\RequirePackage{etoolbox}
\csdef{input@path}{%
 {sty/}
 {figures/}
}

\documentclass[11pt]{book}

\usepackage{page_layout}
\usepackage{couverture}

\usepackage{minitoc}
\dominitoc[n] 

\begin{document}

\frontmatter
	\pagenumbering{roman}
	\NoAutoSpacing
\title{\protect\parbox{\textwidth}{\protect\centering Capturing the Temporal Constraints of Gradual Patterns}}
\author{Dickson Odhiambo OWUOR}

\directorA{Prof. Anne LAURENT}
\directorB{Dr. Joseph Onderi ORERO}

\date{23 Octobre 2020}

\thispagestyle{empty}

\university{UNIVERSITE DE MONTPELLIER}
\univlogo{logo_UM.png}
\univwallpaper{wallpaper.png}
\doctoral{: Information, Structures, Syst\`emes}
\researchunit{: LIRMM}
\researchgroup{: FADO}
\specialisation{ Informatique}

\jury{Anne Laurent}{Professeur}{LIRMM, Université de Montpellier}{Directrice}
\jury{Joseph Onderi Orero}{Ma{\^i}tre de Conférence}{FIT, Strathmore University}{Co-encadrant}
\jury{Nicolas Sicard}{Enseignant-chercheur}{LRIE, l'EFREI}{Examinateur}
\jury{Pascal Poncelet}{Professeur}{LIRMM, Université de Montpellier}{Examinateur}
\jury{Maria Rifqi}{Professeur}{LEMMA, Université Paris II}{Rapporteur}
\jury{Marie-Jeanne Lesot}{Ma{\^i}tre de Conférence, HDR, LIP6}{Sorbonne Université}{Rapporteur}

\phantomsection
\mtcaddchapter[Title]
\maketitle

\begin{dedication}
\begin{flushright}
\null \vspace {\stretch{1}}
\textit{``To my loving parents Leopold and Pamela and, my lovely wife Jackline Bahati...''}
\vspace {\stretch{2}}\null
\end{flushright}
\end{dedication}



\selectlanguage{french}%
\cleardoublepage \phantomsection
\mtcaddchapter[R\'esum\'e]
\begin{abstract}
	\small
	La recherche de motifs fréquents permet d'extraire les corrélations d'attributs par le biais de règles graduelles comme: \textit{``plus il y a de X, plus il y a de Y''}. Ces corrélations sont utiles pour identifier et isoler des relations entre les attributs qui peuvent ne pas être évidentes grâce à des analyses rapides des données. Par exemple, un chercheur peut appliquer une telle recherche pour déterminer quels attributs d'un ensemble de données présentent des corrélations inconnues afin de les isoler pour une exploration plus approfondie ou une analyse. Supposons que le chercheur dispose d'un ensemble de données qui possède les attributs suivants : âge, montant du salaire, du nombre d'enfants et du niveau d'éducation. Un motif graduel extrait peut prendre la forme \textit{``plus le niveau d'éducation est bas, plus le salaire est élevé''}. Étant donné que cette relation est rare, il peut être intéressant pour le chercheur de mettre davantage l'accent sur ce phénomène afin de comprendre. Dans ce travail, nous proposons une technique d'optimisation par des colonies de fourmis qui utilise une approche probabiliste imitant le comportement des fourmis biologiques en cherchant le chemin le plus court pour trouver de la nourriture afin de résoudre des problèmes combinatoires. Nous appliquons la technique d'optimisation des colonies de fourmis afin de générer des candidats des motifs graduels dont la probabilité d'être valide est élevée. Ceci, couplé avec la propriété anti-monotonie, se traduit par le développement d'une méthode efficace. Dans notre deuxième contribution, nous étendons l’extraction de modèles graduels existante à l'extraction de motifs graduels avec un décalage temporel approximatif entre ses attributs affectés. Un tel modèle est appelé motif graduel temporel flou. Cela peut prendre par exemple la forme: \textit{``plus il y a de X, plus il y a de Y \textbf{presque 3 mois plus tard}''} Ces modèles ne peuvent être extraits que de séries de données chronologiques car ils impliquent la présence de l'aspect temporel. Dans notre troisième contribution, nous proposons une donnée modèle de croisement qui permet l'intégration d'implémentations d'algorithmes d'exploration de modèle graduel dans une plateforme Cloud. Cette contribution est motivée par la prolifération des applications IoT dans presque tous les domaines de notre société, ce qui s'accompagne de la fourniture de données chronologiques à grande échelle de différentes sources. Il peut être intéressant pour un chercheur de croiser différentes données de séries chronologiques dans le but d'extraire des motifs graduels temporels des attributs cartographiés. Par exemple un ensemble de données `humidité' peut être temporairement croisé avec un ensemble de données indépendant qui enregistre `Population de mouches', et un schéma peut prendre la forme: \textit{``plus l’humidité est élevée, plus vole presque 2 heures plus tard''}. Notre méthode met l'accent sur l'intégration de l'exploitation des techniques les plus récentes de plate-formes Cloud, car cela facilite l'accès à nos méthodes en allégeant  l’installation et la configuration pour les utilisateurs, permettant ainsi aux utilisateurs de passer plus de temps à se concentrer sur les phénomènes qu'ils analysent.
\end{abstract}

\selectlanguage{english}%
\cleardoublepage \phantomsection
\mtcaddchapter[Abstract]
\begin{abstract}
	Gradual pattern mining allows for extraction of attribute correlations through gradual rules such as: \textit{``the more X, the more Y''}. Such correlations are useful in identifying and isolating relationships among the attributes that may not be obvious through quick scans on a data set. For instance a researcher may apply gradual pattern mining to determine which attributes of a data set exhibit unfamiliar correlations in order to isolate them for deeper exploration or analysis. Assume the researcher has a data set which has the following attributes: age, amount of salary, number of children, and education level. An extracted gradual pattern may take the form \textit{``the lower the education level, the higher the salary''}. Since this relationship is uncommon, it may interest the researcher in putting more focus on this phenomenon in order to understand it. As for many gradual pattern mining approaches, there is a key challenge to deal with huge data sets because of the problem of combinatorial explosion. This problem is majorly caused by the process employed for generating candidate gradual item sets. One way to improve the process of generating candidate gradual item sets involves optimizing this process using a heuristic approach. In this work, we propose an \textit{ant colony optimization} technique which uses a popular probabilistic approach that mimics the behavior biological ants as they search for the shortest path to find food in order to solve combinatorial problems. We apply the ant colony optimization technique in order to generate gradual item set candidates whose probability of being valid is high. This coupled with the anti-monotonicity property, results in the development of a highly efficient \textit{ant}-based gradual pattern mining technique. In our second contribution, we extend an existing gradual pattern mining technique to allow for extraction of gradual patterns together with an approximated temporal lag between the affected gradual item sets. Such a pattern is referred to as a \textit{fuzzy-temporal gradual pattern} and it may take the form: \textit{``the more X, the more Y, \textbf{almost 3 months later}''}. The addition of temporal dimension into the proposed approach makes it even worse regarding combinatorial explosion due to added task of searching for the most relevant time gap. In our third contribution, we propose a data crossing model that allows for integration of mostly gradual pattern mining algorithm implementations into a Cloud platform. This contribution is motivated by the proliferation of \textit{IoT} applications in almost every area of our society and this comes with provision of large-scale time-series data from different sources. It may be interesting for a researcher to cross different time-series data with the aim of extracting temporal gradual patterns from the mapped attributes. For instance a `humidity' data set may be temporally crossed with an unrelated data set that records the `population of flies', and a pattern may take the form: \textit{``the higher the humidity, the higher the number of flies, almost 2 hours later''}. Again, the study emphasizes integration of gradual pattern mining techniques into a Cloud platform because this will facilitate their access on a subscription basis. This alleviates installation and configuration hustles for the users; therefore, it allows them to spend more time focusing on the phenomena they are studying.
	
\end{abstract}
		

\cleardoublepage \phantomsection
\mtcaddchapter[Acknowledgments]
\begin{acknowledgements}
	
	The road to attaining this Ph.D. degree has been intense and it has required a great deal of hard work and unwavering perseverance. I sincerely thank my chief advisor Prof. Anne Laurent for all the good guidance and personal assistance she offered me throughout the Ph.D. journey especially when I worked from LIRMM at Montpellier, France. In addition, I thank my second advisor Dr. Joseph Orero for his constructive recommendations and for creating a favorable working environment for me at Strathmore University (Nairobi, Kenya) which allowed me to put enough focus on the Ph.D. studies.
	
	I render my highest gratitude to Prof. Thomas Runkler (Siemens AG, Munich Germany) for taking his time to mentor me and help me understand some background facts which formed a critical part of the theoretical foundation on which this work is built. I as well thank Mr. Edmond Menya (Strathmore University) and Mr. Olivier Lobry (OREME, Montpellier University) for allowing all our brainstorming sessions and critiquing some parts of this work, since that enabled me to improve on this work.
		
	Next, I extend my sincere gratitude to all the members of the FADO team at LIRMM who went out of their busy schedules, listened to my presentations, understood our work and gave us exceptionally useful feedback. We made use of this feedback to fine-tune this work. I specifically thank Dr. Clement Jonquet for organizing periodic meetings that allowed us to meet and share our research work and experiences.
		
	I express my deep appreciation to Dr. Faaiz Shah for welcoming me to LIRMM on my first visit to Montpellier, France and for ensuring that I completed the relevant documentation that was needed to settle down. Because of him I was able to sort out (in good time) all the logistics pertaining to my registration at Montpellier University for doctoral studies.
	
	Most of the computational experiments performed during this study have been realized with the support of the High Performance Computing Platform \texttt{MESO@LR}, financed by the Occitanie / Pyrénées-Méditerranée Region, Montpellier Mediterranean Metropole and the University of Montpellier.
			
	Finally, I would like to acknowledge that the PhD study in France was thanks to the France Government Scholarship granted through the French Embassy in Kenya. I would like to specifically thank the French Government through the office of Co-operation and Cultural Service (Nairobi, Kenya) and the office of Campus France (Montpellier, France) for their involvement in creating the opportunity for this work to be produced. It is through their financial support and social support that I was able to pursue my doctoral studies at Montpellier University in France.
	
\end{acknowledgements}

	\cleardoublepage \phantomsection \tableofcontents \mtcaddchapter[Table of Contents]
	\cleardoublepage \phantomsection \listoffigures \mtcaddchapter[List of Figures]
	\cleardoublepage \phantomsection \listoftables \mtcaddchapter[List of Tables]


\cleardoublepage \phantomsection
\mtcaddchapter[Acronyms]
\begin{abbreviations}

	\begin{tabular}{ll} 
		\textbf{ACO} & \textbf{A}nt \textbf{C}olony \textbf{O}ptimization\\
		\textbf{CPU} & \textbf{C}entral \textbf{P}rocessing \textbf{U}nit\\
		\textbf{CRUD} & \textbf{C}reate \textbf{R}ead \textbf{U}pdate \textbf{D}elete\\
		\textbf{FuzzTX} & \textbf{F}uzzy \textbf{T}emporal \textbf{C}rossing\\
		\textbf{GP} & \textbf{G}radual \textbf{P}attern\\
		\textbf{GRAANK} & \textbf{GRA}dual r\textbf{ANK}ing\\
		\textbf{GRITE} & \textbf{GR}adual \textbf{IT}emset \textbf{E}xtraction\\
		\textbf{IoT} & \textbf{I}nternet \textbf{o}f \textbf{T}hings\\
		\textbf{JSON} & \textbf{J}ava\textbf{S}cript \textbf{O}bject \textbf{N}otation\\
		\textbf{MQTT} & \textbf{MQ} \textbf{T}elemetry \textbf{T}ransport\\
		\textbf{O$\&$M} & \textbf{O}bervations and \textbf{M}easurements\\
		\textbf{OASIS} & \textbf{O}rganization for \textbf{A}dvancement of \textbf{S}tructured \textbf{I}nformation \textbf{S}tandards \\
		\textbf{ISO} & \textbf{I}nternational \textbf{O}rganization for \textbf{S}tandardization\\
		\textbf{OGC} & \textbf{O}pen \textbf{G}eospatial \textbf{C}onsortium\\
		\textbf{OREME} & \textbf{O}bservatoire de \textbf{RE}cherche \textbf{M}\'editerran\'een de l'\textbf{E}nvironnement\\
		\textbf{SOS} & \textbf{S}ensor \textbf{O}bervation \textbf{S}ervice\\
		\textbf{SWE} & \textbf{S}ensor \textbf{W}eb \textbf{E}nablement\\
		\textbf{T-GRAANK} & \textbf{T}emporal-\textbf{GRAANK}\\
		\textbf{WSN} & \textbf{W}ireless \textbf{S}ensor \textbf{N}etwork\\

	\end{tabular}

\end{abbreviations}

\mainmatter 

\chapter{Introduction}
\label{ch1}
\minitoc \clearpage 
	
	\begin{chapquote}{Sir Isaac Newton \textit{(1642 – 1727)}}
		``If I have seen further it is by standing on the shoulders of Giants''
	\end{chapquote}

	\section{Introduction}
	\label{ch1:introduction}
\begin{figure}[h!]
		\centering
		\includegraphics[width=.64\textwidth]{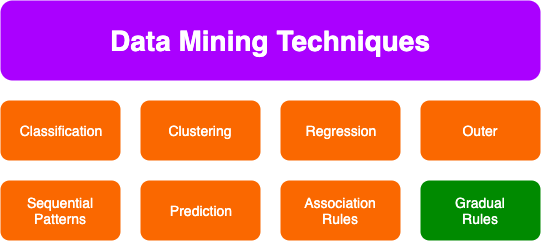}
		\caption{Data mining techniques}
		\label{fig1:data_mining}
	\end{figure}

	``Data mining is a nontrivial process of extracting knowledge from data sets'' \cite{Cios2005}. Experts employ data mining techniques to pull out hidden information from large-scale data sets, which may otherwise take human analysts very long to find. Gradual pattern discovery is a recent extension of the data mining field that allows linguistic expressions of attribute correlations in the form of gradual rules such as \textit{``the more X, the less Y''}, as illustrated in Figure~\ref{fig1:data_mining} \cite{Berzal2007, Di-Jorio2009, Laurent2009, Aryadinata2013, Aryadinata2014a}.
	
	We live in the 21st century where research is characterized by analysis of large-scale data. A great deal of research activities involve correlating different aspects of a data set in order to understand the relationships existing in the observed phenomena. Given such large-scale data, the exploring of every attribute combination for possible correlation proves to be time consuming. However, gradual patterns render the possibility of isolating correlating attributes within a short time \cite{Aryadinata2014a}. 
	
	With that in mind, gradual pattern mining offers many benefits to businesses, companies and other service-oriented domains. For instance in a medical domain, a medical practitioner may carry out tens of tests which may generate data with hundreds of attributes. From this data useful knowledge may be extracted through a gradual rule like: \textit{``the lower the population of organism X, the higher the population of organism Y''}. If this correlation is uncommon, then the practitioner can isolate the two organisms from hundreds of attributes for a deeper analysis. Of course, the need to constantly analyze the performance of existing gradual pattern mining techniques with the aim of improving them should never be abandoned. Researchers in the data mining field should embrace this challenge in order to keep this field active.
		
	Unlike classical mining techniques (such as associative mining) that calculate the frequency of an item set occurrence in a data set, gradual pattern mining calculate the frequency at which gradual attribute correlations occur in a data set. For example in Table~\ref{tab1:sample1}, at all 4 tuples the values of $X$ and $Y$ are both increasing in a subsequent manner. Therefore, a gradual rule may take the form: \textit{``the more X, the more Y''}  \cite{Agrawal1994, Di-Jorio2009}.
		
	\begin{table}[h!]
		\centering
		\begin{tabular}{c l l}
			\toprule
			\textbf{Time} & \textbf{X} & \textbf{Y}\\
			\hline \hline
			13:00 & 1 & 10\\
			13:30 & 2 & 20\\
			14:00 & 3 & 30\\
			14:30 & 4 & 40\\
			\bottomrule
		\end{tabular}
		\caption{A sample timestamped data set}
		\label{tab1:sample1}
	\end{table}
	
	There has been a great deal works for the last 10 years regarding gradual pattern mining. Many of these works do not consider any approach that allows for estimation of the temporal lag among the attributes' correlations. For instance given Table~\ref{tab1:sample1}, classic gradual pattern techniques would extract the pattern \textit{``the more X, the more Y''} and they would not exploit values of \textit{Time} attribute for additional knowledge discovery. It may be possible that a temporal lag may exist between an attribute's change and the impact of that change on another attribute's change. Such a pattern that additionally captures the temporal tendencies between gradual item sets is referred to as a \textit{temporal gradual pattern}.
		
	With respect to Table~\ref{tab1:sample1}, a temporal gradual pattern may take the form: \textit{``the more X, the more Y, \textbf{30 minutes later}''}. Further, temporal gradual pattern mining may be extended to allow for extraction  of \textit{emerging} temporal gradual patterns. A temporal gradual pattern may be said to be \textit{emerging} if its frequency of occurrence varies from one data set to another \cite{Laurent2015}. For the reason that any temporal gradual pattern rely on the temporal-orientation of the data set, their discovery is only possible in \textit{time-series} data sets. 
	
	Time-series data can be defined as a sequence of data points that are temporally-oriented. Sources of time-series data are numerous; for instance it may be obtained from from internal sources (e.g. a data warehouse collecting \textit{IoT} sensor data) or from an external source (e.g. data distributed by government institutions such as weather stations) et cetera.	 Time-series data can also be defined as a \textit{data stream} when it becomes a potentially infinite sequence of precise data points recorded over a period of time \cite{Pitarch2010}. 
	
	Today, \textit{IoT} applications have spread to almost every domain of our society and these applications come with the provision for large-scale time-series data and data streams. Therefore, by adding temporal information to gradual correlations, temporal gradual pattern mining may greatly aid researchers who are interested in identifying the temporal aspects of attribute relationships.
			
	It is also important to mention that due to the increased provision of \textit{IoT}-related time-series data, frameworks and standards such as the (Open Geospatial Consortium) \textit{OGC SensorThings} have emerged to facilitate easy management and sharing of such data among different research institutions \cite{Hicham2018, Liang2016}. These standards aim to integrate sensor data into Spatial Data Infrastructures (SDI) so that they become additional sources of geospatial information besides traditional data types like maps.
	
	SDIs are used to implement frameworks that allow geospatial data to be FAIR (Findable, Accessible, Interoperable and, Reusable). As a result, time-series data offer great potential for integrating real-time observations into geospatial data for richer map visualization. It is now possible for research scientists to remotely monitor environmental sensors, collect and map their readings to specific geographical areas \cite{Hicham2015, Kotsev2018, Ronzhin2019}. With this current trend, the need arises to integrate gradual pattern mining algorithms into \textit{Cloud} platforms that implement such frameworks.
	
	\begin{figure}[h!]
		\centering
		\subfloat[]{%
			\begin{tabular}{c c}
			\toprule
			\textbf{time} & \textbf{room}\\
			& \textbf{temperature}\\
			\hline \hline
			07:00 & 22.0\\
			07:15 & 22.5\\
			07:30 & 23.6\\
			07:45 & 25.0\\
			08:00 & 26.8\\
			\bottomrule
			\end{tabular}
		}%
		$~~~$
		\subfloat[]{%
			\begin{tabular}{c c}
			\toprule
			\textbf{time} & \textbf{number of}\\
			& \textbf{dozing students}\\
			\hline \hline
			07:00 & 0\\
			07:15 & 1\\
			07:30 & 2\\
			07:45 & 6\\
			08:00 & 10\\
			\bottomrule
			\end{tabular}
		}%
		\caption{(a) A sample time-series data set recording the room temperature of a class, and (b) a sample time-series data set recording the number of dozing students}
		\label{fig1:table_samples23}
	\end{figure}
	
	In the research community, great interest has also been expressed regarding \textit{crossing} time-series data from different sources in order to discover new knowledge about phenomena that otherwise could not be discovered by analyzing the individual data sets in isolation. Data crossing enables the matching of different data sets using a predefined criteria and combining their data points to form a new data set \cite{Costa2010, Hicham2015, Hicham2016}. For example, let us assume that both data sets shown in Figure~\ref{fig1:table_samples23} contain data points recorded at simultaneous times. It is easy to see that a new data set may be formed by matching data points that occur at almost similar timestamps as illustrated in Table~\ref{tab1:crossed}.
	
	\begin{table}[h!]
		\centering
		\begin{tabular}{l c c}
			\toprule
			\textbf{time} (approx.) & \textbf{room} & \textbf{number of}\\
			& \textbf{temperature} & \textbf{dozing students}\\
			\hline \hline
			07:00 & 22.0 & 0\\
			07:15 & 22.5 & 1\\
			07:30 & 23.6 & 2\\
			07:45 & 25.0 & 6\\
			08:00 & 26.8 & 10\\
			\bottomrule
			\end{tabular}
		\caption{A new data set formed by crossing 2 data sets temporally}
		\label{tab1:crossed}
	\end{table}
	
\begin{figure}[h!]
		\centering
		\includegraphics[width=\textwidth]{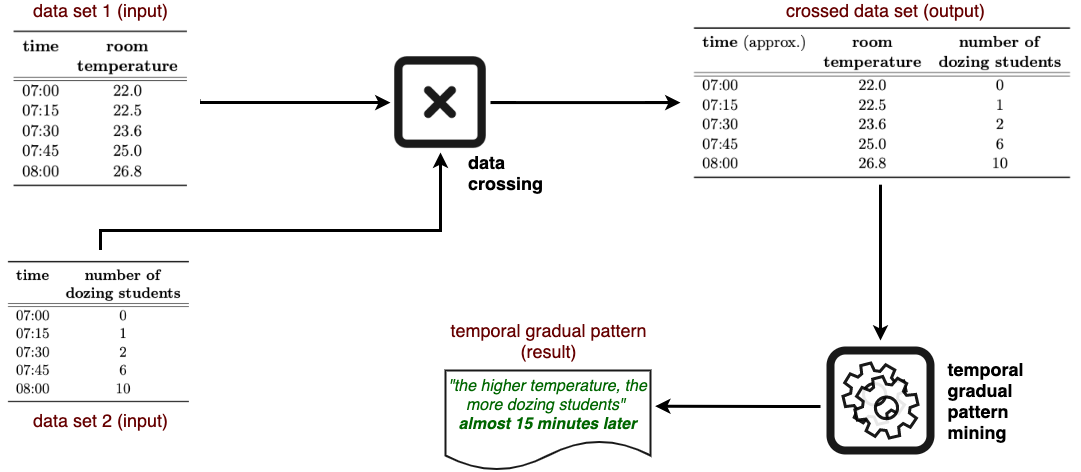}
		\caption{Data crossing model for gradual pattern mining}
		\label{fig1:crossing_model}
	\end{figure}

	Even further, Figure~\ref{fig1:crossing_model} illustrates how the \textit{`room temperature'} time-series data set and the \textit{`dozing students'} time-series data set may be crossed to form a new data set which is mined to extract temporal gradual patterns such as: \textit{``the higher the room temperature, the higher the number of dozing students, 15 minutes later''}. This is an abstract example, but it goes to show that data crossing and gradual pattern mining techniques can be combined to develop a powerful automated data analysis tool.
		
	\section{Problem Statement}
	\label{ch1:problem}
	In the previous section, we mentioned that gradual pattern mining technique is a recent extension in the data mining field and its popularity as a research item has been on the rise since the early 2000s. It is important to highlight that many works on gradual pattern mining approaches employ two main strategies for mining gradual rules: (1) \textit{depth-first search} (DFS) which builds the longest set enumeration tree of gradual candidates and, (2) \textit{breath-first search} (BFS) which generates gradual item set candidates in a level-wise manner. Nonetheless, DFS-based approaches employ recursion (which is computationally complex) to find the longest set enumeration tree and, BFS-based algorithms suffer from the problem of combinatorial candidate explosion when presented with huge data sets.
	
	However, the \textit{anti-monotonicity} property (which holds for gradual item sets) is employed as an efficient pruning strategy in gradual pattern mining. The \textit{anti-monotonicity} property states that: \textit{``if a pattern containing k item-sets is not frequent then all supersets of the pattern are also not frequent''} \cite{Agrawal1994, Di-Jorio2009, Laurent2009}. The \textit{anti-monotonicity} property allows many gradual pattern mining techniques to automatically eliminate gradual item set candidates that are supersets of infrequent item sets or subsets of frequent item sets. In this way, the efficiency of the process of validating candidate gradual item sets is improved.
	
	In reality, the anti-monotonicity property partially solves the problem of combinatorial candidate number explosion that may occur when dealing with large-scale data sets having great numbers of attributes. It should be remembered that for $n$ attributes, there exists $2^{n}$ combinations of gradual item set candidates to consider. For instance, assume $\{A, B\}$ are attributes of a data set. The following candidates are possible: $\{more A, less B\}$, $\{less A, more B\}$, $\{less A, less B\}$, and $\{more A, more B\}$. Therefore, as the number of attributes increase, the number of gradual item set candidates to consider also increase exponentially.
	
	We have also discussed in the previous section how the increasing popularity of \textit{IoT} applications in this century has led to more provision of time-series data. In light of this fact, the need arises to extend gradual pattern mining techniques so that they allow extraction of gradual correlations among the attributes of a time-series data set together with an estimated temporal lag. Such patterns may take the form: \textit{``the more Y, the less Z \textbf{2 years later}''}. Many works have been accomplished regarding gradual pattern mining techniques; however, these works do not take into account the possibility of estimating the temporal lag between gradual item sets.
		
	In addition, it may be interesting to cross time-series data from unrelated sources in order to discover any uncommon relationships. There exists numerous techniques for crossing time-series data from different sources and the most popular techniques rely on a \textit{SQL}-based querying models to achieve this \cite{Wang2013, Boukerche2016, Mrozek2018Dec}. However, many certified frameworks for managing time-series data (such as \textit{OGC SensorThings} framework for managing sensor data) are implemented on APIs that support \textit{NoSQL} querying models.
	
	To shed more light, most research institutions across the world (especially Europe) comply with directives given by a common governing body such as INSPIRE (Infrastructure for Spatial Information in the European Community) \cite{Grellet2017}. One of the most common directive involves encouraging member institutions to use a common certified framework such as OGC SensorThings to manage sensor data \cite{Crommert2004, Kotsev2018}.
	
	It is important to emphasize the adoption of Cloud platforms that implement certified data managing frameworks since consumers of these data (mostly researchers) obviously wish to spend more time studying phenomena than configuring data mining (or analysis) tools. In that case, integrating gradual patterns mining techniques into a Cloud platform such as \textit{Docker} allows researchers to access them on a subscription basis. 
			
	\section{Contributions}
	\label{ch1:contributions}
	With the intention of grappling with the problems discussed in the previous section, the objective of this research study is to address three main gradual pattern mining aspects and they are briefly described as follows.
	
	\subsection{Mining Temporal Gradual Patterns}
	\label{ch1:temporal}
	With regards to exploiting the timestamped attributes in order to estimate temporal lag between extracted gradual item sets, this thesis introduces a \textit{fuzzy} model for estimating time lag as an additional procedure of gradual pattern mining. The process involves transforming a data set through its tuples stepwisely, and applying a fuzzy triangular membership function to estimate time lag. So, a (fuzzy) temporal gradual pattern may take the form \textit{``the more Y, the less Z, \textbf{almost 2 years later}''}.
	
	Moreover, this thesis extends (fuzzy) temporal gradual patterns in order to introduce \textit{emerging} temporal gradual patterns. Emerging patterns (EPs) are item sets whose frequency changes significantly from one data set to another. EPs are described using growth rate, which is the ratio of an EP's frequency support in one data set to its frequency support in another data set \cite{Dong1999b}. Therefore, a temporal gradual emerging pattern my take the form: \textit{``the more Y, the less Z, almost 2 years later''} with a \textbf{frequency support of 0.006} and \textit{``the more Y, the less Z almost, 2 years later''} with a \textbf{frequency support of 0.84}. This pattern has a \textit{growth rate} of $140$.

	\subsection{Optimizing Generation of Candidate Gradual Item Sets}
	\label{ch1:optimize}
\begin{figure}[h!]
		\centering
		\includegraphics[width=.56\textwidth]{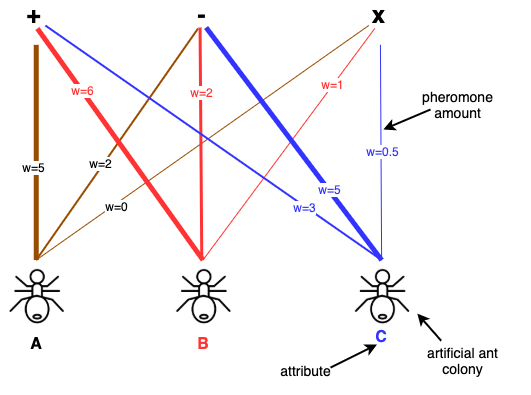}
		\caption{Illustration of ACO for gradual candidate generation. (+ implies attribute is increasing, - implies attribute is decreasing and x implies attribute is irrelevant)}
		\label{fig1:aco_idea}
	\end{figure}

	In this work, we propose a heuristic solution that is based on \textit{ant colony optimization} (ACO) to the problem of (1) combinatorial candidate explosion and, (2) finding longest set tree for the case of gradual pattern mining. For instance, the problem of combinatorial explosion can be solved by optimizing the process candidate generation. The proposed approach involves learning from a given data set the most occurring nature for every attribute (i.e. increasing, decreasing or irrelevant) as illustrated in Figure~\ref{fig1:aco_idea}, and using this knowledge to generate highly probable candidates.
	
	ACO is a popular heuristic approach that imitates the positive feedback reinforcement behavior of biological ants as they search for food: where the more ants following a path, the more pheromones are deposited on that path and, the more appealing that path becomes for being followed by other ants \cite{Dorigo1996, Dorigo2010}. In other words, ACO exploits the behavior of artificial ants in order to search for approximate solutions to discrete optimization problems \cite{Dorigo2019}.

	\subsection{Cloud Integration of Gradual Pattern Mining Techniques}
	\label{ch1:cloud_integration}
	In view of the fact that there exists institutions such as INSPIRE which encourages member institutions to use a common certified frameworks such as OGC SensorThings to manage sensor data, this work is curious about integrating gradual pattern mining techniques into an \textit{OGC SensorThings} framework implementation for the purpose of offering them as additional data analysis tools to research institutions that comply with directives from such governing bodies. Again, the \textit{OGC SensorThings} stands out as one of the best API to interconnect IoT devices, sensor data and applications over the Docker Cloud platform \cite{Crommert2004, Hicham2015, Grellet2017, Kotsev2018}.
	
	In addition, this work introduces a \textit{fuzzy model} that crosses time-series data that can be integrated into an \textit{OGC SensorThings} API implementation. Achieving that integration makes it possible for automatically crossing time-series data from different sources and applying gradual pattern mining algorithms on the crossed data (as illustrated in Figure~\ref{fig1:crossing_model}) within the same Cloud platform.

	\section{Thesis Outline}
	\label{ch1:outline}
	
	The rest of the thesis document is organized as follows:
	
	In Chapter~\ref{ch2}, we describe the preliminary concepts and definitions about gradual patterns, mining techniques for extracting gradual patterns specifically \textit{GRITE} and \textit{GRAANK} approaches and, the traversal strategies employed by existing gradual pattern mining techniques for candidate item set generation. We review literature relating to temporal data mining, data crossing of time-series data from different sources, Cloud platforms into which gradual pattern mining algorithms may be integrated. The concepts discussed in this chapter lays the foundation for the proposed approaches in Chapter~\ref{ch3}, Chapter~\ref{ch4}, Chapter~\ref{ch5}, Chapter~\ref{ch6} and Chapter~\ref{ch7}.
		
	In Chapter~\ref{ch3}, we adopt the essential definitions of gradual patterns from Chapter~\ref{ch2} and propose formal definitions for temporal gradual patterns. In addition to this contribution, we propose a \textit{fuzzy} approach for mining temporal gradual patterns. We develop an algorithm to implement the approach and test on real data. We test the computational performance of the algorithm both when it implements a \textit{traversal} strategy and an \textit{ant}-based strategy for candidate generation and discuss the results we obtained.

	In Chapter~\ref{ch4}, we explore the possibility of employing an \textit{ant colony optimization} technique that uses a probabilistic strategy to generate gradual item set candidates. Further, we propose a representation of generating gradual item set candidates and finding the longest set enumeration tree as optimization problems and, define probabilistic rules through which we may extract gradual patterns more efficiently. Additionally, we develop algorithms that implement this approach to mine for gradual patterns. We compare the computational performance of proposed algorithms against existing gradual pattern mining algorithms and discuss the results we obtained.
			
	In Chapter~\ref{ch5}, we adopt definitions of temporal gradual patterns from Chapter~\ref{ch3} and propose formal definitions for temporal gradual emerging patterns. Further, we describe two different strategies (\textit{border-based} strategy and \textit{ant-based} strategy) for extracting temporal gradual emerging patterns. We develop two algorithms to implement both strategies and test them on real data. We analyze the test results in order to compare the efficiency of both strategies.
	
	\clearpage
	
	In Chapter~\ref{ch6}, we propose a \textit{fuzzy} approach that crosses time-series data from different sources with the ultimate goal of exploiting the crossings for temporal gradual pattern mining. We implement parallel processing on the developed algorithm implementation in order to improve its efficiency. We test the computational performance of the proposed algorithm and discuss the results we obtained.
					
	In Chapter~\ref{ch7}, we describe a software model architecture that integrates the proposed data crossing model implementation together with temporal gradual pattern mining algorithm implementations into a Cloud platform which implements the \textit{OGC SensorThings} API. We develop a \textit{Web application software} that achieves this task as proof of concept.
	
	In Chapter~\ref{ch8}, we conclude this research study and emphasize the applicability of our models in real-life settings. In future perspectives, we discuss the possible research extensions that may be envisaged based on the contributions made by this study.


\chapter{Related Work}
\label{ch2}
\minitoc \clearpage 

	\begin{chapquote}{Lewis Carroll, \textit{Alice in Wonderland}}
		``Begin at the beginning,'' the King said, gravely, ``and go on till you come to an end; then stop.''
	\end{chapquote}

	\section{Introduction}
	\label{ch2:introduction}
	In this chapter we briefly review the history of gradual pattern mining and describe how it emerged from \textit{association rule} mining. We describe the preliminary concepts and formal definitions about gradual patterns. Moreover, we describe existing GRITE and GRAANK approaches and the traversal search strategies that they apply for generating gradual item set candidates. In addition, we review related works: temporal data mining, data crossing and Cloud platforms into which gradual pattern mining algorithms may be integrated.
	
	\section{Gradual Patterns}
	\label{ch2:gp}
	Gradual pattern mining is a research extension in the data mining field and it is a process for extracting gradual correlation knowledge from a numeric data set. Gradual pattern mining seeks to identify gradual dependencies that express correlations between attribute variations in a linguistic manner. For instance a gradual dependency may take the form: \textit{``the more A, the less B''} where A and B are gradual item sets. The gradual dependence between attributes is a type of \textit{tendency} expression formed by \textit{gradual rules} \cite{Eyke2002}. 
	
	To put it another way, gradual pattern mining exploits a quantitative \textit{rule-based} model that allows a simple and expressive way of knowledge representation using the attributes of a data set \cite{Di-Jorio2009, Laurent2009, Aryadinata2013, Owuor2019}. A closely related concept to \textit{gradual rules} that has been developed is \textit{association rules}, which expresses relationships between frequent item sets as association dependencies \cite{Agrawal1993, Agrawal1994}. In fact, gradual rule mining borrows a great deal of its conceptual basis from association rule mining \cite{Agrawal1994, Eyke2002, Laurent2009}. We expound on this in Section~\ref{ch2:gradual_rules}.
	
	\subsection{Association Rules}
	\label{ch2:association_rules}
	Association rule mining was first introduced in \cite{Agrawal1993} and an association rule may take an implicative expression of the form $\{ Y \Rightarrow Z \}$, where $Y$ and $Z$ are item sets. This rule may be interpreted as \textit{``Y implies Z''}, intuitively meaning that in a transactional database: items in set $Y$ tend to also contain items in set $Z$ or the existence of item set $Y$ implies the existence of item set $Z$. 
	
	We describe some formal definitions taken from literature about association rule mining as follows: let $\mathcal{D}$ be a set of transactions,
	
	\textbf{Definition 2.1} \texttt{(Item)}. \textit{An item is a set of literals denoted by $\mathcal{I} = \{i_{1},i_{2},...,i_{n}\}$}. Each transaction $\mathcal{T}$ in $\mathcal{D}$ is a set of items such that $\mathcal{T} \subseteq \mathcal{I}$ and, a transaction $\mathcal{T}$ \textit{contains} $Y$, a set of some items in $\mathcal{I}$ if $Y \subseteq \mathcal{T}$.
	
	\textbf{Definition 2.2} \texttt{(Association Rule)}. \textit{An association rule is an implication of the form $ Y \Rightarrow Z$}, where $Y \subset \mathcal{I}, Z \subset \mathcal{I}$ and $Y \bigcap Z = \emptyset$.
	
	\textbf{Definition 2.3} \texttt{(Confidence)}. \textit{Confidence $c$ is the probability that every transaction $\mathcal{T}$ in data set $\mathcal{D}$ which contains $Y$ also contains $Z$}. Therefore, the association rule $Y \Rightarrow Z$ holds with a confidence of $c$ if $c\%$ of the transactions in $\mathcal{D}$ that contain $Y$ also contain $Z$.
	
	\begin{equation}
		c(Y \Rightarrow Z) = \mathcal{P}(Z \subset \mathcal{T} ~~\vert ~~ Y \subset \mathcal{T})
	\end{equation}
	
	\textbf{Definition 2.4} \texttt{(Support)}. \textit{Support $s$ is the probability that any transaction $\mathcal{T}$ contains both item sets $Y$ and $Z$}. Therefore, the rule $Y \Rightarrow Z$ has a support of $s$ if $s\%$ of the transactions in data set $\mathcal{D}$ contain $Y \bigcup Z$.
	
	\begin{equation}
		s(Y \Rightarrow Z) = \mathcal{P}(Y \subset \mathcal{T} ~~\wedge ~~ Z \subset \mathcal{T})
	\end{equation}
	
	Confidence is also be referred to as the \textit{accuracy} of an association rule and \textit{support} is also referred to as the \textit{frequency} of an association rule. In reality, association rule mining is applied on transactional databases in order to generate all association rules whose confidence and support is greater than user-defined thresholds: \textit{minimum confidence} and a \textit{minimum support} respectively. Henceforth, an association rule can be termed as \textit{frequent} if its support is greater than a respective user-specified threshold \cite{Agrawal1993, Agrawal1994, Eyke2002}.
	
	\textit{Example 2.1.} Let us consider an arbitrary data set $\mathcal{D}_{2.1}$ containing sales transactions of a shopkeeper. Let us set both the \textit{minimum support} and \textit{minimum confidence} to $0.75$.
	
	\begin{table}[h!]
  		\centering
    	\begin{tabular}{c l} 
	  	\toprule
      	\textbf{t-id} & \textbf{item list} \\
      	\hline \hline
      	t101 & bread, milk\\
      	t102 & bread, eggs, milk\\
      	t103 & cheese, milk\\
      	t104 & sugar, milk, coffee, bread\\
      	\bottomrule
    	\end{tabular}
    	\caption{Sample data set $\mathcal{D}_{2.1}$}
    	\label{tab2:sample1}
	\end{table}
	
	Using the transactions in Table~\ref{tab2:sample1}, we may define an association rule: $bread \Rightarrow milk$ which has a support of $75\%$ with a confidence of $100\%$. This may be interpreted as $100\%$ of customers who purchase \textit{bread} also purchase \textit{milk} and the \textit{frequency} of this occurrence $75\%$ of the time. We calculate \textit{support} $s$ by counting all the transactions that have both \textit{bread} and \textit{milk} (i.e. 3) and dividing it by the total number of transactions (i.e. 4). We calculate \textit{confidence} $c$ through dividing the number of transactions that have both \textit{bread} and \textit{milk} (i.e. 3) by the number of transactions that have \textit{bread} (i.e. 3).
	
	\clearpage

	As can be seen, such an analysis may be very useful to the shopkeeper especially when it comes to re-stocking the inventory. It comes as no surprise that applications of association rule mining are more popular in the business domain. These rules aid in decisions that the management or owners of businesses (such as supermarkets) need to make in order to design purchase offers such as including items that are frequently bought together with the aim of maximizing profit \cite{Srikant1997, Aggarwal2015}.
			
	\subsection{Gradual Rules}
	\label{ch2:gradual_rules}
	\cite{Eyke2002} elaborated the theory on quantitative association rule in order to allow for expression of gradual dependence between attributes of a data set in the form of a \textit{fuzzy partition}. The study by \cite{Berzal2007} advocates the same view that transforms attributes as fuzzy linguistic variables in order to form \textit{fuzzy gradual items}. For instance, let us assume we have two attributes: \textit{air temperature} and \textit{mosquito population}.
	
	Three \textit{fuzzy} membership sets $\{low, moderate, high\}$ may be designed for attribute \textit{air temperature} and, fuzzy membership sets $\{small, medium, huge\}$ for attribute \textit{mosquito population} may be designed. Applying the membership degrees of the attributes, a fuzzy gradual association rule may take the form: \textit{``the more air temperature is \textbf{high}, the more mosquito population is \textbf{huge}''}. It is important to realize that a fuzzy gradual item is composed of the attribute together with a corresponding membership to a fuzzy set.
	
	According to \cite{Laurent2009}, association rules can be extended to express causality relationships between pairs of gradual item sets. For instance, let $A_{1}$ be a gradual item: \textit{``the more fast foods''} and, $A_{2}$ be a gradual item: \textit{``the greater the danger of obesity''}. A \texttt{gradual association dependency} may be denoted as $A_{1} \Rightarrow A_{2}$ which may be interpreted as: \textit{``the more the fast food, then the greater the danger of obesity''}. This dependency means that more consumption of fast foods implies an increase in the risk of obesity.
	
\begin{figure}[h!]
		\centering
		\includegraphics[width=.5\textwidth]{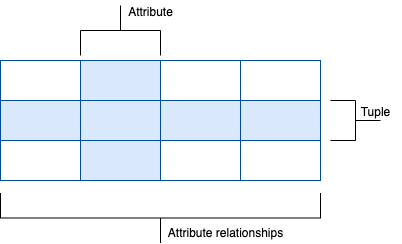}
		\caption{Illustration of attribute dependency}
		\label{fig2:gradual_dependency}
	\end{figure}

	In this study, we hold the same position as \cite{Laurent2009} when it comes to representing gradual item sets since this technique does not require that fuzzy modalities be formed for each gradual item set. As shown in Figure~\ref{fig2:gradual_dependency}, the principal aim of gradual rule mining is to relate the attributes of a data set using a gradual dependency whose quality is measured by the tuple count. 
	
	\subsection{Formal Definitions for Gradual Patterns}
	\label{ch2:definitions}
	We describe formal definitions taken from literature about gradual rule mining as follows:
	
	\textbf{Definition 2.5} \texttt{(Gradual Item)}. \textit{A gradual item is a pair $(i,v)$ where $i$ is an attribute and $v$ is a variation $v \in \lbrace \uparrow,\downarrow \rbrace$. $\uparrow$ stands for an increasing variation while $\downarrow$ stands for a decreasing variation}. 
	
	For example, $(temperature,\uparrow)$ is a gradual item that can be interpreted as ``the higher temperature''.
	
	\textbf{Definition 2.6} \texttt{(Gradual Pattern)}. \textit{A gradual pattern \texttt{GP} is a set of gradual items}.
	
	\begin{equation}
		GP = \lbrace (i_{1},v_{1}), ..., (i_{n},v_{n})\rbrace
	\end{equation}
	
	For example, $\lbrace (temperature,\uparrow),(mosquitoes,\uparrow),(sleep,\downarrow) \rbrace$ is a gradual pattern that can be interpreted as \textit{``the more temperature, the more mosquitoes, the less sleep''}.
	
	\textbf{Definition 2.7} \texttt{(Support)}. \textit{Support \texttt{sup} of a gradual pattern is the ratio of the proportion of tuples that respect the pattern to the total number of tuples}. It must be remembered that deriving the support of gradual patterns involves the ordering of at least two or more tuples, since the patterns are built on the increasing or decreasing nature of attributes \cite{Di-Jorio2009}.
	
	To elaborate further, there are two main approaches for deriving the \textit{support} of a gradual pattern. The compliant subset approach proposed in \cite{Di-Jorio2008, Di-Jorio2009} identifies tuple subsets $\mathcal{D^{*}}$ that can be ordered so that all couples from $\mathcal{D^{*}}$ satisfy the order induced by the pattern. Formally, the support is defined as follows:
	
	\begin{equation} 
		sup(GP) = \frac{max_{\mathcal{D^{*}}\in \mathcal{L}(GP)}\mid \mathcal{D^{*}}\mid}{\mid \mathcal{D}\mid}
	\end{equation}
	
	where $\mathcal{L}(GP)$ denotes the set of all maximal subsets $\mathcal{D^{*}} = {x_{1}, ..., x_{m}} \subseteq \mathcal{D}$ for which there exists a permutation $\pi$ such that $\forall l \in [1, gp-1], x_{\pi l} \preceq_{GP} x_{\pi l+1}.$
	
	The second approach proposed in \cite{Laurent2009} considers the number of tuples that are concordant by exploiting the Kendall's $\tau$ rank correlation. Instead of tuple subsets, it counts the number of tuple couples that satisfy the order induced by the pattern. Therefore, the support is defined by the following formula:
	
	\begin{equation}\label{eqn2:kendalls}
		sup(GP) = \frac{\mid \{ (x, x^{'}) \in \mathcal{D}^{2}/ x \preceq_{GP} x^{'}\} \mid}{\mid \mathcal{D}\mid (\mid \mathcal{D}\mid -1)/2}
	\end{equation}
	
	where $x$ is such that: for attribute $A$, for any object $x \in \mathcal{D}$, $A(x)$ denotes the value $A$ takes for $x$ and $x$ proceeds $x^{'}$.
	
	Another key point to emphasize is that support describes the quality of a gradual pattern and it is measured as the extent to which the pattern holds for a given data set. Therefore, given a user-specified threshold of \textit{minimum support} $\varrho$ a gradual pattern $GP$ is said to be \textit{frequent} if:
	
	\begin{equation}\label{eqn2:frequent}
		sup(GP) \geq \varrho
	\end{equation}
	
	\subsection{Anti-monotonicity Property}
	\label{ch2:antimonotonicity}
	It is easy to observe that the definition of \textit{support} for gradual rule mining is conceptually similar to that for association rule mining, in the sense that both are determined by the proportion of transactions or tuples that  respect the respective rule. \cite{Aggarwal2015} further elaborates association rule \textit{support} by establishing that when an item set $Y$ is contained in a transaction $\mathcal{T}$, all its subsets will also be contained in the transaction. This property is known as the \textit{support monotonicity property}.
	
	\textbf{Property 2.1} \texttt{(Support Monotonicity Property)}. \textit{The support of every subset $X$ of $Y$ is at least equal to that of the support of item set $Y$}.
	
	\begin{equation}
		sup(X) \geq sup(Y) ~~~~ \forall X \subseteq Y
	\end{equation}
	
	\cite{Aggarwal2015} concludes that the \textit{support monotonicity property} implicitly indicates that every subset of a frequent item set is also frequent. This is known as \textit{downward closure property}.
	
	\textbf{Property 2.2} \texttt{(Downward Closure Property)}. \textit{Every subset of a frequent item set is also frequent}.
	
	In the case of gradual pattern mining, if a pattern $GP_{1}$ with gradual item set $\{i_{1}, i_{2}, i_{3} \}$ is not frequent, then it is impossible for a pattern $GP_{2}$ whose gradual items is superset of set $\{i_{1}, i_{2}, i_{3} \}$ to be frequent. This is referred to as the \textit{anti-monotonicity property} of frequent gradual patterns \cite{Aryadinata2013}.
	
	\textbf{Property 2.3} \texttt{(Anti-Monotonicity Property)}. \textit{No frequent gradual pattern containing $n$ attributes can be built over an infrequent gradual pattern containing a subset of these $n$ attributes}. An infrequent gradual pattern is that pattern whose \textit{support} is less than the user-specified threshold as implied in Equation~\eqref{eqn2:frequent}. It is obvious that this property closely resembles the \textit{downward closure property} of association rule mining.
	
	To illustrate using an example, if the gradual pattern $\lbrace (temperature,\uparrow),(mosquitoes,\uparrow),(sleep,\downarrow) \rbrace$ has a \textit{support} that is less than the user-specified threshold, then any gradual pattern that is a superset of this pattern (i.e. $\lbrace (temperature,\uparrow),(mosquitoes,\uparrow),(sleep,\downarrow), (risk,\uparrow) \rbrace$) will also have a \textit{support} that is less than the threshold. 
	
	Altogether, all supersets of a gradual pattern that is infrequent are likewise infrequent and all subsets of a frequent gradual pattern are likewise frequent \cite{Ayouni2010, Aryadinata2014a, Laurent2009, Owuor2019}.
	
	\subsection{Traversal Strategies for Item Set Search}
	\label{ch2:pruning}
	For the purpose of aiding the discussion, we recall the following definitions from literature:
	
	\textbf{Definition 2.9} \texttt{Minimal Frequent Pattern}. \textit{an item set $A$ is said to be minimal if $A$ is frequent and there exists at least one of its superset that is also frequent} \cite{Bayardo1998}. It should be noted that a frequent item set is defined by Equation~\eqref{eqn2:frequent}.
	
	\textbf{Definition 2.10} \texttt{Closed Frequent Pattern}. Let $\mathcal{I}$ be a finite set of objects in data set $\mathcal{D}$. \textit{An item set $C \subseteq \mathcal{I}$ is a closed item set iff: $h(C) = C$ (where $h$ is \textit{Galois closure operator}) \cite{Pasquier1999}.} If $C$ is also \textit{frequent} then it is a closed frequent pattern.
		
	\textbf{Definition 2.11} \texttt{Maximal Frequent Pattern}. \textit{an item set $A$ is said to be maximal if $A$ is frequent and none of its supersets is frequent} \cite{Bayardo1998}.
	
\begin{figure}[h!]
		\centering
		\includegraphics[width=.45\textwidth]{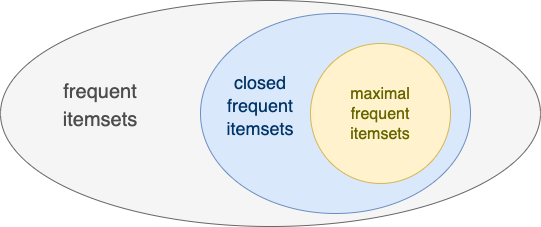}
		\caption{Illustration of frequent item set boundaries}
		\label{fig2:maximal_boundary}
	\end{figure}

	Figure~\ref{fig2:maximal_boundary} illustrates the population of frequent item sets, closed frequent item sets and maximal frequent item sets in a typical data mining data set. It is important to point out that a maximal frequent pattern includes largest possible number of member item sets. For example let $\{ A, B, C, D \}$ be individual frequent item sets, if pattern $\{ ABCD \}$ is frequent then it is a maximal pattern since there can exist no other pattern with more item set members. If pattern $\{ ABC \}$ is also frequent and its support is greater than that of pattern $\{ ABCD \}$, then it is a closed pattern.
		
	Maximal frequent gradual item sets produce the richest knowledge about the relationships of a data set's attributes since they do not miss any member item set that is relevant to the pattern. Consequently, it makes good sense to optimize a gradual pattern mining technique such that it quickly retrieves maximal item sets. This will eliminate the need to search for any subset of the maximal item set, hence improving the efficiency of the technique. 
	
	\subsubsection{Classical Traversal Techniques}
	\label{ch2:classical_traversal}
	
	A lot of research work has been done with the aim of improving the efficiency of mining maximal item sets from data sets. According to \cite{Mabroukeh2010} and \cite{Chand2012}, in association rule mining and sequential pattern mining, researchers have put a lot of effort into optimizing the pruning technique from which maximal or closed item sets are generated. Specifically in the pattern mining realm, algorithms that are based on \textit{depth-first search} (DFS) have proven to be more efficient than algorithms that are based on \textit{breadth-first search} (BFS) since they do not generate useless candidate item sets.
	
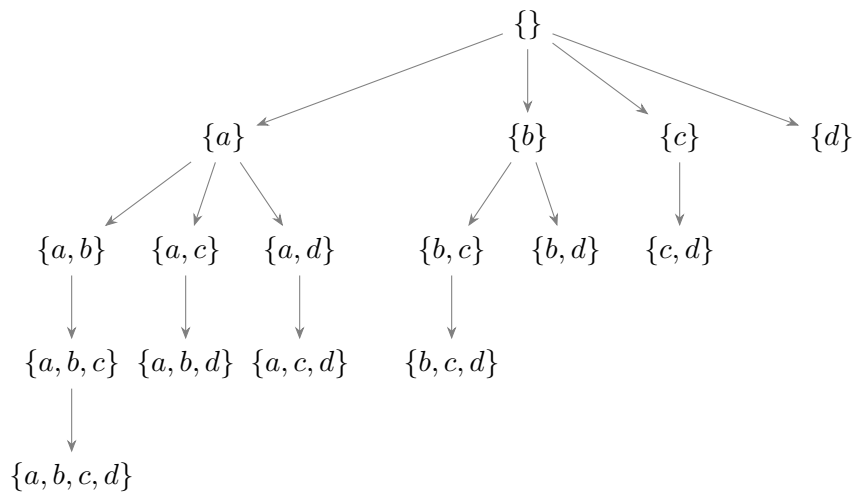
\begin{figure}[h!]
		\centering
		\begin{tikzpicture}
		\begin{scope}
			\node (r0) at (0, 0) {$\{ \}$};
			
			\node (r11) at (-4, -1.5) {$\{a\}$};
			\node (r12) at (0, -1.5) {$\{b\}$};
			\node (r13) at (2, -1.5) {$\{c\}$};
			\node (r14) at (4, -1.5) {$\{d\}$};

			\node (r21) at (-6, -3.0) {$\{a, b\}$};
			\node (r22) at (-4.5, -3.0) {$\{a, c\}$};
			\node (r23) at (-3.0, -3.0) {$\{a, d\}$};
			\node (r24) at (-1, -3.0) {$\{b, c\}$};
			\node (r25) at (0.5, -3.0) {$\{b, d\}$};
			\node (r26) at (2, -3.0) {$\{c, d\}$};

			\node (r31) at (-6, -4.5) {$\{a, b, c\}$};
			\node (r32) at (-4.5, -4.5) {$\{a, b, d\}$};
			\node (r33) at (-3.0, -4.5) {$\{a, c, d\}$};
			\node (r34) at (-1, -4.5) {$\{b, c, d\}$};

			\node (r41) at (-6, -6.0) {$\{a, b, c, d\}$};
		\end{scope}
	
		\begin{scope}
			\path [->] (r0) edge[draw=gray, thin,>={Stealth[gray]}]  (r11);
			\path [->] (r0) edge[draw=gray, thin,>={Stealth[gray]}]  (r12);
			\path [->] (r0) edge[draw=gray, thin,>={Stealth[gray]}]  (r13);
			\path [->] (r0) edge[draw=gray, thin,>={Stealth[gray]}]  (r14);
			
			\path [->] (r11) edge[draw=gray, thin,>={Stealth[gray]}]  (r21);
			\path [->] (r11) edge[draw=gray, thin,>={Stealth[gray]}]  (r22);
			\path [->] (r11) edge[draw=gray, thin,>={Stealth[gray]}]  (r23);
			\path [->] (r12) edge[draw=gray, thin,>={Stealth[gray]}]  (r24);
			\path [->] (r12) edge[draw=gray, thin,>={Stealth[gray]}]  (r25);
			\path [->] (r13) edge[draw=gray, thin,>={Stealth[gray]}]  (r26);

			\path [->] (r21) edge[draw=gray, thin,>={Stealth[gray]}]  (r31);
			\path [->] (r22) edge[draw=gray, thin,>={Stealth[gray]}]  (r32);
			\path [->] (r23) edge[draw=gray, thin,>={Stealth[gray]}]  (r33);
			\path [->] (r24) edge[draw=gray, thin,>={Stealth[gray]}]  (r34);

			\path [->] (r31) edge[draw=gray, thin,>={Stealth[gray]}]  (r41);
		\end{scope}
		\end{tikzpicture}
		\caption{Lattice diagram of possible candidate item sets through a \textit{breadth-first search}}
		\label{fig2:lattice1}
	\end{figure}

	The BFS strategy employs a level-wise candidate generation technique \cite{Agrawal1993}. Figure~\ref{fig2:lattice1} illustrates how candidate item sets may be generated in a level-wise manner from minimal item sets (i.e. $\{a\}$) to a maximal item set (i.e. $\{a, b, c, d\}$). The frequency support of each candidate is calculated and validated only if it surpasses a user-specified threshold.
			
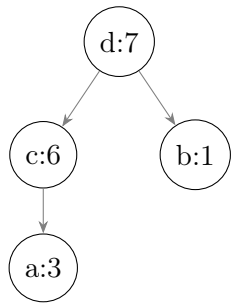
\begin{figure}[h!]
		\centering
		\begin{tikzpicture}
		\begin{scope}[every node/.style={circle, thin, draw}]
			\node (rd) at (0, 0) {d:7};
			\node (rc) at (-1, -1.5) {c:6};
			\node (rb) at (1, -1.5) {b:1};
			\node (ra) at (-1, -3.0) {a:3};
		\end{scope}
	
		\begin{scope}
			\path [->] (rd) edge[draw=gray, thin,>={Stealth[gray]}]  (rc);
			\path [->] (rd) edge[draw=gray, thin,>={Stealth[gray]}]  (rb);
			\path [->] (rc) edge[draw=gray, thin,>={Stealth[gray]}]  (ra);
		\end{scope}
		\end{tikzpicture}
		\caption{Lattice diagram of sample \textit{FP-tree}}
		\label{fig2:lattice2}
	\end{figure}

	The DFS strategy employs a set enumeration tree (also known as a \textit{frequent pattern tree: FP-Tree}) to recursively grow long frequent item sets from short ones \cite{Jiawei2000}. Figure~\ref{fig2:lattice2} illustrates how a maximal pattern $\{d, c, a\}$ is formed by (1) finding the frequency of all single item sets (\texttt{d:7} means \texttt{d} occurs 7 times in the data set: once with \texttt{b} and 6 times with \texttt{c}) and (2) constructing an \textit{FP-Tree} by recursively scanning the data set with ordered transactions.

	Numerous advancements have been made on both \textit{Apriori} and \textit{FP-tree} pattern mining techniques with the aim of improving their computational performance and memory utilization. For example \textit{Max-Miner} is an \textit{Apriori}-based algorithm that employs a \textit{``look ahead''} traversal method instead of the previous \textit{``bottom-up''} traversal method in order to quickly identify maximal item sets as early as possible \cite{Bayardo1998}.
	
	By the same token, \textit{LCM} (Linear time Closed item-set Miner) is a \textit{pattern growth}-based algorithm that uses a set enumeration tree to traverse a depth-first search of frequent closed item sets \cite{Uno2003, Uno2004}. Similarly, \textit{COFI-tree} (Co-Occurrence Frequent Item tree) algorithm proposed by \cite{Hajj2003} is a \textit{pattern growth}-based algorithm that is aimed at reducing memory search space.
	
	\subsubsection{Gradual Rule Traversal Techniques}
	\label{ch2:gradual_traversal}
	With regards to gradual pattern mining, two aspects complicate its mining process: (1) determining frequency support (see Section~\ref{ch2:definitions}) and (2) the \textit{complementary notion} of gradual item sets (for each attribute, there exist two gradual item sets). These two challenging aspects are brought about by the nature of patterns and thus traversal because we need to compare lines. As can be seen in Table~\ref{tab2:sample4} (a) and (b), association rule mining deals with the transactions of a data set while gradual rule mining deals with the attributes of a data set respectively.
	
	\begin{table}[h!]
  		\centering
		\subfloat[]{%
    		\begin{tabular}{l l} 
      		\textbf{id} & \textbf{items}\\
      		\hline \hline
      		t1 & $\{d, a, c, b\}$\\
      		t2 & $\{a, d\}$\\
      		t3 & $\{b, c, a\}$\\
      		t4 & $\{d, a, c\}$\\
      		\bottomrule
    		\end{tabular}}%
    	\qquad \qquad
		\subfloat[]{%
    		\begin{tabular}{l c c c c}
      		\textbf{id} & \textbf{a} & \textbf{b} & \textbf{c} & \textbf{d} \\
      		\hline \hline
      		r1 & 5 & 30 & 43 & 97\\
      		r2 & 6 & 35 & 33 & 86\\
      		r3 & 3 & 40 & 42 & 108\\
      		r4 & 1 & 50 & 49 & 27\\
      		\bottomrule
    		\end{tabular}}%
     		\caption{(a) Sample data set transactions for association rule mining (b) sample numeric data set gradual rule mining}
    		\label{tab2:sample4}
	\end{table}
	
	To elaborate, in association rule mining a single transaction is enough to determine the occurrence frequency of an item set. For instance in Table~\ref{tab2:sample4} (a) given transaction \textit{t3} only, we can tell that item set \texttt{d} is not frequent. In gradual rule mining at least two or more transactions are needed to determine the frequency occurrence of an item set. For instance in Table~\ref{tab2:sample4} (b) given transaction \textit{r3} only, either $(a, \uparrow)$ or $(a, \downarrow)$ is possible. Further in order to mine gradual item sets, the \textit{complementary notion} requires that: for each attribute, there exists two gradual item sets. For instance attribute $a$ creates gradual item sets $(a, \uparrow)$ and $(a, \downarrow)$. This is not the case in association rule mining.
	
	In spite of this, techniques have been developed that allow for gradual rule mining through BFS and DFS strategies. In the case of BFS strategy for gradual rule mining, \cite{Di-Jorio2009, Laurent2009} propose approaches GRITE and GRAANK. We describe these approaches in Section~\ref{ch2:grite} and Section~\ref{ch2:graank} respectively. In the case of DFS strategy for gradual rule mining, \cite{Negrevergne2014} extends LCM presented by \cite{Uno2003, Uno2004} to propose ParaMiner. We describe this approach in Section~\ref{ch2:paraminer}
	
	However, both BFS-based and DFS-based approaches have demerits. BFS-based approaches generate large numbers of candidates when dealing with data sets having huge numbers of attributes. This may lead to a combinatorial candidate explosion which overwhelms the algorithm. DFS's major drawback involves finding the parent node of the set enumeration tree with the longest length. DFS-based techniques employ recursion methods (which have exponential computational complexity) to achieve this.

	\subsection{GRITE Approach}
	\label{ch2:grite}
	\textit{GRITE} is an acronym standing for \textit{GRadual ITemset Extraction} and it is a gradual pattern mining technique proposed by \cite{Di-Jorio2009}. This technique exploits the \textit{anti-monotonicity property} in order to efficiently extract all frequent gradual item sets. GRITE technique employs a \textit{complementary notion} in order generate two gradual items ($a\uparrow ,~ a\downarrow $) for every attribute $a$ of the data set $\mathcal{D}$.
	
	The \textit{complementary notion} advocates the definition that the \textit{frequency support} of complementary gradual item sets is equal as established in Equation~\eqref{eqn2:complementary}. Further, this notion allows the GRITE technique to avoid the consideration of combining complementary gradual items.
	
	\begin{equation}\label{eqn2:complementary}
		sup(\{a,\uparrow\}) = sup(\{a,\downarrow\})
	\end{equation} 
	
	The GRITE technique applies three main approaches in order to extract gradual patterns from a numeric data set:
	\begin{itemize}
		\item application of \textit{binary matrices} in order to represent tuple orders that respect a particular gradual pattern;
		\item application of a \textit{level-wise} ($k-1$\textit{-itemset}) technique to generate ($k$\textit{-itemsets}) as candidates then, applying the bitwise \texttt{AND} operator to \textit{join} gradual item sets;
		\item application of a \textit{precedence graph} method in order to calculate the \textit{frequency support} of gradual patterns.
	\end{itemize}
	
	\textit{Example 2.2.} In order to expound the three approaches, let us consider an arbitrary data set $\mathcal{D}_{2.2}$ containing recordings of atmospheric temperature, atmospheric humidity and number of mosquitoes.
		
	\begin{table}[h!]
  		\centering
    	\begin{tabular}{c c c c} 
	  	\toprule
      	\textbf{id} & \textbf{temperature} & \textbf{humidity} & \textbf{no. of mosquitoes}\\
      	\hline \hline
      	r1 & 30 & .2 & 100\\
      	r2 & 28 & .4 & 300\\
      	r3 & 26 & .5 & 200\\
      	r4 & 26 & .8 & 500\\
      	\bottomrule
    	\end{tabular}
    	\caption{Sample data set $\mathcal{D}_{2.2}$}
    	\label{tab2:sample2}
	\end{table}
	
	\subsubsection{Binary Matrix of Orders}
	\label{ch2:grite_binary}
	As stated earlier, \textit{support} derivation for a gradual pattern involves tracing an ordered list of tuples that respect the pattern. \cite{Di-Jorio2009} proposes a binary representation of these orders, which is defined as follows: let $i$ be a gradual item set, $G_{i}$ be the list of objects respecting it and $T_{G_{i}}$ be the set of tuples in $G_{i}$,
	
	\textbf{Definition 2.8} \texttt{(Binary Matrix of Orders)}. \textit{$G_{i}$ can be represented by a binary matrix $M_{G_{i}}=(m_{a,b})a\in T_{G_{i}}, b\in T_{G_{i}}$, where $m \in \{0,1 \}$}. If there exists an order relation between $a$ and $b$, then the bit corresponding to the tuple of $a$ and the column of $b$ is set to 1, and to 0 otherwise.
	
	For example using Table~\ref{tab2:sample2}, let us consider a gradual $1$\textit{-itemset} $i1 = (temp,\downarrow )$. We have $G_{i1} = \{(r1,r2,r3,r4),(r1,r2,r4,r3) \}$ and $T_{G_{i1}} = \{r1,r2,r3,r4 \}$. The set of orders may be modeled using a binary matrix of size $4\times 4$ as illustrated in Figure~\ref{fig2:binary_matrices1} (a). Figure~\ref{fig2:binary_matrices1} (b) and (c) shows the binary matrices of gradual items $(hum,\uparrow )$ and $(mos,\uparrow )$ respectively.
	
	\begin{figure}[h!]
		\centering
		\subfloat[]{%
			\begin{tabular}{|c|c c c c|}
			\hline
			$\Rsh$ & r1 & r2 & r3 & r4\\
			\hline
			r1 & 0 & 1 & 1 & 1\\
			r2 & 0 & 0 & 1 & 1\\
			r3 & 0 & 0 & 0 & 0\\
			r4 & 0 & 0 & 0 & 0\\
			\hline
			\end{tabular}
			$~~~$ 
			}%
		\subfloat[]{%
			\begin{tabular}{|c|c c c c|}
			\hline
			$\Rsh$ & r1 & r2 & r3 & r4\\
			\hline
			r1 & 0 & 1 & 1 & 1\\
			r2 & 0 & 0 & 1 & 1\\
			r3 & 0 & 0 & 0 & 1\\
			r4 & 0 & 0 & 0 & 0\\
			\hline
			\end{tabular}
			$~~~$ 
			}%
		\subfloat[]{%
			\begin{tabular}{|c|c c c c|}
			\hline
			$\Rsh$ & r1 & r2 & r3 & r4\\
			\hline
			r1 & 0 & 1 & 1 & 1\\
			r2 & 0 & 0 & 0 & 1\\
			r3 & 0 & 1 & 0 & 1\\
			r4 & 0 & 0 & 0 & 0\\
			\hline
			\end{tabular}}%
		\caption{Binary matrices $M_{G_{i1}}$, $M_{G_{i2}}$ and $M_{G_{i3}}$ for gradual items: (a) $i1 = (temp,\downarrow )$, (b) $i2 = (hum,\uparrow )$, (c) $i3 = (mos,\uparrow )$}
		\label{fig2:binary_matrices1}
	\end{figure}
		
	\subsubsection{Level-wise Candidate Generation}
	\label{ch2:grite_candidate}
	The GRITE algorithm implements a level-wise technique that uses $(k-1)$\textit{-itemset} to generate $k$\textit{-itemsets} using a \textit{join} operation which is repeated an exponential number of times. For instance, two \textit{1-itemset} gradual items may be \textit{joined} to generate one \textit{2-itemset} gradual item: $(temp,\downarrow )$ and $(hum,\uparrow )$ may be \textit{joined} to form $\{(temp,\downarrow ), (hum,\uparrow ) \}$. Similarly, four \textit{1-itemset} may be joined to form six \textit{2-itemset} gradual items and so on.
	
	Granted that every gradual \textit{1-itemset} has a bitmap representation of the tuple orders that respect it, bitmap representation of any generated $k$\textit{-itemset} may be obtained by performing a bitwise \texttt{AND} operation of the corresponding \textit{1-itemset} bitmaps. \cite{Di-Jorio2009} proposes a theorem that defines this operation as follows:
		
	\textbf{Theorem 2.1.} \textit{Let $i''$ be a gradual item set generated by joining two gradual item sets $i$ and $i'$. The following relation holds: $M_{G_{i''}} = M_{G_{i}}$ \texttt{AND} $M_{G_{i'}}$}. It comes as no surprise that the theorem relies heavily on the bitwise \texttt{AND} which has a good computational performance.
		
	\begin{figure}[h!]
		\centering
		\subfloat[]{%
			\begin{tabular}{|c|c c c c|}
			\hline
			$\Rsh$ & r1 & r2 & r3 & r4\\
			\hline
			r1 & 0 & 1 & 1 & 1\\
			r2 & 0 & 0 & 1 & 1\\
			r3 & 0 & 0 & 0 & 0\\
			r4 & 0 & 0 & 0 & 0\\
			\hline
			\end{tabular}
			$~~~~~~$ 
			}%
		\subfloat[]{%
			\begin{tabular}{|c|c c c c|}
			\hline
			$\Rsh$ & r1 & r2 & r3 & r4\\
			\hline
			r1 & 0 & 1 & 1 & 1\\
			r2 & 0 & 0 & 0 & 1\\
			r3 & 0 & 0 & 0 & 1\\
			r4 & 0 & 0 & 0 & 0\\
			\hline
			\end{tabular}
			}%
		\caption{Binary matrices $M_{G_{i4}}$ and $M_{G_{i5}}$ for gradual items: (a) $i4 = \{ (temp,\downarrow ), (hum,\uparrow ) \} $, (b) $i5 = \{ (hum,\uparrow ),(mos,\uparrow ) \}$}
		\label{fig2:binary_matrices2}
	\end{figure}
	
	For example using the data set $\mathcal{D}_{2.2}$ in Table~\ref{tab2:sample2}, we can join the \textit{1-itemsets}: $(temp,\downarrow )$, $(hum,\uparrow )$ and $(mos,\uparrow )$ in order to generate two \textit{2-itemset} candidates $\{ (temp,\downarrow ), (hum,\uparrow ) \} $ and $\{ (hum,\uparrow ),(mos,\uparrow ) \}$ and obtain their bitmap representation through a bitwise \texttt{AND} operation as illustrated in Figure~\ref{fig2:binary_matrices2} (a) and (b) respectively.

	\subsubsection{Deriving Frequency Support for GRITE}
	\label{ch2:grite_support}
	For the purpose of deriving the \textit{support} of gradual item sets, the GRITE technique arranges the tuples in a lexicographical order which is characterized by the size of the tuple values. This technique can also be referred to as a \textit{precedence graph} technique. To illustrate this technique we use the following Hasse Diagram: if the value of tuple \textit{r1} is greater than that of \textit{r2}, then the \textit{r1} is placed upward of \textit{r2} and an arrow drawn from \textit{r1} to \textit{r2}.
	
	\cite{Di-Jorio2009} holds the position that given a gradual item set $i$ and its associated binary matrix $M_{G_{i}}$, the \textit{frequency support} \texttt{sup}$(i)$ is derived from the longest list in $M_{G_{i}}$. For instance Figures~\ref{fig2:hase_diagram1} and \ref{fig2:hase_diagram2} show the Hasse Diagrams of gradual items from data set $\mathcal{D}_{2.2}$.
	
	\begin{figure}[h!]
	\centering
	\subfloat[]{%
		\begin{tikzpicture}
		\begin{scope}[every node/.style={circle, thin, draw}]
			\node (r1) at (0, 0) {r1};
			\node (r2) at (0, -1.5) {r2};
			\node (r3) at (-1, -3.0) {r3};
			\node (r4) at (1, -3.0) {r4};
		\end{scope}
	
		\begin{scope}
			\path [->] (r1) edge[draw=gray, thin,>={Stealth[gray]}]  (r2);
			\path [->] (r2) edge[draw=gray, thin,>={Stealth[gray]}]  (r3);
			\path [->] (r2) edge[draw=gray, thin,>={Stealth[gray]}]  (r4);
		\end{scope}
		\end{tikzpicture}
		\qquad \qquad
	}%
	\subfloat[]{%
		\begin{tikzpicture}
		\begin{scope}[every node/.style={circle, thin, draw}]
			\node (r1) at (0, 0) {r1};
			\node (r2) at (0, -1.5) {r2};
			\node (r3) at (0, -3.0) {r3};
			\node (r4) at (0, -4.5) {r4};
		\end{scope}
	
		\begin{scope}
			\path [->] (r1) edge[draw=gray, thin,>={Stealth[gray]}]  (r2);
			\path [->] (r2) edge[draw=gray, thin,>={Stealth[gray]}]  (r3);
			\path [->] (r3) edge[draw=gray, thin,>={Stealth[gray]}]  (r4);
		\end{scope}
		\end{tikzpicture}
		\qquad \qquad
	}%
	\subfloat[]{%
		\begin{tikzpicture}
		\begin{scope}[every node/.style={circle, thin, draw}]
			\node (r1) at (0, 0) {r1};
			\node (r3) at (0, -1.5) {r3};
			\node (r2) at (0, -3.0) {r2};
			\node (r4) at (0, -4.5) {r4};
		\end{scope}
	
		\begin{scope}
			\path [->] (r1) edge[draw=gray, thin,>={Stealth[gray]}]  (r3);
			\path [->] (r3) edge[draw=gray, thin,>={Stealth[gray]}]  (r2);
			\path [->] (r2) edge[draw=gray, thin,>={Stealth[gray]}]  (r4);
		\end{scope}
		\end{tikzpicture}
		\qquad \qquad
	}%
	\caption{Hasse Diagrams for \textit{1-itemset} gradual items: (a) $i1 = (temp,\downarrow )$, (b) $i2 = (hum,\uparrow )$, (c) $i3 = (mos,\uparrow )$}
	\label{fig2:hase_diagram1}
	\end{figure}
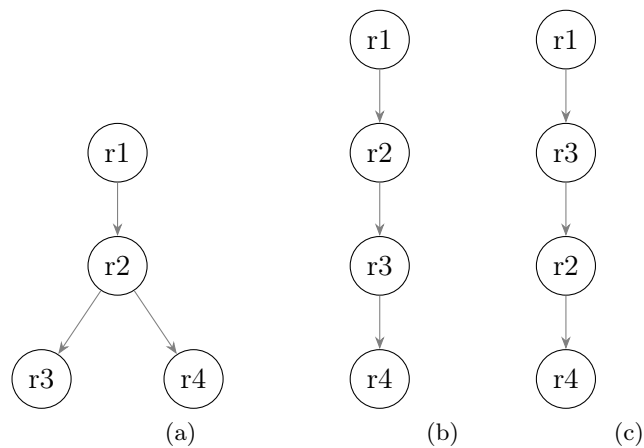

	\begin{figure}[h!]
	\centering
	\subfloat[]{%
		\begin{tikzpicture}
		\begin{scope}[every node/.style={circle, thin, draw}]
			\node (r1) at (0, 0) {r1};
			\node (r2) at (0, -1.5) {r2};
			\node (r3) at (-1, -3.0) {r3};
			\node (r4) at (1, -3.0) {r4};
		\end{scope}
	
		\begin{scope}
			\path [->] (r1) edge[draw=gray, thin,>={Stealth[gray]}]  (r2);
			\path [->] (r2) edge[draw=gray, thin,>={Stealth[gray]}]  (r3);
			\path [->] (r2) edge[draw=gray, thin,>={Stealth[gray]}]  (r4);
		\end{scope}
		\end{tikzpicture}
		\qquad \qquad
	}%
	\subfloat[]{%
		\begin{tikzpicture}
		\begin{scope}[every node/.style={circle, thin, draw}]
			\node (r1) at (0, 0) {r1};
			\node (r2) at (-1, -1.5) {r2};
			\node (r3) at (1, -1.5) {r3};
			\node (r4) at (0, -3.0) {r4};
		\end{scope}
	
		\begin{scope}
			\path [->] (r1) edge[draw=gray, thin,>={Stealth[gray]}]  (r2);
			\path [->] (r1) edge[draw=gray, thin,>={Stealth[gray]}]  (r3);
			\path [->] (r2) edge[draw=gray, thin,>={Stealth[gray]}]  (r4);
			\path [->] (r3) edge[draw=gray, thin,>={Stealth[gray]}]  (r4);
		\end{scope}
		\end{tikzpicture}
		\qquad \qquad
	}%
	\caption{Hasse Diagrams for \textit{2-itemset} gradual items: (a) $i4 = \{ (temp,\downarrow ), (hum,\uparrow ) \} $, (b) $i5 = \{ (hum,\uparrow ),(mos,\uparrow ) \}$}
	\label{fig2:hase_diagram2}
	\end{figure}
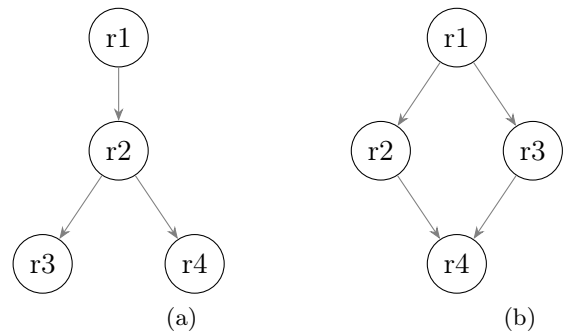
	
	Therefore, the \textit{frequency supports} for the extracted gradual items are as follows: $sup(\{ temp,\downarrow \}) = \dfrac{3}{4}$, $sup(\{ hum,\uparrow \}) = \dfrac{5}{5}$, $sup(\{ mos,\uparrow \}) = \dfrac{5}{5}$, $sup(\{ (temp,\downarrow ), (hum,\uparrow ) \} ) = \dfrac{3}{4}$ and, $sup(\{ (hum,\uparrow ),(mos,\uparrow ) \}) = \dfrac{3}{4}$.

	\subsection{GRAANK Approach}
	\label{ch2:graank}
	\textit{GRAANK} is an acronym standing for \textit{GRAdual rANKing} and it is a gradual pattern mining technique proposed by \cite{Laurent2009}. Similar to GRITE, this technique also exploits the \textit{anti-monotonicity property} in order to efficiently extract all frequent gradual patterns. Again it maintains the \textit{complementary notion} established in the GRITE technique of \textit{frequency support}.
	
	GRAANK technique is an advancement of the GRITE technique; therefore, it similarly applies \textit{binary matrices} for bitmap representation of tuple orders, it also applies a \textit{level-wise} technique for generation of gradual item set candidates, and it also applies the bitwise \texttt{AND} operator to \textit{join} the gradual item sets \cite{Laurent2009}. However, it should be noted that derivation of \textit{frequency support} is implemented differently in the GRAANK technique.
	
	\subsubsection{Deriving Frequency Support for GRAANK}
	\label{ch2:graank_support}
	
	In contrast to the GRITE approach, the GRAANK approach applies a different technique in order to derive the \textit{frequency support} of gradual item sets. It exploits Kendall's $\tau$ rank correlation to evaluate the support of gradual item sets through concordant pairs. Kendall's $\tau$ can be referred to as the frequency of pair-wise inversions and \cite{Laurent2009} defines it as \textit{``the proportion of discordant pairs''}. It is important to note that the gradual item set support definition given in Equation~\eqref{eqn2:kendalls} equals the proportion of concordant pairs of tuples that respect the gradual item set.
	
	Let $n$ be objects to be ranked by $\sigma_{k}$ and $k\in 1, 2$ where $\sigma_{k}(x)$ gives the rank of object x in $\sigma_{k}$ ranking. Then, \textit{concordant pairs} $(i, j)$ are pairs for which rankings agree as follows: either $\sigma_{1}(i) \leq \sigma_{1}(j)$ and $\sigma_{2}(i) \leq \sigma_{2}(j)$, or $\sigma_{1}(i) \geq \sigma_{1}(j)$ and $\sigma_{2}(i) \geq \sigma_{2}(j)$. The sum of ranks for rankings that are not correlated is given by the formula that follows:
	
	\begin{equation}\label{eqn2:rank_sum}
		\dfrac{n(n-1)}{2m}
	\end{equation}
	
	In the case of gradual item set extraction, tuples can be arranged in pairs to form couples which are then tested for respecting the corresponding gradual pattern. It is important to realize that complementary gradual items (i.e. $(i,\uparrow)$ and $(i,\downarrow)$) are distinguished by respectively switching couple numbers (i.e. $(i, j)$ and $(j, i)$. Since none of couple rankings are correlated; then we modify Equation~\eqref{eqn2:rank_sum} as: the total sum of pairs that can be formed is given by $n(n-1)/2$ (where $n$ is the number of tuples in the data set). For example given that the data set in Table~\ref{tab2:sample2} has 4 tuples, we have a maximum of 6 couples which are: $\{ [r1,r2],[r1,r3],[r1,r4],[r2,r3],[r2,r4],[r3,r4] \}$.
	
	Table~\ref{tab2:concordant_list} illustrates a list of concordant couples that respect corresponding gradual item sets and the support which is calculated through dividing the number of concordant couples by the total sum of couples.
	
	\begin{table}[h!]
		\centering
    	\begin{tabular}{l l c} 
	  	\toprule
      	\textbf{item set} & \textbf{list of concordant couples} & \textbf{support}\\
      	\hline \hline
      	$\{ temp,\downarrow \} $ & $\{ [r1,r2],[r1,r3],[r1,r4],[r2,r3],[r2,r4] \}$ & $5/6$\\
      	$\{ hum,\uparrow \}$ & $\{ [r1,r2],[r1,r3],[r1,r4],[r2,r3],[r2,r4],[r3,r4] \}$ & $6/6$\\
      	$\{ mos,\uparrow \}$ & $\{ [r1,r2],[r1,r3],[r1,r4],[r2,r4],[r3,r4] \}$ & $5/6$\\
      	$\{ (temp,\downarrow ), (hum,\uparrow ) \} $ & $\{ [r1,r2],[r1,r3],[r1,r4],[r2,r3],[r2,r4] \}$ & $5/6$\\
      	$\{ (hum,\uparrow ),(mos,\uparrow ) \}$ & $\{ [r1,r2],[r1,r3],[r1,r4],[r2,r4],[r3,r4] \}$ & $5/6$\\
      	\bottomrule
    	\end{tabular}
    	\caption{List of concordant couples and support for some gradual item sets from data set $\mathcal{D}_{2.2}$ in Table~\ref{tab2:sample2}}
    	\label{tab2:concordant_list}
	\end{table}
	
\begin{figure}[h!]
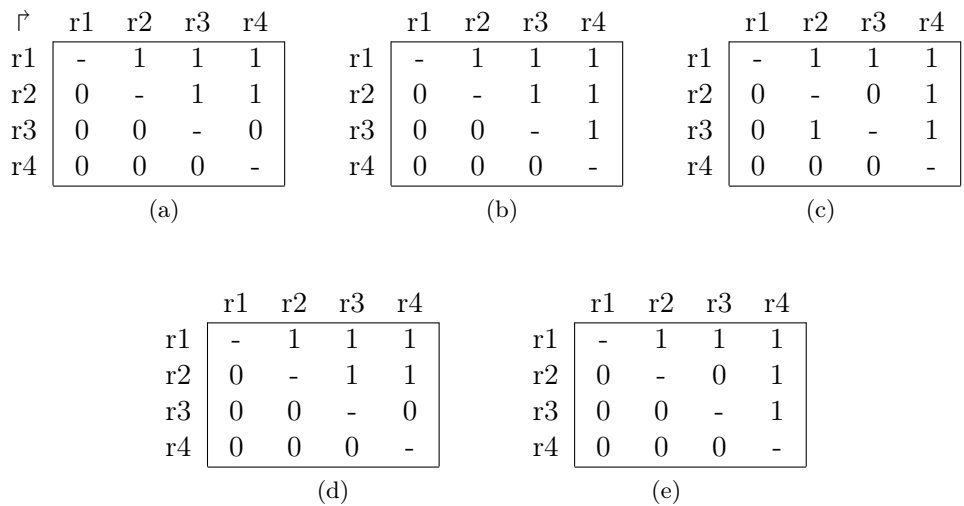

		\centering
		\subfloat[]{%
			\begin{tabular}{c|c c c c|}
			\multicolumn{1}{c}{$\Rsh$} & r1 & r2 & r3 & \multicolumn{1}{c}{r4}\\
			\cline{2-5}
			r1 & - & 1 & 1 & 1\\
			r2 & 0 & - & 1 & 1\\
			r3 & 0 & 0 & - & 0\\
			r4 & 0 & 0 & 0 & -\\
			\cline{2-5}
			\end{tabular}
			$~~~$ 
			}%
		\subfloat[]{%
			\begin{tabular}{c|c c c c|}
			\multicolumn{1}{c}{} & r1 & r2 & r3 & \multicolumn{1}{c}{r4}\\
			\cline{2-5}
			r1 & - & 1 & 1 & 1\\
			r2 & 0 & - & 1 & 1\\
			r3 & 0 & 0 & - & 1\\
			r4 & 0 & 0 & 0 & -\\
			\cline{2-5}
			\end{tabular}
			$~~~$ 
			}%
		\subfloat[]{%
			\begin{tabular}{c|c c c c|}
			\multicolumn{1}{c}{} & r1 & r2 & r3 & \multicolumn{1}{c}{r4}\\
			\cline{2-5}
			r1 & - & 1 & 1 & 1\\
			r2 & 0 & - & 0 & 1\\
			r3 & 0 & 1 & - & 1\\
			r4 & 0 & 0 & 0 & -\\
			\cline{2-5}
			\end{tabular}
			}%
			\medskip \medskip
			
		\subfloat[]{%
			\begin{tabular}{c|c c c c|}
			\multicolumn{1}{c}{} & r1 & r2 & r3 & \multicolumn{1}{c}{r4}\\
			\cline{2-5}
			r1 & - & 1 & 1 & 1\\
			r2 & 0 & - & 1 & 1\\
			r3 & 0 & 0 & - & 0\\
			r4 & 0 & 0 & 0 & -\\
			\cline{2-5}
			\end{tabular}
			$~~~~~~$ 
			}%
		\subfloat[]{%
			\begin{tabular}{c|c c c c|}
			\multicolumn{1}{c}{} & r1 & r2 & r3 & \multicolumn{1}{c}{r4}\\
			\cline{2-5}			
			r1 & - & 1 & 1 & 1\\
			r2 & 0 & - & 0 & 1\\
			r3 & 0 & 0 & - & 1\\
			r4 & 0 & 0 & 0 & -\\
			\cline{2-5}
			\end{tabular}
			}%
		\caption{Binary matrices representing the sets of concordant object pairs for gradual item sets: (a) $i1 = (temp,\downarrow )$, (b) $i2 = (hum,\uparrow )$, (c) $i3 = (mos,\uparrow )$, (d) $i4 = \{ (temp,\downarrow ), (hum,\uparrow ) \} $, (e) $i5 = \{ (hum,\uparrow ),(mos,\uparrow ) \}$}
		\label{fig2:concordant_support}
	\end{figure}

	The Kendall's $\tau$ rank correlation technique applied by the GRAANK approach is more efficient computationally and memory-wise than GRITE's \textit{precedence graph} technique for deriving the frequency support of gradual patterns \cite{Laurent2009}. The concordant couples can be easily represented as a binary matrix as shown in Figure~\ref{fig2:concordant_support} (a), (b), (c), (d) and (e). As can be seen the proportion of concordant couples respecting the corresponding gradual item set can be obtained by summing up all the binary $1$s in the respective matrices.
	
	In the final analysis, GRAANK algorithm performs better than GRITE algorithm for gradual pattern mining. This is because GRAANK algorithm inherits the \textit{binary matrix of orders} technique; therefore benefiting from its efficiency. GRAANK replaces the \textit{precedence graph} technique with the \textit{Kendall's $\tau$ rank correlation} technique whose computation cost for calculating the support of gradual item sets is lower. 
	
	Equally important, both approaches exploit an efficient bitmap representation technique in order to store validated candidate item sets as matrices. Consequently, this reduces the number of multiple scans through the data set to just one initial scan for every attribute since this technique allows representation of resulting gradual item sets through the binary \texttt{AND} operator.
	
	\subsection{ParaMiner Approach}
	\label{ch2:paraminer}
	ParaMiner is a generic approach that exploits parallel processing for mining closed patterns proposed by \cite{Negrevergne2014}.  It extends Linear time Closed itemset Miner (LCM) approach described by \cite{Uno2003, Uno2004} to the problem of gradual pattern mining. 
	
	\textit{Example 2.3.} In order to describe this approach, let us consider a numeric data set $\mathcal{D}_{2.3}$.
	
	\begin{table}[h!]
  		\centering
    	\begin{tabular}{l c c c c}
      		\textbf{id} & \textbf{a} & \textbf{b} & \textbf{c} & \textbf{d} \\
      		\hline \hline
      		r1 & 5 & 30 & 43 & 97\\
      		r2 & 4 & 35 & 33 & 86\\
      		r3 & 3 & 40 & 42 & 108\\
      		r4 & 1 & 50 & 49 & 27\\
      		\bottomrule
    		\end{tabular}
    	\caption{Sample data set $\mathcal{D}_{2.3}$}
    	\label{tab2:sample3}
	\end{table}	
	
	First, ParaMiner encodes a numeric data set into a transactional data set containing variations (i.e. $\uparrow, \downarrow$) of the attributes between each record. For example data set $\mathcal{D}_{2.3}$ can be encoded into a transactional data set as shown in Table~\ref{tab2:transaction1} (a). Second, ParaMiner applies a data set reduction technique to shrink the size of the encoded transaction data set. Table~\ref{tab2:transaction1} (b) illustrates how a data set is reduced by grouping similar gradual items (with weights).
			
	\begin{table}[h!]
		\centering
		\subfloat[]{%
		\begin{tabular}{l c }
      		\textbf{id} & \textbf{item-sets} \\
      		\hline \hline
      		$t_{(r1, r2)}$ & $\{a^{\downarrow}, b^{\uparrow}, c^{\downarrow}, d^{\downarrow} \}$\\
      		$t_{(r1, r3)}$ & $\{a^{\downarrow}, b^{\uparrow}, c^{\downarrow}, d^{\uparrow} \}$\\
      		$t_{(r1, r4)}$ & $\{a^{\downarrow}, b^{\uparrow}, c^{\uparrow}, d^{\downarrow} \}$\\
      		$t_{(r2, r3)}$ & $\{a^{\downarrow}, b^{\uparrow}, c^{\uparrow}, d^{\uparrow} \}$\\
      		$t_{(r2, r4)}$ & $\{a^{\downarrow}, b^{\uparrow}, c^{\uparrow}, d^{\downarrow} \}$\\
      		$t_{(r3, r4)}$ & $\{a^{\downarrow}, b^{\uparrow}, c^{\uparrow}, d^{\downarrow} \}$\\
      		\bottomrule
    	\end{tabular}
    	$~~~$ 
    	}%
		\subfloat[]{%
		\begin{tabular}{l c c}
      		\textbf{tids} & \textbf{weight} & \textbf{item-sets} \\
      		\hline \hline
      		$t_{(r1, r2)}$ & 1 & $\{a^{\downarrow}, b^{\uparrow}, c^{\downarrow}, d^{\downarrow} \}$\\
      		$t_{(r1, r3)}$ & 1 & $\{a^{\downarrow}, b^{\uparrow}, c^{\downarrow}, d^{\uparrow} \}$\\
      		$t_{(r1, r4)}, t_{(r3, r4)}, t_{(r2, r4)}$ & 3 & $\{a^{\downarrow}, b^{\uparrow}, c^{\uparrow}, d^{\downarrow} \}$\\
      		$t_{(r2, r3)}$ & 1 & $\{a^{\downarrow}, b^{\uparrow}, c^{\uparrow}, d^{\uparrow} \}$\\
      		\bottomrule
    	\end{tabular}}%
    	\caption{(a) Transactional encoding of data set $\mathcal{D}_{2.3}$, (b) reduced encoded data set}
    	\label{tab2:transaction1}
	\end{table}
	
	In order to remove infrequent \textit{1-itemset} gradual items, the transactional data set (Table~\ref{tab2:transaction1} (b)) is sorted by item occurrence as shown in Table~\ref{tab2:transactional2} (a). For example, if we set the minimum length of \texttt{tids} to 3, we remove infrequent items as illustrated in Table~\ref{tab2:transactional2} (b).
	
	Third and last, ParaMiner employs a recursive method to find the longest set enumeration tree using the reduced data set (Table~\ref{tab2:transactional2} (b)). Deriving fractional frequency support for the extracted patterns is relatively simple in comparison to GRITE and GRAANK. This support is obtained through dividing the length of the set enumeration tree by the tuple size of the numeric data set. However, it is important to note that the process of encoding a numeric data set into a transactional data set significantly increases its size. For example, if the original numeric data set has $n$ tuples, the encoded data set will have at least $n (n -1) / 2$ tuples.

	\begin{table}[h!]
		\centering
		\subfloat[]{%
		\begin{tabular}{c l}
      		\textbf{item} & \textbf{tids} \\
      		\hline \hline
      		$a^{\downarrow}$ & $\{t_{(r1, r2)}, t_{(r1, r3)}, t_{(r1, r4)}, t_{(r2, r3)},$\\
          	& $t_{(r2, r4)}, t_{(r3, r4)} \}$\\
			$b^{\uparrow}$ & $\{t_{(r1, r2)}, t_{(r1, r3)}, t_{(r1, r4)}, t_{(r2, r3)},$\\
          	& $t_{(r2, r4)}, t_{(r3, r4)} \}$\\   
      		$c^{\uparrow}$ & $\{t_{(r1, r4)}, t_{(r2, r3)}, t_{(r2, r4)}, t_{(r3, r4)} \}$\\
      		$d^{\downarrow}$ & $\{t_{(r1, r2)}, t_{(r1, r4)}, t_{(r2, r4)}, t_{(r3, r4)} \}$\\
      		$c^{\downarrow}$ & $\{t_{(r1, r2)}, t_{(r1, r3)} \}$\\
      		$d^{\uparrow}$ & $\{t_{(r1, r3)}, t_{(r2, r3)} \}$\\  
      		$a^{\uparrow}$ & $\{\emptyset \}$\\
         	$b^{\downarrow}$ & $\{\emptyset \}$\\
      		\bottomrule
    	\end{tabular}}%
		\subfloat[]{%
		\begin{tabular}{c l}
      		\textbf{item} & \textbf{tids} \\
      		\hline \hline
      		$a^{\downarrow}$ & $\{t_{(r1, r2)}, t_{(r1, r3)}, t_{(r1, r4)},$\\
          	& $t_{(r2, r3)}, t_{(r2, r4)}, t_{(r3, r4)} \}$\\
			$b^{\uparrow}$ & $\{t_{(r1, r2)}, t_{(r1, r3)}, t_{(r1, r4)},$\\
          	& $t_{(r2, r3)}, t_{(r2, r4)}, t_{(r3, r4)} \}$\\   
      		$c^{\uparrow}$ & $\{t_{(r1, r4)}, t_{(r2, r3)}, t_{(r2, r4)},$\\
      		& $t_{(r3, r4)} \}$\\
      		$d^{\downarrow}$ & $\{t_{(r1, r2)}, t_{(r1, r4)}, t_{(r2, r4)},$\\
          	& $t_{(r3, r4)} \}$\\
      		\bottomrule
    	\end{tabular}
			}%
		\caption{(a) sorted items by occurrence and, (b) sorted reduced transactional data set}
    	\label{tab2:transactional2}
	\end{table}
	
	One contribution of this study entails proposing a heuristic solution that is based on \textit{ant colony optimization} to the problem of (1) combinatorial candidate explosion and, (2) finding longest set tree for the case of gradual pattern mining.

	\section{Temporal Data Mining}
	\label{ch2:temporal}
	Another contribution of this research study involves proposing and describing a \textit{fuzzy model} that extends the GRAANK approach in order to extract temporal gradual rules from numeric data sets. A temporal gradual rule may take the form: \textit{``the higher the temperature, the more mosquitoes \textbf{almost 4 hours later}''}. In this section, we review related literature concerning temporal data mining.
	
	\cite{Roddick1999} state that temporal data mining deals with the analysis of events by one or more dimensions of time. They distinguish temporal data mining into two main fields: one concerns extracting similar patterns within the same or among different time sequences, this is referred to as \textit{trend analysis}; the other concerns extracting causal relationships among temporally oriented events.
	
	\textit{Trend analysis} was introduced by \cite{Agrawal1995} as they tried to solve the problem of \textit{`absence of time constraint'} in their proposed algorithm \textit{AprioriAll} for extracting sequential patterns \cite{Chand2012}. For example this problem may be represented by the scenario that follow:
	
	\textit{a shop does not care if someone bought `bread', followed by `bread and jam' three weeks later; they may want to specify that a customer should support a sequential pattern only if adjacent elements occur within a specified interval, say 3 days. Therefore, for a customer to support this pattern, the customer should have bought `bread and jam' within 3 days of buying `bread'.}
	
	\cite{Srikant1996} proposed \textit{Generalized Sequential Pattern} (GSP) that utilizes a user-specified \textit{time gap} to generate candidates for temporally-oriented frequent patterns. \textit{GSP} is 5 times faster than \textit{AprioriAll} since it generates less candidates. However, \cite{Masseglia2009} proposed \textit{Graph for Time Constraint} (GTC) that is more efficient than GSP because it handled time constraints prior to and separately from the counting step of the data sequence.
	
	It should be noted that for both GSP and GTC, a pattern is considered to be relevant, if its item sets occur at a minimum time interval determined by a user-specified \textit{sliding-window size}. We shy away from this approach since it locks out any frequent pattern whose time constraint is larger or smaller than the specified \textit{window size}. For example given the previous scenario, since the \textit{window size} is set at \textit{`3 days'} any element occurring slightly above 3 days (i.e. 3 days and 4 hours) is not considered.
		
	The latter field that deals with discovering causal relationships may be conceptualized by a gradual pattern that correlates the causal effect among the attributes of a data set. \cite{Fiot2009} proposed an \textit{`trend evolution'} approach that extracted gradual trends which are represented in the form: \textit{``An increasing number butter purchases during a short period is frequently followed by a purchase of cheese a few hours later.”} This approach involves converting a numeric database into a fuzzy membership degree database (referred to as a \textit{trend database}) using a user-specified fuzzy partition.
	
	\textit{Example 2.4.} Let us consider an example adopted from \cite{Fiot2009}.
	
	\begin{table}[h!]
  		\centering
    	\begin{tabular}{c c c c c}
	  	\toprule
      	& & \textbf{oil price} & \textbf{oil consumption} & \textbf{car sales}\\
        & \textbf{month} & ($\$$ per 100 L) & (KL per day) & (K) \\
      	\hline \hline
      	d1 & Feb & 112 & 330 & 150\\
      	d2 & Mar & 115 & 315 & 145\\
      	d3 & Apr & 125 & 300 & 143\\
      	d4 & May & 120 & 320 & 140\\
      	\bottomrule
    	\end{tabular}
    	\caption{Sample numeric data sequence}
    	\label{tab2:sample5}
	\end{table}
	
	Specifically, the \textit{trend database} is generated from the \textit{evolution procedure} that is applied on two tuples that are related to the same attribute within the original data set. For example Table~\ref{tab2:trend_db} is generated by applying a user-defined \textit{fuzzy set partition} on Table~\ref{tab2:sample5}. We comment that the \textit{evolution procedure} may be improved by the avoidance of relating all the tuples of the data set by themselves, since this may lead to a combinatorial explosion in the number of \textit{evolved} (or generated) tuples.
	
	\begin{table}[h!]
  		\centering
    	\begin{tabular}{c c c c c}
	  	\toprule
      	\textbf{month} & \textbf{price}, high & \textbf{consumption}, medium & \textbf{sales}, low &\\
      	\hline \hline
      	Feb & .1 &  & & $r^{1}$\\
      	Mar & .2 & .3 & .4 & $r^{2}$\\
      	Apr & .4 & .5 & .7 & $r^{3}$\\
      	May & 1 & .2 & .3 & $r^{4}$\\
      	\bottomrule
    	\end{tabular}
    	\caption{Generated trend data sequence}
    	\label{tab2:trend_db}
	\end{table}
	
	In conclusion, the \textit{trend evolution} technique represents the fuzziness of temporal correlations among frequent item sets by using a linguistic format (i.e. \textit{``a few hours later''}). However, we propose an approach in Chapter~\ref{ch3} that represents the temporal fuzziness of frequent gradual item sets through a numeric format (i.e. \textit{``almost 3 hours later''}).
	
	\section{Data Crossing for Pattern Mining}
	\label{ch2:data_crossing}
	The third contribution of this study encompasses proposing and describing a \textit{fuzzy model} that crosses time-series data from different sources in order to exploit the crossings for gradual pattern mining. In this section, we review existing techniques concerning crossing time-series data.
	
	To begin with, data crossing may be defined as: 
	
	\textit{``a process that enables the matching of different data sets using a pre-defined criteria and combining their data points to form a new data set''} \cite{Hicham2015, Hicham2016, Costa2010}.
		
	It should be remembered that the main reason for crossing different data set is to attempt to discover hidden correlations that otherwise cannot be discovered by analyzing the individual data sets in isolation. Several previous work exist related to crossing data from different sensors in a WSN \textit{(Wireless Sensor Network)} via \textit{query processing} \cite{Vaidehi2011, Gonccalves2012, Wang2013}.
	
	\textit{Query processing} allows users to retrieve information about sensed data values by issuing pre-defined queries to the storage engine. In fact, the nature of storage engine used by a WSN determines which type of \textit{query processing} is applicable \cite{Gonccalves2012, Wang2013}. According to \cite{Gonccalves2012}, there are 3 groups of storage models: \textit{local storage} model where sensors keep their on data in a distributed manner, \textit{external storage} model, where all sensors store data in a common database in a centralized manner, and \textit{data-centric} storage model which combines both models.
	
	It is important to highlight that \textit{query processing} relies on declarative languages (e.g. \texttt{SQL} - Structured Query Language) in all these models. \texttt{Q1} shows an example of a query. Therefore, most research work relate to increasing the efficiency of \textit{query processing} in either a distributed or a centralized \textit{WSN} model \cite{Vaidehi2011, Gonccalves2012, Boukerche2016}. However, many works on these techniques do not consider crossing data sets by approximating the \textit{`date-time'} attribute.
		
	\begin{Verbatim}
		Q1:	SELECT MAX HUMIDITY
			   FROM SENSOR_DATA
			   WHERE SAMPLE INTERVAL = 1 min
	\end{Verbatim}
	
	However, the emergence of the term \textit{fuzzy join}, which enables declarative languages such as \texttt{SQL} to generate views based on textual similarity (i.e. through approximations). A fuzzy-search model is proposed in \cite{Mrozek2018Oct} and  \cite{Mrozek2018Dec} which extends \texttt{U-SQL} to allow \textit{fuzzy join} approximations on numeric data sets. This is implemented by representing attribute values as fuzzy members and assigning them to a linguistic variable.
	
	\begin{figure}[h!]
    	\centering
		\includegraphics[width=.5\textwidth] {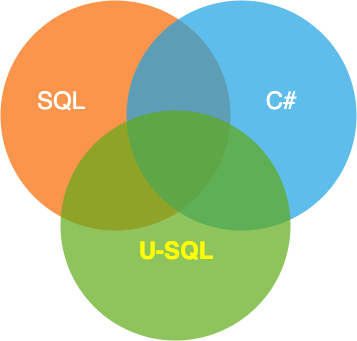}
		\caption{Relationship between \texttt{U-SQL} and C$\#$ $\&$ \texttt{SQL}}
		\label{fig2:usql}
	\end{figure}
	
	\texttt{U-SQL}\footnote{\url{https://docs.microsoft.com/en-us/u-sql/}} is big data query language created by Microsoft for (Azure Data Lake Analytics) ADLA service and it is a combination of \texttt{SQL} and \texttt{C$\#$} languages as illustrated in Figure~\ref{fig2:usql} (\texttt{Q2} shows a sample \texttt{U-SQL} excerpt). It is important to note that \texttt{$@$RESULT} and \texttt{$@$SENSOR$\_$DATA} are variables and \texttt{$@$SENSOR$\_$DATA} may include a separate code excerpt which extracts data from different sources. \texttt{U-SQL} can be applied on the following data storage engines: Azure Data Lake Storage, Azure Blob Storage, and Azure \texttt{SQL} DB, Azure \texttt{SQL} Data Warehouse. It should be underlined that currently it is extremely difficult to apply \texttt{U-SQL} on non-Microsoft storage engines such as \textit{Postgres}.

	\begin{Verbatim}
		Q2:	@RESULT =
			   SELECT temperature, humidity, COUNT(*) AS DataSize
			   FROM @SENSOR_DATA
			   FETCH FIRST 10 ROWS;	
	\end{Verbatim}

	However, data management frameworks especially the \textit{OGC SensorThings} are built on top of NoSQL models. Therefore, SQL -based or U-SQL-based querying models are difficult to integrate into such frameworks. To provide one solution to this problem, we propose a model that is based on a NoSQL fuzzy model. This allows for our technique for crossing unrelated data sets to be easy to integrate into \textit{OGC SensorThings} API which is built on \textit{Postgres} data storage engine and its querying model does not support \texttt{U-SQL}. 
		
\begin{figure}[h!]
    	\centering
		\includegraphics[width=.6\textwidth] {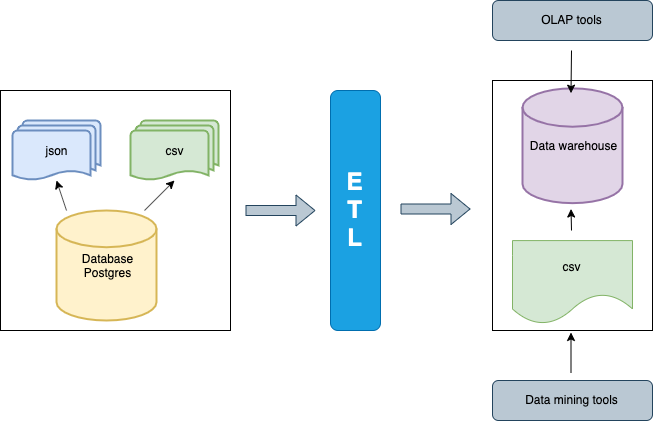}
		\caption{Illustration of ETL process}
		\label{fig2:etl_process}
	\end{figure}

	\cite{Costa2010} proposes another model that extracts, transforms and loads (ETL) into a data warehouse different time-series data collected in a \textit{WSN}, as illustrated in Figure~\ref{fig2:etl_process}. In this work, they demonstrate how to transform the data by normalizing the data types of its attributes including \textit{date-time} before loading it into a data warehouse.
	
	The drawback with the normalization technique proposed by \cite{Costa2010} is the problem of \textit{perfect matching} since it merges \textit{tuples} based on the large \textit{date-time} period and discards values with small granularity. Under those circumstances crossing is only possible if the largest \textit{date-time} values match perfectly.

	In this research study, we propose a \textit{fuzzy} model in Chapter~\ref{ch6} that will extract time-series data from unrelated sources, transform them using a fuzzy membership function so that crossing is possible through estimation even when the \textit{date-time} values do not match. Additionally, we apply this model on an \textit{OGC SensorThings} API implementation since it offers provision for numerous unrelated time-series data.
		
	\section{OGC SensorThings Framework}
	\label{ch2:ogc}
	OGC\footnote{\url{https://www.opengeospatial.org/}} is an international consortium consisting of over 530 companies, government agencies, research organizations and universities driven to make geospatial data and services \textit{FAIR} (Findable, Accessible, Interoperable and Reusable). The OGC SensorThings API is an open standard that is built on top of the OGC Sensor Web Enablement (SWE) and International Organization for Standardization Observation and Measurement (ISO/OGC O$\&$M) data model \cite{Crommert2004, Liang2016}.

	The SensorThings API is specifically designed for constrained \textit{IoT} (Internet of Things) devices; therefore, it employs the use of efficient technologies such as RESTful (Representational State Transfer) API, JSON (JavaScript Object Notation) encoding, MQTT protocol, OASIS OData (Open Data protocol) and flexible URL conventions. The OGC SensorThings framework is an advancement of the OGC SOS (Sensor Observation Service) framework. In contrast to its predecessor, the SensorThings framework is specifically designed for resource-constrained IoT devices and web-based platforms \cite{Liang2016}.
		
\begin{figure}[h!]
    	\centering
		\includegraphics[width=\textwidth] {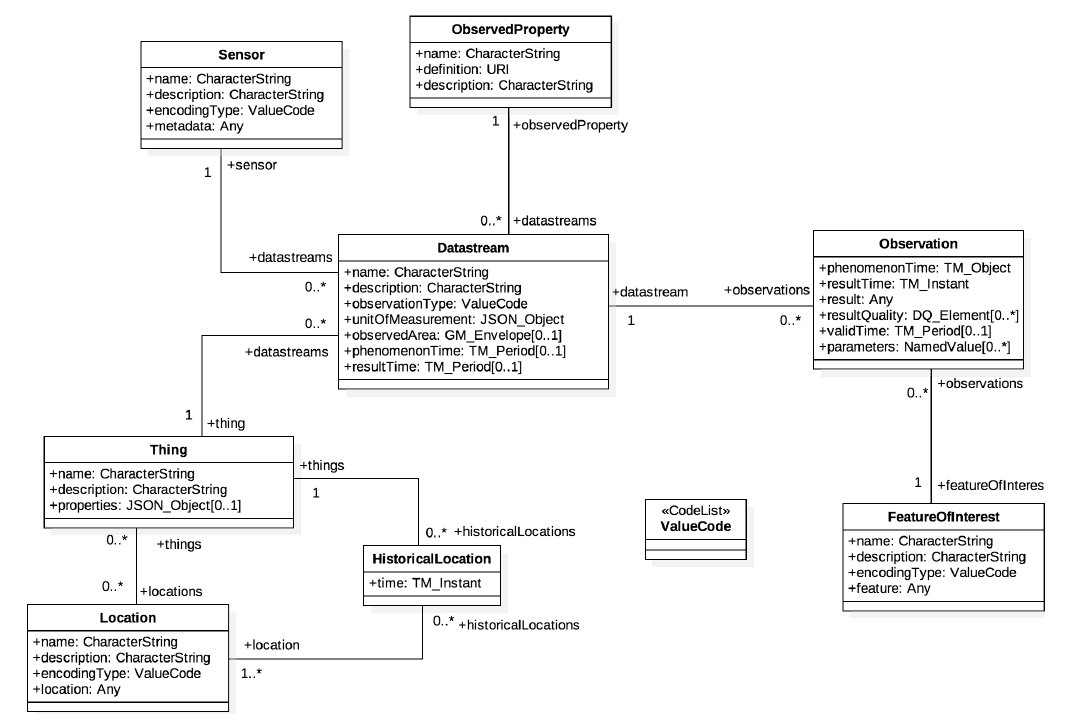}
		\caption{The 8 sensing entities that make OGC SensorThings framework \cite{Liang2016}}
		\label{fig2:ogc_entities}
	\end{figure}

	The SensorThings API is composed of 2 parts: (1) the \textit{Sensing} part and (2) the \textit{Tasking} part. The Tasking part provides a standard way for parameterizing (also called tasking) of IoT devices such as: sensors and actuators, smart cities in-situ platforms, wearable devices, drones, autonomous vehicles and so forth. This research study is interested in the \textit{Sensing} part which allows IoT devices and applications to perform CRUD operations through HTTP requests. The \textit{Sensing} part is designed based on the ISO/OGC O$\&$M data model and it defines 8 entities for IoT sensing applications. Figure~\ref{fig2:ogc_entities} depicts the 8 entities together with their properties and relationships \cite{Hicham2015, Liang2016}.
	
	According to \cite{Liang2016} the 8 entities are defined as follows:
	
	\textbf{(1) Thing:} with regard to IoT, a thing is an object of the physical world or the information world (i.e. virtual) that is capable of being identified and integrated into communication networks. The properties of the \texttt{Thing} entity include: name, description, and properties.
	
	\textbf{(2) Location:} is a geographical description that locates the \texttt{Thing} or the \texttt{Things} it  associates with. This entity is defined as the last known location of the \texttt{Thing}. It should be noted that a \texttt{Thing}'s \texttt{Location} may be similar to the \texttt{Thing}'s \texttt{Observation}'s \texttt{FeatureOfInterest}. The properties of the \texttt{Location} entity include: name, description, encoding type and location (whose type is defined by the encoding type property).
	
	\textbf{(3) HistoricalLocation:} provides the times of the last known and previous locations of the \texttt{Thing} entity. This entity has one property, that is time which records the time of the \texttt{Thing}'s known \texttt{Location}.

	\textbf{(4) Datastream:} this entity groups a collection of \texttt{Observations} measuring the same \texttt{ObservedProperty} and produced by the same \texttt{Sensor}. The properties of the \texttt{Datastream} entity include: name, description, unit of measurement, observation type, observed area, phenomenon time and result time.
	
	\textbf{(5) Sensor:} is an instrument that observes a property or phenomenon with the aim of producing an estimated value of the phenomenon. The properties of the \texttt{Sensor} entity include: name, description, encoding type and meta-data.
	
	\textbf{(6) ObservedProperty:} this entity specifies the phenomenon of an \texttt{Observation}. The properties of the \texttt{ObservedProperty} entity include: name, definition and description.
	
	\textbf{(7) Observation:} this entity defined as the act of measuring or determining the value of a phenomenon. The properties of the \texttt{Observation} entity include:  phenomenon time, result, result time, result quality, valid time and parameters.
		
	\textbf{(8) FeatureOfInterest:} this entity is described together with the \texttt{Observation} entity in this way: an \texttt{Observation} leads to a phenomenon being assigned a value. A phenomenon can also be described as a \textit{`property of feature'} or the \texttt{FeatureOfInterest} of the \texttt{Observation}. The \texttt{FeatureOfInterest} entity has the following properties: name, description, encoding type and feature.
	
\begin{figure}[h!]
    	\centering
		\includegraphics[width=.75\textwidth] {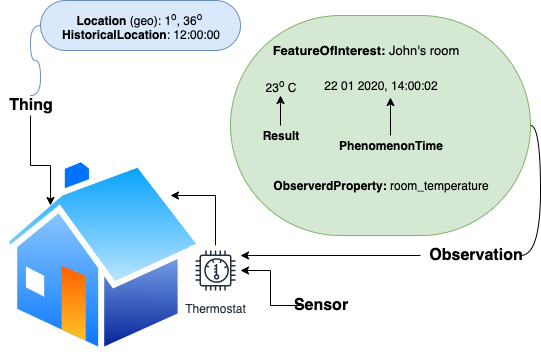}
		\caption{An illustration of the sensing entities in real life}
		\label{fig2:ogc_demo}
	\end{figure}

	For example if we are interested in observing the room temperature of a building as a phenomenon, we may install a WI-FI connected thermostat inside the building. The sensing entities of our experiment are illustrated in Figure~\ref{fig2:ogc_demo}.
	
	According to \cite{Liang2016}, the OGC SensorThings API offers three main benefits listed as follows:
	\begin{enumerate}
		\item it allows the proliferation of new high value services with lower overhead of development and wider reach,
		\item it lowers the risks, time and cost across a full IoT product cycle and,
		\item it simplifies the connections between devices-to-devices and devices-to-applications.
	\end{enumerate}
	
	Popular certified software implementations of the SensorThings API are accomplished: (1) in \texttt{Java} language by \textit{$52^{o}$ North Initiative for Geospatial Open Source Software}\footnote{\url{https://52north.org/}} and (2) in \textit{Google's} \texttt{golang (GO)} language by \textit{GOST (golang SensorThings) IoT Platform}\footnote{\url{https://www.gostserver.xyz/}}. When the SensorThings API is implemented on a Cloud platform, it exposes a service document resources (which offers the 8 entity sets) that allows its clients to navigate the entities in a hyper-media driven manner.
	
	All things considered, this research study recommends the \textit{GOST} software implementation of the SensorThings API because it is utilizes the \textit{Docker} Cloud platform. The \textit{Docker} Cloud platform provides good support integrating our gradual pattern mining algorithms which are implemented in \texttt{Python} language.

	\section{Cloud Platforms for Gradual Pattern Mining}
	\label{ch2:cloud}
	This research study's fourth contribution includes describing a software architecture model which integrates gradual pattern mining algorithms into a \textit{Cloud platform} that implements an \textit{OGC SensorThings API}. The SensorThings API provides an efficient framework for managing and sharing unrelated \textit{time-series} sensor data. For this reason the study emphasizes the implementation of the API in order to exploit the availability of time-series data from different sources for extraction of temporal gradual patterns.
	
	It is important to highlight that existing algorithm implementations of any version of GRITE and GRAANK is either in \texttt{Python, C++, Java or R} language. These implementation languages are preferred because they allow for parallel computing which improves performances of the algorithms. However, these algorithm implementations are not scalable therefore in this section, we examine Cloud platforms that may support integration of such algorithm implementations. We also discuss the benefits that may come with such a Cloud integration.
		
	To begin with, there exist numerous formal definitions of Cloud computing. \cite{Bernstein2014} provides two of these definitions that are internationally accepted:
	
	(1) \textit{``Cloud computing is form of resources pooling where a provider's computing resources are pooled to serve multiple consumers using a multi-tenant model, with different physical and virtual resources dynamically assigned and re-assigned according to customer demand''},
	
	(2) \textit{``Cloud computing is a form of rapid elasticity where computing capabilities can be elastically provisioned and released, in some cases automatically, to scale rapidly and inward commensurate with demand''}.
	
\begin{figure}[h!]
    	\centering
		\includegraphics[width=.64\textwidth] {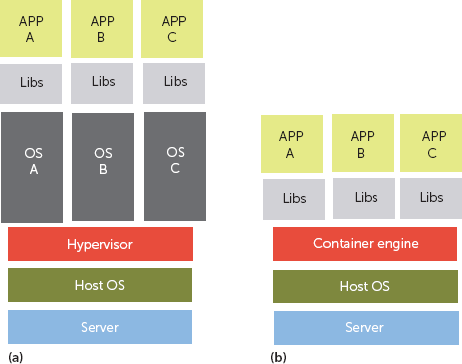}
		\caption{Comparison of (a) hypervisor and (b) container-based deployments \cite{Bernstein2014}}
		\label{fig2:containers_vms}
	\end{figure}

	Both definitions involve provision of computing services to a customer rapidly and on demand basis. There exists two fundamental virtualization technologies in Cloud computing for providing such services: the \textit{hypervisor} and the \textit{container}. The hypervisor technology provides virtual machines which can execute directly on \textit{bare metal} hardware or on an additional software layer. A container is a virtualization technology that avails protected portions of the operating system (i.e. containers virtualize the operating system) \cite{Merkel2014, Bernstein2014, Anderson2015}. Figure~\ref{fig2:containers_vms} illustrates the difference between these two technologies.
	
	In the recent years the popularity of the container virtualization technology has exceeded that of hypervisor virtualization technology by a large margin. The increase in popularity is largely attributed to the benefits provided by \textit{Docker} platform, which implements a container virtualization technology \cite{Merkel2014}. Different from its hypervisor-based counterparts, \textit{Docker} is comes with light-weight computing resource which allows consumers to enjoy improvements in computation speed and memory performance.
	
	\textit{Docker} is an open source project that was started in early 2013 offering a systematic deployment of Linux applications inside portable \textit{containers}. \textit{Docker} containers provide hardware abstraction for both developers and users since multiple containers can independently run on the same \textit{Operating System} without affecting each other. This feature make allows developers to build and test their applications on any environment and execute it on all platforms without any modification \cite{Merkel2014, Bernstein2014, Anderson2015}. 
		
	Commercially, there are many providers of Cloud computing services. These services are categorized into three main models: (1) Infrastructure-as-a-service (IaaS), (2) Platform-as-a-service (PaaS) and (3) Software-as-a-service (SaaS). \textit{Microsoft Azure} and \textit{Amazon Web Service} are two of the largest Cloud computing services providers. Both of them allow for deployment of \textit{Docker} through their platforms and they provide it as PaaS.
	
\begin{figure}[h!]
    	\centering
		\includegraphics[width=.64\textwidth] {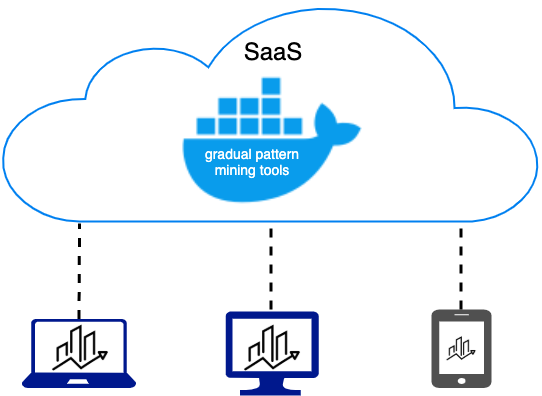}
		\caption{Illustration of SaaS for GP mining algorithms}
		\label{fig2:saas}
	\end{figure}

	It is important to highlight that integrating gradual pattern mining algorithms into a \textit{Docker} platform enables the access of these algorithms through a SaaS (Software-as-a-service) model as illustrated in Figure~\ref{fig2:saas}. In a SaaS distribution model, software is centrally hosted and licensed on a subscription basis. This introduces a flexibility that spares users the agony of spending hours trying to install analysis software \cite{Joshi2018}.
		
	In addition, integration of gradual pattern mining algorithms into the Cloud allows users to utilize servers that are more secure, reliable and flexible than on-premises servers at a cheaper subscription price. Again due to replicability, server downtime is almost non-existence. Computing resources can be scaled up or down depending on the demand of the gradual pattern mining algorithm.

	All in all, this study aims to avail gradual pattern mining on a light-weight platform that is easily configurable by different research observatories (such as OREME\footnote{\url{https://oreme.org/}} ) and institutions around the World. For this reason \textit{Docker} is the most suitable Cloud platform to achieve this because of the benefits mentioned above and the \textit{GOST} implementation of the OGC SensorThings API is recommended because it provides a functional \textit{Docker} Web application. In Chapter~\ref{ch7}, we present a system architecture that is built on top of \textit{Docker} OGC SensorThings API. 
	
	\newpage
	
	\section{Summary}
	\label{ch2:summary}
	To sum up, we began by describing association rules and how gradual rules are built on top of association rules. 
	
	We presented definitions for gradual rules and we discussed two (\textit{breadth-first} and \textit{depth-first}) traversal techniques for item set search in association rule mining. In the case of gradual rule mining, the \textit{breadth-first search} technique is easier to implement than the \textit{depth-first search} due to the complexity presented in (1) determining frequency support of gradual item sets and (2) the \textit{complementary notion} of gradual item sets. Next, we described three popular existing approaches (GRITE GRAANK and ParaMiner) for mining gradual rules. As a result we propose a heuristic solution to the problem of combinatorial candidate explosion and the problem of finding parent node of an FP-Tree.
		
	We review existing literature relating to temporal pattern mining. We discuss two existing techniques for mining temporal frequent item sets: the first one employs a \textit{sliding-window size} technique; the second one employs an \textit{evolution trend} with a user-specified fuzzy partition. Many of the existing approaches do not represent temporal gradual patterns using approximated numeric values. Therefore, we propose an approach that employs a fuzzy model to extract gradual patterns of the form: \textit{``the more X, the more Y, \textbf{almost 3 minutes later}''}.
	
	We review existing literature concerning crossing time-series data from different sources. We discover that \textit{query processing} is the most popular approach. However, it relies on a \texttt{U-SQL}-based \textit{fuzzy-join} model for crossing unrelated data sets in a Microsoft Azure data lake environment. We intend to construct a crossing model that may be integrated into an OGC SensorThings framework whose querying model does not support \texttt{U-SQL}.
	
	Finally, we review existing Cloud technologies that best suit integration for gradual pattern mining algorithm implementations. We recommend the \textit{Docker} Cloud platform and the \textit{GOST} certified implementation of the OGC SensorThings API since they will be easier for research institutions to adopt. We propose a system architecture model to demonstrate proof of concept.
	


\chapter{Temporal Gradual Patterns}
\label{ch3}
\minitoc \clearpage 

	\begin{chapquote}{Henry Ford \textit{(1863 – 1947)}}
		``Whether you think you can or think you can't, you are right''
	\end{chapquote}

	\section{Introduction}
	\label{ch3:introduction}
	In this chapter, we propose and describe a fuzzy modality that extends the GRAANK approach (which is described in Section~\ref{ch2:graank}) in order to extract fuzzy-temporal gradual patterns. We propose definitions for temporal gradual patterns and use them to describe the proposed temporal gradual pattern mining technique referred to as T-GRAANK (Temporal-GRAANK). We implement an algorithm for the proposed technique and test it on real data.

	\section{Temporal Gradual Patterns}
	\label{ch3:tgp}
	We begin by recalling that gradual pattern mining allows for retrieval of correlations between attributes of a numeric data set through rules such as: \textit{``the more exercise, the less stress''}. However, it is possible that a temporal lag may exist between changes in the attributes and their impact on others, existing gradual pattern mining techniques do not take this aspect into account. We illustrate the possibility of this temporal aspect of gradual item sets in the example that follows.

	\textit{Example 3.1.} We consider an arbitrary data set $D_{4.1}$ in Table~\ref{tab3:sample1}, containing the number of physical exercises that a person performed along with the stress levels together with the corresponding dates.

	\begin{table}[h!]
  		\centering
    	\begin{tabular}{c c c c} 
	  	\toprule
	  	\multicolumn{4}{c}{Correlation between exercise and stress level}\\
      	\hline
      	\textbf{id} & \textbf{date} & \textbf{activity} & \textbf{stress}\\
      	&  (day/month) & (exercise) & \textbf{level}\\
      	\hline \hline
      	r1 & 01/06 & 1 & 4\\
      	r2 & 02/06 & 2 & 2\\
      	r3 & 03/06 & 2 & 3\\
      	r4 & 04/06 & 1 & 5\\
      	r5 & 05/06 & 3 & 1\\
      	\bottomrule
    	\end{tabular}
    	\caption{Sample data set $D_{4.1}$}
    	\label{tab3:sample1}
	\end{table}

	For instance in Table~\ref{tab3:sample1}, we may extract a gradual pattern of the form: $\lbrace (exercise, \uparrow), (stress, \downarrow) \rbrace _{sup = 3}$, the support is 3 because tuples $<$r1, r2, r5$>$ can be ordered successively to match the gradual rule. As can be seen, the extracted gradual pattern has not utilized the \textit{`date'} attribute to include it in the representation of retrieved knowledge. It may be interesting to include the extraction of this attribute in the gradual pattern mining technique in order to show the time specificity of the extracted gradual pattern.

	In this chapter, we propose and describe a fuzzy approach that handles such situations in order to retrieve patterns such as: \textit{``the more exercise, the less stress \textbf{almost 1 month later}''}. In order to allow for extraction of this kind of patterns, we propose three main steps: (1) transform the data set in a step-wise manner in order to retrieve \textit{`time differences'}, (2) apply a modified GRAANK algorithm implementation on the transformed data set in order to extract gradual patterns and (3) apply a fuzzy model on the \textit{`time differences'} in order to estimate a time lag between the extracted gradual patterns.

	\subsection{Proposed Definitions for Temporal Gradual Patterns}
	\label{ch3:definitions}
	For the purpose of describing temporal gradual patterns, we propose the definitions that follow:

	\textbf{Definition 3.1.} \texttt{Time Lag.} \textit{A time lag $\alpha t$ is the amount of time $t$ that elapses before or after the changes in a one gradual item affects the changes in another gradual item.}

	\begin{equation}
		\alpha \in \{=+, =-, \approx +, \approx - \}
	\end{equation}
	where `$=+t$' implies \textit{`exactly $t$ time later'}, `$=-t$' implies \textit{`exactly $t$ time before'}, `$\approx + t$' implies \textit{`almost $t$ time later'} and, `$\approx - t$' implies \textit{`almost $t$ time before'}. For example `$\approx + 5 sec$' is a time lag that may interpreted as \textit{`almost 5 seconds later'}. The \textit{time tag}'s value $t$ is derived from the formula that follows:

	\begin{equation}
		t = \mathfrak{m} \in \mathbb{M}
	\end{equation}
	where $\mathfrak{m}$ is the center value of a fuzzy triangular membership function of the fuzzy set $\mathbb{M}$ which is composed of time differences deduced from a \textit{date-time} attribute as: $\mathbb{M} = \{|r_{1} - r_{1+s}|, ..., |r_{n} - r_{n+s}|\}$. We fully describe $\mathbb{M}$ in Section~\ref{ch3:transformation} and, $\mathfrak{m}$ in Section~\ref{ch3:fuzzy_model}.

	For example given a \textit{date-time} sequence: $\{3, 3, 4, 5, 5\}$, the center of a triangular membership function spanning that universe may be taken to be $4$.

	\textbf{Definition 3.2.} \texttt{Temporal Gradual Item.} \textit{A temporal gradual pattern \texttt{g} can be defined as a gradual pattern that includes a crisp time lag}. A temporal gradual item is made up of two parts: a gradual item and a crisp time lag.

	\begin{equation}
		g = (i, v)_{\alpha t}
	\end{equation}
	where $(i, v)$ is a gradual item and $\alpha t$ is a time lag where $\alpha \in \lbrace =+, =- \rbrace$ such that `$=+t$' implies a time lag of $t$ later and,  `$=-t$' implies a time lag of $t$ earlier. For example $(stress, \downarrow)_{=+2weeks}$ is a temporal gradual item interpreted as the ``the less stress 2 weeks later''.

	\textbf{Definition 3.3.} \texttt{Fuzzy Temporal Gradual Item.} \textit{A temporal gradual pattern \texttt{g$_{f}$} can be defined as a gradual pattern that includes a fuzzy time lag}. A temporal gradual item is made up of two parts: a gradual item and a fuzzy time lag.

	\clearpage 

	\begin{equation}
		g_{f} = (i, v)_{\alpha t}
	\end{equation}
	where $\alpha \in \lbrace \approx +, \approx - \rbrace$ such that `$\approx + t$' implies a time lag of almost $t$ later and, `$\approx - t$' implies a time lag of almost $t$ earlier. For example $(exercise, \uparrow)_{\approx - 1week}$ is a fuzzy-temporal gradual item that can be interpreted as the ``the more exercise almost 1 week earlier''.

	\textbf{Definition 3.4.} \texttt{Reference Gradual Item.} \textit{A reference gradual item $(i_{0},v_{0})$ is a gradual item from which other temporal gradual items are varied with time.} It should be understood that a reference gradual item is specified by the user, it cannot be the `date-time' attribute and it is part of fuzzy-temporal gradual pattern or a temporal gradual pattern. Definitions 3.6 and 3.7 elaborate this definition.

	\textbf{Definition 3.5.} \texttt{Temporal Gradual Pattern.} \textit{A temporal gradual pattern \texttt{TGP} is a set of temporal gradual items with one the reference gradual item set}.

	\begin{equation}
		TGP = \lbrace (i_{0},v_{0}), (i_{1},v_{1})_{\alpha t_{1}}, ..., (i_{n},v_{n})_{\alpha t_{n}}\rbrace
	\end{equation}
	where $(i_{0},v_{0})$ is the reference gradual item and $(i_{n},v_{n})_{\alpha t_{n}}$ are temporal gradual items.

	\textbf{Definition 3.6.} \texttt{Fuzzy-Temporal Gradual Pattern.} \textit{A fuzzy temporal gradual pattern \texttt{TGP}$_{f}$ is a set of fuzzy temporal gradual items with one the reference gradual item set}.

	\begin{equation}
		TGP_{f} = \lbrace (i_{0},v_{0}), (i_{1},v_{1})_{\alpha t_{1}}, ..., (i_{n},v_{n})_{\alpha t_{n}}\rbrace
	\end{equation}

	For example $\lbrace (jogging,\uparrow),(walking,\uparrow),(stress,\downarrow)_{\approx +2weeks} \rbrace$ is a fuzzy temporal gradual item set that can be interpreted as ``the more jogging, the more walking, the less stress almost 2 weeks later''.

	\textbf{Definition 3.7.} \texttt{Representativity.} \textit{The representativity \texttt{rep} of a fuzzy-temporal gradual pattern is the ratio of the number of transformed tuples to the number of all tuples.}

	\begin{equation}
		rep(TGP_{f}) = \frac{|R'|}{|R|}
	\end{equation}
	where $R'$ is tuple subsets in the transformed data set and, $R$ is tuple subsets in original data set.

	Given a threshold of minimum representativity $\delta$, a fuzzy-temporal gradual pattern \texttt{TGP}$_{f}$ is said to be relevant if:

	\begin{equation}
		rep(TGP_{f}) \geq \delta
	\end{equation}

	Definition 3.7 also holds for \textit{temporal gradual patterns}.

	\section{Mining Temporal Gradual Patterns}
	\label{ch3:mining}
	Our goal is to transform a time-series data set into a temporal format that allows for extraction of gradual patterns with the corresponding time lag information. We propose a fuzzy model that uses a membership function to estimate the time lag of a temporal gradual pattern. We describe these three approaches in the sections that follow.

	\subsection{Data Transformation Model}
	\label{ch3:transformation}
	In this section, we demonstrate how a typical time-series data set can be transformed into a temporal format in order to allow for extraction of \textit{fuzzy temporal gradual patterns} based on a user-specified \textit{minimum representativity threshold}.

	Let $D$ be a time-series data set with a set of attributes $\{a_{t},a_{1}, a_{2}, ..., a_{k} \}$: $a_{t}$ is `date-time' attribute and every attribute $a_{k}$ is composed of a set of tuples $R = \{r_{1}, r_{2},..., r_{n} \}$. We provide a function notation $f$ for data transformations in Equation~\eqref{eqn3:transform}, which requires two inputs: $a_{r}$ a user-specified reference attribute, $s$ a transformation step derived from the minimum representativity threshold.

	\begin{equation}\label{eqn3:transform}
		D^{s} = \{ f(s, a_{r}, D) ~\vert~ s \in \mathbb{N} \} = \{a _{k} ^{s} \}
	\end{equation}
	such that:
	\[
		a _{k} ^{s} =
		\begin{cases}
			a_{k} = \{|r_{1} - r_{1+s}|,..., |r_{n} - r_{n+s}|\}  \qquad if~ a_{k} == a_{t}\\
			a_{k} = \{r_{1},..., r_{n-s} \} \qquad \qquad \qquad \qquad \; if~ a_{k} == a_{r}\\
			a_{k} = \{r_{1+s},..., r_{n}\} \qquad \qquad \qquad \qquad \; otherwise
		\end{cases}\notag
	\]
	where $D^{s}$ is a set of transformed data sets; $\mathbb{N} = 1,...,\mathbb{Z}$ and $\mathbb{Z} \approx |R| (1 - \delta)$, $R$ is all tuple subsets in data set $D$.

	We propose the pseudocode steps that follow for transforming time-series data sets (the algorithm is shown in Algorithm~\ref{alg4:transform}):

	\begin{enumerate}
		\item Calculate maximum number of integer transformation \texttt{steps} using the \textit{minimum representativity threshold}.
		\item For every transformation \texttt{step}, transform the data set in a step-wise manner with respect to the \textit{reference attribute} as elaborated in  Equation~\eqref{eqn3:transform}.
		\item On every transformed data set, apply the T-GRAANK algorithm to mine for fuzzy-temporal gradual patterns. We discuss this algorithm in Section~\ref{ch3:tgraank}.
		\item Repeat steps 2 and 3 until all transformation \texttt{steps} are exhausted.
	\end{enumerate}

	\textit{Example 3.2.} We consider an arbitrary data set containing the number of hours a person spent performing physical exercises together with the stress levels after irregular number of days, as shown in Table~\ref{tab3:sample2} (a).

	\begin{table}[h!]%
  		\centering
  		\footnotesize
  		\subfloat[]{
    	\begin{tabular}{c c c c}
	  	\toprule
      	\textbf{id} & \textbf{date} & \textbf{exercise} & \textbf{stress}\\
      	&  (day/month) & (hours) & \textbf{levels}\\
      	\hline \hline
      	r1 & 01/06 & 1 & 4\\
      	r2 & 04/06 & 2 & 2\\
      	r3 & 05/06 & 3 & 3\\
      	r4 & 10/06 & 1 & 2\\
      	r5 & 12/06 & 3 & 3\\
      	\bottomrule
    	\end{tabular}}%
    	\qquad
    	\subfloat[]{
    	\begin{tabular}{c|c|c|c}
	  		\toprule
      		\textbf{id} & \textbf{days \textit{lag}} & \textbf{exercise} & \textbf{stress}\\
      		& ($r_{n}-r_{n+1}$) & ($r_{n}$) & ($r_{n+1}$)\\
      		\hline \hline
      		t1 & 3 & 1 & 2\\
      		t2 & 1 & 2 & 3\\
      		t3 & 5 & 3 & 2\\
      		t4 & 2 & 1 & 3\\
      		t5 & - & - & -\\
      		\bottomrule
    		\end{tabular}
    	}
    	\caption{(a) A sample data set $D_{4.2}$, (b) transformed data set $D'_{4.2}$}
    	\label{tab3:sample2}
	\end{table}

	If we take attribute \textit{`exercise'} as the \textit{reference attribute} ($a_{r}$) and a minimum representativity threshold of $0.8$, then the attributes may be applied to the transformation function $f$ to generate a transformed data set as illustrated in Table~\ref{tab3:sample2} (b).

	After this transformation, we mine the Table~\ref{tab3:sample2} (b) for gradual patterns. For instance the longest path that match gradual pattern
	$\lbrace (exercise, \uparrow), (stress,\downarrow) \rbrace$ is $<$t2, t3$>$, so the support is $\frac{2}{4}$. We observe that a representativity of $0.8$ constitutes 4 out of 5 tuples. On the negative side, there is a decrease in the \textit{representativity} of the data as we progress our transformations to larger steps. On the positive side, \textit{representativity} has a less significant effect on large data sets because of their great number of tuples.

	Next, we approximate a \textit{time lag} for the extracted pattern $\lbrace (exercise, \uparrow), (stress,\downarrow) \rbrace$ using the values of attribute \textit{`days lag'} (i.e. $\{ 3, 1, 5, 2\}$). It is important to note that our interest is to estimate \textit{time lag} using tuples from which we derived our pattern, in this case $<$t2, t3$>$. In order to approximate the most relevant \textit{time lag}, we apply a fuzzy model which is described in the section that follows.

	\begin{algorithm}[h!]
	\scriptsize
	\caption{Transforming Time-Series Data Sets}
	\label{alg4:transform}
	\SetKwFunction{append}{append}\SetKwFunction{totalRows}{totalRows}
	\SetKwFunction{totalCols}{totalCols}\SetKwFunction{TGRAANK}{T-GRAANK}
	\SetKwInOut{Input}{\textbf{Input}}\SetKwInOut{Output}{\textbf{Output}}

	\Input{$D -$ data set, $refColumn -$ reference column, $minSup -$ minimum support, $minRep -$ minimum representativity}
	\Output{$TGP -$ set of (fuzzy) Temporal Gradual Patterns}

	\BlankLine
	$S_{Max} \leftarrow$ maximum number of steps w.r.t $minRep$\;
	$R_{Max} \leftarrow $\totalRows($D$)\;
	\For{$s\leftarrow 1$ \KwTo $S_{Max}$}{
		\For{$i\leftarrow 0$ \KwTo $(R_{Max}-s)$}{
			$d \leftarrow tblCell_{[i+s]} - tblCell_{[i]}$\tcc*[H]{column with time}
			$tempRow \leftarrow$\append($refColumn$)\;
			$C_{Max} \leftarrow$ \totalCols($Row_{[i]})$\;
			\For{$j\leftarrow 1$ \KwTo $C_{Max}$}{

				\If{$Column_{[j]} \not =$ $(refColumn$ or $timeColumn)$}{
					$tempRow \leftarrow$\append($tblCell_{[j][i+s]}$)\;
				}
			}
			$D^{'} \leftarrow$\append($tempRow$)\;
			$T_{d} \leftarrow$\append($d$)\;
		}
		$tgps \leftarrow$ \TGRAANK{$D^{'}, T_{d}, minSup$}\;
		$TGP \leftarrow$\append($tgps$)\;
	}
	\Return $TGP$\;
	\end{algorithm}

	\subsection{Building the Fuzzy Model}
	\label{ch3:fuzzy_model}
	Generally, there exists a great number of membership modalities that one can build functions from (for instance triangular, trapezoidal, Gaussian among others), and it is very difficult to determine which one will fit the data set perfectly. However, it is enough to pick modalities that span the whole universe and remain scalable \cite{Zadeh1965, Schockaert2008, Ayouni2010, Mandal2012}.

\begin{figure}[h!]
	\centering
    \subfloat[]{%
		\begin{tikzpicture}
		\begin{axis}[
				height=5cm, width=7.5cm,
				ylabel=membership degree,
				no marks, axis lines*=left, clip=false,
  				ymin=0,
  				xticklabels=\empty,
 				yticklabels=\empty,
 				grid = major, enlargelimits = false]
    	\addplot[very thick, color=cyan] coordinates{(0,0) (2,0) (4,0) (6,3.96) (8,0) (10,0)};
    	\node[below] at (axis cs:4, 0)  {$x_{min}$};
    	\node[below] at (axis cs:8, 0)  {$x_{max}$};
		\end{axis}
		\end{tikzpicture}}%
		\qquad \qquad
	\subfloat[]{%
		\begin{tikzpicture}
		\begin{axis}[
				no markers, domain=0:10, samples=100,
				height=5cm, width=7.5cm,
				ylabel=membership degree,
				no marks, axis lines*=left, clip=false,
  				ymin=0,
  				xticklabels=\empty,
 				yticklabels=\empty,
 				grid = major, enlargelimits = false]
    	\addplot[very thick,color=cyan] {gauss(6,1)};
    	\addplot+[domain=2:4, draw=red, fill=red!100] {gauss(6,1)} \closedcycle;
    	\addplot+[domain=8:10, draw=red, fill=red!100] {gauss(6,1)} \closedcycle;
    	\node[below] at (axis cs:4, 0)  {$x_{min}$};
    	\node[below] at (axis cs:8, 0)  {$x_{max}$};
		\end{axis}
		\end{tikzpicture}}%
    \caption{(a) Triangular MF, (b) Gaussian MF}
    \label{fig3:membership_fxn}
	\end{figure}
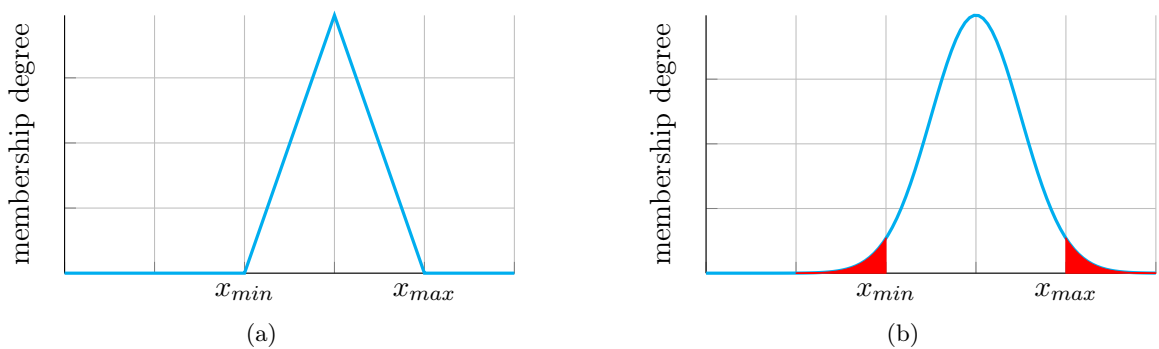

	Since our goal is to represent the fuzziness of the \textit{`time lag'} numerically, we focus on identifying a membership function (MF) that allows the peak position to accommodate a single numerical value. This automatically eliminates MFs whose peaks accommodate a range of values (such as the trapezoidal MF) and leaves us two options shown in Figure~\ref{fig3:membership_fxn}.

	\pgfmathdeclarefunction{gauss}{2}{%
		\pgfmathparse{1/(#2*sqrt(2*pi))*exp(-((x-#1)^2)/(2*#2^2))}%
	}

	Equally important, our approach seeks to approximate the peak position (also called \textit{TRUE center}), minimum and maximum extremes of a MF, such that the MF includes a majority of members in the population of calculated \textit{time differences}. For this reason, the triangular MF is better justified for this study than the Gaussian MF, since the latter allows false values outside the minimum and maximum extremes to have small membership degrees (illustrated by shaded areas in Figure~\ref{fig3:membership_fxn}b).

	\subsubsection{Slide, Re-calculate Technique for Time Lag Estimation}
	\label{ch3:fuzzy_slide}
	The \textit{TRUE} center of a distribution is established when the largest proportion of members are closely spaced around it \cite{Montgomery2003}. In light of this, we propose the algorithm in Algorithm~\ref{alg4:slide} which initially takes the median as the center and slides the MF step-wisely to left or/and right until we find this value.

	We point out that our \textit{``slide re-calculate''} technique is conjectural. However, through simulation we observed that its accuracy is greatly improved when an optimum \textit{``slice''} (or gap) is chosen by which the MF is slid. In our case, at $line4:$ $slice=0.1~of~Median$. This implies that at worst, the algorithm will slide a total of 20 times: 10 times to the left and 10 times to the right. Each slide is equivalent to a single loop iteration.

	\clearpage

	\begin{algorithm}[h!]
	\small
	\caption{Fuzzy Slide, Re-calculate Pseudocode}
	\label{alg4:slide}
	\SetKwFunction{append}{append}\SetKwFunction{quartile}{quartile}
	\SetKwFunction{fuzzTrimf}{fuzzTrimf}\SetKwFunction{countAverage}{countAverage}
	\SetKwInOut{Input}{\textbf{Input}}\SetKwInOut{Output}{\textbf{Output}}

	\Input{$selTs -$ selected time-lags, $allTs -$ all time-lags, $minSup -$ minimum support}
	\Output{$q2 -$ approximated time lag value, $sup -$ support}

	\BlankLine
	$q1 \leftarrow$ \quartile($1, allTs$), $q2 \leftarrow$ \quartile($2, allTs$), and $q3 \leftarrow$ \quartile($3, allTs$)\;
	$boundaries \leftarrow$ \append($q1, q2, q3$)\;
	$center \leftarrow$ \quartile($2, selTs$)\;
	$left, right \leftarrow False$, $slice \leftarrow (0.1 * q2)$, and $sup \leftarrow 0$\;
	\While{$sup < minSup$}{
		$memberships \leftarrow$ \fuzzTrimf($selTs, boundaries$)\;
		$sup \leftarrow$ \countAverage($memberships$)\;
		\eIf{$sup >= minSup$}{
			\Return $q2, sup$\;
		}{
			\uIf{$left~is~False$}{
				\eIf{$center <= q2$}{
					$q1 \leftarrow (q1 - slice)$, $q2 \leftarrow (q2 - slice)$, $q3 \leftarrow (q3 - slice)$\;
					$boundaries \leftarrow$ \append($q1, q2, q3$)\;
				}{
					$left \leftarrow True$\;
				}
			}\uElseIf{$right~is~False$}{
				\eIf{$center >= q2$}{
					$q1 \leftarrow (q1 + slice)$, $q2 \leftarrow (q2 + slice)$, $q3 \leftarrow (q3 + slice)$\;
					$boundaries \leftarrow$ \append($q1, q2, q3$)\;
				}{
					$right \leftarrow True$\;
				}
			}\Else{
				\Return $False, False$\;
			}
		}
	}
	\end{algorithm}

\begin{figure}[h!]
    	\centering
    	\subfloat[]{%
		\begin{tikzpicture}
		\begin{axis}[
				height=4cm,width=7.5cm,
				xlabel=days,
				ylabel=membership,
				no marks,
				axis lines=middle,
  				clip=false,
  				ymin=0,ymax=1,
  				xmin=0,xmax=6,
  				ytick={0,0.5,1},
  				xtick={0,...,6},
  				legend pos=north west]
    	\addplot[color=cyan] coordinates{(0,0) (1,0) (2.5,1) (5,0) (6,0)};
    	\draw [thin, dashed, draw=cyan]
        	(axis cs: 2.5,0) -- (axis cs: 2.5,1);
		\end{axis}
		\end{tikzpicture}}%
	\subfloat[]{
		\begin{tikzpicture}
		\begin{axis}[
				height=4cm, width=7.5cm,
				xlabel=days,
				ylabel=,
				no marks,
				axis lines=middle,
  				clip=false,
  				ymin=0,ymax=1,
  				xmin=0,xmax=5,
  				ytick={0,0.5,1},
  				xtick={0,...,5},
  				legend pos=north west]
    	\addplot[color=cyan] coordinates{(0,0) (1.5,1) (4,0) (5,0)};
    	\draw [thin, dashed, draw=cyan]
        	(axis cs: 1.5,0) -- (axis cs: 1.5,1);
		\end{axis}
		\end{tikzpicture}}%
    \caption{(a) Membership function for $r_{n+1}$, (b) modified membership function for $r_{n+1}$}
    \label{fig3:initial_mf}
	\end{figure}
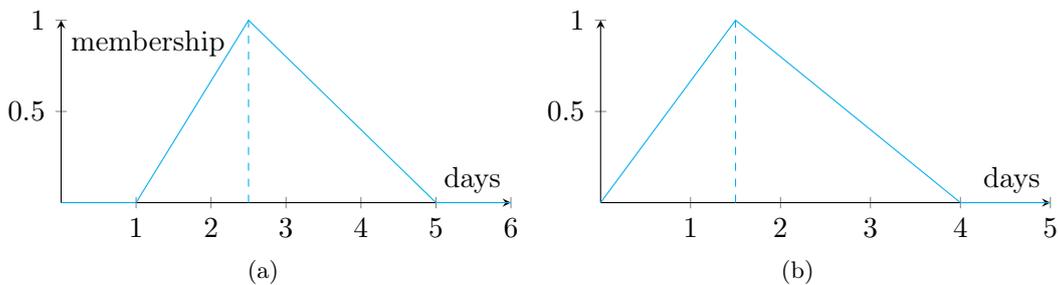

	As an illustration, Figure~\ref{fig3:initial_mf} (a) shows the MF for the data set in Table~\ref{tab3:sample2} (b). Applying the MF to the \textit{`days lag'} population $\{ 3, 1, 5, 2\}$ with respect to members of path $<$t2, t3$>$:  $\{ 1, 5\}$, we generate the fuzzy data set: $\{ ((1, 0), (5, 0)\}$. Therefore, the membership degree support for `$\approx 2.5days$' is $\frac{0}{2}$.

	As it can be seen, the problem may be that the MF in Figure~\ref{fig3:initial_mf} (a) is either be too narrow or is pivoted on a wrong medial value. We shy away from widening the MF boundaries since this increases the size of the universe. However, if we slide the MF to the left by \textit{``$1day$ slice''} as shown in Figure~\ref{fig3:initial_mf} (b) we now observe the fuzzy set with respect to path $<$t2, t3$>$ becomes: $\{(1,.8), (5,0) \}$. Altogether, pattern: $\lbrace (exercise, \uparrow), (stress, \downarrow)_{\approx + 1.5days} \rbrace$ has a support of $\frac{2}{4}$, a \textit{representativity} of $\frac{4}{5}$ and a \textit{time lag}: `$\approx +1.5days$' whose support is $\frac{1}{2}$.

	\subsubsection{Large Data Sets With Outliers}
	\label{ch3:large_fuzzy}
	The data set in \textit{Example 3.2} is small and the distribution of \textit{time differences} between the tuples tend to be almost uniform. However, most real life data sets are huge and the \textit{time differences} may have one or more outliers. Under those circumstances, designing a membership function that spans the entire universe of \textit{time lags} may lead to extremely false approximations of the actual \textit{time lag}.

\begin{figure}[h!]
		\centering
		\begin{tikzpicture}
		\begin{axis}[
				height=5cm, width=10cm,
				xlabel= ,
				ylabel= ,
				no marks,
				axis lines=middle,
  				clip=false,
  				ymin=0,ymax=1,
  				xmin=0,xmax=12,
  				ytick={0,0.5,1},
  				xtick={2,4,6,8,10},
  				xticklabels={Min,Q1,Median,Q3,Max},
  				legend pos=north west]
    	\addplot[color=cyan] coordinates{(2,0) (6,1) (10,0)};
    	\addplot[color=brown] coordinates{(4,0) (6,1) (8,0)};
    	\draw [thin, dashed, draw=red]
        	(axis cs: 6,0) -- (axis cs: 6,1);
		\end{axis}
		\end{tikzpicture}
    \caption{Triangular membership function for large data sets}
    \label{fig3:large_mf}
	\end{figure}
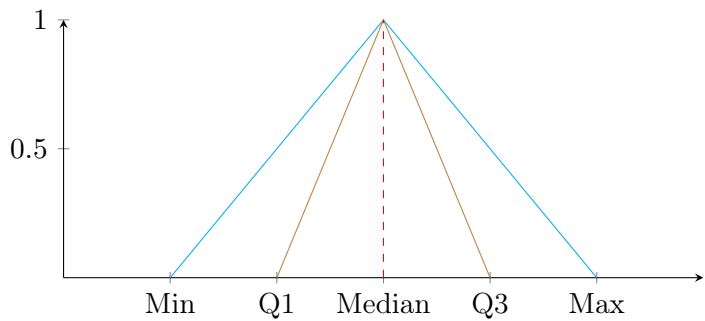

	We propose the use of quartiles or percentiles in order to narrow the span of the MF, so that the extreme outliers to the left and/or right of the universe are ignored as illustrated in Figure~\ref{fig3:large_mf}. Once the MF has been determined, the successive steps for approximating \textit{time lag} are similar to the ones in Section~\ref{ch3:fuzzy_slide}.

	\subsection{T-GRAANK Technique}
	\label{ch3:tgraank}
	We recall that the overall procedure for extracting temporal gradual patterns involves 3 steps: (1) transforming the data set (2) applying a variant of GRAANK technique to fetch gradual patterns from the transformed data set and (3) applying a fuzzy model to estimate the time lag of gradual item sets. In this section, we describe the second step. 

	T-GRAANK denotes Temporal-GRAANK since it modifies the GRAANK technique proposed in \cite{Laurent2009}. The algorithm applies a \textit{breadth-wise search} strategy to generate gradual item set candidates as shown in Algorithm~\ref{alg4:tgraank}. It should be noted that the methods at $line 12$ and $line 13$ are implemented by Algorithm~\ref{alg4:slide}.

	\begin{algorithm}[h!]
	\small
	\caption{$T-GRAANK$ Algorithm}
	\label{alg4:tgraank}
	\SetKwFunction{APRIORIgen}{APRIORIgen}\SetKwFunction{calculateSupport}{calculateSupport}
	\SetKwFunction{concordantPositions}{concordantPositions}\SetKwFunction{timeDifferences}{timeDifferences}
	\SetKwFunction{buildTriMembership}{buildTriMembership}\SetKwFunction{fuzzyFunc}{fuzzyFunc}
	\SetKwFunction{append}{append}
	\SetKwInOut{Input}{\textbf{Input}}\SetKwInOut{Result}{\textbf{Result}}

	\Input{$D^{'} -$ transformed data set, $T_{d} -$ time differences, $minSup -$ minimum support}
	\Result{$TGP -$ set of (fuzzy) temporal gradual patterns}
	\BlankLine
	\ForEach{Attribute $A$ in $D^{'}$}{
		$G \leftarrow$ build concordance matrices $A^{\uparrow}$ and $A^{\downarrow}$\;
	}
	$G^{'}\leftarrow$ \APRIORIgen($G$) \tcc*[H]{generates frequent gradual item-set candidates}
	\ForEach{Candidate $C$ in $G^{'}$}{
		$sup \leftarrow$ \calculateSupport($C$)\;
		\eIf{$sup < minSup$}{
			discard $C$\;
		}{
			$pos_{indices}\leftarrow$ \concordantPositions($C_{pairs}$)\;
			$time_{lags}\leftarrow$ \timeDifferences($pos_{indices},T_{d}$)\;
			$boundaries \leftarrow$ \buildTriMembership($T_{d}$)\;
			$t_{lag}\leftarrow$ \fuzzyFunc($time_{lags},boundaries$)\;
			$TGP \leftarrow$\append($C, t_{lag}$)\;
		}
	}
	\Return $TGP$\;
	\end{algorithm}

	\section{Experiments}
	\label{ch3:experiments}
	In this section, we present an experimental study of computational performance of our proposed T-GRAANK approach for mining temporal gradual patterns. We implement T-GRAANK algorithm in \texttt{Python} language.

	\subsection{Source Code}
	\label{ch3:source_code}
	The \texttt{Python} source code for T-GRAANK algorithm is available at our GitHub repository \url{https://github.com/owuordickson/t-graank.git}.

	\subsection{Computational Complexity}
	\label{ch3:complexity}
	In this section, we derive the asymptotic computational time complexity of the T-GRAANK algorithm shown in Algorithm~\ref{alg4:tgraank} using the \texttt{Big-O} analysis technique.

	It is important to mention that since our proposed approach is a variant of the GRAANK approach proposed in \cite{Laurent2009}, it similarly benefits from the binary matrix representation. However, the T-GRAANK technique has a higher computational complexity since it processes multiple \textit{transformed data sets} whose computation is approximately equivalent to the combined computation of successively repeated GRAANK operations.

	\clearpage

	We derive the asymptotic complexity of each algorithm by focusing on the number loops and assuming that other steps are computationally relatively constant. Therefore, in Table~\ref{tab3:complexity}: $n$ denotes number of tuples, $m$ denotes number of columns, $r$ denotes \textit{representativity}, $C$ denotes constant statements \cite{Vaz2017}.

	\begin{table}[h!]
  		\centering
    	\begin{tabular}{c c c}
	  	\toprule
      	\textbf{Algorithm} & \textbf{Asymptotic Complexity} & \textbf{Bound}\\
      	\hline \hline
      	Algorithm~\ref{alg4:slide} & $\mathcal{X}=20(C)+C$ & $O(20)$ \\
      	Algorithm~\ref{alg4:tgraank} & $\mathcal{Y}=n^{4}+n^{3}+\mathcal{X}+C$ & $O(n^{4})$ \\
       	Algorithm~\ref{alg4:transform} & $\mathcal{Z}=n(1-r)(nm+\mathcal{Y}+C)$ & $O(n^{5})$ \\
      	\bottomrule
    	\end{tabular}
    	\caption{Asymptotic time complexity of T-GRAANK algorithm}
    	\label{tab3:complexity}
	\end{table}

	From Table~\ref{tab3:complexity}, it should be noted that the complexities of Algorithms~\ref{alg4:slide} and \ref{alg4:tgraank} are successively nested in Algorithm~\ref{alg4:transform}. Therefore, the overall time complexity is given by: $f(\mathcal{Z})$ whose upper bound is slightly greater than $O(n^{5})$ which deduces the worst-case performance of the algorithm.

	\subsection{Data Set Description}
	\label{ch3:dataset}
	We test the computational performance of the T-GRAANK algorithm implementation in order to determine the behavior of the Algorithm with respect to a user-specified \textit{minimum representativity} and \textit{minimum support} thresholds in Section~\ref{ch3:performance}. We execute the T-GRAANK algorithm on a synthetic data set with 50 tuples and 3 attributes. The test runs were performed on a 2.9 GHz Intel Core i7 MacBook Pro 2012 model, with 8 GB 1600 MHz DDR3 RAM.

	We further test the computational performance of the T-GRAANK algorithm by applying it on a larger `Power Consumption' data set, obtained from \texttt{UCI Machine Learning Repository} \cite{Dua2019}. This is a time-series numerical data set with 9 attributes and 2075259 tuples has  describes the electric power consumption in one household (located in Sceaux, France) in terms of active power, voltage and global intensity with a one-minute sampling rate between 2006 and 2010. We performed test runs on 4 node instances of a HPC (High Performance Computing) \textbf{Meso@LR} platform\footnote{\url{https://meso-lr.umontpellier.fr}} each made up of 14 cores and 128GB of RAM.

	In order to prove the applicability of the T-GRAANK technique, we performed two separate tasks: (1) harvest NDVI data from a satellite positioned over 4 regions in Kenya; (2) obtain rainfall amount data about the 4 regions from a weather report and apply T-GRAANK on the data in order to determine if they will match the conclusion obtained from analyzing the NDVI data. The aim is to confirm the conclusions made in \cite{Davenport1993}, that the NDVI (Normalized Difference Vegetation Index) is a sensitive indicator of the inter-annual variability of rainfall in the East African region. Data employed in this use case come from the \textit{data-cube} Open Source framework tool provided by \url{https://www.opendatacube.org} and the results were confirmed using data from the Kenya Meteorological Service repository \url{http://www.meteo.go.ke/index.php?q=archive}. We present the results in Section~\ref{ch3:use_case}.

	\subsection{Experiment Results}
	\label{ch3:results}
	In this section, we present results of our experimental study of 3 data sets. First, we show computational performance results and, the extracted patterns results when T-GRAANK algorithm is applied on a synthetic data set and UCI power consumption data set. These results are available at: \url{https://github.com/owuordickson/meso-hpc-lr/tree/master/results/tgps/uci}. Second, we show results of a use case example.

	\subsubsection{Computational Performance Results}
	\label{ch3:performance}
\begin{figure}[h!]
	\centering
    \subfloat[]{%
		\begin{tikzpicture}
		\begin{axis}[
			height=6.5cm, width=7.5cm,
			grid=both,
  			grid style={line width=.1pt, draw=gray!20},
    		major grid style={line width=.2pt, draw=gray!50},
  			axis lines=middle,
  			minor tick num=5,
    		xmin=0.5, xmax=1,
    		ymin=0, ymax=4000,
    		xlabel={Minimum Threshold},
    		ylabel={Time (ms)},
    		legend pos=north east,
    		xlabel style={at={(axis description cs:0.5,-0.1)},anchor=north},
  			ylabel style={at={(axis description cs:-0.15,0.5)},rotate=90, anchor=south}
  			]
  			\addplot[smooth, mark=*, color=blue] table[x=Sup, y=Runtime]{uci_sup.dat};
  	  	\addplot[smooth, mark=square, color=red] table[x=Rep, y=Runtime]{uci_rep.dat};
			\legend{min-sup ($\sigma$), min-rep $(\delta)$};
		\end{axis}
		\end{tikzpicture}}%
	\subfloat[]{%
		\begin{tikzpicture}
		\begin{axis}[
			height=6.5cm, width=7.5cm,
			grid=both,
  			grid style={line width=.1pt, draw=gray!20},
    		major grid style={line width=.2pt, draw=gray!50},
  			axis lines=middle,
  			minor tick num=5,
    		xmin=0, xmax=1,
    		ymin=0, ymax=1250,
    		xlabel={Minimum Threshold},
    		ylabel={Time (ms)},
    		legend pos=north east,
    		xlabel style={at={(axis description cs:0.5,-0.1)},anchor=north},
  			ylabel style={at={(axis description cs:-0.15,0.5)},rotate=90, anchor=south}
  			]
			\addplot[smooth, color=blue, mark=*]coordinates {
    			(0.1,1166.2)(0.3,886.8)(0.5,753.4)(0.6,717.1)(0.8,256.1)(0.9,239.6)
    		};
    		\addplot[smooth, color=red, mark=square]coordinates {
    				(0.1,1200.9)(0.3,955.1)(0.5,753.4)(0.6,572.4)(0.9,159.2)
    		};
    		\legend{min-sup ($\sigma$), min-rep $(\delta)$}
		\end{axis}
		\end{tikzpicture}}%
    \caption{(red graph) Plot of run-time against \textit{minimum representativity} (min-rep) with min-sup held constant at 0.9. (blue graph) Plot of run-time against \textit{minimum support} (min-sup) with min-rep held constant at 0.9. (a)  UCI data set: 9 attributes, 10k tuples on 14 HPC CPU cores and, (b) Synthetic data set: 3 attributes, 50 tuples on 4 CPU cores.}
    \label{fig3:long_runtime}
	\end{figure}
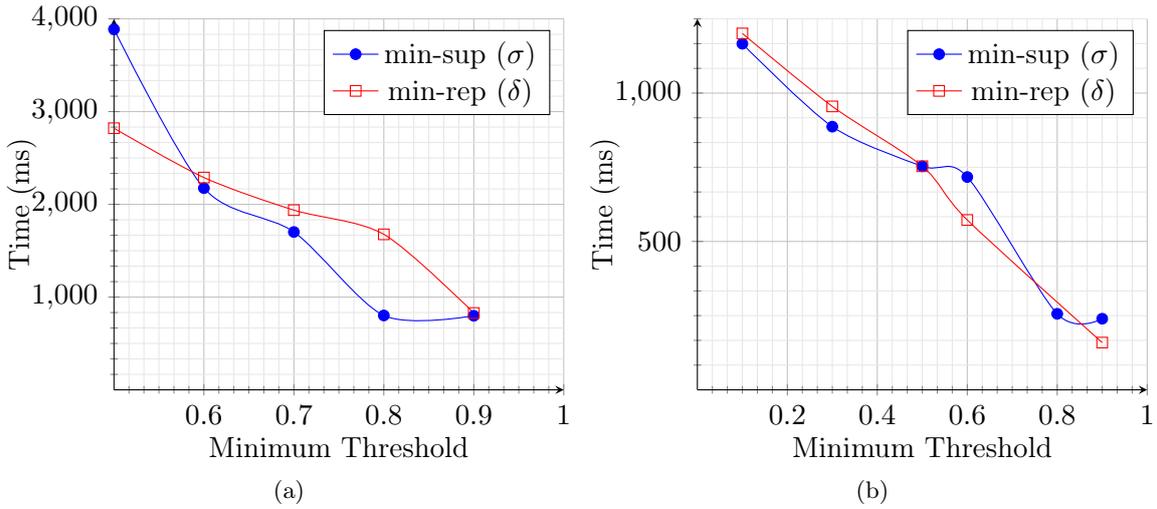

	The run time performance of T-GRAANK is shown in Figure~\ref{fig3:long_runtime}. In both Figure~\ref{fig3:long_runtime} (a) and (b), it can be seen that as \textit{representativity} threshold decreases, there is an increase in \textit{run time}. This behavior is due to the fact that the number of data set transformations to be mined is inversely proportional to the minimum representativity threshold; therefore, as representativity decreases the number of transformed data sets increase and this increases the run-time. It can also be observed that as minimum support threshold decreases, the \textit{run time} increases. This is because as support reduces, the number of gradual pattern candidates increase; consequently, increasing the run-time.

	T-GRAANK algorithm is an extension of GRAANK algorithm proposed in \cite{Laurent2009}. On one hand, the T-GRAANK algorithm is more computationally intensive than GRAANK algorithm. On the other hand, the increase in computations can be justified by the fact that new knowledge about \textit{temporal tendencies} is extracted which was not possible previously.

	\clearpage

	\subsubsection{Extracted Temporal Gradual Patterns}
	\label{ch3:extracted_tgps}
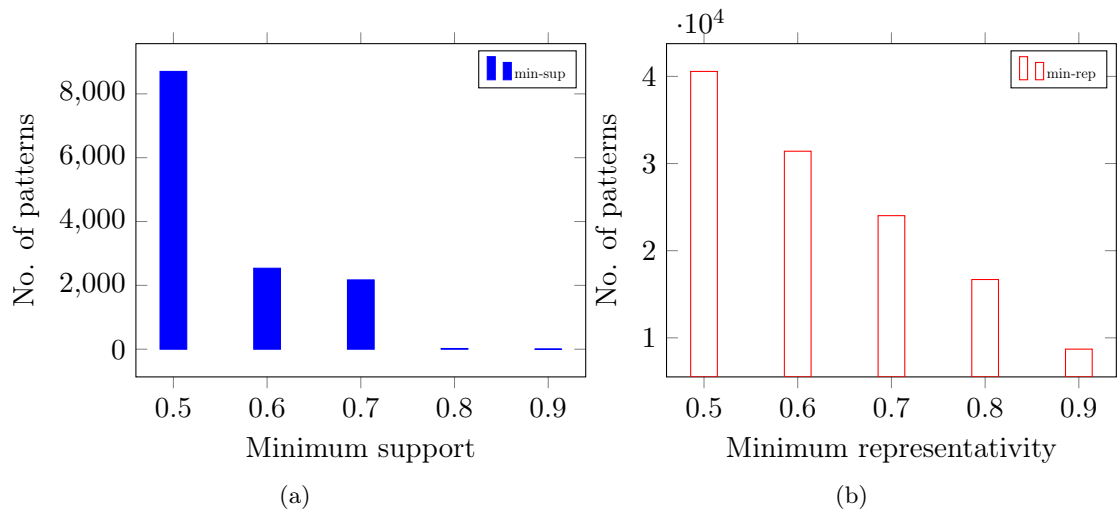
\begin{figure}[h!]
		\centering
		\subfloat[]{%
			\begin{tikzpicture}
  			\begin{axis}[
  				height=6cm, width=7.5cm,
  				ybar,
    			symbolic x coords={0.5, 0.6, 0.7, 0.8, 0.9},
            	xtick=data,
   				xlabel=Minimum support,
   				ylabel=No. of patterns,
   				legend style={nodes={scale=0.5}},
 				]
			\addplot[color=blue, fill=blue] table[x=Sup, y=Patterns]{uci_sup.dat};
			\legend{min-sup}
  			\end{axis}
  			\end{tikzpicture}}%
  		\subfloat[]{%
  			\begin{tikzpicture}
  			\begin{axis}[
  				height=6cm, width=7.5cm,
  				ybar,
    			symbolic x coords={0.5, 0.6, 0.7, 0.8, 0.9},
            	xtick=data,
   				xlabel=Minimum representativity,
   				ylabel=No. of patterns,
   				legend style={nodes={scale=0.5}},
 				]
			\addplot[color=red] table[x=Rep, y=Patterns]{uci_rep_5.dat};
			\legend{min-rep}
  			\end{axis}
  			\end{tikzpicture}}%
  	\caption{Patterns extracted from the UCI data set: (a) no. of patterns vs min-sup with min-rep held constant at 0.9, (b) no. of patterns vs min-rep with min-sup held constant at 0.5.}
  	\label{fig3:extracted_tgps}
	\end{figure}

	In Figure~\ref{fig3:extracted_tgps} (a), the number of temporal gradual patterns reduce significantly as minimum support increases. This is because as minimum support increases, it demands only gradual patterns whose quality surpass the respective threshold. In Figure~\ref{fig3:extracted_tgps} (b), the number of extracted patterns significantly reduce as minimum representativity increases. One reason for this is that as representativity increases fewer transformed data sets are generated; consequently, the algorithm is provided with fewer transformed data sets from which it can extract temporal gradual patterns.

	\subsubsection{Use Case Example}
	\label{ch3:use_case}
	In order to test applicability of the T-GRAANK technique, we performed two separate tasks related to weather and compared their results. The aim of the use case is to confirm the conclusions of \cite{Davenport1993}, that the NDVI (Normalized Difference Vegetation Index) is a sensitive indicator of the inter-annual variability of rainfall in the East African region.

	\begin{table}[h!]
  		\centering
    	\begin{tabular}{c c c}
	  	\toprule
      	\textbf{town} & \multicolumn{2}{c}{\textbf{amount}}\\
      	 & 2013 & 2015\\
      	\hline \hline
      	MAK & 104 & 75\\
      	WAJ & 49 & 69\\
      	ELD & 174 & 200\\
      	NRB & 44 & 223\\
      	\bottomrule
    	\end{tabular}
    	\caption{Rainfall distribution in Kenya}
    	\label{tab3:rainfall_distribution}
	\end{table}

	\clearpage

	In the first task, we retrieved the historical (October-December) rainfall distribution amounts of 4 towns in Kenya from a weather report in \cite{KenyaReport2014,KenyaReport2016}, shown in Table~\ref{tab3:rainfall_distribution}. Here, we chose two observable patterns: $\lbrace(MAK,\downarrow),(WAJ,\uparrow)\rbrace_{=+2years}$ and $\lbrace(ELD,\uparrow),(NRB,\uparrow)\rbrace_{=+2years}$.

	In the second task, we first generated NDVI data (for year 2013 and 2015) from LANDSAT 7 satellite images over Kenya using a novel tool known as \texttt{data-cube}\footnote{\url{https://www.opendatacube.org}}. Data-cube is a \texttt{Python-based} platform for the expanded use of satellite data in an Open Source framework. The generated NDVI data is a time-series data with 4 attributes (Makueni, Wajir, Eldoret and Nairobi) and 13 tuples (containing NDVI values of the respective 4 towns). Lastly in the second task, we applied our proposed approach on the NDVI data and we obtained the results shown in Table~\ref{tab3:results_ndvi}. As can be seen, the patterns generated by our algorithm match the selected patterns in Table~\ref{tab3:rainfall_distribution}; except for pattern $\lbrace WAJ+,MAK-\rbrace$, where the \textit{time lag} is slightly less.

	\begin{table}[h!]
  		\centering
    	\begin{tabular}{l l l c}
	  	\toprule
      	\textbf{Ref. Item} & \textbf{Pattern : Sup} & \textbf{Time Lag : Sup} & \textbf{Rep}\\
      	\hline \hline
      	NRB & $\lbrace ELD+,NRB+\rbrace : 0.666$ & $\approx +1.999yrs: 1.0$& $50\%$\\
      	 & $\lbrace WAJ+,NRB+,MAK+\rbrace : 0.666$ & $\approx +1.999yrs: 1.0$& $50\%$\\\\
      	 \hdashline \\
      	WAJ & $\lbrace ELD+,WAJ+\rbrace : 0.600$ & $\approx +1.223yrs: 0.5$& $62.5\%$\\
      	 & $\lbrace WAJ+,MAK-\rbrace : 0.600$ & $\approx +1.747yrs: 0.5$& $62.5\%$\\
      	\bottomrule
    	\end{tabular}
    	\caption{NDVI fuzzy-temporal gradual pattern results}
    	\label{tab3:results_ndvi}
	\end{table}

	We emphasize that it is difficult to get clear satellite images between short intervals due to cloud coverage; therefore, the \textit{data-cube} tool runs an algorithm that re-creates the image based on previous images. It may be for this reason that the time approximation for the pattern $\lbrace WAJ+,MAK-\rbrace$ is slightly less than 2 years.

	\section{Summary}
	\label{ch3:summary}
	In this chapter, we propose an approach for extending the existing GRAANK algorithm in order to extract fuzzy temporal gradual patterns. This approach combines two main concepts: (1) a fuzzy model  for estimating temporal tendencies of the patterns and (2) a gradual pattern mining technique for extracting the temporal gradual patterns.

	It is important to mention that we also test the T-GRAANK algorithm on another time-series data set and present the results in Chapter~\ref{ch4:tgraank_comparison}. We do not show these results in this chapter because they include a comparative study of ACO-TGRAANK algorithm which is built on top of an Ant Colony Optimization approach that is introduced in Chapter~\ref{ch4}.
	





\chapter{Ant Colony Optimization for Gradual Pattern Mining}
\label{ch4}
\minitoc \clearpage 

	\begin{chapquote}{Jack Kerouac, \textit{The Dharma Bums}}
		``One day I will find the right words, and they will be simple''
	\end{chapquote}

	\section{Introduction}
	\label{ch4:introduction}
	In this chapter, we propose a heuristic solution that is based on \textit{ant colony optimization} to the problem of (1) combinatorial candidate explosion and, (2) finding the longest set tree for the case of gradual pattern mining. We use definitions in Chapter~\ref{ch2:definitions} to aid discussions in this chapter. Finally, we perform experiments to analyze the performance of our proposed algorithms.

	\section{Ant Colony Optimization}
	\label{ch4:aco}
	Gradual pattern mining is a technique in the data mining realm that maps correlations between attributes of a data set as gradual dependencies. A gradual dependency may take a form of \textit{``the more Attribute$_{K}$, the less Attribute$_{L}$"}. In order to formulate such rules, a candidate rule is generated from respective attribute combinations and its quality tested against a user-specified frequency support threshold.
	
	Many gradual pattern mining approaches extend either a \textit{breadth-first search} (BFS) or \textit{depth-first search} (DFS) strategy for mining gradual item sets. In this chapter, we propose an \textit{ant colony optimization} (ACO) strategy that uses a probabilistic approach to improve efficiency of both BFS-based and DFS-based approaches for mining gradual item sets. ACO, as originally described by \cite{Dorigo1996}, is a general-purpose heuristic approach for optimizing various combinatorial problems. It exploits the behavior of a colony of artificial ants in order to search for approximate solutions to discrete optimization problems \cite{Smith2001, Silva2002, Blum2005, Runkler2005, Dorigo2019}. The application areas for ACO are vast; for instance in the telecommunication domain, \cite{Mong2003} employed it to optimally load balance circuit-switched networks.

	ACO imitates the positive feedback reinforcement behavior of biological ants as they search for food: where the more ants following a path, the more chemical pheromones are deposited on that path and, the more appealing that path becomes for being followed by other ants \cite{Dorigo1996, Dorigo2010}. 
	
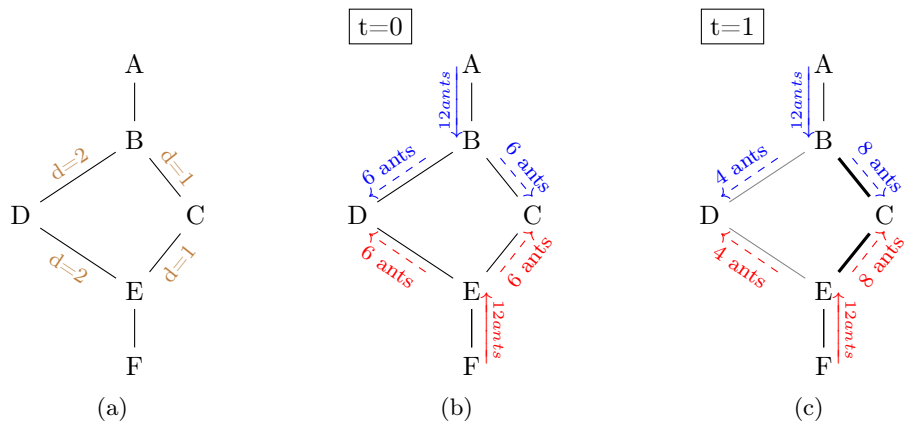
\begin{figure}[h!]
		\centering
		\small
  		\subfloat[]{%
		\begin{tikzpicture}
		\begin{scope}
			\node (n1) at (0,1) {A};
			\node (n2) at (0,0) {B};
			\node (n3) at (0.8,-1) {C};
			\node (n4) at (-1.5,-1) {D};
			\node (n5) at (0,-2) {E};
			\node (n6) at (0,-3) {F};
		\end{scope}
	
		\begin{scope}
			\path (n1) edge[draw=black] (n2);
			\draw[-] (n2) edge  node[sloped, above, brown] {\scriptsize d=1} (n3);
			\draw[-] (n2) edge  node[sloped, above, brown] {\scriptsize d=2} (n4);
			\draw[-] (n3) edge  node[sloped, below, brown] {\scriptsize d=1} (n5);
			\draw[-] (n4) edge  node[sloped, below, brown] {\scriptsize d=2} (n5);
			\path (n5) edge[draw=black] (n6);
		\end{scope}
		\end{tikzpicture}
		}%
		$\qquad \qquad$
  		\subfloat[]{%
		\begin{tikzpicture}
		
		\begin{scope}[every node/.style={thin, draw}]
			\node (t) at (-1.2, 1.5) {t=0};
		\end{scope}
		
		\begin{scope}
			\node (n1) at (0,1) {A};
			\node (n2) at (0,0) {B};
			\node (n3) at (0.8,-1) {C};
			\node (n4) at (-1.5,-1) {D};
			\node (n5) at (0,-2) {E};
			\node (n6) at (0,-3) {F};
		\end{scope}
	
		\begin{scope}
			\path (n2) edge[draw=black] (n3);
			\path (n2) edge[draw=black] (n4);
			\path (n3) edge[draw=black] (n5);
			\path (n4) edge[draw=black] (n5);
			\path (n5) edge[draw=black] (n6);

			\draw[-] (n2) -- node[sloped, above, blue] {\scriptsize $\xleftarrow{12 ants}$}  (n1);
			\draw[dashed, blue] ([xshift=2.5ex]n2.south) edge[->]  node[sloped, above, blue] {\scriptsize 6 ants} (n3.north);
			\draw[dashed, blue] ([xshift=-4ex]n2.south) edge[->]  node[sloped, above, blue] {\scriptsize 6 ants} ([xshift=1ex]n4.north);
			
			\draw[-] (n5) -- node[sloped, above, red] {\scriptsize $\xleftarrow{12 ants}$}  (n6);
			\draw[dashed, red] ([xshift=2.5ex]n5.north) edge[->]  node[sloped, below, red] {\scriptsize 6 ants} ([yshift=0.2ex]n3.south);
			\draw[dashed, red] ([xshift=-4ex]n5.north) edge[->]  node[sloped, below, red] {\scriptsize 6 ants} ([xshift=1ex]n4.south);
		\end{scope}
		
		\end{tikzpicture}
		}%
		$\qquad \qquad$
  		\subfloat[]{%
		\begin{tikzpicture}
		
		\begin{scope}[every node/.style={thin, draw}]
			\node (t) at (-1.2, 1.5) {t=1};
		\end{scope}
		
		\begin{scope}
			\node (n1) at (0,1) {A};
			\node (n2) at (0,0) {B};
			\node (n3) at (0.8,-1) {C};
			\node (n4) at (-1.5,-1) {D};
			\node (n5) at (0,-2) {E};
			\node (n6) at (0,-3) {F};
		\end{scope}
	
		\begin{scope}
			\path (n2) edge[draw=black, very thick] (n3);
			\path (n2) edge[draw=gray, very thin] (n4);
			\path (n3) edge[draw=black, very thick] (n5);
			\path (n4) edge[draw=gray, very thin] (n5);
			\path (n5) edge[draw=black] (n6);

			\draw[-] (n2) -- node[sloped, above, blue] {\scriptsize $\xleftarrow{12 ants}$}  (n1);
			\draw[dashed, blue] ([xshift=2.5ex]n2.south) edge[->]  node[sloped, above, blue] {\scriptsize 8 ants} (n3.north);
			\draw[dashed, blue] ([xshift=-4ex]n2.south) edge[->]  node[sloped, above, blue] {\scriptsize 4 ants} ([xshift=1ex]n4.north);
			
			\draw[-] (n5) -- node[sloped, above, red] {\scriptsize $\xleftarrow{12 ants}$}  (n6);
			\draw[dashed, red] ([xshift=2.5ex]n5.north) edge[->]  node[sloped, below, red] {\scriptsize 8 ants} ([yshift=0.2ex]n3.south);
			\draw[dashed, red] ([xshift=-4ex]n5.north) edge[->]  node[sloped, below, red] {\scriptsize 4 ants} ([xshift=1ex]n4.south);
		\end{scope}
		
		\end{tikzpicture}
		}%
    	\caption{An example of artificial ants: (a) initial paths with weighted distances, (b) at time $t=0$ there is no pheromone intensity on any path; so ants choose all paths with equal probability and, (c) at time $t=1$ the pheromone intensity is stronger on the shorter paths; therefore more ants prefer these paths.}
  	\label{fig4:artificial_ants}
  	\end{figure}

	\textit{Example 4.1.} We consider a sample graph of artificial ants moving along on the edges of nodes A, B, C, D, E and F as shown in Figure~\ref{fig4:artificial_ants}. 
	
	In order to make an accurate interpretation of ant colony system, assume that the paths between nodes B and D, D and E have longer lengths than paths between nodes B and C, C and E (as indicated by the weighted distances in Figure~\ref{fig4:artificial_ants}a). Let us consider what happens at regular discrete time intervals: $t=0,1,2...$ . Suppose that 12 new ants come to node B from A and, 12 new ants come to node E from F at each time interval. Each ant travels at a speed of \textit{2 weighted distance per time interval} and, that by moving along at time $t$ it deposits pheromones of intensity 1, which completely and instantaneously evaporates in the middle of the successive time interval $(t+1, t+2)$.

	At time $t=0$ no path has any pheromone intensity and, there are have 12 new ants at node B and 12 new ants at node E. The ants select between nodes C and D randomly; therefore, on average 6 ants from each node move towards C and 6 ants towards D - as shown in Figure~\ref{fig4:artificial_ants}b.
	
	At time $t=1$ there are 12 new ants at node B and 12 new ants at node E. The 12 new ants at node B will find a pheromone intensity of 6 on the path that leads to node D deposited by the 6 ants that used it previously from node B and, a pheromone intensity of 12 on the path that leads to node C deposited by 6 ants that used it coming from node B and 6 that arrived from node E. Therefore, the ants will find the path to node C more desirable than that to node D. The probability of the ants choosing to move towards node C is 2/3, while that of moving towards node D is 1/3 and, 8 ants will move towards node C and 4 ants towards node D. The same is true for the 12 new ants at node E, as shown in Figure~\ref{fig4:artificial_ants}c.

	\subsection{Preliminary Mathematical Notations}
	\label{ch4:preliminary}
	In this section, we introduce formal mathematical notations taken from literature concerning ant colony system (ACS). We use the traveling salesman problem (TSP)  \cite{Junger1995} in order to describe these preliminary mathematical notations \cite{Dorigo1996, Dorigo2010}. 
	
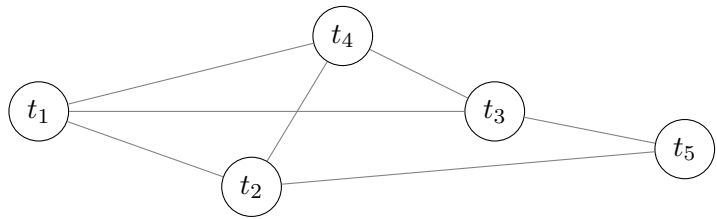
\begin{figure}[h!]
	\centering
	\begin{tikzpicture}
	\begin{scope}[every node/.style={circle, thin, draw}]
		\node (n1) at (0,1) {$t_{1}$};
		\node (n2) at (2.8,0) {$t_{2}$};
		\node (n3) at (6,1) {$t_{3}$};
		\node (n4) at (4,2) {$t_{4}$};
		\node (n5) at (8.5,0.5) {$t_{5}$};
	\end{scope}
	
	\begin{scope}
		\path (n1) edge[draw=gray] (n2);
		\path (n1) edge[draw=gray] (n3);
		\path (n1) edge[draw=gray] (n4);
		\path (n2) edge[draw=gray] (n4);
		\path (n2) edge[draw=gray] (n5);
		\path (n4) edge[draw=gray] (n3);
		\path (n3) edge[draw=gray] (n5);
	\end{scope}
	\end{tikzpicture}
	\caption{A sample TSP towns graph}
	\label{fig4:tsp_sample}
	\end{figure}

	Given a set of $n$ towns, the TSP problem can be stated as the problem of finding the shortest route that visits each town once. The path between town $i$ and $j$ may be represented as $d_{i,j}$ and, an instance of the TSP may also be presented as a graph composed of nodes and edges as shown in Figure~\ref{fig4:tsp_sample}.
	
	\newpage
	Let $\langle b_{i}(t) \mid i = 1,2,...n \rangle$ be the number of ants in town $i$ at time $t$ and, $q = \sum _{i=1} ^{n} b_{i}(t)$ be the total number of ants. Each ant is a simple agent that:
	\begin{itemize}
		\item chooses a town to go to based on the probability that is a function of the town distance and amount of pheromone present on the connecting edge,
		\item is prohibited from returning to already visited towns until a tour visit of all towns is completed (this is managed by a \textit{tabu list}) and,
		\item deposits pheromone trails on each visited edge $(i,j)$, once it completes the tour.
	\end{itemize}
	
	Let $\tau _{i,j}(t)$ be the \textit{pheromone intensity} on edge $(i,j)$ at time $t$ and, each ant at time $t$ chooses the next town to visit, where it will be at time $t+1$. We define $q$ movements made by $q$ ants in the interval $(t, t+1)$ a single \textit{iteration} of the ACS algorithm. Therefore, every $n$ iterations of the algorithm each ant completes a visit of all the towns (tour cycle) and, at this point the pheromone intensity is updated according to the formula in Equation~\eqref{eqn4:p_update}.
	
	\begin{equation}\label{eqn4:p_update}
		\tau _{i,j} (t+n) = \rho ~.~  \tau _{i,j} (t) +  \Delta \tau _{i,j}
	\end{equation}
	where $\rho$ (usually $<1$) is a coefficient such that $(1 - \rho)$ represents the evaporation of pheromone intensity between time $t$ and $t+n$,
	
	\begin{equation}
		\Delta \tau _{i,j} = \sum _{k=1} ^{q} \Delta \tau _{i,j} ^{k}\notag
	\end{equation}
	where $\Delta \tau _{i,j} ^{k}$ is the quantity per unit of length of pheromone substance laid on edge $(i, j)$ by the $kth$ ant between time $t$ and $t+1$ and it is given by:
	
	\begin{equation}
		 \Delta \tau _{i,j} ^{k} = 
		 \begin{cases}
		 	\frac{C}{L_{k}} \qquad $if $ kth $ ant uses path$ (i,j) $ in its tour (between time $ t $ and $ t+n)\\
			0 \qquad \; ~ otherwise
		 \end{cases}\notag
	\end{equation}
	where $C$ is a constant and $L_{k}$ is the tour length of the $kth$ ant.
	
	In order to satisfy the constraint that an ant visits all the $n$ towns, each ant is associated with a \textit{tabu list} $(tabu_{k})$ that stores the towns already visited by the ant up to time $t$ and forbids it from re-visiting them until the \textit{tour cycle} is complete. At the end of the tour cycle, the tabu list is used to compute the distance of the path followed by the ant then, it is emptied and, the next tour cycle begins.
	
	We introduce the term \textit{visibility} ($\eta _{i, j} = 1/d_{i, j}$) which we use to define the probability $p_{i, j} ^{k} (t)$ of the $kth$ ant moving from town $i$ to $j$ as shown in Equation~\eqref{eqn4:rule_orig}.
	
	\begin{equation}\label{eqn4:rule_orig}
		p_{i, j} ^{k} (t) = 
		\begin{cases}
			\frac{[\tau _{i, j} (t)]^{\alpha} ~.~ [\eta _{i, j}]^{\beta}}{\sum _{k \in allowed_{k}} [\tau _{i, k} (t)]^{\alpha} ~.~ [\eta _{i, k}]^{\beta}} \qquad $if $ j\in allowed_{k}\\
			0 \qquad \qquad \qquad \qquad \qquad  \qquad \; ~ otherwise
		\end{cases}
	\end{equation}
	where $allowed_{k} = \{N - tabu_{k} \}$ and $\alpha$ and $\beta$ are parameters that control the relative importance of the pheromone intensity against visibility.
	
	Therefore, the probability $p_{i, j} ^{k} (t)$ is a trade-off between visibility $\eta _{i, j}$ (which instructs that the closest town should be selected with highest probability - implements a greedy constructive heuristic) and, pheromone intensity at time $t$  $\tau _{i, j} (t)$ (which instructs that the edge $(i, j)$ with most ant traffic should be selected with highest probability - implementing an autocatalytic process).
	
	According to \cite{Hartmann2008}, Ant colony optimization (ACO) utilizes a set of artificial ants to probabilistically contrive solutions $\mathcal{S}$ through a collective memory, \textit{pheromones} stored in matrix $\mathcal{T}$, together with a problem specific heuristic $\eta$. In this chapter, we will consider one variant of ACO called `\texttt{MAX-MIN} ant system' \cite{Stutzle2000} to optimize both BFS and DFS for the case of gradual pattern mining.
	
	\subsection{ACO for BFS Candidate Generation}
	\label{ch4:aco_bfs}
	In this section we describe \textit{ACO-GRAANK} technique which is an optimized version of the \textit{GRAANK} technique described in Chapter~\ref{ch2:graank}. The \textit{ACO-GRAANK} technique inherits the efficient bitwise binary representation of concordant tuple couples that respect a gradual item set, but utilizes an \textit{ant}-based technique for generating candidate item sets. In order to apply ACO to the GRAANK approach, we need to identify a suitable heuristic representation of the gradual item set candidate generation problem that can be solved using an ant colony system.
	
	\begin{table}[h!]
  		\centering
      	\caption{Sample data set $\mathcal{D}_{4.1}$ with 3 attributes: $n = \{a, b, c \}$.}
    	\begin{tabular}{l c c c}
      		\textbf{id} & \textbf{a} & \textbf{b} & \textbf{c}\\
      		\hline \hline
      		r1 & 5 & 30 & 43\\
      		r2 & 4 & 35 & 33\\
      		r3 & 3 & 40 & 42\\
      		r4 & 1 & 50 & 49\\
      		\bottomrule
    		\end{tabular}
    	\label{tab4:sample1}
	\end{table}
	
	Given a set of $n = \{ a_{1},a_{2},... \}$ attributes of a data set $\mathcal{D}$ (as shown in Table~\ref{tab4:sample1}), gradual BFS techniques seek to find \textit{frequent} gradual patterns $M = \{ m_{1}, m_{2},...,m_{k} \}$ (where $\forall m \in M: m \subseteq n$) by generating and testing numerous candidates. It is important to recall that a non-trivial gradual pattern $m_{k}$ is set of at least 2 gradual items and, a singleton gradual item is a pair composed of an attribute and \textit{variation} (increasing/decreasing) - see Chapter~\ref{ch2:definitions}. 
	
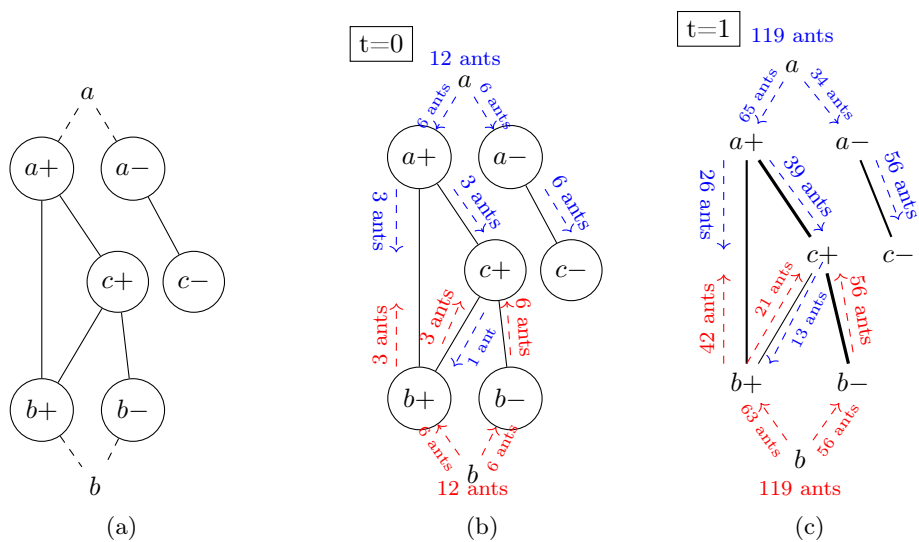
\begin{figure}[h!]
		\centering
		\small
  		\subfloat[]{%
		\begin{tikzpicture}
		\begin{scope}[circlestyle/.style={circle, thin, draw}]
			\node (nu) at (-0.6,1) {$a$};
			
			\node[circlestyle] (n11) at (-1.2,0) {$a+$};
			\node[circlestyle] (n12) at (0,0) {$a-$};

			\node[circlestyle] (n21) at (-0.2,-1.5) {$c+$};
			\node[circlestyle] (n22) at (0.8,-1.5) {$c-$};
			
			\node[circlestyle] (n31) at (-1.2,-3.2) {$b+$};
			\node[circlestyle] (n32) at (0,-3.2) {$b-$};
			
			\node (nd) at (-0.5,-4.2) {$b$};
		\end{scope}
	
		\begin{scope}
			\path (nu) edge[draw=black, dashed] (n11);
			\path (nu) edge[draw=black, dashed] (n12);
			
			\draw[-] (n11) edge (n31);
			\draw[-] (n11) edge (n21);
			\draw[-] (n12) edge (n22);
			\draw[-] (n21) edge (n31);
			\draw[-] (n21) edge (n32);
			
			\path (n31) edge[draw=black, dashed] (nd);
			\path (n32) edge[draw=black, dashed] (nd);
			
		\end{scope}
		\end{tikzpicture}
		}%
		$\qquad \qquad$
  		\subfloat[]{%
		\begin{tikzpicture}
		
		\begin{scope}[every node/.style={thin, draw}]
			\node (t) at (-1.7, 1.5) {t=0};
		\end{scope}
		
		\begin{scope}[circlestyle/.style={circle, thin, draw}]
			\node (nu) at (-0.6,1) {$a$};
			\node[above, blue] at (-0.6, 1.1) {\scriptsize 12 ants};
			
			\node[circlestyle] (n11) at (-1.2,0) {$a+$};
			\node[circlestyle] (n12) at (0,0) {$a-$};

			\node[circlestyle] (n21) at (-0.2,-1.5) {$c+$};
			\node[circlestyle] (n22) at (0.8,-1.5) {$c-$};
			
			\node[circlestyle] (n31) at (-1.2,-3.2) {$b+$};
			\node[circlestyle] (n32) at (0,-3.2) {$b-$};
			
			\node (nd) at (-0.5,-4.2) {$b$};
			\node[above, red] at (-0.5,-4.6) {\scriptsize 12 ants};
		\end{scope}
	
		\begin{scope}
			\path (nu) edge[draw=blue, dashed, ->] node[blue, sloped, above]{\tiny 6 ants} (n11);
			\path (nu) edge[draw=blue, dashed, ->] node[blue, sloped, above]{\tiny 6 ants} (n12);
			
			\draw[-] (n11) edge (n31);
			\draw[-] (n11) edge (n21);
			\draw[-] (n12) edge (n22);
			\draw[-] (n21) edge (n31);
			\draw[-] (n21) edge (n32);
			
			\path (nd) edge[draw=red, dashed, ->] node[red, sloped, below]{\tiny 6 ants} (n31);
			\path (nd) edge[draw=red, dashed, ->] node[red, sloped, below]{\tiny 6 ants} (n32);
			
			\draw[dashed, blue] ([xshift=-2ex]n11.south) edge[->]  node[sloped, below, blue] {\scriptsize 3 ants} ([yshift=10ex, xshift=-2ex]n31.north);
			\draw[dashed, blue] ([xshift=3ex]n11.south) edge[->]  node[sloped, above, blue] {\scriptsize 3 ants} ([yshift=0.6ex, xshift=-1ex]n21.north);
			\draw[dashed, blue] ([xshift=3ex]n12.south) edge[->]  node[sloped, above, blue] {\scriptsize 6 ants} ([yshift=0.6ex]n22.north);
			
			\draw[dashed, blue] ([xshift=-1ex, yshift=-1ex]n21.south) edge[->]  node[sloped, below, blue] {\tiny 1 ant} ([xshift=3ex]n31.north);

			\draw[dashed, red] ([xshift=-2ex]n31.north) edge[->]  node[sloped, above, red] {\scriptsize 3 ants} ([yshift=-10ex, xshift=-2ex]n11.south);
			\draw[dashed, red] ([yshift=2ex, xshift=2ex]n31.north) edge[->]  node[sloped, above, red] {\scriptsize 3 ants} ([xshift=-3ex]n21.south);
			\draw[dashed, red] ([yshift=1ex]n32.north) edge[->]  node[sloped, above, red] {\scriptsize 6 ants} ([xshift=1ex]n21.south);
			
		\end{scope}
				
		\end{tikzpicture}
		}%
		$\qquad$
  		\subfloat[]{%
		\begin{tikzpicture}
		
		\begin{scope}[every node/.style={thin, draw}]
			\node (t) at (-1.7, 1.5) {t=1};
		\end{scope}
		
		\begin{scope}[circlestyle/.style={cycle, thin, draw}]
			\node (nu) at (-0.6,1) {$a$};
			\node[above, blue] at (-0.6, 1.2) {\scriptsize 119 ants};
			
			\node (n11) at (-1.2,0) {$a+$};
			\node (n12) at (0.2,0) {$a-$};

			\node (n21) at (-0.2,-1.5) {$c+$};
			\node (n22) at (0.8,-1.5) {$c-$};
			
			\node (n31) at (-1.2,-3.2) {$b+$};
			\node (n32) at (0.2,-3.2) {$b-$};
			
			\node (nd) at (-0.5,-4.2) {$b$};
			\node[above, red] at (-0.5,-4.8) {\scriptsize 119 ants};
		\end{scope}
	
		\begin{scope}
			\path (nu) edge[draw=blue, dashed, ->] node[blue, sloped, above]{\tiny 65 ants} (n11);
			\path (nu) edge[draw=blue, dashed, ->] node[blue, sloped, above]{\tiny 34 ants} (n12);
			
			\draw[thick] (n11) edge (n31);
			\draw[very thick] (n11) edge (n21);
			\draw[thick] (n12) edge (n22);
			\draw[] (n21) edge (n31);
			\draw[very thick] (n21) edge (n32);
			
			\path (nd) edge[draw=red, dashed, ->] node[red, sloped, below]{\tiny 63 ants} (n31);
			\path (nd) edge[draw=red, dashed, ->] node[red, sloped, below]{\tiny 56 ants} (n32);
			
			\draw[dashed, blue] ([xshift=-2ex]n11.south) edge[->]  node[sloped, below, blue] {\scriptsize 26 ants} ([yshift=10ex, xshift=-2ex]n31.north);
			\draw[dashed, blue] ([xshift=2ex]n11.south) edge[->]  node[sloped, above, blue] {\scriptsize 39 ants} ([yshift=0.5ex]n21.north);
			\draw[dashed, blue] ([xshift=2ex]n12.south) edge[->]  node[sloped, above, blue] {\scriptsize 56 ants} ([yshift=1.5ex]n22.north);
			
			\draw[dashed, blue] ([yshift=1ex]n21.south) edge[->]  node[sloped, below, blue] {\tiny 13 ants} ([xshift=2ex]n31.north);

			\draw[dashed, red] ([xshift=-2ex]n31.north) edge[->]  node[sloped, above, red] {\scriptsize 42 ants} ([yshift=-10ex, xshift=-2ex]n11.south);
			\draw[dashed, red] ([yshift=0ex]n31.north) edge[->]  node[sloped, above, near end, red] {\tiny 21 ants} ([xshift=-2ex]n21.south);
			\draw[dashed, red] ([yshift=1ex]n32.north) edge[->]  node[sloped, above, red] {\scriptsize 56 ants} ([xshift=1.5ex]n21.south);
			
		\end{scope}		
		\end{tikzpicture}
		}%
    	\caption{\footnotesize An example of artificial ants for BFS gradual pattern mining: (a) initial paths, (b) at time $t=0$ there is no pheromone intensity on any path; so ants choose paths with equal probability, and (c) at time $t=1$ the pheromone intensity is stronger on paths with cheapest gradual variations; therefore, ants prefer these paths.}
  	\label{fig4:artificial_ants_bfs}
  	\end{figure}

	\textit{Example 4.2.} We consider a sample graphs of artificial ants moving on the edges of gradual items $a+, a-, b+, b-, c+$ and, $c-$ (where $+$ denotes `increasing', $-$ denotes `decreasing' and minimum support threshold $\varrho = 0.5$) as shown in Figure~\ref{fig4:artificial_ants_bfs}. For the sake of simplicity, we remove all edges connecting to nodes (or gradual items) whose \textit{frequency support} $< \varrho$ and, assume that $support(\{a+, c+ \}) > \varrho$ but $support(\{a+, c+, b+ \}) < \varrho$.

	In order to make an accurate interpretation of the ACS, assume that the distances of all node edges are equal. Let us consider what happens at regular discrete time intervals: $t=0,1,2,...$ . Suppose that 12 new ants come to nodes $a$ and $b$ respectively at time $t=0$ and another 119 new ants to nodes $a$ and $b$ at time $t=1,2,...$ . Each ant travels at a speed of 2 \textit{edge distances per time interval} and, that by moving along at time $t$ it deposits pheromones of intensity 1 only if all the nodes visited up to that time $t$ form a gradual item set whose $support \geq \varrho$. The pheromone instantaneously evaporates in the middle of the successive time interval $(t+1, t+2)$.

	At time $t=0$ no path has any pheromone intensity; therefore, the 12 news ants at nodes $a$ and $b$ choose all paths  with equal probability as shown in Figure~\ref{fig4:artificial_ants_bfs}b. Since $support(\{a+, c+, b+ \}) < \varrho$ the ant from $a$ that chose edges between nodes $a+$, $c+$ and $b+$ and the 3 ants from $b$ that chose edges between nodes $b+$, $c+$ and $c+$ do not deposit any pheromone on any of these edges.
	
	At time $t=1$ 119 new ants arrive at nodes $a$ and $b$. The 119 new ants at node $a$ will find a pheromone intensity 6 on the path that leads to $a-$ deposited by ants that used it previously from node $a-$ and, a pheromone intensity of 15 at node $a+$: 6 on the path that leads to $b+$ (3 deposited by ants that departed from $a+$ and 3 deposited by ants that arrived from $b+$), 9 on the path that leads to $c+$ (3 deposited by ants that departed from $a+$ and 6 deposited by ants that arrived from $c+$). The probability of the ants choosing to move towards node $a-$ is 6/21, that of moving towards node $a+$ is 15/21, that of moving towards node $b+$ is 6/15 and, that of moving towards $c+$ 9/15. The same is true for the new 119 ants at node $b$, as shown in Figure~\ref{fig4:artificial_ants_bfs}c. Therefore, the best variation routes (or candidates) are: $\{a+,c+,b-\},\{a+,b+\},\{a-,c-\}$.
	
	\subsubsection{Proposed Mathematical Notations of ACS for GPCG Problem}
	\label{ch4:aco_bfs_notations}
	The gradual pattern candidate generation (GPCG) problem can be stated as: 
	
	\textit{``the problem of finding the cheapest variation routes that connect all gradual items that satisfy the minimum frequency requirement once.''}
	
	In gradual pattern mining, a potential candidate must not include any \textit{infrequent} singleton gradual item set (anti-monotonicity property - see Chapter\ref{ch2:antimonotonicity}). Therefore, when apply ant colony system (ACS) to the case of GPCG problem, we define each node of the ACS graph as a singleton gradual item derived from its attribute and the variation edge distance $d_{i, j}$ between nodes $i, j$ is given by formula in Equation~\eqref{eqn4:bfs_distance}:
	
	\begin{equation}\label{eqn4:bfs_distance}
		d_{i, j} = 
		 \begin{cases}
		 	1 \qquad \quad $if $ sup(i) \geq \varrho $ and $ sup(j) \geq \varrho\\
			\infty \qquad \; otherwise
		 \end{cases}
	\end{equation}
	where $sup(i)$ is the frequency support of gradual item set $\{i\}$, $\varrho$ is a user-specified minimum support threshold - see Chapter~\ref{ch2:definitions}. The possibility of gradual item $(i, j)$ being frequent is either visible `1' or not visible `$\infty$'.
	
	Let $\langle b_{i}(t) \mid i = 1,2,...n \rangle$ be the number of ants at node $i$ at time $t$ and, $q = \sum _{i=1} ^{n} b_{i}(t)$ be the total number of ants. Each ant is a simple agent that:
	\begin{itemize}
		\item chooses a node to go to based on the probability that is a function of the amount of pheromone present on the connecting edge,
		\item is prohibited from returning to already visited nodes until a tour visit of all nodes is completed (this is managed by a \textit{tabu list}) and,
		\item deposits pheromone trails on all visited edges if support of all visited nodes $sup(\mathbb{N}) \geq \varrho$, once it completes the tour. This is because a gradual candidate is formed by combining all the visited nodes into a set; and, if the candidate is not frequent then its path is also not appealing. Lastly, every edge $(i, j)$ is removed if: $sup(i) < \varrho$ and $sup(j) < \varrho$.
	\end{itemize}

	Let $\tau _{i,j}(t)$ be the \textit{pheromone intensity} on edge $(i,j)$ at time $t$ and, each ant at time $t$ chooses the next node to visit, where it will be at time $t+1$. We define $q$ movements made by $q$ ants in the interval $(t, t+1)$ a single \textit{iteration} of the ACS algorithm. Therefore, every $n$ iterations of the algorithm each ant completes a visit of all valid nodes (tour cycle) and, at this point the pheromone intensity is updated according to the formula in Equation~\eqref{eqn4:p_update_bfs}.
	
	\begin{equation}\label{eqn4:p_update_bfs}
		 \tau _{i, j} (t+n) = 
		 \begin{cases}
		 	 \rho ~.~  \tau _{i, j} (t) +  \Delta \tau _{i, j} \qquad $if $ sup(\mathbb{N}_{n}) \geq \varrho\\
			0 \qquad \qquad \qquad \qquad \quad otherwise
		 \end{cases}
	\end{equation}
	where $\rho$ and $\Delta \tau _{i, j}$ are similar to Equation~\eqref{eqn4:p_update} and, $\mathbb{N}_{n}$ is the set of all nodes visited by the ant at end of $n$ iterations
	
	\begin{equation}
		\mathbb{N}_{n} = \{i_{t}\}_{t=0} ^{t=n}\notag
	\end{equation}
	
	Similarly, each ant is associated with a tabu list $tabu_{k}$ that stores the nodes already visited up to time $t$ and forbids the ant from revisiting them until the tour cycle is complete. In addition, the tabu list also keeps track of node edges which do not meet the minimum frequency support requirement. Since we modify notation $d_{i, j}$ in Equation~\eqref{eqn4:bfs_distance}, visibility ($\eta _{i, j} = 1/d_{i, j}$) is consequently affected. The probability $p_{i, j} ^{k} (t)$ of the $kth$ ant moving from node $i$ to $j$ becomes:
	
	\begin{equation}\label{eqn4:rule_bfs}
		p_{i, j} ^{k} (t) = 
		\begin{cases}
			\frac{[\tau _{i, j} (t)]^{\alpha}}{\sum _{k \in allowed_{k}} [\tau _{i, k} (t)]^{\alpha}} \qquad $if $ j\in allowed_{k} $ and $ (sup(i) \geq \varrho ~\&~ sup(j) \geq \varrho)\\
			0 \qquad \qquad \qquad \qquad \quad otherwise
		\end{cases}
	\end{equation}
	\begin{center}
	where $allowed_{k}$ and $\alpha$ are similar to Equation~\eqref{eqn4:rule_orig}.
	\end{center}
	
	Therefore, the probability $p_{i, j} ^{k} (t)$ is a function of pheromone intensity at time $t$: $\tau _{i, j} (t)$ (which instructs that the edge $(i, j)$ with most ant traffic should be selected with highest probability - implementing an autocatalytic process). In gradual pattern mining \textit{visibility} between nodes $i$ and $j$ can be compared to the possibility of forming a candidate by combining node $i$ and $j$; and, it is either `visible' to the ant if both nodes are \textit{frequent} or `invisible' to the ant if one of the nodes is \textit{infrequent}. So, $\eta _{i, j}$ is denoted as either 1 or 0 for `visible' or `invisible' respectively.
		
	\subsection{ACO for DFS FP-Tree Search}
	\label{ch4:aco_dfs}
	In this section, we describe ACO-ParaMiner technique which is an optimized version of the existing DFS-based ParaMiner technique proposed by \cite{Negrevergne2014} (see Chapter~\ref{ch2:paraminer}). ACO-ParaMiner inherits the technique of transactional encoding from ParaMiner, but utilizes an \textit{ant-based} technique to find the longest \textit{frequent pattern tree} (FP-Tree). Similarly, for the purpose of applying ACO to a DFS-based approach, we identify a suitable heuristic representation to the problem of finding the parent node of the longest FP-tree.
	
	\begin{table}[h!]
		\centering
		\caption{(a) Sample data set $\mathcal{D}_{4.2}$ and, (b) its corresponding sorted reduced transactional data set when minimum length of \texttt{tid} is 3 - see Chapter~\ref{ch2:paraminer}.}
		\subfloat[]{%
		\begin{tabular}{l c c c c}
      		\textbf{id} & \textbf{a} & \textbf{b} & \textbf{c} & \textbf{d} \\
      		\hline \hline
      		r1 & 5 & 30 & 43 & 97\\
      		r2 & 4 & 35 & 33 & 86\\
      		r3 & 3 & 40 & 42 & 108\\
      		r4 & 1 & 50 & 49 & 27\\
      		\bottomrule
    	\end{tabular}
    	$~~~$ 
    	}%
		\subfloat[]{%
		\begin{tabular}{c l}
      		\textbf{item} & \textbf{tids} \\
      		\hline \hline
      		$a^{\downarrow}$ & $\{t_{(r1, r2)}, t_{(r1, r3)}, t_{(r1, r4)},$\\
          	& $t_{(r2, r3)}, t_{(r2, r4)}, t_{(r3, r4)} \}$\\
			$b^{\uparrow}$ & $\{t_{(r1, r2)}, t_{(r1, r3)}, t_{(r1, r4)},$\\
          	& $t_{(r2, r3)}, t_{(r2, r4)}, t_{(r3, r4)} \}$\\   
      		$c^{\uparrow}$ & $\{t_{(r1, r4)}, t_{(r2, r3)}, t_{(r2, r4)},$\\
      		& $t_{(r3, r4)} \}$\\
      		$d^{\downarrow}$ & $\{t_{(r1, r2)}, t_{(r1, r4)}, t_{(r2, r4)},$\\
          	& $t_{(r3, r4)} \}$\\
      		\bottomrule
    	\end{tabular}
		}%
    	\label{tab4:sample2}
	\end{table}
		
	Given a set of $n=\{a_{1}, a_{2}, ..., a_{k}\}$ attributes of a data set where each attribute has a set of tuples $a_{k} = \{r_{1}, r_{2}, ... \}$ (as shown in Table~\ref{tab4:sample2}a), gradual DFS techniques seek to find \textit{frequent} gradual patterns $M = \{ m_{1}, m_{2},... \}$ (where $\forall m \in M: m \subseteq n$) by encoding the data set into a transactional data set (as shown in Table~\ref{tab4:sample2}b) and recursively searching the transactional data set for the longest FP-Tree - see Chapter~\ref{ch2:paraminer}.
	
\begin{figure}[h!]
		\centering
		\small
  		\subfloat[]{%
		\begin{tikzpicture}
		\begin{scope}
			\node (n1) at (0,1) {};
			\node (n2) at (0,0) {$r_{1}$};
			\node (n3) at (0.8,-1) {$r_{4}$};
			\node (n4) at (-1.5,-1) {$r_{3}$};
			\node (n5) at (0,-2) {$r_{2}$};
			\node (n6) at (0,-3) {};
		\end{scope}
	
		\begin{scope}
			\path (n1) edge[draw=black, dashed] (n2);
			\draw[-] (n2) edge  node[sloped, above, brown] {\scriptsize d=0.25} (n3);
			\draw[-] (n2) edge  node[sloped, above, brown] {\scriptsize d=0.5} (n4);
			\draw[-] (n3) edge  node[sloped, below, brown] {\scriptsize d=0.25} (n5);
			\draw[-] (n4) edge  node[sloped, below, brown] {\scriptsize d=0.33} (n5);
			\path (n5) edge[draw=black, dashed] (n6);
		\end{scope}
		\end{tikzpicture}
		}%
		$\qquad \qquad$
  		\subfloat[]{%
		\begin{tikzpicture}
		
		\begin{scope}[every node/.style={thin, draw}]
			\node (t) at (-1.2, 1.5) {t=0};
		\end{scope}
		
		\begin{scope}
			\node (n1) at (0,1) {};
			\node (n2) at (0,0) {$r_{1}$};
			\node (n3) at (0.8,-1) {$r_{4}$};
			\node (n4) at (-1.5,-1) {$r_{3}$};
			\node (n5) at (0,-2) {$r_{2}$};
			\node (n6) at (0,-3) {};
		\end{scope}
			
		\begin{scope}
			\path (n2) edge[draw=black] (n3);
			\path (n2) edge[draw=black] (n4);
			\path (n3) edge[draw=black] (n5);
			\path (n4) edge[draw=black] (n5);

			\draw[blue] (n2) edge[dashed, <-]  node[sloped, below, blue] {\tiny 12 ants} (n1);
			\draw[dashed, blue] ([xshift=2.5ex]n2.south) edge[->]  node[sloped, above, blue] {\scriptsize 6 ants} (n3.north);
			\draw[dashed, blue] ([xshift=-4ex]n2.south) edge[->]  node[sloped, above, blue] {\scriptsize 6 ants} ([xshift=1ex]n4.north);
			
			\draw[red] (n6) edge[dashed, ->]  node[sloped, below, red] {\tiny 12 ants} (n5);
			\draw[dashed, red] ([xshift=2.5ex]n5.north) edge[->]  node[sloped, below, red] {\scriptsize 6 ants} ([yshift=0.2ex]n3.south);
			\draw[dashed, red] ([xshift=-4ex]n5.north) edge[->]  node[sloped, below, red] {\scriptsize 6 ants} ([xshift=1ex]n4.south);
		\end{scope}
		
		\end{tikzpicture}
		}%
		$\qquad \qquad$
  		\subfloat[]{%
		\begin{tikzpicture}
		
		\begin{scope}[every node/.style={thin, draw}]
			\node (t) at (-1.2, 1.5) {t=1};
		\end{scope}
		
		\begin{scope}
			\node (n1) at (0,1) {};
			\node (n2) at (0,0) {$r_{1}$};
			\node (n3) at (0.8,-1) {$r_{4}$};
			\node (n4) at (-1.5,-1) {$r_{3}$};
			\node (n5) at (0,-2) {$r_{2}$};
			\node (n6) at (0,-3) {};
		\end{scope}
	
		\begin{scope}
			\path (n2) edge[draw=black, very thick] (n3);
			\path (n2) edge[draw=gray, very thin] (n4);
			\path (n3) edge[draw=black, very thick] (n5);
			\path (n4) edge[draw=gray, very thin] (n5);

			\draw[blue] (n2) edge[dashed, <-]  node[sloped, below, blue] {\tiny 12 ants} (n1);
			\draw[dashed, blue] ([xshift=2.5ex]n2.south) edge[->]  node[sloped, above, blue] {\scriptsize 8 ants} (n3.north);
			\draw[dashed, blue] ([xshift=-4ex]n2.south) edge[->]  node[sloped, above, blue] {\scriptsize 4 ants} ([xshift=1ex]n4.north);
			
			\draw[red] (n6) edge[dashed, ->]  node[sloped, below, red] {\tiny 12 ants} (n5);
			\draw[dashed, red] ([xshift=2.5ex]n5.north) edge[->]  node[sloped, below, red] {\scriptsize 8 ants} ([yshift=0.2ex]n3.south);
			\draw[dashed, red] ([xshift=-4ex]n5.north) edge[->]  node[sloped, below, red] {\scriptsize 4 ants} ([xshift=1ex]n4.south);
		\end{scope}
		
		\end{tikzpicture}
		}%
    	\caption{\footnotesize An example of artificial ants for DFS: (a) initial paths with distances, (b) at time $t=0$ there is no pheromone intensity on any path; so ants choose all paths with equal probability and, (c) at time $t=1$ the pheromone intensity is stronger on the shorter paths; therefore more ants prefer these paths.}
  	\label{fig4:artificial_ants_dfs}
  	\end{figure}
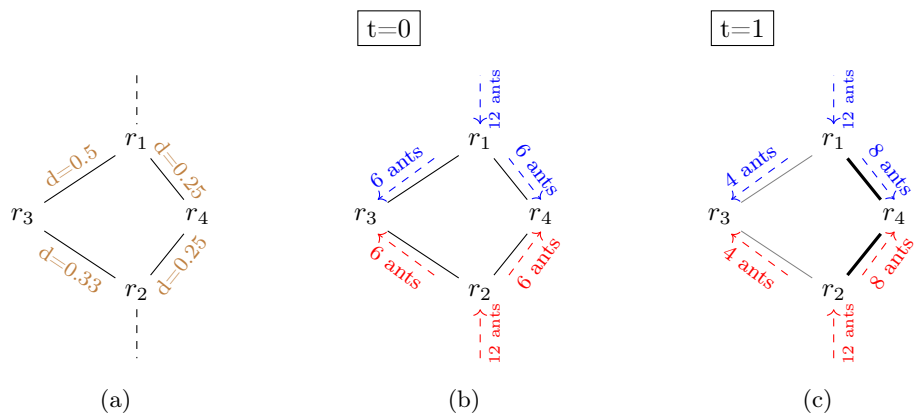

	\textit{Example 4.3.} We consider a sample graphs of artificial ants moving on the edges of nodes (or tuples) $r_{1}, r_{2}, r_{3}$ and $r_{4}$ as shown in Figure~\ref{fig4:artificial_ants_dfs}. For the case of DFS in gradual pattern mining, we propose to use the occurrence count of tuples in the encoded transactional data set to determine the length of distance between nodes (i.e. $d_{i, j} = \frac{1}{1 + \sum (r_{i}, r_{j})_{count}}$).
		
  	In order to make an accurate interpretation of ant colony system, assume that the paths between nodes $r_{1}$ and $r_{3}$, $r_{3}$ and $r_{2}$ have longer lengths than paths between nodes $r_{1}$ and $r_{4}$, $r_{4}$ and $r_{2}$ (as indicated by the distances in Figure~\ref{fig4:artificial_ants_dfs}a). Let us consider what happens at regular discrete time intervals: $t=0,1,2...$ . Suppose that 12 new ants come to node $r_{1}$ and, 12 new ants come to node $r_{2}$ at each time interval. Each ant travels at a speed of 0.5 \textit{distance per time interval} and, that by moving along at time $t$ it deposits pheromones of intensity 1, which completely and instantaneously evaporates in the middle of the successive time interval $(t+1, t+2)$.

	At time $t=0$ no path has any pheromone intensity and, there are have 12 new ants at node $r_{1}$ and 12 new ants at node $r_{2}$. The ants select between nodes $r_{4}$ and $r_{3}$ randomly; therefore, on average 6 ants from each node move towards $r_{4}$ and 6 ants towards $r_{3}$ - as shown in Figure~\ref{fig4:artificial_ants_dfs}b.
	
	At time $t=1$ there are 12 new ants at node $r_{1}$ and 12 new ants at node $r_{2}$. The 12 new ants at node $r_{1}$ will find a pheromone intensity of 6 on the path that leads to node $r_{3}$ deposited by the 6 ants that used it previously from node $r_{1}$ and, a pheromone intensity of 12 on the path that leads to node $r_{4}$ deposited by 6 ants that used it coming from node $r_{1}$ and 6 that arrived from node $r_{2}$. Therefore, the ants will find the path to node $r_{4}$ more desirable than that to node $r_{3}$. The probability of the ants choosing to move towards node $r_{4}$ is 2/3, while that of moving towards node $r_{3}$ is 1/3 and, 8 ants will move towards node $r_{4}$ and 4 ants towards node $r_{3}$. The same is true for the 12 new ants at node $r_{2}$, as shown in Figure~\ref{fig4:artificial_ants_dfs}c. Therefore, the most appealing FP-Tree is $\{(r_{1}, r_{4}), (r_{2}, r_{4})\}$ which both appear in transactions with gradual items $\{ b^{\uparrow}, c^{\uparrow}, d^{\downarrow} \} \Leftrightarrow \{b+, c+, d-\}$.

	\subsubsection{Proposed Mathematical Notations of ACS for GPFP-Tree Problem}
	\label{ch4:aco_dfs_notations}
	The gradual pattern FP-Tree (GPFP-Tree) problem can be stated as: 
		
	\textit{``the problem of finding the longest FP-Tree from which frequent gradual patterns can be constructed.''}
	
	In the case of DFS for gradual pattern mining, we inherit the concept of encoding data sets into transactional data sets and harness this to represent the GPFP-Tree problem as a slightly different version of the TSP problem. Given a set of $n$ tuples of a data set, we seek to find the cheapest route that visits the most tuples once. The path between tuples $i$ and $j$ may be represented as $d_{i, j}$ and an instance of the GPFP-Tree may also be represented as a graph composed of tuples (or nodes) and edges. Each edge distance is derived from the formula:
	
	\begin{equation}\label{eqn4:dfs_distance}
		d_{i, j} = \frac{1}{1 + \sum (r_{i}, r_{j})_{count}}
	\end{equation}
	where $\sum (r_{i}, r_{j})_{count}$ is the occurrence count of tuple pair $(r_{i}, r_{j})$ in an encoded transactional data set.
	
	Let $\langle b_{i}(t) \mid i = 1,2,...n \rangle$ be the number of ants in node $i$ at time $t$ and, $q = \sum _{i=1} ^{n} b_{i}(t)$ be the total number of ants. Each ant is a simple agent that:
	\begin{itemize}
		\item chooses a node to go to based on the probability that is a function of the node distance and amount of pheromone present on the connecting edge,
		\item is prohibited from returning to already visited nodes until a tour visit of all allowed nodes is completed (this is managed by a \textit{tabu list}) and,
		\item deposits pheromone trails on each visited edge $(i, j)$, once it completes the tour.
	\end{itemize}
	
	Let $\tau _{i, j}(t)$ be the \textit{pheromone intensity} on edge $(i, j)$ at time $t$ and, each ant at time $t$ chooses the next node to visit, where it will be at time $t+1$. We define $q$ movements made by $q$ ants in the interval $(t, t+1)$ a single \textit{iteration} of the ACS algorithm. Therefore, every $n$ iterations of the algorithm each ant completes a visit of all the nodes (tour cycle) and, at this point the pheromone intensity is updated according to the formula in Equation~\eqref{eqn4:p_update_dfs}.
	
	\begin{equation}\label{eqn4:p_update_dfs}
		\tau _{i, j} (t+n) = \rho ~.~  \tau _{i, j} (t) +  \Delta \tau _{i, j}
	\end{equation}
	\begin{center}
		where $\rho$ and $\Delta \tau _{i, j}$ are similar to Equation~\eqref{eqn4:p_update}.
	\end{center}
	
	Similar to the original ACS, each ant is associated with a \textit{tabu list} $(tabu_{k})$ that stores the nodes already visited by the ant up to time $t$ and forbids it from re-visiting them until the \textit{tour cycle} is complete. At the end of the tour cycle, the tabu list is used to compute the distance of the path followed by the ant then, it is emptied and, the next tour cycle begins. The probability $p_{i, j} ^{k} (t)$ of the $kth$ ant moving from node $i$ to $j$ as shown in Equation~\eqref{eqn4:rule_dfs}.
	
	\begin{equation}\label{eqn4:rule_dfs}
		p_{i, j} ^{k} (t) = 
		\begin{cases}
			\frac{[\tau _{i, j} (t)]^{\alpha} ~.~ [\eta _{i, j}]^{\beta}}{\sum _{k \in allowed_{k}} [\tau _{i, k} (t)]^{\alpha} ~.~ [\eta _{i, k}]^{\beta}} \qquad $if $ j\in allowed_{k}\\
			0 \qquad \qquad \qquad \qquad \qquad  \qquad \; ~ otherwise
		\end{cases}
	\end{equation}
	\begin{center}
		where $allowed_{k}$ and $\alpha$ and $\beta$ are similar to Equation~\eqref{eqn4:rule_orig}.
	\end{center}
	
	Therefore, in the GPFP-Tree problem probability $p_{i, j} ^{k} (t)$ is a trade-off between $\eta _{i, j}$ and $\tau _{i, j} (t)$. Put differently, it a trade-off between a greedy constructive heuristic and an autocatalytic process.
	
	\subsection{Convergence Proof}
	\label{ch4:convergence}
	There exist two main ant colony optimization techniques: (1) \textit{Ant Colony System} (ACS) proposed in \cite{Dorigo1997} and (2) \texttt{MAX--MIN} \textit{Ant System} (MMAS) proposed in \cite{Stutzle2000}. We recommend the MMAS over ACS for the problem of gradual pattern mining since it offers a better performance for finding an optimal solution for either a combinatorial problem or finding the longest FP-Tree problem.

	In the case of BFS gradual pattern mining we wish to find an \textit{optimal solution} (which is a valid maximal gradual item set) that updates the matrix such that the preceding generated solutions are either subsets of or similar to the optimal solution. Such a characteristic may also be referred to as a \textit{Convergence Property}. The study of \cite{Stutzle2002} illustrates a \textit{convergence proof} that applied directly to MMAS. The proof holds that:

	 \textit{``for any small constant $\epsilon > 0$ and for a sufficiently large number of algorithm iterations $t$, the probability of finding an optimal solution at least once is $\mathcal{P}*(t) \geq 1 - \epsilon$ and that the probability tends to $1$ for $t\rightarrow \infty$.''}

	Further, \cite{Stutzle2002} establishes that after an \textit{optimal solution} has been found, it takes a limited number of algorithm iterations for the pheromone trails that belong to the found optimal solution to grow higher than any other pheromone trail. With regard to BFS gradual pattern mining, this implies that the values of the pheromone matrix will no longer change significantly after such iterations. Therefore, we propose that this \textit{convergence property} of ACO techniques can be harnessed to determine the limit of algorithm iterations for generating gradual item set candidates. With regard to DFS gradual pattern mining, after an \textit{optimal parent node} is found, it takes a limited number of algorithm iterations for the pheromone trails corresponding to the optimal parent node to grow higher than other pheromone trails. Therefore, the algorithm should generate similar nodes subsequently.

	However, the proofs do not hint at any time required to find an optimal gradual item set solution. It is equally important to point out that for both ACS and MMAS approaches, if no ant finds a better \textit{``best-so-far''} trail within a finite number of iterations, all ants tend to construct a similar pheromone trail. This may also be referred to as the \textit{local minima phenomenon} \cite{Satukitchai2015}. In the case of BFS gradual pattern mining, this may imply that a non-optimal gradual item set solution may be found to be optimal if no ant finds the best maximal item set solution within a limited number of iterations.

	\section{Proposed ACO-based Approaches}
	\label{ch4:aco_approaches}
	In this section, we present two approaches: ACO-GRAANK and ACO-ParaMiner which optimize GRAANK and ParaMiner algorithms respectively. Both algorithms are variants of \textit{``MAX-MIN ant system''} \cite{Stutzle2000}.
	
	\subsection{ACO-GRAANK Approach}
	\label{ch4:aco_graank}
	The chief aim of gradual pattern mining approaches is to extract (if possible maximal) gradual item sets whose quality surpass a user-specified threshold. Classical GRITE and GRAANK techniques employ a level-wise search (BFS) traversal and join strategy to build maximal gradual item sets by combining minimal item sets. This strategy has a high time complexity especially when presented with a huge number of minimal gradual item sets.

	On the contrary, it can be shown that a heuristic approach can generate, with extremely high efficiency, gradual item set candidates whose probability of being valid is high \cite{Kalpana2008, Haifeng2016}. Moreover, this eliminates the repetition that comes with combining minimal item sets. In this chapter, we present an \textit{ant}-based approach that guides artificial ants to find highly probably valid gradual item set candidates.

	In order to represent the gradual pattern mining problem as a combinatorial problem, we take the position that a gradual item set may also be referred to as a \texttt{pattern solution}. In that case all possible gradual item set solutions ($\mathcal{S}_{n}$) are admissible and can be generated based on the pheromone matrix ($\mathcal{T}_{a,j}$). Notwithstanding, all the gradual item sets in a generated solution will be evaluated and the solution updated with only valid item sets.

	First we mention that initially there exists an equal chance for any attribute $A$ (of data set $\mathcal{D}$) to either increase ($+$) or decrease ($-$) or be irrelevant ($\times$). As the algorithm acquires more knowledge about valid patterns, the possibilities of the 3 options are adjusted accordingly. We propose an artificial pheromone matrix as shown in Equation~\eqref{eqn4:p-matrix_graank}. The matrix contains knowledge about pheromone proportions of the 3 gradual options for each attribute.
		
	\begin{equation}
	\label{eqn4:p-matrix_graank}
		\mathcal{T}_{a, j} = q \times 3
	\end{equation}
	\begin{center}
		where $q$: number of attributes, $a=1,...,q$ and, $j\in \{+, -, \times\}$
	\end{center}
	
	At the beginning all artificial pheromones $p_{a, j}$ in matrix $\mathcal{T}_{a, j}$ are initialized to 1, then they are updated as follows at time interval $(t, t+1)$:
	\begin{itemize}
		\item every generated gradual item set solution is evaluated and only valid solutions are used to update the artificial pheromone matrix. Invalid solutions are stored and used to reject their supersets;
		\item in a given valid solution, each gradual item set is used to update the corresponding artificial pheromone $p_{a, j}(t)$ (where $j$ is either $+$ or $-$) by formula: $p_{a, j}(t) = p_{a, j}(t) + 1$.
	\end{itemize}

	The probability rule $\mathcal{P}_{a, j} (t)$ is given by calculating the proportions of its artificial pheromones $p_{a*}$, as shown in Equations~\eqref{eqn4:acograank_rule}. 
	
	\begin{equation}\label{eqn4:acograank_rule}
		\mathcal{P}_{a, j} (t) = \frac{p_{a, j}}{\sum _{k \in \{+, -, \times \}} p_{a, k} (t)}
	\end{equation}
		
	
	

	Algorithm~\ref{alg4:aco_graank} illustrates an implementation of ACO-GRAANK technique. As can be seen, the main steps are:
	
	\begin{enumerate}
		\item Build binary matrices of 1-itemset from input data set $\mathcal{D}$ (see Chapter~\ref{ch2:graank}).
		\item Generate gradual item sets from pheromone matrix $\tau_{aj}$ and validate each generated item set by comparing its support against the minimum support $\varrho$.
		\item If the generated pattern is valid, use gradual items to update the pheromone matrix.
		\item Repeat steps 2-3 until the algorithm starts to generate similar gradual item sets (see Convergence Proof in Section~\ref{ch4:convergence}).
	\end{enumerate}
	
	\begin{algorithm}[h!]
	\caption{ACO-GRAANK algorithm}
	\label{alg4:aco_graank}
	\SetKwInOut{Input}{\textbf{Input}}
	\SetKwInOut{Output}{\textbf{Output}}	
	\SetKwFunction{genGP}{gen-gp-soln}
	\SetKwFunction{buildBin}{bin-gp-matrices}
	\SetKwFunction{continue}{continue}
	\SetKwFunction{evalSol}{evaluate-gp}
	\SetKwFunction{updateP}{update-pheromones}

	\Input{$\mathcal{D}-$ numeric data set, $\sigma-$ minimum support threshold}
	\Output{W - winner gradual patterns}
	
	$W\leftarrow \emptyset,~L\leftarrow \emptyset, ~\tau \leftarrow 1$
	\tcc*[H]{$W-$ winner set, $L-$ loser set, $\tau -$ P-matrix}
	$\mathcal{B}^{*} \leftarrow $ \buildBin($\mathcal{D}$)
	\tcc*[H]{$\mathcal{B}^{*}-$ binary representations}
	$repeated \leftarrow 0$\;
	\While{$(repeated == 0)$}{
		$gp_{gen} \leftarrow$ \genGP($A, \tau$)
		\tcc*[H]{$A-$ attributes of $\mathcal{D}$}
		\uIf{$(gp_{gen} \in W)$}{
			$repeated \leftarrow 1$\;
		}\uElseIf{$(gp_{gen} \supseteq L) ~OR~ (gp_{gen} \subset W)$}{
			\continue\;
		}\uElse{
			$supp \leftarrow $ \evalSol($gp_{gen}, \mathcal{B}^{*}$)\;
			\uIf{$(supp \geq \sigma)$}
			{
				$W \leftarrow W \cup gp_{gen}$\;
				$\tau \leftarrow$ \updateP{$gp_{gen}, \tau$}\;
			}\uElse{
				$L \leftarrow L \cup gp_{gen}$\;
			}
		}	
	}
	\Return $W$\;	
	\end{algorithm}
	
	\clearpage
	
	\subsection{ACO for Temporal Gradual Pattern Mining}
  	\label{ch4:aco_tgraank}
	ACO-TGRAANK implementation based on ACO-GRAANK technique proposed in this Chapter. ACO-TGRAANK denotes Ant Colony Optimization Temporal-GRAANK algorithm since it modifies the T-GRAANK technique proposed in Chapter~\ref{ch3}.

	\begin{algorithm}[h!]
	\caption{$ACO-TGRAANK$ Algorithm}
	\label{alg4:acotgraank}
	\SetKwFunction{ACOgen}{ACO-GRAANK}
	\SetKwFunction{calculateSupport}{calculateSupport}
	\SetKwFunction{concordantPositions}{concordantPositions}\SetKwFunction{timeDifferences}{timeDifferences}
	\SetKwFunction{buildTriMembership}{buildTriMembership}\SetKwFunction{fuzzyFunc}{fuzzyFunc}
	\SetKwFunction{append}{append}
	\SetKwInOut{Input}{\textbf{Input}}\SetKwInOut{Result}{\textbf{Result}}

	\Input{$D^{'} -$ transformed data-set, $T_{d} -$ time differences, $minSup -$ minimum support}
	\Result{$TGP -$ set of (fuzzy) temporal gradual patterns}
	\BlankLine

	$W^{'}\leftarrow$ \ACOgen($D^{'}, minSup$) \tcc*[H]{generates gradual item-set solutions}
	\ForEach{Pattern $C$ in $W^{'}$}{
		$pos_{indices}\leftarrow$ \concordantPositions($C_{pairs}$)\;
		$time_{lags}\leftarrow$ \timeDifferences($pos_{indices},T_{d}$)\;
		$boundaries \leftarrow$ \buildTriMembership($T_{d}$)\;
		$t_{lag}\leftarrow$ \fuzzyFunc($time_{lags},boundaries$)\;
		$TGP \leftarrow$\append($C, t_{lag}$)\;
	}
	\Return $TGP$\;
	\end{algorithm}
	
	Instead of applying a breadth-first search strategy to generate temporal gradual patterns (i.e. applied in T-GRAANK approach), this algorithm applies the ACO-GRAANK approach as shown at $line~1$ in Algorithm~\ref{alg4:acotgraank}. The rest of the steps in the algorithm remain the same as the ones described in the T-GRAANK approach in Chapter~\ref{ch3:mining}.
	
	\subsection{ACO-ParaMiner Approach}
	\label{ch4:aco_paraminer}
	The principal aim of DFS-based approaches is to extract frequent closed item sets. ParaMiner extends LCM (Linear time Closed item-set Miner) proposed by \cite{Uno2004} to the case of gradual pattern mining. ParaMiner derives its efficiency from (1) a data set reduction process which reduces the size of input the algorithm has to process and (2) parallelizing the recursive function of finding the parent node of the longest FP-Tree. 
	
	Despite this, it should be remembered that for the case of gradual pattern mining, encoding of a numeric data set into a transactional data set is required (see Chapter~\ref{ch2:gradual_traversal}). In reality, the size of encoded transactional data set is almost squared the size of the original data set. For example, if the original numeric data set has $n$ tuples, the encoded data set will have $n (n -1) / 2$ tuples. This surge in the input data set size impacts the efficiency of the approach negatively.
	
	We propose the following ACO optimizations to the ParaMiner approach (described in Chapter~\ref{ch2:paraminer}): (1) in addition to reducing the transactional data set by combining similar item set transactions and removing infrequent ones, we construct a \texttt{cost matrix} of all corresponding nodes. (2) we replace the parallelized recursive function with a non-recursive heuristic function that quickly learns the longest FP-tree with the help of the \textit{cost matrix}.
	
	\clearpage
	
	To illustrate, we use data set $\mathcal{D}_{4.3}$ in Table~\ref{tab4:sample3}. In order to remove infrequent items, the transactional data set is sorted by item occurrence as shown in Table~\ref{tab4:transactional2} (b). If we set the minimum length of \texttt{tids} to 3, we remove infrequent items as illustrated in Table~\ref{tab4:trans_costmatrix} (a).
	
	\begin{table}[h!]
  		\centering
      	\caption{Sample data set $\mathcal{D}_{4.3}$}
    	\begin{tabular}{l c c c c}
      		\textbf{id} & \textbf{a} & \textbf{b} & \textbf{c} & \textbf{d} \\
      		\hline \hline
      		r1 & 5 & 30 & 43 & 97\\
      		r2 & 4 & 35 & 33 & 86\\
      		r3 & 3 & 40 & 42 & 108\\
      		r4 & 1 & 50 & 49 & 27\\
      		\bottomrule
    		\end{tabular}
    	\label{tab4:sample3}
	\end{table}

	\begin{table}[h!]
		\centering
		\caption{(a) example of a reduced transactional data set, (b) sorted items by occurrence}
		\subfloat[]{%
		\begin{tabular}{l c l}
      		\textbf{tids} & \textbf{w} & \textbf{item-sets} \\
      		\hline \hline
      		$t_{(r1, r2)}$ & 1 & $\{a^{\downarrow}, b^{\uparrow}, c^{\downarrow}, d^{\downarrow} \}$\\
      		$t_{(r1, r3)}$ & 1 & $\{a^{\downarrow}, b^{\uparrow}, c^{\downarrow}, d^{\uparrow} \}$\\
      		$t_{(r1, r4)}, t_{(r2, r4)}$ & 3 & $\{a^{\downarrow}, b^{\uparrow}, c^{\uparrow}, d^{\downarrow} \}$\\
      		$t_{(r3, r4)}$ & &\\
      		$t_{(r2, r3)}$ & 1 & $\{a^{\downarrow}, b^{\uparrow}, c^{\uparrow}, d^{\uparrow} \}$\\
      		\bottomrule
    	\end{tabular}}%
		\subfloat[]{%
		\begin{tabular}{c l}
      		\textbf{item} & \textbf{tids} \\
      		\hline \hline
      		$a^{\downarrow}$ & $\{t_{(r1, r2)}, t_{(r1, r3)}, t_{(r1, r4)}, t_{(r2, r3)},$\\
          	& $t_{(r2, r4)}, t_{(r3, r4)} \}$\\
			$b^{\uparrow}$ & $\{t_{(r1, r2)}, t_{(r1, r3)}, t_{(r1, r4)}, t_{(r2, r3)},$\\
          	& $t_{(r2, r4)}, t_{(r3, r4)} \}$\\   
      		$c^{\uparrow}$ & $\{t_{(r1, r4)}, t_{(r2, r3)}, t_{(r2, r4)}, t_{(r3, r4)} \}$\\
      		$d^{\downarrow}$ & $\{t_{(r1, r2)}, t_{(r1, r4)}, t_{(r2, r4)}, t_{(r3, r4)} \}$\\
      		$c^{\downarrow}$ & $\{t_{(r1, r2)}, t_{(r1, r3)} \}$\\
      		$d^{\uparrow}$ & $\{t_{(r1, r3)}, t_{(r2, r3)} \}$\\  
      		$a^{\uparrow}$ & $\{\emptyset \}$\\
         	$b^{\downarrow}$ & $\{\emptyset \}$\\
      		\bottomrule
    	\end{tabular}
			}%
    	\label{tab4:transactional2}
	\end{table}
	
	\begin{table}[h!]
		\centering
		\caption{(a) sorted reduced transactional data set, (b) corresponding \texttt{cost matrix}}
		\subfloat[]{%
		\begin{tabular}{c l}
      		\textbf{item} & \textbf{tids} \\
      		\hline \hline
      		$a^{\downarrow}$ & $\{t_{(r1, r2)}, t_{(r1, r3)}, t_{(r1, r4)},$\\
          	& $t_{(r2, r3)}, t_{(r2, r4)}, t_{(r3, r4)} \}$\\
			$b^{\uparrow}$ & $\{t_{(r1, r2)}, t_{(r1, r3)}, t_{(r1, r4)},$\\
          	& $t_{(r2, r3)}, t_{(r2, r4)}, t_{(r3, r4)} \}$\\   
      		$c^{\uparrow}$ & $\{t_{(r1, r4)}, t_{(r2, r3)}, t_{(r2, r4)},$\\
      		& $t_{(r3, r4)} \}$\\
      		$d^{\downarrow}$ & $\{t_{(r1, r2)}, t_{(r1, r4)}, t_{(r2, r4)},$\\
          	& $t_{(r3, r4)} \}$\\
      		\bottomrule
    	\end{tabular}
    	$~~~$ 
    	}%
		\subfloat[]{%
		\begin{tabular}{|c|c c c c|}
			\hline
			$\Rsh$ & r1 & r2 & r3 & r4\\
			\hline
			r1 & 1 & $1/4$ & $1/3$ & $1/5$\\
			r2 & 1 & 1 & $1/4$ & $1/5$\\
			r3 & 1 & 1 & 1 & $1/5$\\
			r4 & 1 & 1 & 1 & 1\\
			\hline
			\end{tabular}}%
    	\label{tab4:trans_costmatrix}
	\end{table}
	
	\newpage
	We define a \texttt{cost matrix}: $\mathcal{C}_{i,j} = n\times n$ (where $n$ is number of tuples in the original numeric data set i.e. Table~\ref{tab4:sample3}) and initialize it to 1. Using the sorted and reduced transactional data set in Table~\ref{tab4:trans_costmatrix} (a), we update elements of \texttt{cost matrix} that correspond to the inverse of occurrence count of \textbf{tids} as shown in Equation~\eqref{eqn4:c-matrix}. Table~\ref{tab4:trans_costmatrix} (b) illustrates the \texttt{cost matrix} of Table~\ref{tab4:trans_costmatrix} (a). In this case we observe that \textbf{tids} with more occurrences have least costs; therefore, they are better candidates for parent nodes.
	
	\begin{equation}\label{eqn4:c-matrix}
		\mathcal{C}_{i, j} = \frac{1}{1 + \sum (ri, rj)_{count}}
	\end{equation}
	\begin{center}
		where $(ri, rj)_{count}$ is the occurrence count of tid pair $(ri, rj)$.
	\end{center}
	We define an artificial pheromone matrix as shown in Equation~\eqref{eqn4:p-matrix_paraminer}. This matrix contains knowledge about the pheromone proportions of every node of the data set.
	
	\begin{equation}\label{eqn4:p-matrix_paraminer}
		\Gamma_{i, j} = n \times n
	\end{equation}
	\begin{center}
		where n: number of tuples in numeric data set
	\end{center}
		
	At the beginning all artificial pheromones $p_{i, j}$ in matrix $\Gamma_{i, j}$ are initialized to 1, then they are updated as follows at time interval $(t, t+1)$:
	
	\begin{itemize}
		\item every generated node, is used to retrieve the corresponding attributes from the sorted reduced transactional data set. A set intersection of the \textbf{tids} of all these attributes provides an FP-Tree whose length is tested against the specified threshold;
		\item if the length of the FP-tree surpasses the specified threshold, the pheromones corresponding to the \textbf{tids} are incremented by 1: $p_{i,j} = p_{i,j}(t) + 1$;
		\item if the length of the FP-Tree falls below the specified threshold, the pheromones corresponding to the \textbf{tids} are evaporated by a factor $\epsilon$: $(1 - \epsilon) p_{i, j}(t)$.
	\end{itemize}
	
	We propose a probabilistic rule that allows us to learn a parent node of the longest FP-Tree in a non-recursive manner. This rule is shown in Equation~\eqref{eqn4:acoparaminer_rule}. 
	
	\begin{equation}\label{eqn4:acoparaminer_rule}
		\mathcal{P}_{i, j}(t) = \frac{p_{i, j} (1 / \mathcal{C}_{i, j})}{\sum _{k=1} ^{n} p_{i, j} ^{k} (t) (1 / \mathcal{C}_{i, j} ^{k})}
	\end{equation}
	
	\clearpage
	
	Algorithm~\ref{alg4:aco_paraminer} illustrates an implementation of ACO-Paraminer technique. As can be seen, the main steps are:
	
	\begin{enumerate}
		\item Encode a numeric data set $\mathcal{D}$ into a transactional data set and reduce it by combining similar item sets and removing infrequent item sets. Use the reduced transactional data set $\mathcal{D}_{red}$ to construct a Cost Matrix $\mathcal{C}$.
		\item Generate nodes that have high probability of being parent nodes using the Cost Matrix and Pheromone Matrix $\Gamma_{i,j}$. 
		\item For each generated node, retrieve corresponding attributes from $\mathcal{D}_{red}$ and determine the intersection of their \texttt{tids}. If the length of the resulting intersection surpasses the minimum length threshold $\varsigma$, then update the pheromones.
		\item Repeat steps 2-3 until the algorithm starts to generate similar nodes.
	\end{enumerate}
		
	\begin{algorithm}[h!]
	\caption{ACO-ParaMiner algorithm}
	\label{alg4:aco_paraminer}
	\SetKwInOut{Input}{\textbf{Input}}
	\SetKwInOut{Output}{\textbf{Output}}	
	\SetKwFunction{encData}{encode-data}
	\SetKwFunction{reduce}{reduce-data}
	\SetKwFunction{genNode}{gen-node}
	\SetKwFunction{getTIDs}{get-tids}
	\SetKwFunction{intersect}{intersection}
	\SetKwFunction{updateP}{update-pheromones}
	\SetKwFunction{evapoP}{evaporate-pheromones}

	\Input{$\mathcal{D}-$ numeric data set, $\varsigma-$ minimum length threshold}
	\Output{$GP-$ gradual patterns}
	
	$N\leftarrow \emptyset,~\mathcal{C}\leftarrow 1, ~\Gamma \leftarrow 1$
	\tcc*[H]{$N-$ node set, $\mathcal{C}-$  C-matrix, $\Gamma-$ P-matrix}
	$GP \leftarrow \emptyset, ~repeated \leftarrow 0$\;
	$\mathcal{D}_{enc} \leftarrow$ \encData($\mathcal{D}$)
	\tcc*[H]{$\mathcal{D}_{enc}-$ encoded data set}
	$\mathcal{D}_{red}, \mathcal{C} \leftarrow$ \reduce($\mathcal{D}_{enc}, \mathcal{C}$)
	\tcc*[H]{$\mathcal{D}_{red}-$ reduced data set}
	
	\While{$(repeated == 0)$}{
		$n_{i,j} \leftarrow$ \genNode($\mathcal{C}, \Gamma$)\;
		\uIf{$(n_{i,j} \not \in N)$}{
			$T^{*}, gp \leftarrow$ \getTIDs($n_{i, j}, \mathcal{D}_{red}$)
			\tcc*[H]{$T^{*}-$ TIDs, $gp-$ gradual pattern}
			$len \leftarrow$ \intersect($T^{*}$)\;
			\uIf{$(len \geq \varsigma)$}
			{
				$N \leftarrow N \cup n_{i, j}$\;
				$GP \leftarrow GP \cup gp$\;
				$\Gamma \leftarrow$ \updateP($T^{*}, \Gamma$)\;
			}\uElse{
				$\Gamma \leftarrow$ \evapoP($T^{*}, \Gamma, \epsilon$)\;
			}
		}\uElse{
			$repeated \leftarrow 1$\;
		}
	}

	\Return $GP$\;
	\end{algorithm}
	
	\clearpage
	\newpage
	\section{Experiments}
	\label{ch4:experiments}
	In this section, we present an experimental study of computational performance of our algorithms. We implemented BFS-based GRAANK (as described in \cite{Laurent2009}) and ACO-GRAANK; and DFS-based ParaMiner (as described in \cite{Negrevergne2014}) and ACO-ParaMiner algorithms in \texttt{Python} language. All experiments were conducted on a (High Performance Computing) HPC platform \textbf{Meso@LR}\footnote{\url{https://meso-lr.umontpellier.fr}}. We used one node made up of 14 cores and 128GB of RAM.
	
	\subsection{Source Code}
	\label{ch4:source_code}
	The \texttt{Python} source code of all the 4 algorithms is available at: \begin{footnotesize}\url{https://github.com/owuordickson/ant-colony-gp.git}\end{footnotesize}. Since \cite{Negrevergne2014} does not provide a \texttt{Python} implementation of ParaMiner, we extended a \texttt{Python} source code of LCM \begin{footnotesize}(\url{https://github.com/scikit-mine.git})\end{footnotesize} to implement our ParaMiner as described by \cite{Negrevergne2014}.

	\subsection{Data Set Description}
	\label{ch4:dataset}

	\begin{table}[h!]
	\small
	\centering
	\caption{Experiment data sets}
	\begin{tabular}{|c|c|c|c|c|}
		\hline
		Data set & $\#$tuples & $\#$attributes & Timestamped & Domain\\
		\hline \hline
		Breast Cancer (B\&C) & 116 & 10 & No & Medical\\
		Cargo 2000 (C2K) & 3942 & 98 & No & Transport\\
		Directio (Buoys) & 948000 & 21 & Yes & Coastline\\
		Power Consumption (UCI) & 2075259 & 9 & Yes & Electrical\\
		\hline
	\end{tabular}
	\label{tab4:dataset}
	\end{table}
	
	Table~\ref{tab4:dataset} shows the features of the numerical data sets used in the experiments for evaluating the computational performance of the algorithms. The `Breast Cancer' data set, obtained from \texttt{UCI Machine Learning Repository} \cite{Patricio2018}, is composed of 10 quantitative predictors and binary variable indicating the presence or absence of breast cancer. The predictors are recordings of anthropometric data gathered from the routine blood analysis of 116 participants (of whom 64 have breast cancer and 52 are healthy).
	
	The `Cargo 2000' data set,  obtained from \texttt{UCI Machine Learning Repository} \cite{Metzger2015}, describes 98 tracking and tracing events that span 5 months of transport logistics execution. The `Power Consumption' data set, obtained from \texttt{UCI Machine Learning Repository} \cite{Dua2019}, describes the electric power consumption in one household (located in Sceaux, France) in terms of active power, voltage and global intensity with a one-minute sampling rate between 2006 and 2010.
	
	The `Directio' data set is one of 4 data sets obtained from OREMES’s data portal\footnote{\url{https://data.oreme.org}} that recorded swell sensor signals of 4 buoys near the coast of the Languedoc-Roussillon region in France between 2012 and 2019 \cite{Bouchette2019}. 
	\clearpage
		
	\subsection{Experiment Results}
	\label{ch4:results}
	In this section, we present results of our experimental study on the 4 data sets described in Section~\ref{ch4:dataset} using implemented algorithms: GRAANK, ParaMiner, ACO-GRAANK, ACO-ParaMiner. The results reveal behaviors of the 4 algorithms when applied on different data sets that vary in tuple and attribute size. We use these results to compare computational performances of the 4 algorithms in Section~\ref{ch4:comparative_1} and extracted gradual patterns in Section~\ref{ch4:consistent_gps}. We discuss the results in Section~\ref{ch4:discussion}. All experiment results can be obtained from:\begin{footnotesize} \url{https://github.com/owuordickson/meso-hpc-lr/tree/master/results/gps/14cores}\end{footnotesize}.
	
	\subsubsection{Comparative Computational Experiments}
	\label{ch4:comparative_1}
	This experiment compares the computational runtime and memory usage of algorithms ACO-GRAANK, ACO-ParaMiner, GRAANK, ParaMiner when applied on data sets: B\&C, C2K, Buoys, UCI.
		
	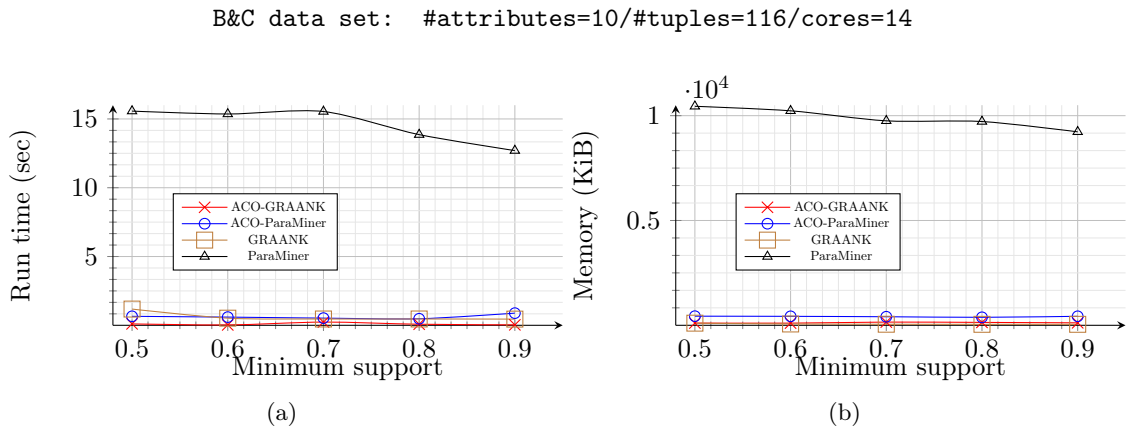
\begin{figure}[h!]
		\centering
		\small
		\texttt{B\&C data set: \#attributes=10/\#tuples=116/cores=14}\par\medskip
  		\subfloat[]{%
		\begin{tikzpicture}
  			\begin{axis}[
  				height=4.5cm, width=7.5cm,
  				grid=both,
  				grid style={line width=.1pt, draw=gray!20},
    			major grid style={line width=.2pt, draw=gray!50},
  				axis lines=middle,
  				minor tick num=5,
    			ymin=0, ymax=16,
    			xmin=0.48, xmax=0.95,
    			xtick={0.5, 0.6, 0.7, 0.8, 0.9},
   				xlabel=Minimum support, 
   				ylabel=Run time (sec),
  				xlabel style={at={(axis description cs:0.5,-0.1)},anchor=north},
  				ylabel style={at={(axis description cs:-0.15,0.5)},rotate=90, anchor=south},
 	 			legend style={at={(0.5,0.6)}, nodes={scale=0.5}},
 				]
			\addplot[smooth, mark=x, mark size=3pt, color=red] table[x=Support, y=ACO_GRAANK]{b_c_runtime.dat};
  	  		\addlegendentry{ACO-GRAANK}

  			\addplot[smooth, mark=o, color=blue] table[x=Support, y=ACO_LCM]{b_c_runtime.dat};
			\addlegendentry{ACO-ParaMiner}
			
			\addplot[smooth, mark=square, mark size=3pt, color=brown] table[x=Support, y=GRAANK]{b_c_runtime.dat};
  	  		\addlegendentry{GRAANK}

  			\addplot[smooth, mark=triangle, color=black] table[x=Support, y=LCM]{b_c_runtime.dat};
			\addlegendentry{ParaMiner}
  			\end{axis}
  		\end{tikzpicture}}%
  		\subfloat[]{%
		\begin{tikzpicture}
  			\begin{axis}[
  				height=4.5cm, width=7.5cm,
  				grid=both,
  				grid style={line width=.1pt, draw=gray!20},
    			major grid style={line width=.2pt, draw=gray!50},
  				axis lines=middle,
  				minor tick num=5,
    			ymin=0, ymax=10500,
    			xmin=0.48, xmax=0.95,
    			xtick={0.5, 0.6, 0.7, 0.8, 0.9},
   				xlabel=Minimum support, 
   				ylabel=Memory (KiB),
  				xlabel style={at={(axis description cs:0.5,-0.1)},anchor=north},
  				ylabel style={at={(axis description cs:-0.15,0.5)},rotate=90, anchor=south},
  	 	 		legend style={at={(0.5,0.6)}, nodes={scale=0.5}},
 				]

			\addplot[smooth, mark=x, mark size=3pt, color=red] table[x=Support, y=ACO_GRAANK]{b_c_memory.dat};
  	  		\addlegendentry{ACO-GRAANK}

  			\addplot[smooth, mark=o, color=blue] table[x=Support, y=ACO_LCM]{b_c_memory.dat};
			\addlegendentry{ACO-ParaMiner}
			
			\addplot[smooth, mark=square, mark size=3pt, color=brown] table[x=Support, y=GRAANK]{b_c_memory.dat};
  	  		\addlegendentry{GRAANK}

  			\addplot[smooth, mark=triangle, color=black] table[x=Support, y=LCM]{b_c_memory.dat};
			\addlegendentry{ParaMiner}
  			\end{axis}
  		\end{tikzpicture}}%
    	\caption{Breast Cancer (B\&C) data set: (a) plot of run time against minimum support threshold and, (b) plot of memory usage against minimum support.}
  		\label{fig4:exp_comparison1}
  	\end{figure}
  	
  	Figure~\ref{fig4:exp_comparison1} (a) shows the runtime and, Figure~\ref{fig4:exp_comparison1} (b) shows the memory usage of these 4 algorithms when applied on B\&C data set. We observe that ParaMiner (black triangular curve) has the slowest runtime followed (in order) by GRAANK (brown square curve), ACO-ParaMiner (blue circle curve) and, ACO-GRAANK (red cross curve). Again ParaMiner has the highest memory usage followed (in order) by ACO-ParaMiner, ACO-GRAANK and, GRAANK.

  	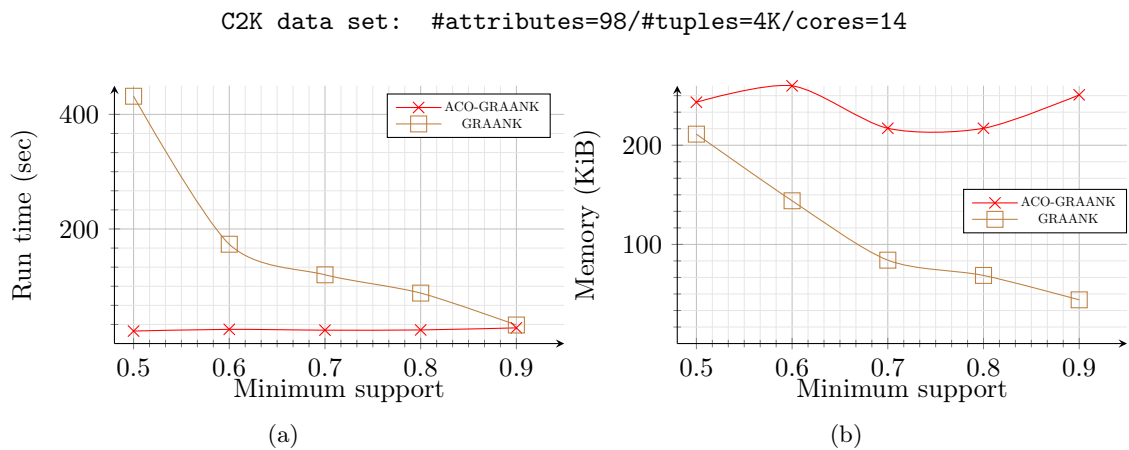
\begin{figure}[h!]
		\centering
		\small
		\texttt{C2K data set: \#attributes=98/\#tuples=4K/cores=14}\par\medskip
  		\subfloat[]{%
		\begin{tikzpicture}
  			\begin{axis}[
  				height=5cm, width=7.5cm,
  				grid=both,
  				grid style={line width=.1pt, draw=gray!20},
    			major grid style={line width=.2pt, draw=gray!50},
  				axis lines=middle,
  				minor tick num=5,
    			ymin=0, ymax=450,
    			xmin=0.48, xmax=0.95,
    			xtick={0.5, 0.6, 0.7, 0.8, 0.9},
   				xlabel=Minimum support, 
   				ylabel=Run time (sec),
  				xlabel style={at={(axis description cs:0.5,-0.1)},anchor=north},
  				ylabel style={at={(axis description cs:-0.15,0.5)},rotate=90, anchor=south},
 	 			legend style={nodes={scale=0.5}},
 	 			legend pos=north east
 				]

			\addplot[smooth, mark=x, mark size=3pt, color=red] table[x=Support, y=ACO_GRAANK_L]{c2k_runtime.dat};
  	  		\addlegendentry{ACO-GRAANK}
			
			\addplot[smooth, mark=square, mark size=3pt, color=brown] table[x=Support, y=GRAANK_L]{c2k_runtime.dat};
  	  		\addlegendentry{GRAANK}
  			\end{axis}
  		\end{tikzpicture}}%
  		\subfloat[]{%
		\begin{tikzpicture}
  			\begin{axis}[
  				height=5cm, width=7.5cm,
  				grid=both,
  				grid style={line width=.1pt, draw=gray!20},
    			major grid style={line width=.2pt, draw=gray!50},
  				axis lines=middle,
  				minor tick num=5,
    			ymin=0, ymax=260,
    			xmin=0.48, xmax=0.95,
    			xtick={0.5, 0.6, 0.7, 0.8, 0.9},
   				xlabel=Minimum support, 
   				ylabel=Memory (KiB),
  				xlabel style={at={(axis description cs:0.5,-0.1)},anchor=north},
  				ylabel style={at={(axis description cs:-0.15,0.5)},rotate=90, anchor=south},
   	 	 		legend style={at={(1,0.6)}, nodes={scale=0.5}},
 				]

			\addplot[smooth, mark=x, mark size=3pt, color=red] table[x=Support, y=ACO_GRAANK_L]{c2k_memory.dat};
  	  		\addlegendentry{ACO-GRAANK}
			
			\addplot[smooth, mark=square, mark size=3pt, color=brown] table[x=Support, y=GRAANK_L]{c2k_memory.dat};
  	  		\addlegendentry{GRAANK}
			\end{axis}
  		\end{tikzpicture}}%
    	\caption{Cargo 2000 (C2K) data set: (a) plot of run time against minimum support threshold and, (b) plot of memory usage against minimum support.}
  		\label{fig4:exp_comparison2a}
  	\end{figure}
  	
\begin{figure}[h!]
		\centering
		\small
		\texttt{C2K data set: \#attributes=98/\#tuples=200/cores=14}\par\medskip
  		\subfloat[]{%
		\begin{tikzpicture}
  			\begin{axis}[
  				height=5cm, width=7.5cm,
  				grid=both,
  				grid style={line width=.1pt, draw=gray!20},
    			major grid style={line width=.2pt, draw=gray!50},
  				axis lines=middle,
  				minor tick num=5,
    			ymin=0, ymax=60,
    			xmin=0.48, xmax=0.95,
    			xtick={0.5, 0.6, 0.7, 0.8, 0.9},
   				xlabel=Minimum support, 
   				ylabel=Run time (sec),
  				xlabel style={at={(axis description cs:0.5,-0.1)},anchor=north},
  				ylabel style={at={(axis description cs:-0.15,0.5)},rotate=90, anchor=south},
 	 			legend style={nodes={scale=0.5}},
 	 			legend pos=north east
 				]

			\addplot[smooth, mark=x, mark size=3pt, color=red] table[x=Support, y=ACO_GRAANK]{c2k_runtime.dat};
  	  		\addlegendentry{ACO-GRAANK}

  			\addplot[smooth, mark=o, color=blue] table[x=Support, y=ACO_LCM]{c2k_runtime.dat};
			\addlegendentry{ACO-ParaMiner}
			
			\addplot[smooth, mark=square, mark size=3pt, color=brown] table[x=Support, y=GRAANK]{c2k_runtime.dat};
  	  		\addlegendentry{GRAANK}

  			\end{axis}
  		\end{tikzpicture}}%
  		\subfloat[]{%
		\begin{tikzpicture}
  			\begin{axis}[
  				height=5cm, width=7.5cm,
  				grid=both,
  				grid style={line width=.1pt, draw=gray!20},
    			major grid style={line width=.2pt, draw=gray!50},
  				axis lines=middle,
  				minor tick num=5,
    			ymin=0, ymax=900,
    			xmin=0.48, xmax=0.95,
    			xtick={0.5, 0.6, 0.7, 0.8, 0.9},
   				xlabel=Minimum support, 
   				ylabel=Memory (KiB),
  				xlabel style={at={(axis description cs:0.5,-0.1)},anchor=north},
  				ylabel style={at={(axis description cs:-0.15,0.5)},rotate=90, anchor=south},
   	 	 		legend style={at={(0.5,0.7)}, nodes={scale=0.5}},
 				]

			\addplot[smooth, mark=x, mark size=3pt, color=red] table[x=Support, y=ACO_GRAANK]{c2k_memory.dat};
  	  		\addlegendentry{ACO-GRAANK}

  			\addplot[smooth, mark=o, color=blue] table[x=Support, y=ACO_LCM]{c2k_memory.dat};
			\addlegendentry{ACO-ParaMiner}
			
			\addplot[smooth, mark=square, mark size=3pt, color=brown] table[x=Support, y=GRAANK]{c2k_memory.dat};
  	  		\addlegendentry{GRAANK}

			\end{axis}
  		\end{tikzpicture}}%
    	\caption{Cargo 2000 (C2K) data set: (a) plot of run time against minimum support threshold and, (b) plot of memory usage against minimum support.}
  		\label{fig4:exp_comparison2b}
  	\end{figure}
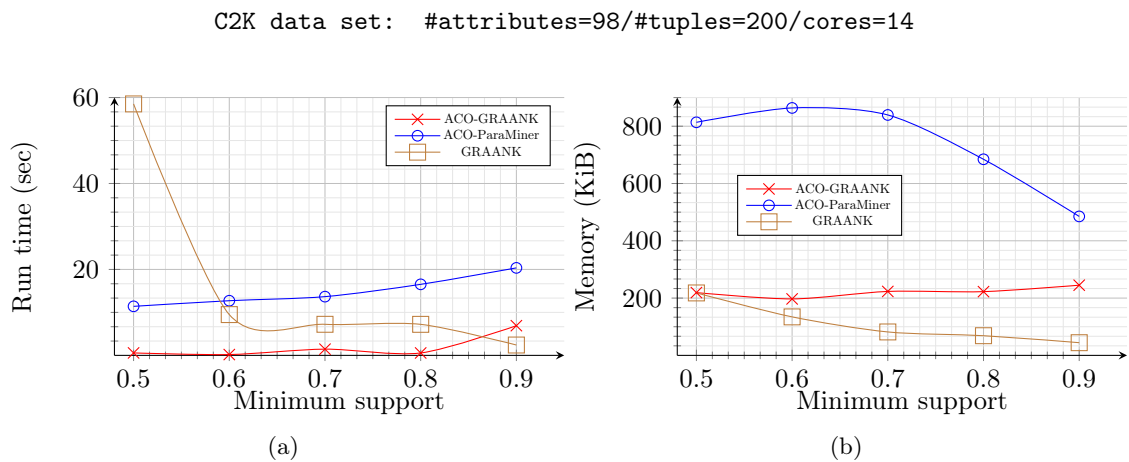

  	Figure~\ref{fig4:exp_comparison2a} (a) and (b) show runtime and memory usage of ACO-GRAANK and GRAANK on C2K data set. We observe that GRAANK has a slower runtime than ACO-GRAANK that reduces as the minimum support increases but, it has a lower memory usage than ACO-GRAANK. Using C2K data set with 3942 tuples, ParaMiner and ACO-ParaMiner yield \textit{Memory Error} outputs; so, we use C2K data set with 200 tuples as shown in Figure~\ref{fig4:exp_comparison2b}. 
  	
  	In Figure~\ref{fig4:exp_comparison2b} (a), ParaMiner has a runtime greater than 72 hours and it does not appear in the graph plot. GRAANK has the slowest runtime at support 0.5 which reduces significantly below ACO-ParaMiner's runtime and, ACO-GRAANK has the fastest runtime. In Figure~\ref{fig4:exp_comparison2b} (b), ACO-ParaMiner has the slowest runtime followed (in order) by ACO-GRAANK and GRAANK.
  	
  	Figure~\ref{fig4:exp_comparison3a} (a) and (b) show the runtime and memory usage of ACO-GRAANK and GRAANK algorithms on Buoys data set with 15402 tuples. For supports 0.5 and 0.6 GRAANK yields a \textit{Out of Memory Error} output and no runtime. For ACO-GRAANK runtime and memory usage slightly increase with support. 
  	
	\pgfplotsset{select coords between index/.style 2 args={
    	x filter/.code={
        	\ifnum\coordindex<#1\def\pgfmathresult{}\fi
        	\ifnum\coordindex>#2\def\pgfmathresult{}\fi
    	}
	}}
  	\begin{figure}[h!]
		\centering
		\small
		\texttt{Buoys data set: \#attributes=21/\#tuples=15K/cores=14}\par\medskip
  		\subfloat[]{%
		\begin{tikzpicture}
  			\begin{axis}[
  				height=5cm, width=7.5cm,
  				grid=both,
  				grid style={line width=.1pt, draw=gray!20},
    			major grid style={line width=.2pt, draw=gray!50},
  				axis lines=middle,
  				minor tick num=5,
    			ymin=0, ymax=560,
    			xmin=0.48, xmax=0.95,
    			xtick={0.5, 0.6, 0.7, 0.8, 0.9},
   				xlabel=Minimum support, 
   				ylabel=Run time (sec),
  				xlabel style={at={(axis description cs:0.5,-0.1)},anchor=north},
  				ylabel style={at={(axis description cs:-0.15,0.5)},rotate=90, anchor=south},
  				legend style={at={(0.5,0.6)}, nodes={scale=0.5}},
 				]

			\addplot[smooth, mark=x, mark size=3pt, color=red] table[x=Support, y=ACO_GRAANK_L]{buoys_runtime.dat};
  	  		\addlegendentry{ACO-GRAANK}
			
			\addplot[smooth, mark=square, mark size=3pt, color=brown, select coords between index={2}{4}] table[x=Support, y=GRAANK_L]{buoys_runtime.dat};
  	  		\addlegendentry{GRAANK}
  			\end{axis}
  		\end{tikzpicture}}%
  		\subfloat[]{%
		\begin{tikzpicture}
  			\begin{axis}[
  				height=5cm, width=7.5cm,
  				grid=both,
  				grid style={line width=.1pt, draw=gray!20},
    			major grid style={line width=.2pt, draw=gray!50},
  				axis lines=middle,
  				minor tick num=5,
    			ymin=0, ymax=250,
    			xmin=0.48, xmax=0.95,
    			xtick={0.5, 0.6, 0.7, 0.8, 0.9},
   				xlabel=Minimum support, 
   				ylabel=Memory (KiB),
  				xlabel style={at={(axis description cs:0.5,-0.1)},anchor=north},
  				ylabel style={at={(axis description cs:-0.15,0.5)},rotate=90, anchor=south},
 	 			legend style={nodes={scale=0.5}},
 	 			legend pos=north east
 				]

			\addplot[smooth, mark=x, mark size=3pt, color=red] table[x=Support, y=ACO_GRAANK_L]{buoys_memory.dat};
  	  		\addlegendentry{ACO-GRAANK}
			
			\addplot[mark=square, mark size=3pt, color=brown, select coords between index={1}{4}] table[x=Support, y=GRAANK_L]{buoys_memory.dat};
  	  		\addlegendentry{GRAANK}

  			\end{axis}
  		\end{tikzpicture}}%
    	\caption{Directio (Buoys) data set: (a) plot of run time against minimum support threshold and, (b) plot of memory usage against minimum support.}
  		\label{fig4:exp_comparison3a}
  	\end{figure}
  	
  	ParaMiner and ACO-ParaMiner both yield \textit{Memory Error} outputs on Buoys data set with 15402 tuples; so, we use a data set with 200 tuples as shown in Figure~\ref{fig4:exp_comparison3b}. In Figure~\ref{fig4:exp_comparison3b} (a) and (b), ParaMiner has the slowest runtime and memory usage while ACO-GRAANK, GRAANK and ACO-ParaMiner have relatively low runtimes and memory usages. 
  	
  	\begin{figure}[h!]
		\centering
		\small
		\texttt{Buoys data set: \#attributes=21/\#tuples=200/cores=14}\par\medskip
  		\subfloat[]{%
		\begin{tikzpicture}
  			\begin{axis}[
  				height=5cm, width=7.5cm,
  				grid=both,
  				grid style={line width=.1pt, draw=gray!20},
    			major grid style={line width=.2pt, draw=gray!50},
  				axis lines=middle,
  				minor tick num=5,
    			ymin=0, ymax=350,
    			xmin=0.48, xmax=0.95,
    			xtick={0.5, 0.6, 0.7, 0.8, 0.9},
   				xlabel=Minimum support, 
   				ylabel=Run time (sec),
  				xlabel style={at={(axis description cs:0.5,-0.1)},anchor=north},
  				ylabel style={at={(axis description cs:-0.15,0.5)},rotate=90, anchor=south},
 	 			legend style={nodes={scale=0.5}},
 	 			legend pos=north east
 				]

			\addplot[smooth, mark=x, mark size=3pt, color=red] table[x=Support, y=ACO_GRAANK]{buoys_runtime.dat};
  	  		\addlegendentry{ACO-GRAANK}

  			\addplot[smooth, mark=o, color=blue] table[x=Support, y=ACO_LCM]{buoys_runtime.dat};
			\addlegendentry{ACO-ParaMiner}
			
			\addplot[smooth, mark=square, mark size=3pt, color=brown] table[x=Support, y=GRAANK]{buoys_runtime.dat};
  	  		\addlegendentry{GRAANK}

  			\addplot[smooth, mark=triangle, color=black] table[x=Support, y=LCM]{buoys_runtime.dat};
			\addlegendentry{ParaMiner}
  			\end{axis}
  		\end{tikzpicture}}%
  		\subfloat[]{%
		\begin{tikzpicture}
  			\begin{axis}[
  				height=5cm, width=7.5cm,
  				grid=both,
  				grid style={line width=.1pt, draw=gray!20},
    			major grid style={line width=.2pt, draw=gray!50},
  				axis lines=middle,
  				minor tick num=5,
    			ymin=0, ymax=56000,
    			xmin=0.48, xmax=0.95,
    			xtick={0.5, 0.6, 0.7, 0.8, 0.9},
   				xlabel=Minimum support, 
   				ylabel=Memory (KiB),
  				xlabel style={at={(axis description cs:0.5,-0.1)},anchor=north},
  				ylabel style={at={(axis description cs:-0.15,0.5)},rotate=90, anchor=south},
   	 	 		legend style={at={(0.5,0.6)}, nodes={scale=0.5}},
 				]

			\addplot[smooth, mark=x, mark size=3pt, color=red] table[x=Support, y=ACO_GRAANK]{buoys_memory.dat};
  	  		\addlegendentry{ACO-GRAANK}

  			\addplot[smooth, mark=o, color=blue] table[x=Support, y=ACO_LCM]{buoys_memory.dat};
			\addlegendentry{ACO-ParaMiner}
			
			\addplot[smooth, mark=square, mark size=3pt, color=brown] table[x=Support, y=GRAANK]{buoys_memory.dat};
  	  		\addlegendentry{GRAANK}

  			\addplot[smooth, mark=triangle, color=black] table[x=Support, y=LCM]{buoys_memory.dat};
			\addlegendentry{ParaMiner}
  			\end{axis}
  		\end{tikzpicture}}%
    	\caption{Directio (Buoys) data set: (a) plot of run time against minimum support threshold and, (b) plot of memory usage against minimum support.}
  		\label{fig4:exp_comparison3b}
  	\end{figure}
  	
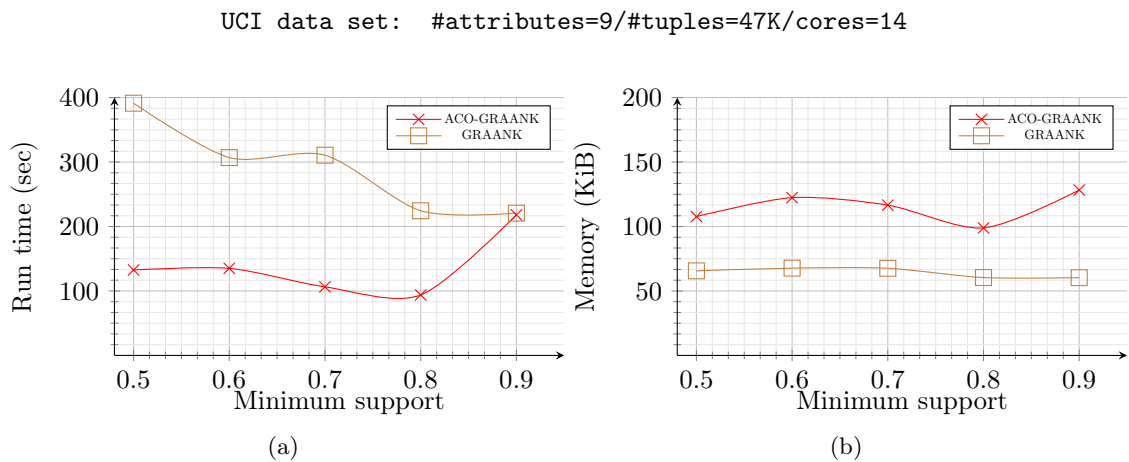
\begin{figure}[h!]
		\centering
		\small
		\texttt{UCI data set: \#attributes=9/\#tuples=47K/cores=14}\par\medskip
  		\subfloat[]{%
		\begin{tikzpicture}
  			\begin{axis}[
  				height=5cm, width=7.5cm,
  				grid=both,
  				grid style={line width=.1pt, draw=gray!20},
    			major grid style={line width=.2pt, draw=gray!50},
  				axis lines=middle,
  				minor tick num=5,
    			ymin=0, ymax=400,
    			xmin=0.48, xmax=0.95,
    			xtick={0.5, 0.6, 0.7, 0.8, 0.9},
   				xlabel=Minimum support, 
   				ylabel=Run time (sec),
  				xlabel style={at={(axis description cs:0.5,-0.1)},anchor=north},
  				ylabel style={at={(axis description cs:-0.15,0.5)},rotate=90, anchor=south},
 	 			legend style={nodes={scale=0.5}},
 	 			legend pos=north east
 				]

			\addplot[smooth, mark=x, mark size=3pt, color=red] table[x=Support, y=ACO_GRAANK_L]{uci_runtime.dat};
  	  		\addlegendentry{ACO-GRAANK}
			
			\addplot[smooth, mark=square, mark size=3pt, color=brown] table[x=Support, y=GRAANK_L]{uci_runtime.dat};
  	  		\addlegendentry{GRAANK}
			\end{axis}
  		\end{tikzpicture}}%
  		\subfloat[]{%
		\begin{tikzpicture}
  			\begin{axis}[
  				height=5cm, width=7.5cm,
  				grid=both,
  				grid style={line width=.1pt, draw=gray!20},
    			major grid style={line width=.2pt, draw=gray!50},
  				axis lines=middle,
  				minor tick num=5,
    			ymin=0, ymax=200,
    			xmin=0.48, xmax=0.95,
    			xtick={0.5, 0.6, 0.7, 0.8, 0.9},
   				xlabel=Minimum support, 
   				ylabel=Memory (KiB),
  				xlabel style={at={(axis description cs:0.5,-0.1)},anchor=north},
  				ylabel style={at={(axis description cs:-0.15,0.5)},rotate=90, anchor=south},
 	 			legend style={nodes={scale=0.5}},
 	 			legend pos=north east
 				]

			\addplot[smooth, mark=x, mark size=3pt, color=red] table[x=Support, y=ACO_GRAANK_L]{uci_memory.dat};
  	  		\addlegendentry{ACO-GRAANK}
  	  					
			\addplot[smooth, mark=square, mark size=3pt, color=brown] table[x=Support, y=GRAANK_L]{uci_memory.dat};
  	  		\addlegendentry{GRAANK}
  			\end{axis}
  		\end{tikzpicture}}%
    	\caption{Power consumption (UCI) data set: (a) plot of run time against minimum support and, (b) plot of memory usage against minimum support.}
  		\label{fig4:exp_comparison4a}
  	\end{figure}

\begin{figure}[h!]
		\centering
		\small
		\texttt{UCI data set: \#attributes=9/\#tuples=1K/cores=14}\par\medskip
  		\subfloat[]{%
		\begin{tikzpicture}
  			\begin{axis}[
  				height=5cm, width=7.5cm,
  				grid=both,
  				grid style={line width=.1pt, draw=gray!20},
    			major grid style={line width=.2pt, draw=gray!50},
  				axis lines=middle,
  				minor tick num=5,
    			ymin=0, ymax=300,
    			xmin=0.48, xmax=0.95,
    			xtick={0.5, 0.6, 0.7, 0.8, 0.9},
   				xlabel=Minimum support, 
   				ylabel=Run time (sec),
  				xlabel style={at={(axis description cs:0.5,-0.1)},anchor=north},
  				ylabel style={at={(axis description cs:-0.15,0.5)},rotate=90, anchor=south},
 	 			legend style={at={(0.5,0.6)}, nodes={scale=0.5}},
 				]

			\addplot[smooth, mark=x, mark size=3pt, color=red] table[x=Support, y=ACO_GRAANK]{uci_runtime.dat};
  	  		\addlegendentry{ACO-GRAANK}

  			\addplot[smooth, mark=o, color=blue] table[x=Support, y=ACO_LCM]{uci_runtime.dat};
			\addlegendentry{ACO-ParaMiner}
			
			\addplot[smooth, mark=square, mark size=3pt, color=brown] table[x=Support, y=GRAANK]{uci_runtime.dat};
  	  		\addlegendentry{GRAANK}

  			\addplot[smooth, mark=triangle, color=black] table[x=Support, y=LCM]{uci_runtime.dat};
			\addlegendentry{ParaMiner}
			\end{axis}
  		\end{tikzpicture}}%
  		\subfloat[]{%
		\begin{tikzpicture}
  			\begin{axis}[
  				height=5cm, width=7.5cm,
  				grid=both,
  				grid style={line width=.1pt, draw=gray!20},
    			major grid style={line width=.2pt, draw=gray!50},
  				axis lines=middle,
  				minor tick num=5,
    			ymin=0, ymax=310000,
    			xmin=0.48, xmax=0.95,
    			xtick={0.5, 0.6, 0.7, 0.8, 0.9},
   				xlabel=Minimum support, 
   				ylabel=Memory (KiB),
  				xlabel style={at={(axis description cs:0.5,-0.1)},anchor=north},
  				ylabel style={at={(axis description cs:-0.15,0.5)},rotate=90, anchor=south},
 	   	 	 	legend style={at={(0.5,0.6)}, nodes={scale=0.5}},
 				]

			\addplot[smooth, mark=x, mark size=3pt, color=red] table[x=Support, y=ACO_GRAANK]{uci_memory.dat};
  	  		\addlegendentry{ACO-GRAANK}

  			\addplot[smooth, mark=o, color=blue] table[x=Support, y=ACO_LCM]{uci_memory.dat};
			\addlegendentry{ACO-ParaMiner}
			
			\addplot[smooth, mark=square, mark size=3pt, color=brown] table[x=Support, y=GRAANK]{uci_memory.dat};
  	  		\addlegendentry{GRAANK}

  			\addplot[smooth, mark=triangle, color=black] table[x=Support, y=LCM]{uci_memory.dat};
			\addlegendentry{ParaMiner}
  			\end{axis}
  		\end{tikzpicture}}%
    	\caption{Power consumption (UCI) data set: (a) plot of run time against minimum support and, (b) plot of memory usage against minimum support.}
  		\label{fig4:exp_comparison4b}
  	\end{figure}
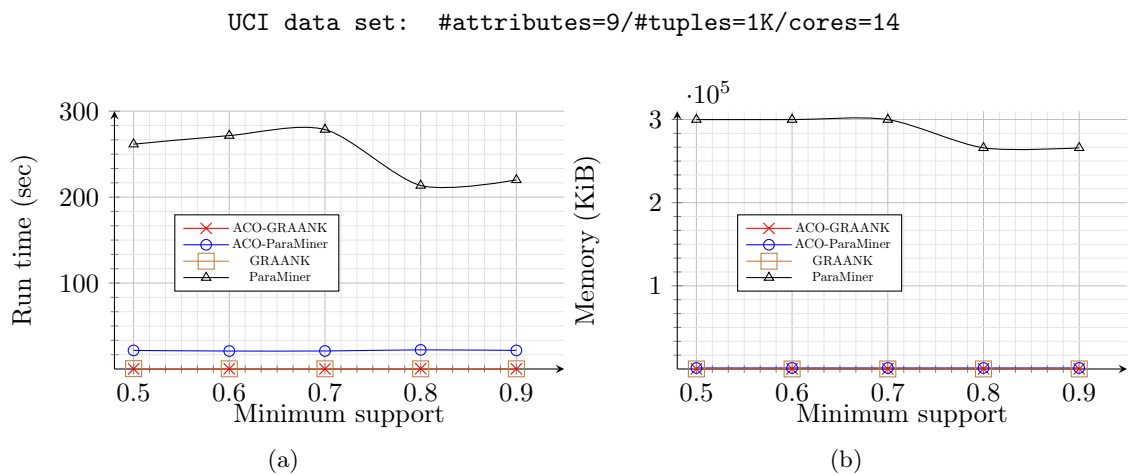

  	Figure~\ref{fig4:exp_comparison4a} (a) and (b) show the runtime and memory usage of ACO-GRAANK and GRAANK algorithms on UCI data set with 46774 tuples. ACO-GRAANK has a lower runtime than GRAANK but, a higher memory usage. ParaMiner and ACO-ParaMiner both yield \textit{Memory Error} outputs on UCI data set with 46774 tuples; so, we use a UCI data set with 1000 tuples as shown in Figure~\ref{fig4:exp_comparison4b}. In Figure~\ref{fig4:exp_comparison4b} (a) we observe that ParaMiner has the slowest runtime followed by ACO-ParaMiner and GRAANK and ACO-GRAANK have relatively low runtimes. In Figure~\ref{fig4:exp_comparison4b} (b), we observe that ParaMiner has the highest memory usage and, ACO-GRAANK, ACO-ParaMiner, GRAANK have relatively low memory usages.
  	
	\newpage	
	\subsubsection{Consistent Gradual Patterns}
	\label{ch4:consistent_gps}
	This experiment analyzes gradual patterns extracted by algorithms ACO-GRAANK, ACO-ParaMiner, GRAANK, ParaMiner from data sets: B\&C, C2K with 200 tuples, Buoys with 200 tuples, UCI with 1000 tuples.
	
	\begin{table}[h!]
	\small
	\centering
	\caption{Consistent gradual patterns}
	\begin{tabular}{|l|l|}
		\hline
		Data set & Consistent gradual patterns\\
		\hline \hline
		B\&C & $\{(Insulin, \downarrow), (HOMA, \downarrow)\}$, support: 0.94\\
		\hline
		C2K & $\{(i1\_rcf\_1\_p, \uparrow), (i1\_rcf\_1\_e, \uparrow)\}$, support: 0.837\\
		\hline
		Buoys & $\{(Tz, \downarrow), (Tav, \downarrow)\}$, support: 0.945\\
		\hline
		UCI & $\{(Global\_activepower, \uparrow), (Global\_intensity, \uparrow)\}$, support: 0.954 \\
		\hline
	\end{tabular}
	\label{tab4:consistent_gs}
	\end{table}
	
	\clearpage
	
\begin{figure}[h!]
		\centering
		\small
  		\subfloat[]{%
  		\begin{tikzpicture}
  			\begin{axis}[
  				height=4cm, width=6.5cm,
  				ybar,			
    			symbolic x coords={0.5, 0.6, 0.7, 0.8, 0.9},
    			bar width=.1cm,
            	xtick=data,
   				xlabel=Minimum support, 
   				ylabel=No. of patterns,
   				legend style={nodes={scale=0.5}},
 				]
			\addplot table[x=Support, y=ACO_GRAANK]{b_c_patterns.dat};
			\addplot table[x=Support, y=ACO_LCM]{b_c_patterns.dat};
			\addplot table[x=Support, y=GRAANK]{b_c_patterns.dat};
			\addplot table[x=Support, y=LCM]{b_c_patterns.dat};
			\legend{ACO-GRAANK, ACO-ParaMiner, GRAANK, ParaMiner}
  			\end{axis}
  		\end{tikzpicture}}%
  		\subfloat[]{%
  		\begin{tikzpicture}
  			\begin{axis}[
  				height=4cm, width=6.5cm,
  				ybar,			
    			symbolic x coords={0.5, 0.6, 0.7, 0.8, 0.9},
      			bar width=.1cm,
            	xtick=data,
   				xlabel=Minimum support, 
   				ylabel=No. of patterns,
   				legend style={nodes={scale=0.5}},
 				]
			\addplot table[x=Support, y=ACO_GRAANK]{c2k_patterns.dat};
			\addplot table[x=Support, y=ACO_LCM]{c2k_patterns.dat};
			\addplot table[x=Support, y=GRAANK]{c2k_patterns.dat};
			\legend{ACO-GRAANK, ACO-ParaMiner, GRAANK}
  			\end{axis}
  		\end{tikzpicture}}%
  		\\
  		\subfloat[]{%
  		\begin{tikzpicture}
  			\begin{axis}[
  				height=4cm, width=6.5cm,
  				ybar,			
    			symbolic x coords={0.5, 0.6, 0.7, 0.8, 0.9},
    			bar width=.1cm,
            	xtick=data,
   				xlabel=Minimum support, 
   				ylabel=No. of patterns,
   				legend style={nodes={scale=0.5}},
 				]
			\addplot table[x=Support, y=ACO_GRAANK]{buoys_patterns.dat};
			\addplot table[x=Support, y=ACO_LCM]{buoys_patterns.dat};
			\addplot table[x=Support, y=GRAANK]{buoys_patterns.dat};
			\addplot table[x=Support, y=LCM]{buoys_patterns.dat};
			\legend{ACO-GRAANK, ACO-ParaMiner, GRAANK, ParaMiner}
			\end{axis}
  		\end{tikzpicture}}%
  		\subfloat[]{%
  		\begin{tikzpicture}
  			\begin{axis}[
  				height=4cm, width=6.5cm,
  				ybar,			
    			symbolic x coords={0.5, 0.6, 0.7, 0.8, 0.9},
    			bar width=.1cm,
            	xtick=data,
   				xlabel=Minimum support, 
   				ylabel=No. of patterns,
   				legend style={at={(0.5,1)}, anchor=north, legend columns=-1, nodes={scale=0.3}},
 				]
			\addplot table[x=Support, y=ACO_GRAANK]{uci_patterns.dat};
			\addplot table[x=Support, y=ACO_LCM]{uci_patterns.dat};
			\addplot table[x=Support, y=GRAANK]{uci_patterns.dat};
			\addplot table[x=Support, y=LCM]{uci_patterns.dat};
  			\end{axis}
  		\end{tikzpicture}}%

    	\caption{Bar graph of number of patterns against minimum support for data sets: (a) B\&C, (b) C2K, (c) Buoys and, (d) UCI}
  		\label{fig4:exp_patterns}
  	\end{figure}
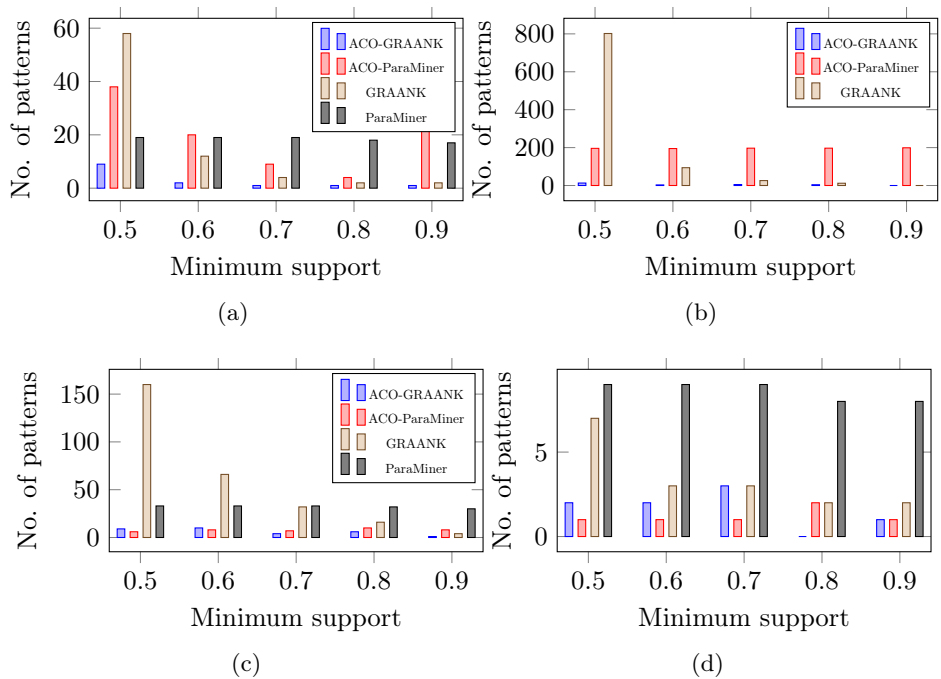

	Table~\ref{tab4:consistent_gs} shows consistent gradual patterns extracted by all algorithms from the 4 data sets whose support is high. It is important to remember that the higher the value of support, the better the quality of the gradual pattern. Figure~\ref{fig4:exp_patterns} compares the number of extracted gradual patterns by all algorithms.
	
	In Figure~\ref{fig4:exp_patterns} (a), B\&C data set: we observe that GRAANK extracts the most pattern at support 0.5 followed (in order) by ACO-ParaMiner, ParaMiner and ACO-GRAANK. For other supports ParaMiner and ACO-ParaMiner have the most extracted patterns. In Figure~\ref{fig4:exp_patterns} (b), C2K data set with 200 tuples: we observe that with exemption of support 0.5, ACO-ParaMiner extracts more patterns than GRAANK and ACO-GRAANK.
	
	In Figure~\ref{fig4:exp_patterns} (c), Buoys data set with 200 tuples: we observe that GRAANK extracts the most pattern for supports 0.5 and 0.6 relative to ACO-ParaMiner, ACO-GRAANK and ParaMiner. For the other supports ParaMiner extracts the most patterns. In Figure~\ref{fig4:exp_patterns} (d), UCI data set with 1000 tuples: we observe that ParaMiner extracts the most patterns followed (in order) by GRAANK, ACO-GRAANK and ACO-ParaMiner.
	
	\subsubsection{Comparative Computational Experiment: T-GRAANK vs ACO-TGRAANK}
  	\label{ch4:tgraank_comparison}
  	The aim of this experiment is to determine which algorithm implementation performs better: (1) T-GRAANK implementation based on GRAANK technique proposed in \cite{Laurent2009} or (2) ACO-TGRAANK implementation based on ACO-GRAANK technique proposed in Section~\ref{ch4:aco_tgraank}. We applied the Buoys data set on both T-GRAANK and ACO-TGRAANK algorithms in order to compare their performances.
	
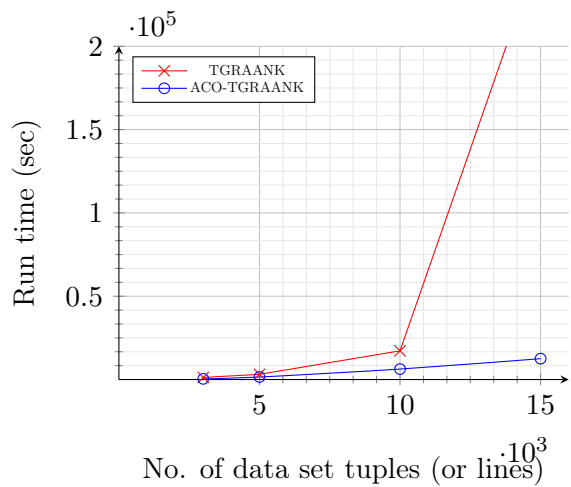
\begin{figure}[h!]
  	\centering
  	\texttt{Buoys: attributes=11/cpus=28/tuples=3k-15k/min-sup=0.5/min-rep=0.9995}\par\medskip
  	\begin{tikzpicture}
    		\begin{axis}[
  				height=6cm, width=7.5cm,
  				scaled x ticks=base 10:-3,
  				grid=both,
  				grid style={line width=.1pt, draw=gray!20},
    			major grid style={line width=.2pt, draw=gray!50},
  				axis lines=middle,
  				minor tick num=5,
    			ymin=0, ymax=200000,
    			xmin=0, xmax=16000,
    			xtick={5000, 10000, 15000},
   				xlabel=No. of data set tuples (or lines), 
   				ylabel=Run time (sec),
  				xlabel style={at={(axis description cs:0.5,-0.2)},anchor=north},
  				ylabel style={at={(axis description cs:-0.15,0.5)},rotate=90, anchor=south},
 	 			legend style={nodes={scale=0.5}},
 	 			legend pos=north west
 				]
			\addplot[mark=x, mark size=3pt, color=red] table[x=Size, y=Runtime]{tgraank.dat};
  	  		\addlegendentry{TGRAANK}

  			\addplot[smooth, mark=o, color=blue] table[x=Size, y=Runtime]{aco_tgraank.dat};
			\addlegendentry{ACO-TGRAANK}
  			\end{axis}
    	\end{tikzpicture}
  		\caption{Buoys data set: graph plot of run time against size of data set for T-GRAANK and ACO-TGRAANK algorithms}
    	\label{fig4:tgraank_compare}
    \end{figure}

  Using the results we obtained, we plot the graphs in Figure~\ref{fig4:tgraank_compare}. We mention that at the 15k-line data set the T-GRAANK algorithm exceeded the limit on the total run time we set (\textit{3 days}) for each job allocation in the HPC node. We observe that the run time performance of ACO-TGRAANK algorithm implementation is better than that of the T-GRAANK algorithm especially for large data sets. From this, we can make the conclusion that the ACO-based technique is more efficient at extracting temporal gradual patterns than classical technique especially when dealing with huge data sets.

	\subsection{Discussion of Results}
	\label{ch4:discussion}
	
	\subsubsection{Memory Usage}
	\label{ch4:discuss_memory}
	We observe that ParaMiner and ACO-ParaMiner have higher requirements for memory size than GRAANK and ACO-GRAANK. This phenomenon explains why ParaMiner and ACO-ParaMiner (1) yield \textit{Memory Error} outputs for relatively large data sets and, (2) have slower runtimes for relatively small data sets. As described in Section~\ref{ch2:gradual_traversal}, ParaMiner is based on a DFS strategy whose drawbacks (for the case of gradual pattern mining) are as follows:
	\begin{enumerate}
		\item a numeric data set has to be encoded into a transactional data set before DFS can be applied. This significantly increases the size of the data set and consequently size of usage memory required;
		\item DFS employs a recursive approach to find all the longest paths. However, we notice from the experiment results that ACO-ParaMiner has improved runtime since it uses a non-recursive heuristic approach to achieve this.
	\end{enumerate}
	
	For the case of GRAANK and ACO-GRAANK, their relatively low memory usages can be largely attributed to the modeling of tuple orders using a binary matrix. However, ACO-GRAANK has higher memory usage requirement than GRAANK since it generates numerous and random maximal gradual items.

	\subsubsection{Execution Runtime}
	\label{ch4:discuss_runtime}
	We observe that generally BFS-based algorithms (GRAANK and ACO-GRAANK) have relatively faster runtimes than DFS-based algorithms. This mainly is due to low memory usage requirements explained in the previous section. We also observe that ACO-GRAANK has relatively faster runtimes than GRAANK because it employs a heuristics approach that learns high quality gradual patterns quicker than the classical level-wise approach.
	
	\subsubsection{Extracted Gradual Patterns}
	\label{ch4:discuss_gps}
	It is important to recall from Chapter~\ref{ch2:pruning} that \textit{maximal} and \textit{closed} patterns provide the richest knowledge of attribute correlations because such patterns combine all the attributes of the data set that have gradual relationships. Having mentioned that, we observe that generally the number of patterns extracted by ParaMiner is relatively constant against different supports because it searches for all closed patterns (which are more rich in terms of attribute correlation knowledge) \cite{Negrevergne2014}, while the number of patterns extracted by GRAANK varies significantly with support since it searches for all patterns that surpass the support threshold. However, ACO-GRAANK searches for most maximal patterns (which are fewer but provide more rich knowledge of attribute correlations) and, ACO-ParaMiner searches for most frequent closed patterns.

	\section{Summary}
	\label{ch4:summary}
	In this chapter, we describe an ant colony optimization technique for BFS-based and DFS-based approaches for mining gradual patterns. Given the computational results in Section~\ref{ch4:experiments}, we establish that ACO-GRAANK and ACO-ParaMiner algorithms out-perform classical GRAANK and ParaMiner algorithms and, they mine fewer but high quality maximal and closed patterns respectively. Additionally, we establish that ACO-TGRAANK algorithm implementation which extends the ACO strategy proposed in this Chapter is the best performing algorithm implementation for mining temporal gradual patterns.
	
















\chapter{Temporal Gradual Emerging Patterns}
\label{ch5}
\minitoc \clearpage 

	\begin{chapquote}{William Lawrence Bragg \textit{(1890 – 1971)}}
		``The important thing in science is not so much to obtain new facts as to discover new ways of thinking about them''
	\end{chapquote}

	\section{Introduction}
	\label{ch5:introduction}
	In this chapter, we extend temporal gradual patterns to introduce and describe temporal gradual emerging patterns. We propose to extend a border-based manipulation technique to the case of mining temporal gradual emerging patterns. In addition, we extend ant colony optimization strategy described in Chapter~\ref{ch3} to propose a more efficient technique for constructing temporal gradual emerging patterns.

	\section{Emerging Patterns}
	\label{ch5:ep}
	Emerging patterns (EPs) are item sets whose frequency increases significantly from one data set to another. EPs are described using \textit{growth rate}, which is the ratio of an EP's frequency support in one data set to its frequency support in another data set. For instance, suppose a car shop in 2017 had 200 purchases of $\{FOG\_LAMPS, BATTE-$ $RY, TYRES\}$ out of 1000 transactions, and in 2018 it had 500 such purchases out of 1000 transactions. This purchase is an EP with a \textit{growth rate} of $2.5$ from the year 2017 to 2018.

	More specifically, an EP is present if its \textit{growth rate} across data sets is larger than a given specified minimal numerical threshold. EPs can be applied to discover distinctions that exist amongst a collection of data sets with classes such as \textit{``hot vs cold''}, \textit{``poisonous vs edible''}. In other words, EPs are a powerful tool for capturing discriminating characteristics between the classes of different data sets \cite{Dong1999a, Jinyan2001a, Kotagiri2003}.
	
	Mining gradual emerging patterns (GEPs) aims at identifying distinctions between numerical data sets in terms of attribute correlations \cite{Laurent2015}. For instance, given \textit{`windspeed'} and \textit{`atmospheric temperature'} as attributes of two numerical data sets, a GEP may take the form: \textit{``the higher the windspeed, the lower the temperature''} with a frequency support of $0.01$ in one data set, and \textit{``the higher the windspeed, the lower the temperature''} with a frequency support of $0.8$ in another data set. However, many correlations appear over time, as for instance it is the case when phenomena appear after some meteorological situation due to latency. Many previous works have not taken this into account. 
		
	In this chapter, we introduce a new pattern category called temporal gradual emerging patterns (TGEPs), which extends temporal gradual patterns (TGPs), for knowledge discovery in numeric timestamped data sets. It is important to mention that apart from the study presented in Chapter~\ref{ch3}, many other research works exist that allow for transformation of a single timestamped data set into numerous data sets based on \textit{date-time} attribute. Therefore, we may define TGEPs as: \textit{temporal gradual item sets whose frequency supports increase significantly between transformed data sets}. TGEP mining unearths a new possibility for discovering gradual trends with respect to time in timestamped numerical data sets.

	For instance a TGEP may take the form: \textit{``the higher the windspeed, the lower the temperature, almost 3 minutes later''} with a frequency support of $0.005$ in the first transformed data set, and \textit{``the higher the windspeed, the lower the temperature, almost 3 minutes later''} with a frequency support of $0.75$ in the second transformed data set. Most often, apart from the significant increase in frequency support, it may also be possible for a drift of time gap to additionally occur. For instance the second transformed data set may produce the pattern: \textit{``the higher the windspeed,  the lower the temperature, almost 7 minutes later''} with a frequency support of $0.75$. 
	
	The latter delivers a more meaningful pattern knowledge from the data set, since it may also be interpreted as the amount of time that elapses before/after a temporal gradual item emerges. This pattern has a \textit{growth rate} of $150$ after approximately \textit{4 minutes}. Apart from this simple example, numerous real-life timestamped data sets from unrelated sources may be crossed and mined for interesting TGEPs. In spite of this, it should be emphasized that TGEPs are extracted from transformed data sets. As illustrated in Chapter~\ref{ch3:transformation}, a timestamped data set may be transformed into numerous data sets each of which can be mined for TGPs. 
	
	For this same reason, the complexity of dealing with more than 2 transformed data sets when extracting TGEPs arises. The process of mining all TGPs from each of these data sets and comparing the patterns against each other to identify emerging ones proves to be computationally time-consuming. Consequently, we introduce contributions in the sections that follow that propose and describe 2 more efficient approaches for mining TGEPs.

	\subsection{Preliminary Concepts and Notations}
	\label{ch5:preliminary}
	In order to describe TGEPs, we recall the definitions of gradual patterns as given in Chapter~\ref{ch2:definitions}. Additionally in this section, we provide some preliminary definitions of emerging patterns (as given by \cite{Dong1999b, Jinyan2001b, Kotagiri2003}) and gradual emerging patterns (as given by \cite{Laurent2015}).
	
	In the case of emerging patterns, assume a data set $\mathcal{D}$ is defined by item set $I = \{i_{1}, i_{2}, ..., i_{n} \}$ and it consists of transactions $\{t_{1}, t_{2}, ..., t_{n} \}$. Every transaction is a subset of item set $I$ (as shown in Figure~\ref{fig5:sample_ep} (a)).

	\textit{Example 5.1.} We consider two data sets: $(\mathcal{D}_{1})$ in Figure~\ref{fig5:sample_ep} (a), and $(\mathcal{D}_{2})$ in Figure~\ref{fig5:sample_ep} (b).
		
	\begin{figure}[h!]
		\centering
		\small
		\subfloat[]{%
		\begin{tabular}{c l}
			\toprule
			\textbf{transaction} & \textbf{items}\\
			\hline \hline
			t1 & bread, milk, sugar\\
			t2 & eggs, milk\\
			t3 & cheese, bread\\
			t4 & butter, sugar\\
			\bottomrule
		\end{tabular}
		}%
		$~~$
		\subfloat[]{%
		\begin{tabular}{c l}
			\toprule
			\textbf{transaction} & \textbf{items}\\
			\hline \hline
			t1 & bread, milk, sugar\\
			t2 & eggs, bread, milk\\
			t3 & bread, milk, sugar, cheese\\
			t4 & bread, milk, sugar, eggs\\
			\bottomrule
		\end{tabular}
		}%
		\caption{Two sample data sets containing transactions}
		\label{fig5:sample_ep}
	\end{figure}
	
	\textbf{Definition 5.1.} \texttt{Emerging Pattern.} \textit{An emerging pattern $(EP)$ is a set of items that appear less frequently in transactions of one data set $\mathcal{D}_{1}$ and more frequently in transactions of another data set $\mathcal{D}_{2}$.} 
	
	For example, (as illustrated in Figure~\ref{fig5:sample_ep}) $\{bread, milk, sugar\}$ is an emerging item set since its frequency occurrence count in transactions (or frequency support) is significantly greater in data set $\mathcal{D}_{2}$ than in data set $\mathcal{D}_{1}$.
	
	\textbf{Definition 5.2.} \texttt{Growth Rate.} Growth rate \textit{$(gr)$ of an emerging pattern $(EP)$ is the ratio of frequency support of the pattern in data set $\mathcal{D}_{2}$ to another data set $\mathcal{D}_{1}$.} The quality of an emerging pattern is measured by growth rate.
	
	For example, given data sets $\mathcal{D}_{1}$ and $\mathcal{D}_{2}$, the \textit{growth rate} of item set $EP$ in favour of data set $\mathcal{D}_{2}$ is given as follows:

	\begin{equation}
		gr(EP) = \dfrac{sup(EP)_{\mathcal{D}_{2}}}{sup(EP)_{\mathcal{D}_{1}}}
	\end{equation}

	Therefore, given a user-specified minimum numerical threshold $\rho$, a pattern $EP$ is said to be emerging only if:

	\begin{equation}
		gr(EP) \geq \rho
	\end{equation}

	It is important to emphasize that growth rate is derived from frequency support of the involved patterns \cite{Dong1999b, Dong2005}. Similar to transactional frequent pattern mining, frequency support as a quality measure for extracted patterns also applies to gradual pattern mining. It is for this reason that the concept of emerging patterns can be extended to the case of gradual pattern mining \cite{Laurent2015}. Therefore, a gradual emerging pattern may be defined as follows:

	\textbf{Definition 5.3.} \texttt{Gradual Emerging Pattern.} \textit{A gradual emerging pattern $(GEP)$ is a set of gradual item sets whose support $sup(GEP)_{\mathcal{D}_{g2}} > sup(GEP)_{\mathcal{D}_{g1}}$.} Where $\mathcal{D}_{g}$ is a numeric data set whose attributes hold singleton values/items per transaction (i.e. Table~\ref{tab5:tgep_sample} in Section~\ref{ch5:tgep}).
		
	For example, $GP = \lbrace (rain,\uparrow),(wind,\uparrow) \rbrace$ is a gradual emerging pattern if support $sup(GP)$ is significantly greater in data set $\mathcal{D}_{g2}$ than in data set $\mathcal{D}_{g1}$.	 It should be underlined that gradual patterns cannot be extracted from data sets whose attributes hold a set of items/values (i.e. Figure~\ref{fig5:sample_ep}) due to tuple pairing that occurs during extraction.

	\section{Temporal Gradual Emerging Patterns}
	\label{ch5:tgep}
	In this section, we seek to describe temporal gradual emerging patterns (TGEPs). We begin by briefly recalling the extraction process of temporal gradual patterns (TGPs) (see Chapter~\ref{ch3} for full description) because an emerging TGP makes up a TGEP. TGP mining extends gradual pattern mining in order to additionally estimate the time lag that may exist between gradual item sets. For instance, a TGP may take the form \textit{``the higher X, the higher Y, \textbf{almost 3 months later}''}. 
	
	\textit{Example 5.2.} Let us consider a data set $(\mathcal{D}_{g})$ shown in Table~\ref{tab5:tgep_sample}.

	\begin{table}[h!]
	\centering
	\small
	\begin{tabular}{c c c c}
	  	\toprule
      	\textbf{id} & \textbf{date} & \textbf{rain} & \textbf{wind}\\
      	&  (day/month) & (mm) & (km/h)\\
      	\hline \hline
      	r1 & 01/06 & 10 & 42\\
      	r2 & 04/06 & 13 & 21\\
      	r3 & 05/06 & 18 & 35\\
      	r4 & 10/06 & 10 & 21\\
      	r5 & 12/06 & 18 & 35\\
      	\bottomrule
    \end{tabular}
    \caption{Sample data set of rainfall amount and wind speed recorded at different dates.}
    \label{tab5:tgep_sample}
	\end{table}
	
	The first process in the extraction of TGPs is to \textit{transform} a timestamped numerical data set step-wisely into a temporal format. This process requires that one attribute be selected as a \textit{reference attribute} and \textit{transformation step} be set. For example, (using data set in Table~\ref{tab5:tgep_sample}) if \textit{`rain'} is selected as the reference attribute and a transformation step be set at $1$; then the data set ($\mathcal{D}_{g}$) in this table will be transformed into data set ($\mathcal{D}'_{g}$) in Table~\ref{tab5:transformed} such that: 
	\begin{itemize}
		\item every tuple ($r_{n}$) of \textit{`rain'} attribute is mapped to tuple ($r_{n+1}$) of \textit{`wind'} attribute, and 
		\item date difference is calculated as ($r_{n} - r_{n+1}$).
	\end{itemize}
		
	\begin{table}[h!]
		\centering
		\small
		\begin{tabular}{c|c|c|c}
	  		\toprule
      		\textbf{id} & \textbf{date \textit{diff}} & \textbf{rain} & \textbf{wind}\\
      		& ($r_{n}-r_{n+1}$) & ($r_{n}$) & ($r_{n+1}$)\\
      		\hline \hline
      		t1 & 3 & 10 & 21\\
      		t2 & 1 & 13 & 35\\
      		t3 & 5 & 18 & 21\\
      		t4 & 2 & 10 & 35\\
      		t5 & - & - & -\\
      		\bottomrule
    	\end{tabular}
    	\caption{Data set $(\mathcal{D}'_{g})$ transformed from data set $\mathcal{D}_{g}$ by \textit{step:} $s = 1$.}
    	\label{tab5:transformed}
	\end{table}
	
	It is observable that a transformation step of $1$, leads to generation of a transformed data set $(\mathcal{D}'_{g})$ that represents $4$ out of the $5$ tuples of the original data set $(\mathcal{D}_{g})$. Therefore, $\mathcal{D}'_{g}$ has a \textit{representativity} of $0.8$. Representativity \textit{$(rep)$ of a TGP is the ratio of tuple size in a transformed data set ($\mathcal{D}'_{g}$) to tuple size in the original data set ($\mathcal{D}_{g}$).} Therefore, given a  minimum numerical threshold $\delta$, a $TGP$ is relevant only if:

	\begin{equation}
		rep(TGP) \geq \delta
	\end{equation}

	The second process in the mining of TGPs is to extract gradual patterns from the \textit{transformed data set} and approximate a \textit{time lag} associated with the extracted gradual patterns.
		
	As shown above, TGPs are extracted from \textit{transformed data sets}. The value of the specified minimum representativity $(\delta)$ determines the number of transformation steps; consequently, the number of transformed data sets. It may be the case that a particular TGP occurs frequently in more than one transformed data set; as a result, it becomes an \textit{emerging TGP}. Under those circumstance, we may define a TGEP as follows.

	\textbf{Definition 5.4.} \texttt{Temporal Gradual Emerging Pattern.} \textit{A temporal gradual emerging pattern $(TGEP)$ is a set of temporal gradual patterns that appear more frequently in one transformed data set $\mathcal{D}'_{g}$ and less frequently in another transformed data set $\mathcal{D}''_{g}$.}

	\section{Border-based Discovery of TGEPs}
	\label{ch5:border_tgep}
	In this section, we propose an approach that exploits border manipulation for extraction of TGEPs. First, we describe how border manipulation technique may be applied to the case of frequent item sets and gradual item sets. Finally, we propose to extend it to the case of temporal gradual item sets.
	
	\subsection{Border Representation of Frequent Item Sets}
	\label{ch5:frequent_border}

	The border-based approach was introduced by \cite{Dong1999b} and it offers condensed representation and efficient manipulation of large interval closed item sets. In considerations of clearly describing the border-based approach for discovery of TGEPs, we formalize the notion of interval-closed sets and provide the definition of a border as given by \cite{Jinyan2001a, Dong1999b, Dong1999a, Laurent2015}.

	\textbf{Property 5.1.} \texttt{Interval Closed Sets}. Collections of sets $\mathcal{C}$ are said to be interval closed if: $\forall X,Z \in \mathcal{C}; ~\forall Y$ such that $X \subseteq Y \subseteq Z$, it also holds that $Y \in \mathcal{C}$. Such sets are also referred to as \textit{convex} sets.

	\textbf{Definition 5.5.}	 \texttt{Border.} \textit{A border is an ordered pair $< \mathcal{L}, \mathcal{R} >$, (where $\mathcal{L}$ \texttt{- left-hand bound of the border} and $\mathcal{R}$ \texttt{- right-hand bound of the border}) if: (a) each of $\mathcal{L}$ and $\mathcal{R}$ is an antichain\footnote{A collection of sets $\mathcal{C}$ is an \textit{antichain} if $\forall X,Y \in \mathcal{C}, X \not\subseteq Y$ and $Y \not\subseteq X$} collection of sets, and (b) each element of $\mathcal{L}$ is a subset of some element in $\mathcal{R}$ and each element of $\mathcal{R}$ is a superset of some element in $\mathcal{L}$}.

	For example, a collection of sets $[ \mathcal{L}, \mathcal{R} ]$ may be represented by (or is said to have) a border $< \mathcal{L}, \mathcal{R} >$, where $[ \mathcal{L}, \mathcal{R} ] = \{Y \mid \exists X \in \mathcal{L}, \exists Z \in \mathcal{R}$ such that $X \subseteq Y \subseteq Z \}$.
		
	For the purpose of relating Property 5.1 to Definition 5.5, \cite{Dong1999b} presents three main propositions: (a) the collection of all large item sets with respect to a minimum threshold $(\sigma)$ is interval closed, (b) each interval-closed collection $\mathcal{C}$ of sets has a unique border $< \mathcal{L}, \mathcal{R} >$, where $\mathcal{L}$ is the collection of minimal item sets in $\mathcal{C}$ and $\mathcal{R}$ is the collection of maximal item sets, and (c) the collection of large item sets with respect to a minimum threshold $(\sigma)$ in a data set has a \textit{left-rooted} border.

	To clarify, a border $< \mathcal{L}, \mathcal{R} >$ is called \textit{left-rooted} if $\mathcal{L}$ is a singleton set (i.e. the left-hand bound is $\{ \emptyset \}$ and its right-hand bound is the collection of maximal item sets) and it is \textit{right-rooted} if $\mathcal{R}$ is a singleton set. As a result of these propositions, \cite{Dong1999b} illustrates that an efficient maximal set pattern mining algorithm (i.e. Max-Miner) may be used to extract collections of large item sets from data sets which in-turn provide \textit{left-rooted} borders for each data set with respect to a minimum support threshold.

	All in all, \cite{Dong1999b} demonstrates how an efficient discovery of left-rooted borders from two data sets (i.e. $\mathcal{D}_{1}$ and $\mathcal{D}_{2}$) through Max-Miner algorithm, followed by a repetitive border differential procedure (implemented as BORDER-DIFF algorithm) on the two borders allows for extraction of emerging patterns. The overall algorithm is known as MBD-LLBORDER algorithm.

	\subsection{Border Representation of Gradual Item Sets}
	\label{ch5:gradual_border}

	It should be noted that the MBD-LLBORDER algorithm cannot be applied directly to the case of gradual item sets, since the nature of gradual item sets is slightly different from that of classical frequent item sets. Unlike frequent item sets, (in terms of tuple transactions and gradual items) gradual item sets in both cases deal with pairs and not singletons. We demonstrate this in Chapter~\ref{ch2:gp}.

	As demonstrated in Section~\ref{ch2:gp}: (a) deriving the frequency support of a gradual pattern involves ordering tuples in concordant pairs; and (b) discovering any meaningful correlation knowledge among attributes of a data set involves extracting non-trivial gradual patterns composed of at least $2$ gradual items. It should be remembered that the border representation of large item sets presented by \cite{Dong1999b} (see Section~\ref{ch5:frequent_border}) is most suitable for classic frequent item sets that are each composed of singleton items.

	For example, in the classic item set case a $4$\textit{-length} pattern $\{A,B,C,D\}$ may fully be decomposed into its $4$ items: $\{A\}$, $\{B\}$, $\{C\}$, $\{D\}$. In the gradual item set case a $4$\textit{-length} pattern $\{(A,\uparrow), (B,\downarrow), (C,\uparrow), (D,\downarrow) \}$ at best may be decomposed into its $6$ gradual items: $\{(A,\uparrow),(B,\downarrow)\}, \{(A,\uparrow),(C,\uparrow)\}, \{(A,\uparrow),(D,\downarrow)\}, \{(B,\downarrow),(C,\uparrow)\}, \{(B,\downarrow),(D,\downarrow)\}, \{(C,\uparrow),(D,\downarrow)\}$.

	Nevertheless, \cite{Laurent2015} identifies two properties of gradual patterns that allow for border representation of gradual patterns. They are: (a) a collection of frequent gradual patterns is interval-closed, and (b) a collection of frequent gradual patterns may be represented as a left-rooted border $< \{ \emptyset \}, \mathcal{R} >$, where $\mathcal{R}$ is the set of maximal gradual item sets in the collection broken down into gradual items of length $2$.
		
	With reference to the first identified property of gradual patterns, it derives explicitly from \textit{anti-monotonicity} feature of gradual patterns. The anti-monotonicity property states that: \textit{``no frequent gradual pattern containing $n$ items can be built over an infrequent gradual pattern containing a subset of these $n$ items.''} For instance, if a maximal gradual pattern $\{(A,\uparrow), (B,\downarrow), (C,\uparrow)\}$ is frequent (support surpasses the minimum specified threshold), then all subsets of this pattern are also frequent \cite{Di-Jorio2008, Owuor2019}.
		
	With reference to the second identified property of gradual patterns, it derives from \textit{pairings} that come with mining non-trivial gradual patterns. Consequently, \cite{Laurent2015} proposes a subsequent representation of maximal gradual item sets into its smaller gradual items of length $2$. That is to say, a maximal gradual pattern of length $k$ may be re-represented by a set of $k(k-1)/2$ gradual items of length $2$.

	\subsection{Border Representation of Temporal Gradual Item Sets}
	\label{ch5:temporal_border}

	Temporal gradual patterns (TGPs) (as described in Section~\ref{ch5:tgep}) are simply gradual patterns that have been improved to indicate an estimated \textit{time lag} among the gradual item sets. In fact \cite{Owuor2019} presents T-GRAANK approach which extends the GRAANK approach (presented in \cite{Laurent2009} for mining gradual patterns) for mining TGPs.

	It should be clarified that the distinctive feature of approximated time lag of TGPs is majorly introduced by the \textit{data set transformation} process. As an advantage, the transformation process allows for generation of multiple transformed data sets from the original timestamped data set using a specified \textit{representativity} threshold \cite{Owuor2019}. The novel idea of harnessing this process so as to mine a single data set for \textit{emerging TGPs} (or TGEPs) is quite interesting. In fact, we propose Algorithm~\ref{alg5:border_ep}, BT-GRAANK stands for MBD-LLBORDER Temporal GRAdual rANKing.
	
	\begin{algorithm}[h!]
	\caption{\textit{BT-GRAANK} algorithm}
	\label{alg5:border_ep}
	\small

	\SetKwFunction{border}{\textbf{MBD-LL}BORDER}
	\SetKwFunction{maximal}{maximal}
	\SetKwFunction{transform}{transform}
	\SetKwFunction{tgraank}{extract-tgps}
	\SetKwInOut{Input}{\textbf{Input}}
	\SetKwInOut{Output}{\textbf{Output}}

	\Input{$D -$ data set, $refCol -$ reference column, $\sigma -$ minimum support, $\delta -$ minimum representativity}
	\Output{$TGEPs -$ TGEPs represented as borders}

	$\mathcal{D}^{*}, T_{d}^{*} \leftarrow$ \transform($D, \delta, refCol$)
	\tcc*[H]{$\mathcal{D}-$ transformed data set, $T_{d}-$ time differences, $*$ denotes multiple}

	$tgps^{*} \leftarrow$ \tgraank($\mathcal{D}^{*}, T_{d}^{*}, \sigma$)\;

	$leftBdr^{*} \leftarrow$ \maximal($tgps^{*}$)\tcc*[H]{maximal TGPs as left-rooted borders}

	$TGEPs \leftarrow$ \border($leftBdr^{*}$)\;

	\Return $TGEPs$\;
	\end{algorithm}
	
	\begin{enumerate}
		\item Build multiple transformed data sets $\mathcal{D}'_{g}, \mathcal{D}''_{g}, ..., \mathcal{D}^{*}_{g}$ from a timestamped data set $\mathcal{D}_{g}$ from a specified representativity threshold $\delta$ (see Section~\ref{ch5:tgep}).
		\item Extract all TGPs from each transformed data set with a specified support threshold $\sigma$.
		\item Construct border representations of all maximal TGPs from the transformed data sets as described in Section~\ref{ch5:gradual_border}.
		\item Apply a modified MBD-LLBORDER algorithm (using modified a union operator) to the borders obtained in Step 3 (two borders at a time).
	\end{enumerate}
	
	It is important to note that this border-based strategy is an efficient technique for mining TGEPs; however, the search space grows exponentially with respect to the number of attributes \cite{Garcia-Vico2018, Owuor2019}. Therefore, this strategy is not suitable for discovering patterns in huge data sets.
	
	\section{Ant-based Discovery of TGEPs}
	\label{ch5:aco_tgep}
	In this section, first we introduce an alternative approach that is based on ant colony optimization (ACO) for mining gradual patterns (GPs) and temporal gradual patterns (TPGs), and second we extend this approach in order to mine for temporal gradual emerging patterns (TGEPs).
	
	\subsection{ACO for TGEP Mining}
	\label{ch5:aco_gp}
	In Chapter~\ref{ch4}, we have proposed an ACO strategy that uses a probabilistic approach to efficiently generate gradual item set candidates from a data set's attributes. Ant colony optimization (ACO), as originally described by \cite{Dorigo1996}, is a general-purpose heuristic approach for optimizing various combinatorial problems. It exploits the behavior of a colony of artificial ants in order to search for approximate solutions to discrete optimization problems.

	In order to apply ACO to the problem of gradual pattern mining, we have described in Chapter~\ref{ch4:aco}: (1) a suitable representation of the gradual item set candidate generation problem, (2) a probabilistic rule $\mathcal{P}$ for generating solutions $\mathcal{S}_{n}$, (3) a technique for updating the pheromone matrix $\mathcal{T}_{a,j}$, (4) a convergence proof for confirming that this approach finds an optimal gradual pattern from a data set.
	
	It should be remembered that not only may the ACO strategy be applied to the case of gradual pattern mining but also to the case of TGP mining. As described in Chapter~\ref{ch3:mining}, TGP mining involves two main processes: (1) transforming a timestamped data set into multiple data sets using a specified representativity threshold, and (2) applying a modified GRAANK algorithm that extracts gradual patterns together with an approximated time lag. Therefore, the modified GRAANK algorithm is substitute for a modified ACO-based algorithm.

	For the purpose of applying ACO strategy to the case of TGEPs, we propose a modification of the technique for updating the pheromone matrix $\mathcal{T}_{a,j}$ shown in Equation~\eqref{eqn5:p_matrix}. The matrix contains knowledge about pheromone proportions of the 3 gradual options for each attribute. In fact, it is through the pheromone matrix that the algorithm learns how to generate highly probably valid gradual item set candidates.

	\begin{equation}\label{eqn5:p_matrix}
		\mathcal{T}_{a, j} = q \times 3
	\end{equation}
	\begin{center}
		where $q$: number of attributes, $a=1,...,q$ and, $j\in \{+, -, \times\}$
	\end{center}
		
	At the beginning all artificial pheromones $p_{a, j}$ in $\mathcal{T}_{a, j}$ are initialized to 1, then we modify the manner in which the pheromone matrix is updated as follows at time interval $(t, t+1)$: 
	\begin{itemize}
		\item Every generated gradual item set solution is evaluated and only valid solutions are used to update the artificial pheromone matrix. Invalid solutions are stored with aim of using them to reject their supersets.
		\item In a given valid solution, each gradual item set is used to update the corresponding artificial pheromone $p_{a, j} (t)$ (where $j$ is either $+$ or $-$) using Equation~\eqref{eqn5:update_p}. 
		\item Again using the same valid solution, for every attribute that does not have a gradual item set appearing in the solution - we update the corresponding irrelevant pheromone $p_{a, j}$ (where $j$ is $\times$) using Equation~\eqref{eqn5:update_p}.
	\end{itemize}

	\begin{equation}\label{eqn5:update_p}
		p_{a, j}(t) = p_{a, j}(t) + sup(Sol)
	\end{equation}
	\begin{center}
		where $sup(Sol)$ is the frequency support of the generated pattern solution
	\end{center}
	
	It should be highlighted that the proposed modification only affects Equation~\eqref{eqn5:update_p}. Previously in Chapter~\ref{ch4:aco_graank} pheromone $p_{a, j}$ is incremented by $1$; here, it is incremented by $sup(Sol)$. It is important to point out that as long as pheromones are incremented, the ACO technique effectively learns how to generate highly probable gradual item set candidates. However, since the value of $sup(Sol)$ is always less than $1$; the ACO technique takes a little longer to converge, consequently slightly reducing its efficiency.
	
	Equally important, the proposed modification allows for accumulation of the \textit{frequency supports} of validated temporal gradual patterns. Therefore, matrix $\mathcal{T}_{a, j}$ may also be referred to as \textit{support-based} pheromone matrix. We exploit this matrix for constructing TGEPs in the section that follows.

	\subsection{Growth-rate Manipulation for Mining TGEPs}
	\label{ch5:aco_gr}
	It should be noted that applying the proposed ACO-based approach on a data set extracts numerous GPs or TGPs, but only one pheromone matrix for every data set or transformed data set respectively (see Equation~\eqref{eqn5:p_matrix}). As illustrated in the Section~\ref{ch5:aco_gp}, the values of the matrix depend on the patterns extracted since each valid pattern increments it by its support (see Equation~\eqref{eqn5:update_p}). In that case, a single matrix cumulatively stores support values of all extracted patterns. The matrix can be normalized using the number of algorithm iterations determined by the \textit{convergence property}.
	
	The definitions given in Section~\ref{ch5:ep} about emerging patterns (EPs) and their growth-rate validates the idea that: if two data sets each provide its \textit{support-based} pheromone matrix, then dividing the two matrices element-wisely generates a \textit{growth-rate matrix}. Through division, the growth-rate matrix reduces any irrelevant EPs to zero but allows for construction of relevant EPs. To this end, there exists no reason to keep the gradual patterns previously extracted.
	
	\textit{Example 5.3.} Let \textit{$\{$Rain, Wind, Temperature$\}$} be attributes of data sets $\mathcal{D}_{1}$ and $\mathcal{D}_{2}$. $\mathcal{P}_{1}$ and $\mathcal{P}_{2}$ (shown in Figure~\ref{fig5:p_matrices}) be the pheromone matrices of data sets $\mathcal{D}_{1}$ and $\mathcal{D}_{2}$ respectively. 
	
	\begin{figure}[h!]
		\centering
		\subfloat[]{%
			\begin{tabular}{c|c c c|}
			\multicolumn{1}{c}{} & $\uparrow$ & $\downarrow$ & \multicolumn{1}{c}{$\times$}\\
			\cline{2-4}
			Rain & .8 & 0 & 0\\
			Wind & 0 & .9 & 0\\
			Temp & 0 & 0 & .85\\
			\cline{2-4}
			\end{tabular}
			$~~~~~$ 
			}%
		\subfloat[]{%
			\begin{tabular}{c|c c c|}
			\multicolumn{1}{c}{} & $\uparrow$ & $\downarrow$ & \multicolumn{1}{c}{$\times$}\\
			\cline{2-4}
			Rain & .4 & 0 & 0\\
			Wind & 0 & .6 & 0\\
			Temp & 0 & .75 & 0\\
			\cline{2-4}
			\end{tabular}
			}%
		\caption{(a) normalized support values of $\mathcal{P}_{1}$, and (b) normalized support values of $\mathcal{P}_{2}$}
		\label{fig5:p_matrices}
	\end{figure}

	A growth-rate matrix in favor of $\mathcal{P}_{1}$ is shown in Figure~\ref{fig5:gr_matrix1}. As can be deduced from the growth-rate matrix in Figure~\ref{fig5:gr_matrix1}, we may construct a GEP $\{(Rain, \uparrow), (Wind, \downarrow)\}$ with a growth-rate of at least $1.5$ from data set $\mathcal{D}_{1}$ to $\mathcal{D}_{2}$.

	\begin{figure}[h!]
	\centering
	\begin{tabular}{c|c c c|}
		\multicolumn{1}{c}{} & $\uparrow$ & $\downarrow$ & \multicolumn{1}{c}{$\times$}\\
		\cline{2-4}
		Rain & 2 & 0 & 0\\
		Wind & 0 & 1.5 & 0\\
		Temp & 0 & 0 & $\infty$\\
		\cline{2-4}
	\end{tabular}
	\caption{Growth-rate matrix from pheromone matrix $\mathcal{P}_{1}$ to $\mathcal{P}_{2}$}
	\label{fig5:gr_matrix1}
	\end{figure}

	In this chapter, our main aim is to extend the ACO strategy to the case of TGEP mining and compare its performance to the border-based strategy. Although mining TGPs using an ACO-based approach is easily achievable (see Section~\ref{ch5:aco_gp}), constructing TGEPs from growth-rate matrices is difficult since the matrices do not provide information about associated time-lags. For this reason, we propose an additional \textit{time-lag matrix} which is updated with approximated time-lags of validated patterns every time the \textit{pheromone matrix} is updated with support values of these patterns. Finally, the combined content of the growth-rate matrix and the time-lag matrix allow for the construction of TGEPs.

	\textit{Example 5.4.} Let $TGP_{1} = \{(Rain^{\uparrow}, Wind^{\downarrow})_{\approx +2 mins},sup=0.8 \}$ be extracted from a transformed data set $\mathcal{D}'_{g}$, and $TGP_{2} = \{(Rain^{\uparrow}, Wind^{\downarrow})_{\approx +6 mins},$ $sup=0.4 \}$ be extracted from a transformed data set $\mathcal{D}''_{g}$. Figure~\ref{fig5:tgp_matrices} shows the support pheromone matrices and the corresponding time-lag matrices for the transformed data sets.
		
	\begin{figure}[h!]
		\centering
		\subfloat[]{%
			\begin{tabular}{c|c c c|}
			\multicolumn{1}{c}{} & $\uparrow$ & $\downarrow$ & \multicolumn{1}{c}{$\times$}\\
			\cline{2-4}
			Rain & .8 & 0 & 0\\
			Wind & 0 & .8 & 0\\
			Temp & 0 & 0 & .8\\
			\cline{2-4}
			\end{tabular}
			$~~~~~$ 
		}%
		\subfloat[]{%
			\begin{tabular}{c|c c c|}
			\multicolumn{1}{c}{} & $\uparrow$ & $\downarrow$ & \multicolumn{1}{c}{$\times$}\\
			\cline{2-4}
			Rain & $+2 mins$ & 0 & 0\\
			Wind & 0 & $+2 mins$ & 0\\
			Temp & 0 & 0 & 0\\
			\cline{2-4}
			\end{tabular}
		}%
		\\
		\subfloat[]{%
			\begin{tabular}{c|c c c|}
			\multicolumn{1}{c}{} & $\uparrow$ & $\downarrow$ & \multicolumn{1}{c}{$\times$}\\
			\cline{2-4}
			Rain & .04 & 0 & 0\\
			Wind & 0 & .04 & 0\\
			Temp & 0 & 0 & .04\\
			\cline{2-4}
			\end{tabular}
			$~~~~~$ 
		}%
		\subfloat[]{%
			\begin{tabular}{c|c c c|}
			\multicolumn{1}{c}{} & $\uparrow$ & $\downarrow$ & \multicolumn{1}{c}{$\times$}\\
			\cline{2-4}
			Rain & $+6 mins$ & 0 & 0\\
			Wind & 0 & $+6 mins$ & 0\\
			Temp & 0 & 0 & 0\\
			\cline{2-4}
			\end{tabular}
		}%

		\caption{(a) pheromone matrix for $\mathcal{D}'_{g}$, (b) time-lag matrix for $\mathcal{D}'_{g}$, (c) pheromone matrix $\mathcal{D}''_{g}$, and (d) time-lag matrix for $\mathcal{D}''_{g}$, }
		\label{fig5:tgp_matrices}
	\end{figure}
	
	A growth-rate matrix in favor of the support-based pheromone matrix of transformed data set $\mathcal{D}'_{g}$ is shown in Figure~\ref{fig5:gr_matrix2}. This growth-rate matrix is mapped element-wisely onto time-lag matrices of $\mathcal{D}'_{g}$ and $\mathcal{D}''_{g}$ in order to eliminate irrelevant time-lag elements; in this case none of the elements are irrelevant.
			
	\begin{figure}[h!]
	\centering
	\begin{tabular}{c|c c c|}
		\multicolumn{1}{c}{} & $\uparrow$ & $\downarrow$ & \multicolumn{1}{c}{$\times$}\\
		\cline{2-4}
		Rain & 20 & 0 & 0\\
		Wind & 0 & 20 & 0\\
		Temp & 0 & 0 & 20\\
		\cline{2-4}
	\end{tabular}
	\caption{Growth-rate matrix from pheromone matrix of data set $\mathcal{D}'_{g}$ to $\mathcal{D}''_{g}$}
	\label{fig5:gr_matrix2}
	\end{figure}
	
	As can be deduced by combining growth-rate matrix in Figure~\ref{fig5:gr_matrix2} and time-lag matrices in Figure~\ref{fig5:tgp_matrices} (b) and (d), we may construct a TGEP $\{(Rain, \uparrow), (Wind, \downarrow)\}$ with a growth-rate of $20$ after approximately $4$ minutes. All things considered, we propose Algorithm~\ref{alg5:trenc}, TRENC stands for Temporal gRadual Emerging aNt Colony optimization.

	\begin{algorithm}[h!]
	\caption{\textbf{TRENC} algorithm}
	\label{alg5:trenc}

	\SetKwFunction{acograd}{aco-matrices}

	\SetKwFunction{division}{gen-growthrate}
	\SetKwFunction{build}{construct}

	\SetKwFunction{transform}{transform}

	\SetKwInOut{Input}{\textbf{Input}}
	\SetKwInOut{Output}{\textbf{Output}}

	\Input{$D -$ data set, $refCol -$ reference column, $\sigma -$ minimum support, $\delta -$ minimum repsentatitvity}
	\Output{$TGEPs -$ TGEPs in JSON format}

	$\mathcal{D}^{*}, T_{d}^{*} \leftarrow$ \transform($D, \delta, refCol$)
	\tcc*[H]{$\mathcal{D}-$ transformed data set, $T_{d}-$ time differences, $*$ denotes multiple}

	$\mathcal{P}^{*}, \mathcal{T}^{*} \leftarrow$ \acograd($D^{*}, T_{d}^{*}, \sigma$)
	\tcc*[H]{$\mathcal{P}-$ pheromone matrix, $\mathcal{T}-$ time-lag matrix }

	$\mathcal{G}^{*} \leftarrow$ \division($\mathcal{P}[x]$, $\mathcal{P}^{*}$)\tcc*[H]{$\mathcal{G}-$ growth-rate matrix, $x-$ user-specified w.r.t preferred transformed data set}

	$TGEPs \leftarrow$ \build($\mathcal{G}^{*}, \mathcal{T}^{*}$)\;

	\Return $TGEPs$\;
	\end{algorithm}

	\begin{enumerate}
		\item Build multiple transformed data sets $\mathcal{D}'_{g}, \mathcal{D}''_{g}, ..., \mathcal{D}^{*}_{g}$ from a timestamped data set $\mathcal{D}_{g}$ using a specified representativity threshold $\delta$ (see Section~\ref{ch5:tgep}).
		\item From each transformed data set, build a normalized support pheromone matrix along with corresponding time-lag matrices.
		\item Generate growth-rate matrices from the pheromone matrices obtained in Step 2 (two pheromone matrices at a time).
		\item Combine each growth-rate matrix with the two corresponding time-lag matrices to construct TGEPs.
	\end{enumerate}
	
	\clearpage
		
	\section{Experiments}
	\label{ch5:experiments}
	In this section, we implement the border-based BT-GRAANK algorithm (described in Section~\ref{ch5:border_tgep}) and the ant-based TRENC algorithm (described in Section~\ref{ch5:aco_tgep}) for mining TGEPs and analyze their computational performances. All experiments were conducted on a (High Performance Computing) HPC platform \textbf{Meso@LR}\footnote{\url{https://meso-lr.umontpellier.fr}}. We used one node made up of 112 cores and 128GB of RAM.

	\subsection{Source Code}
	\label{ch5:source_code}
	The \texttt{Python} source code of our proposed algorithms are available at our GitHub repository: \url{https://github.com/owuordickson/trenc.git}.

	\subsection{Data Set Description}
	\label{ch5:dataset}
	Table~\ref{tab5:dataset} shows the features of the data sets used in the experiments for evaluating the computational performance of our proposed algorithms.

	\begin{table}[h!]
	\centering
	\small
	\caption{Experiment data sets}
	\begin{tabular}{|l|c|c|c|c|}
		\hline
		Data set & $\#$tuples & $\#$attributes & Timestamped & Domain\\
		\hline \hline
		Buoys (Directio) & 6121 & 21 & Yes & Coastline\\
		Power Consumption (UCI) & 10001 & 9 & Yes & Electrical\\
		\hline
	\end{tabular}
	\label{tab5:dataset}
	\end{table}

	The `Power Consumption' (UCI) data set, obtained from \texttt{UCI Machine Learning Repository} \cite{Dua2017}, describes the electric power consumption in one household (located in Sceaux, France) in terms of active power, voltage and global intensity with a one-minute sampling rate between December 2006 and November 2010.

	The `Directio' data set is one of 4 data sets obtained from OREMES’s data portal\footnote{\url{https://data.oreme.org}} that recorded swell sensor signals of 4 buoys near the coast of the Languedoc-Roussillon region in France between 2012 and 2019 \cite{Bouchette2019}. These data sets can be retrieved from: \url{https://github.com/owuordickson/trenc/tree/master/data}.

	\subsection{Experiment Results}
	\label{ch5:results}
	In this section, we present the results of our experimental study on the two data sets (described in Section~\ref{ch5:dataset}) using our proposed algorithms with a minimum growth-rate threshold $\rho = 1.0$. These results reveal that the two algorithms behave differently when applied on different data sets (especially if they vary in number of attributes). We use these results to analyze and compare the computational efficiency and parallel efficiency of the algorithms as presented in Section~\ref{ch5:comparative_1} and Section~\ref{ch5:comparative_2} respectively. We discuss these results in Section~\ref{ch5:discussion}. All the experiment results can be obtained from: \url{https://github.com/owuordickson/meso-hpc-lr/tree/master/results/tgeps/112cores}.

	\subsubsection{Comparative Experiments: computational efficiency}
	\label{ch5:comparative_1}
	This experiment compares the run-time and number of temporal gradual emerging patterns (TGEPs) extracted by TRENC and BT-GRAANK from data sets \textit{UCI} and \textit{Directio}. We mention that the minimum representativity threshold is set at \texttt{0.99} so that very few transformations are applied on the original data sets, which improves the quality of TGEPs (see Section~\ref{ch5:tgep}).

	\begin{figure}[h!]
		\centering
		\small
		\texttt{UCI data set: \#attributes=9/\#tuples=10K/cores=56/min-rep=0.99}\par\medskip
  		\subfloat[]{%
		\begin{tikzpicture}
  			\begin{axis}[
  				height=5cm, width=7.5cm,
  				grid=both,
  				grid style={line width=.1pt, draw=gray!20},
    			major grid style={line width=.2pt, draw=gray!50},
  				axis lines=middle,
  				minor tick num=5,
    			ymin=0, ymax=275,
    			xmin=0.48, xmax=0.95,
    			xtick={0.5, 0.6, 0.7, 0.8, 0.9},
   				xlabel=Minimum support,
   				ylabel=Run time (sec),
  				xlabel style={at={(axis description cs:0.5,-0.1)},anchor=north},
  				ylabel style={at={(axis description cs:-0.15,0.5)},rotate=90, anchor=south},
 	 			legend style={nodes={scale=0.5}},
 	 			legend pos=north east
 				]

			\addplot[smooth, mark=x, mark size=3pt, color=red] table[x=Sup, y=Time]{border_uci_56.dat};
  	  		\addlegendentry{BT-GRAANK}

  			\addplot[smooth, mark=o, color=blue] table[x=Sup, y=Time]{trenc_uci_56.dat};
			\addlegendentry{TRENC}
  			\end{axis}
  		\end{tikzpicture}}%
  		\subfloat[]{%
  		\begin{tikzpicture}
  			\begin{axis}[
  				height=4.7cm, width=7.5cm,
  				ybar,
    			symbolic x coords={0.5, 0.6, 0.7, 0.8, 0.9},
            	xtick=data,
   				xlabel=Minimum support,
   				ylabel=No. of patterns,
   				legend style={nodes={scale=0.5}},
 				]
			\addplot table[x=Sup, y=Patterns]{border_uci_56.dat};
			\addplot table[x=Sup, y=Patterns]{trenc_uci_56.dat};
			\legend{BT-GRAANK, TRENC}
  			\end{axis}
  		\end{tikzpicture}}%
    	\caption{UCI data set $(\rho = 1.0)$: (a) plot of run time against minimum support threshold and, (b) bar graph of number of patterns against minimum support threshold.}
  		\label{fig5:tgep_comparison1}
  	\end{figure}
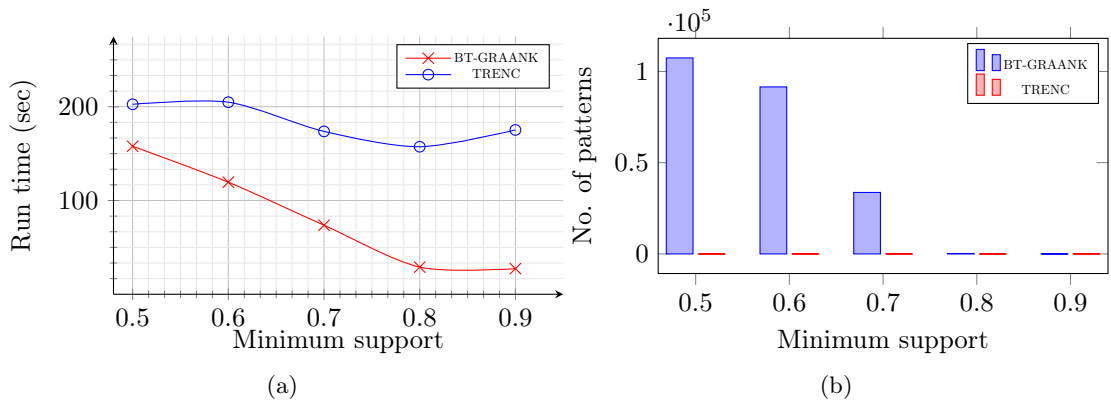

  	Figure~\ref{fig5:tgep_comparison1} (a) shows run-time and, Figure~\ref{fig5:tgep_comparison1} (b) shows the number of TGEPs extracted  by TRENC and BT-GRAANK algorithms when applied on the \textit{UCI} data set. We observe that the run-time of BT-GRAANK (blue curve) is lower than that of TRENC (red curve) and it reduce significantly (as well as the number of TGEPs) as support threshold is increased. For the case of TRENC, the run-time and number of TGEPs are almost constant.

  	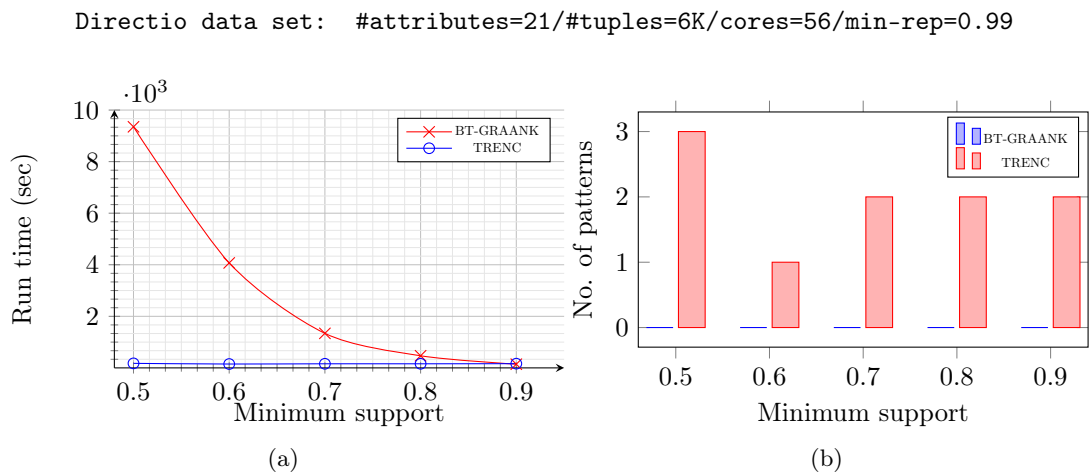
\begin{figure}[h!]
		\centering
		\small
		\texttt{Directio data set: \#attributes=21/\#tuples=6K/cores=56/min-rep=0.99}\par\medskip
  		\subfloat[]{%
		\begin{tikzpicture}
  			\begin{axis}[
  				height=5cm, width=7.5cm,
  				scaled y ticks=base 10:-3,
  				grid=both,
  				grid style={line width=.1pt, draw=gray!20},
    			major grid style={line width=.2pt, draw=gray!50},
  				axis lines=middle,
  				minor tick num=5,
    			ymin=0, ymax=10000,
    			xmin=0.48, xmax=0.95,
    			xtick={0.5, 0.6, 0.7, 0.8, 0.9},
   				xlabel=Minimum support,
   				ylabel=Run time (sec),
  				xlabel style={at={(axis description cs:0.5,-0.1)},anchor=north},
  				ylabel style={at={(axis description cs:-0.15,0.5)},rotate=90, anchor=south},
 	 			legend style={nodes={scale=0.5}},
 	 			legend pos=north east
 				]

			\addplot[smooth, mark=x, mark size=3pt, color=red] table[x=Sup, y=Time]{border_directio_56.dat};
  	  		\addlegendentry{BT-GRAANK}

  			\addplot[smooth, mark=o, color=blue] table[x=Sup, y=Time]{trenc_directio_56.dat};
			\addlegendentry{TRENC}
  			\end{axis}
  		\end{tikzpicture}}%
  		\subfloat[]{%
  		\begin{tikzpicture}
  			\begin{axis}[
  				height=4.7cm, width=7.5cm,
  				ybar,
    			symbolic x coords={0.5, 0.6, 0.7, 0.8, 0.9},
            	xtick=data,
   				xlabel=Minimum support,
   				ylabel=No. of patterns,
   	 	 		legend style={nodes={scale=0.5}},
 				]
			\addplot table[x=Sup, y=Patterns]{border_directio_56.dat};
			\addplot table[x=Sup, y=Patterns]{trenc_directio_56.dat};
			\legend{BT-GRAANK, TRENC}
  			\end{axis}
  		\end{tikzpicture}}%
    	\caption{Directio data set $(\rho = 1.0)$: (a) plot of run time against minimum support threshold and, (b) bar graph of number of patterns against minimum support threshold.}
  		\label{fig5:tgep_comparison2}
  	\end{figure}
  	
  	Figure~\ref{fig5:tgep_comparison2} (a) shows run-time performance and, Figure~\ref{fig5:tgep_comparison2} (b) shows the number of TGEPs extracted  by TRENC and Border-TGRAANK algorithms when applied on the \textit{Directio} data set. In this instance, BT-GRAANK (in comparison to TRENC) has the highest run-time (which reduces) and fewest TGEPs as the support threshold is increased. Again, the run-time and number of extracted of TGEPs are almost constant for the case of TRENC.
	
  	It should be remembered that support threshold plays an important role in determining the quality and quantity of extracted frequent patterns (see Chapter~\ref{ch2:gp}). The higher the threshold, the higher the quality of the patterns; consequently, trivial patterns are ignored. This explains why the run-time and number of TGEPs reduce significantly for the case of BT-GRAANK. Concerning TRENC, it is based on ACO strategy (see Chapter~\ref{ch4:aco_graank}) which mines for maximal patterns first. For this particular case, it may seem that the maximal patterns mined are of high quality; therefore, it seems that the support threshold has little effect on the quantity of TGEPs.

	\subsubsection{Comparative Experiments: parallel efficiency}
	\label{ch5:comparative_2}
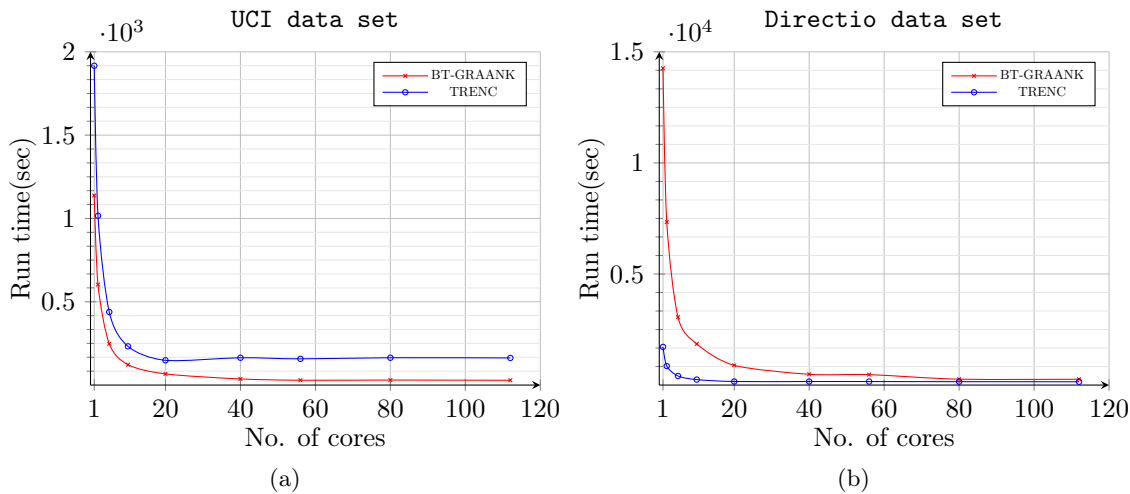
\begin{figure}[h!]
		\centering
		\small
  		\subfloat[]{%
		\begin{tikzpicture}
  			\begin{axis}[
  				title= \texttt{UCI data set},
  				height=6cm, width=7.5cm,
  				scaled y ticks=base 10:-3,
  				grid=both,
  				grid style={line width=.1pt, draw=gray!20},
    			major grid style={line width=.2pt, draw=gray!50},
  				axis lines=middle,
  				minor tick num=5,
  				ymin=0,ymax=2000,
  				xmin=0,xmax=120,
  				xtick={1,20,40,60,80,100,120},
  				xlabel=No. of cores, ylabel=Run time(sec),
  				xlabel style={at={(axis description cs:0.5,-0.1)},anchor=north},
  				ylabel style={at={(axis description cs:-0.1,0.5)},rotate=90,anchor=south},
 				legend style={nodes={scale=0.5}},
 				legend pos=north east]

 			\addplot[smooth, mark=x, mark size=1pt, color=red] table[x=Cores, y=Time]{border_uci.dat};
  			\addlegendentry{BT-GRAANK}
 			\addplot[smooth, mark=o, mark size=1pt, color=blue] table[x=Cores, y=Time]{trenc_uci.dat};
  	  		\addlegendentry{TRENC}

  			\end{axis}
  		\end{tikzpicture}}%
  		\subfloat[]{%
  		\begin{tikzpicture}
  			\begin{axis}[
  				title= \texttt{Directio data set},
  				height=6cm, width=7.5cm,
  				grid=both,
  				grid style={line width=.1pt, draw=gray!20},
    			major grid style={line width=.2pt, draw=gray!50},
  				axis lines=middle,
  				minor tick num=5,
  				ymin=0,ymax=15000,
  				xmin=0,xmax=120,
  				xtick={1,20,40,60,80,100,120},
  				xlabel=No. of cores, ylabel=Run time(sec),
  				xlabel style={at={(axis description cs:0.5,-0.1)},anchor=north},
  				ylabel style={at={(axis description cs:-0.1,0.5)},rotate=90,anchor=south},
 				legend style={nodes={scale=0.5}},
 				legend pos=north east]

 			\addplot[smooth, mark=x, mark size=1pt, color=red] table[x=Cores, y=Time]{border_directio.dat};
  			\addlegendentry{BT-GRAANK}
 			\addplot[smooth, mark=o, mark size=1pt, color=blue] table[x=Cores, y=Time]{trenc_directio.dat};
  	  		\addlegendentry{TRENC}

  			\end{axis}
  		\end{tikzpicture}}%
  		\caption{Plot of run time versus no. of cores on data sets (a) UCI and, (b) Directio}
		\label{fig5:parallel1}
	\end{figure}

	This experiment compares the run-time of BT-GRAANK and TRENC algorithms against different number of CPU cores on data sets \textit{UCI} and \textit{Directio}. 	In Figure~\ref{fig5:parallel1}, the run-time of both algorithms reduce as the number of cores increase.

	We use these results to analyze their multiprocessing behavior using the \texttt{speedup} and \texttt{parallel efficiency} performance measures. (1) \texttt{Speedup S(n)} may be defined as: \textit{``the ratio of the execution time of a single processor to the execution time of $n$ processors''} ($S(n) = T_{1}/T_{n}$). (2) \texttt{Parallel efficiency E(n)} may be defined as: \textit{``the average utilization of $n$ processors''} ($	E(n) = S(n)/n$) \cite{Eager1989}.
	
	\clearpage

	\begin{figure}[h!]
		\centering
		\small
		\texttt{UCI data set: \#attributes=9/\#tuples=10K/min-rep=0.99/min-sup=0.8}\par\medskip
  		\subfloat[]{%
  		\begin{tikzpicture}
  			\begin{axis}[
  				height=6cm, width=7.5cm,
  				grid=both,
  				grid style={line width=.1pt, draw=gray!20},
    			major grid style={line width=.2pt, draw=gray!50},
  				axis lines=middle,
  				minor tick num=5,
  				ymin=0,ymax=115,
  				xmin=0,xmax=115,
  				xtick={1,20,40,60,80,100,120},
    			ytick={1,20,40,60,80,100,120},
  				xlabel=No. of cores, ylabel=Speedup,
  				xlabel style={at={(axis description cs:0.5,-0.1)},anchor=north},
  				ylabel style={at={(axis description cs:-0.15,0.5)},rotate=90,anchor=south},
 				legend style={nodes={scale=0.5}},
 				legend pos=north west]

  			\addplot[smooth, mark=x, mark size=1pt, color=red] table[x=Cores, y=Speedup]{border_uci_sp.dat};
  			\addlegendentry{BT-GRAANK}
  			\addplot[smooth, mark=o, mark size=1pt, color=blue] table[x=Cores, y=Speedup]{trenc_uci_sp.dat};
  	  		\addlegendentry{TRENC}
  	  		\addplot[dashed, color=brown] table[x=Cores, y=Speedup]{speedup_1.dat};

  			\end{axis}
  		\end{tikzpicture}}%
  		\subfloat[]{%
  		\begin{tikzpicture}
  			\begin{axis}[
  				height=6cm, width=7.5cm,
  				grid=both,
  				grid style={line width=.1pt, draw=gray!20},
    			major grid style={line width=.2pt, draw=gray!50},
  				axis lines=middle,
  				minor tick num=5,
  				ymin=0,ymax=1.2,
  				xmin=0,xmax=115,
  				ytick={0,.5,1},
  				xtick={1,20,40,60,80,100,120},
  				xlabel=No. of cores, ylabel=Parallel efficiency,
  				xlabel style={at={(axis description cs:0.5,-0.1)},anchor=north},
  				ylabel style={at={(axis description cs:-0.15,0.5)},rotate=90,anchor=south},
 				legend style={nodes={scale=0.5}},
 				legend pos=north east]

  			\addplot[smooth, mark=x, mark size=1pt, color=red] table[x=Cores, y=Efficiency]{border_uci_sp_eff.dat};
  			\addlegendentry{BT-GRAANK}
  			\addplot[smooth, mark=o, mark size=1pt, color=blue] table[x=Cores, y=Efficiency]{trenc_uci_sp_eff.dat};
  	  		\addlegendentry{TRENC}
  	  		\addplot[thick, dashed, color=brown] table[x=Cores, y=Efficiency]{sp_efficiency_1.dat};

  			\end{axis}
  		\end{tikzpicture}}%
  		\caption{UCI data set $(\rho = 1.0)$: (a) plot of speed up versus number of cores (b) plot of parallel efficiency versus number of cores}
  		\label{fig5:parallel2}
  	\end{figure}
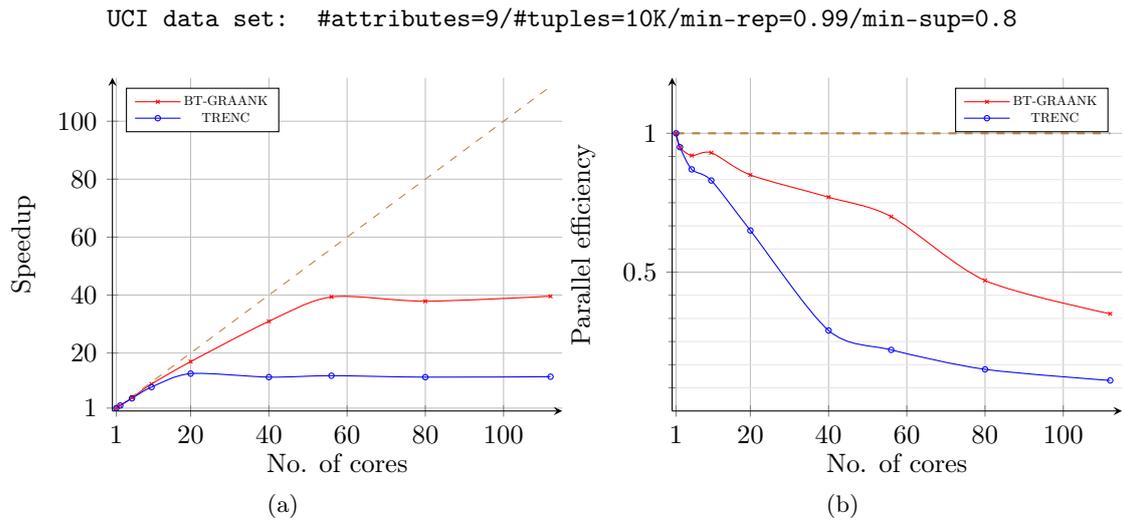
	
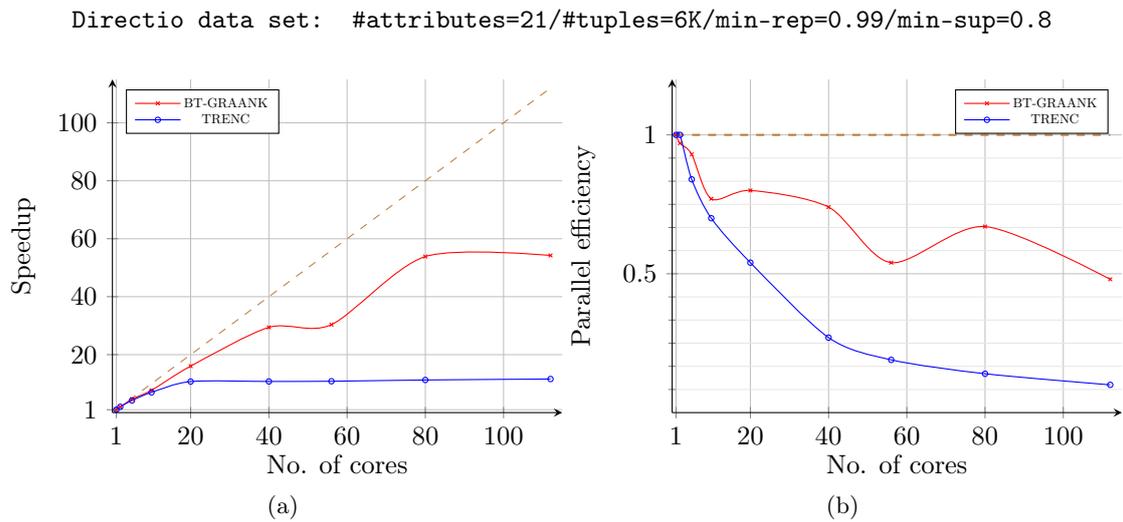
\begin{figure}[h!]
		\centering
		\small
		\texttt{Directio data set: \#attributes=21/\#tuples=6K/min-rep=0.99/min-sup=0.8}\par\medskip
  		\subfloat[]{%
  		\begin{tikzpicture}
  			\begin{axis}[
  				height=6cm, width=7.5cm,
  				grid=both,
  				grid style={line width=.1pt, draw=gray!20},
    			major grid style={line width=.2pt, draw=gray!50},
  				axis lines=middle,
  				minor tick num=5,
  				ymin=0,ymax=115,
  				xmin=0,xmax=115,
  				xtick={1,20,40,60,80,100,120},
    			ytick={1,20,40,60,80,100,120},
  				xlabel=No. of cores, ylabel=Speedup,
  				xlabel style={at={(axis description cs:0.5,-0.1)},anchor=north},
  				ylabel style={at={(axis description cs:-0.15,0.5)},rotate=90,anchor=south},
 				legend style={nodes={scale=0.5}},
 				legend pos=north west]

  			\addplot[smooth, mark=x, mark size=1pt, color=red] table[x=Cores, y=Speedup]{border_directio_sp.dat};
  			\addlegendentry{BT-GRAANK}
  			\addplot[smooth, mark=o, mark size=1pt, color=blue] table[x=Cores, y=Speedup]{trenc_directio_sp.dat};
  	  		\addlegendentry{TRENC}
  	  		\addplot[dashed, color=brown] table[x=Cores, y=Speedup]{speedup_1.dat};

  			\end{axis}
  		\end{tikzpicture}}%
  		\subfloat[]{%
  		\begin{tikzpicture}
  			\begin{axis}[
  				height=6cm, width=7.5cm,
  				grid=both,
  				grid style={line width=.1pt, draw=gray!20},
    			major grid style={line width=.2pt, draw=gray!50},
  				axis lines=middle,
  				minor tick num=5,
  				ymin=0,ymax=1.2,
  				xmin=0,xmax=115,
  				ytick={0,.5,1},
  				xtick={1,20,40,60,80,100,120},
  				xlabel=No. of cores, ylabel=Parallel efficiency,
  				xlabel style={at={(axis description cs:0.5,-0.1)},anchor=north},
  				ylabel style={at={(axis description cs:-0.15,0.5)},rotate=90,anchor=south},
 				legend style={nodes={scale=0.5}},
 				legend pos=north east]

  			\addplot[smooth, mark=x, mark size=1pt, color=red] table[x=Cores, y=Efficiency]{border_directio_sp_eff.dat};
  			\addlegendentry{BT-GRAANK}
  			\addplot[smooth, mark=o, mark size=1pt, color=blue] table[x=Cores, y=Efficiency]{trenc_directio_sp_eff.dat};
  	  		\addlegendentry{TRENC}
  	  		\addplot[thick, dashed, color=brown] table[x=Cores, y=Efficiency]{sp_efficiency_1.dat};

  			\end{axis}
  		\end{tikzpicture}}%
  		\caption{Directio data set $(\rho = 1.0)$: (a) plot of speed up versus number of cores (b) plot of parallel efficiency versus number of cores}
  		\label{fig5:parallel3}
  	\end{figure}

  	In Figure~\ref{fig5:parallel2} (a), we observe that BT-GRAANK (in comparison to TRENC) has a highest Speedup and, in Figure~\ref{fig5:parallel2} (b) has a highest parallel efficiency when both are applied on the UCI data set. Again, in Figure~\ref{fig5:parallel3} (a), we observe that BT-GRAANK (in comparison to TRENC) has a highest Speedup and, in Figure~\ref{fig5:parallel3} (b) has a highest parallel efficiency when both are applied on the Directio data set. However, it should be observed, from Figure~\ref{fig5:parallel1} (b), that the run-time of BT-GRAANK is higher than that of TRENC on data set Directio.

	\subsubsection{Consistent Temporal Gradual Emerging Patterns (TGEPs)}
	\label{ch5:consistent_tgeps}
	In this section, we present the consistent TGEPs (see Section~\ref{ch5:tgep} for Definition) extracted from the two data sets described above.

	\begin{table}[h!]
	\small
	\centering
	\caption{Consistent TGEPs extracted from data sets UCI and Directio}
	\begin{tabular}{|l|l|}
		\hline
		Data set & Consistent TGEPs $(\rho = 1.0)$\\
		\hline \hline
		Buoys (Directio) & $\{(Tz, \uparrow), (Hm0, \uparrow)\}$: growth-rate 1.0 \\&after every 2.5 hours\\
		\hline
		Power Consumption (UCI) & $\{(activepower, \uparrow), (voltage, \downarrow)\}$: growth-rate 1.04 \\&after every 24 hours\\
		\hline
	\end{tabular}
	\label{tab5:consistent_tgeps}
	\end{table}

	\subsection{Discussion of Results}
	\label{ch5:discussion}

	\subsubsection{Computation Run-time Complexity}
	\label{ch5:discuss_runtime}
	Unlike BT-GRAANK, we observe that the run-time for TRENC on both data sets is almost constant. As described in Section~\ref{ch5:aco_gp} is based on a heuristic technique for efficiently generating maximal gradual item sets. Therefore, the run-time required for extracting patterns through this technique is determined by how long the pheromone matrix takes to converge. Again, this property enables TRENC to have a lower run-time than BT-GRAANK (see Section~\ref{ch5:temporal_border}) when executed on a data set with a large number of attributes. For example since data set Directio has more attributes than data set UCI, there is a significant increase in run-time for the case BT-GRAANK and a relatively small change for the case of TRENC (see Figure~\ref{fig5:tgep_comparison1} (a) and Figure~\ref{fig5:tgep_comparison2} (a).

	Finally, we observe that BT-GRAANK has a higher Speedup and parallel efficiency than TRENC for both data sets as shown in Figure~\ref{fig5:parallel2} and Figure~\ref{fig5:parallel3}. Majorly, this is due the fact that TRENC's run-time is almost constant despite the variations in support threshold and number of cores. This implies the advantage that TRENC can extract high quality TGEPs using few processors and at any support threshold.

	\subsubsection{Extracted TGEPs}
	\label{ch5:discuss_tgeps}
	We observe that BT-GRAANK extracts more TGEPs than TRENC from the UCI data set. It should be emphasized that BT-GRAANK identifies borders from two maximal items. For this reason, numerous borders are used to construct few TGEPs (see Section~\ref{ch5:temporal_border}). In fact, we discover that both algorithms identify similar consistent TGEPs from the UCI data set as shown in Table~\ref{tab5:consistent_tgeps}.
	
	\section{Summary}
	\label{ch5:summary}
	In this work, first we have introduced the concept of temporal gradual emerging patterns; second, we have proposed two strategies for mining temporal gradual emerging patterns; third, we have proposed an experimental computational performance comparison including a parallel implementation of these two approaches on a HPC supercomputer. Finally, we recommend ant-based strategy as the most suitable strategy for mining temporal gradual emerging patterns especially when dealing with huge data sets.






\def\removeleadingzeros#1{\if0#1 \expandafter\else#1\fi}

\def\transformtime#1:#2:#3!{
\pgfkeys{/pgf/fpu=true,/pgf/fpu/output format=fixed}
\pgfmathparse{\removeleadingzeros#1*3600-\pgfkeysvalueof{/pgfplots/timeplot zero}*3600+\removeleadingzeros#2*60+\removeleadingzeros#3}
\pgfkeys{/pgf/fpu=false}
}

\pgfplotsset{
timeplot zero/.initial=0,
timeplot/.style={
    x coord trafo/.code={\expandafter\transformtime##1!},
    x coord inv trafo/.code={%
        \pgfkeys{/pgf/fpu=true,/pgf/fpu/output format=fixed}
        \pgfmathsetmacro\hours{floor(##1/3600)+\pgfkeysvalueof{/pgfplots/timeplot zero}}
        \pgfmathsetmacro\minutes{floor((##1-(\hours-\pgfkeysvalueof{/pgfplots/timeplot zero})*3600)/60)}
        \pgfmathsetmacro\seconds{##1-floor(##1/60)*60}
        \def\pgfmathresult{\pgfmathparse{mod(\hours,60)<10?"0":{},int(mod(\hours,60))}\pgfmathresult:\pgfmathparse{mod(\minutes,60)<10?"0":{},int(mod(\minutes,60))}\pgfmathresult:\pgfmathparse{mod(\seconds,60)<10?"0":{},int(mod(\seconds,60))}\pgfmathresult}
        \pgfkeys{/pgf/fpu=false}
    },
scaled x ticks=false,
xticklabel=\tick
}
}


\chapter{Data Crossing for Gradual Pattern Mining}
\label{ch6}
\minitoc \clearpage 

	\begin{chapquote}{Ray Bradbury, \textit{Zen in the Art of Writing}}
		``You must stay drunk on writing so reality cannot destroy you''
	\end{chapquote}

	\section{Introduction}
	\label{ch6:introduction}
	In this chapter, we propose and describe a fuzzy model for crossing unrelated time-series data sets with the ultimate goal of exploiting them for temporal gradual pattern mining. By using a fuzzy model, our proposed approach becomes more robust than other crisp models that could miss a phenomenon because of small data variations. We develop an algorithm that implements our proposed model and we test it on real data. In addition, we apply parallel processing on the algorithm implementation and measure its computational performance. We specifically test our model on numeric time-series data sets so as to extract temporal gradual patterns afterwards.
		
	\section{Crossing Time-series Data}
	\label{ch6:fuzztx}
\begin{figure}[h!]
		\centering
		\includegraphics[width=.6\textwidth]{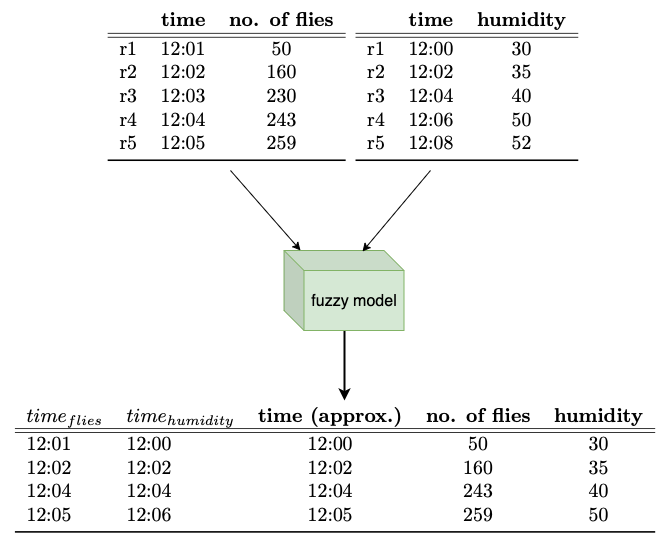}
		\caption{Illustration of crossing `no. of flies' and `humidity' data sets using a fuzzy model}
		\label{fig6:crossing_model}
	\end{figure}

	Today, with the proliferation of \textit{Internet of Things} (IoT) applications in almost every area of our society comes the trouble of deducing relevant information from real-time time-series data (from different sources) for decision making. A possible solution to this may be \textit{data crossing}. We recall the definition of data crossing given in Section~\ref{ch2:data_crossing} as \textit{``a process that enables the matching of different data sets using a pre-defined criteria and combining their data points to form a new data set''}. Figure~\ref{fig6:crossing_model} illustrates the process of crossing two time-series data sets to form one data set through a fuzzy model.

	\textit{Example 6.1.} We consider two sample time-series data sets shown in Table~\ref{tab6:sample1}.
	
	\begin{table}[h!]
  		\centering
		\subfloat[]{%
    		\begin{tabular}{l c c} 
      		 & \textbf{time} & \textbf{no. of flies} \\
      		\hline \hline
      		r1 & 12:01 & 50\\
      		r2 & 12:02 & 160\\
      		r3 & 12:03 & 230\\
      		r4 & 12:04 & 243\\
      		r5 & 12:05 & 259\\
      		\bottomrule
    		\end{tabular}}%
    	$~~~~~$
		\subfloat[]{%
    		\begin{tabular}{l c c}
      		 & \textbf{time} & \textbf{humidity} \\
      		\hline \hline
      		r1 & 12:00 & 30\\
      		r2 & 12:02 & 35\\
      		r3 & 12:04 & 40\\
      		r4 & 12:06 & 50\\
      		r5 & 12:08 & 52\\
      		\bottomrule
    		\end{tabular}}%
    		\caption{(a) Sample of population of flies time-series data (b) sample of humidity time-series data}
    		\label{tab6:sample1}
	\end{table}
	
	The \textit{date-time} attribute reveals how closely simultaneous the occurrence of data points of the 2 data sets are. For example, time-series data sets having most of their data points occurring at almost the same time, when crossed, yield a data set that maps almost all the data points of the individual sets.

	In this example, we notice that there exists a degree of fuzziness for any time interval that matches respective tuples in both data sets. For instance if we use a triangular membership function and we pick `$1200$ \textit{hours}' as the center of this function - the membership degrees (MDs) of humidity data set's \textit{`time'} attribute may be approximated as: $\{(1200, 1.0), (1202, 0.8)$, $(1204, 0.6),(1208, 0.4)$, $(1208, 0.2) \}$.

	Similarly, the MDs of the number of flies data set's \textit{`time'} attribute may be approximated as: $\{(1201, 0.9)$, $(1202, 0.8)$, $(1203, 0.7)$, $(1204, 0.6)$, $(1205, 0.5) \}$. Therefore, for any center that we pick between `$1200$ \textit{hours}' and `$1208$ \textit{hours}', the MD in the population of \textit{time} attribute decreases from closest value to the furthest value. This interesting (MD) feature can be harnessed and applied on a fuzzy model that may cross time-series data from different sources. We describe this model in the section that follows.

	\subsection{Building the Fuzzy Model}
	\label{ch6:fuzzy_model}
	In this section, we construct a fuzzy model for crossing time-series data sets from different sources. We cross them with the intention of extracting temporal gradual patterns.
	
	In \textit{Fuzzy sets} theory, there exists a great number of membership functions that one can apply on a data set for fitting purposes \cite{Zadeh1965, Ayouni2010, Mandal2012}. In this chapter, we pick a triangular membership function (MF) so that we can order the MDs of \textit{date-time} population with reference to a single-value center. Automatically, this eliminates any MF whose center includes more than one value.
	
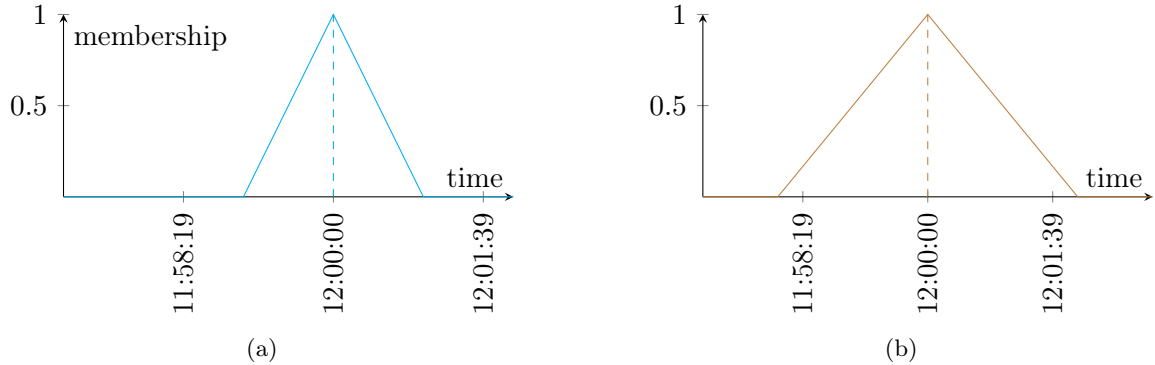
\begin{figure}[h!]
		\subfloat[]{%
		\begin{tikzpicture}
		\begin{axis}[
				height=4cm,
				width=7.5cm,
				xlabel=time,
				ylabel=membership,
				enlargelimits = false,
				no marks,
				axis lines=middle,
  				clip=false,
  				ymin=0,ymax=1,
  				ytick={0,0.5,1},
  				timeplot, timeplot zero=0,
    			xticklabel style={rotate=90, anchor=near xticklabel},
  				legend pos=north west]
  		\addplot+[color=cyan] table[x index=0, y index=1] {fuzzy1.dat};
    	\draw [thin, dashed, draw=cyan] (axis cs: 12:00:00,0) -- (axis cs: 12:00:00,1);
		\end{axis}
		\end{tikzpicture}
		}%
		\qquad \qquad
		\subfloat[]{%
		\begin{tikzpicture}
		\begin{axis}[
				height=4cm,
				width=7.5cm,
				xlabel=time,
				ylabel=\empty,
				enlargelimits = false,
				no marks,
				axis lines=middle,
  				clip=false,
  				ymin=0,ymax=1,
  				ytick={0,0.5,1},
  				timeplot, timeplot zero=0,
    			xticklabel style={rotate=90, anchor=near xticklabel},
  				legend pos=north west]
  		\addplot+[color=brown] table[x index=0, y index=1] {fuzzy2.dat};
    	\draw [thin, dashed, draw=brown] (axis cs: 12:00:00,0) -- (axis cs: 12:00:00,1);
		\end{axis}
		\end{tikzpicture}
		}
		\caption{(a) Membership function for temperature time-series data (b) membership function for humidity time-series data}
    	\label{fig6:membership}
	\end{figure}

	It is important to mention that we pick a triangular MF over the Gaussian MF since it is simpler to implement and, moreover our interest is not in fitting the data set perfectly \cite{Mandal2012}. For instance, it is easy to construct an initial triangular MF for the \textit{date-time} population of each time-series data by using the minimum value as the center and the smallest difference in the population to calculate the minimum and maximum extremes as shown in Figure~\ref{fig6:membership} (a) and (b). 
	
	Using these 2 triangular MFs, we can build a model that crosses tuples based on MDs. We exploit this idea and propose the \textit{procedure} that follows: (see proposed Algorithm~\ref{alg6:fuzztx} for pseudocode)
	
	\begin{enumerate}
		\item select the triangular MF with the largest boundaries
		\item apply the MF on each data set's \textit{date-time} population to pick the \textit{tuple-index} of the value with the largest (MD)
		\item use the \textit{tuple-index} to retrieve and cross tuples
		\item slide the MF positively by its boundary, and repeat all steps from step 1 until the population is exhausted
	\end{enumerate}
		
	As an illustration, we apply the pseudo-code steps on the data sets in Table~\ref{tab6:sample1} (a) and (b) to obtain the data set in Table~\ref{tab6:crossed} (b).

	\begin{table}[h!]
  		\centering
			\subfloat[]{%
    		\begin{tabular}{c c c}
      		\textbf{center} & \textbf{humidity} & \textbf{flies} \\
      	 	& \textbf{index} & \textbf{index} \\
      	 	& (max. MD) & (max. MD) \\
      		\hline \hline
      		12:00 & r1 ($1.0$) & r1 ($0.9$)\\
      		12:02 & r2 ($1.0$) & r2 ($1.0$)\\
      		12:04 & r3 ($1.0$) & r4 ($1.0$)\\
      		12:06 & r4 ($1.0$) & r5 ($0.9$)\\
      		\bottomrule
    		\end{tabular}}%
    		$~~~$
			\subfloat[]{%
    		\begin{tabular}{c c c}
    		&&\\
    		&&\\
      		\textbf{time} & \textbf{humidity} & \textbf{no. of flies} \\
      		\hline \hline
      		12:00 & 30 & 50\\
      		12:02 & 35 & 160\\
      		12:04 & 40 & 243\\
      		12:06 & 50 & 259\\
      		\bottomrule
    		\end{tabular}}%
    	\caption{(a) Tuple indices of humidity and pop. of flies data sets after applying steps 1 and 2 (b) crossed data set after applying steps 3 and 4}
    	\label{tab6:crossed}
	\end{table}

	Having crossed the 2 data sets, it is possible to apply gradual pattern mining techniques especially T-GRAANK (Temporal GRAdual rANKing) (see Chapter~\ref{ch3:mining}) to extract the gradual correlation between humidity and population of flies. For instance a relevant pattern may be: $\{(humidity,\uparrow ), (flies,\uparrow )_{\approx +2mins} \}$ which may be interpreted as \textit{``the higher the humidity, the higher number of flies, almost 2 minutes later''}.

	\subsection{FuzzTX Algorithm}
	\label{ch6:fuzztx}
	In this section, we present Algorithm~\ref{alg6:fuzztx} which implements the FuzzTX model described in Section~\ref{ch6:fuzzy_model}. In this algorithm, we first extract the \textit{date-time} values from each time-series data set and store them in individual arrays ($line~4$). The smallest difference between elements in each \textit{date-time} array is added to the boundary array $\mathcal{B}$ ($line~8$). This array is used to determine the boundaries of the MF ($line~10$). Next, we build a triangular MF that initially starts from the smallest \textit{date-time} value and it is positively slid by a specific boundary until it is equal to the largest \textit{date-time} value ($line~13$).

	\begin{algorithm}[h!]
	\small
	\caption{\textit{FuzzTX} algorithm}
	\label{alg6:fuzztx}
	\SetKwFunction{extract}{ExtractTime}
	\SetKwFunction{min}{min}
	\SetKwFunction{max}{max}
	\SetKwFunction{del}{Delete}
	\SetKwFunction{break}{Break}
	\SetKwFunction{mindiff}{MinDiff}
	\SetKwFunction{buildmf}{BuildTriMF}
	\SetKwFunction{fuzztx}{FuzzyX}
	\SetKwInOut{Input}{\textbf{Input}}
	\SetKwInOut{Output}{\textbf{Output}}

	\Input{time-series data sets $DS^{*}$}
	\Output{data set $\mathcal{D}$}

	$\mathcal{B} \leftarrow \emptyset,~ \mathcal{D} \leftarrow \emptyset$\;
	$T_{min},~T_{max}$\;

	\For{$ds$ in $DS^{*}$}{
		$T_{arr} \leftarrow$ \extract{$ds$}\;
		$T_{max} \leftarrow$ \max{$T_{arr}$} \tcc*[l]{iff greater}
		$T_{min} \leftarrow$ \min{$T_{arr}$} \tcc*[l]{iff lesser}
		$min_{d} \leftarrow$ \mindiff{$T_{arr}$}\;
				$\mathcal{B} \leftarrow \mathcal{B} \cup min_{d}$\;
	}

	$bound_{sel} \leftarrow$ \max{$\mathcal{B}$} \tcc*[l]{largest boundary}
	$t \leftarrow T_{min}$\;
	\While{$t \leq T_{max}$}{
		$mf \leftarrow$ \buildmf{$t-bound_{sel},~t,~t+bound_{sel}$}\;
		\For{$T_{arr}$ of $each~ds~in~DS^{*}$}{
			$index \leftarrow$ \max{\fuzztx{$mf,~T_{arr}$}} \tcc*[l]{index with largest membership degree}
			\uIf{$index$}{
				$x_{tuple} \leftarrow x_{tuple} \cup ds_{tuple}[index]$\;
				\del{$ds_{tuple}[index]$}\;}
			\uElse{
				$x_{tuple} \leftarrow False$\;
				\break{}\;
			}
		}
		\If{$x_{tuple}$}{
			$\mathcal{D}_{tuples} \leftarrow \mathcal{D}_{tuples} \cup x_{tuple}$\;
		}
		$t \leftarrow t + bound_{sel}$\;
	}
	\Return	$\mathcal{D}$
	\end{algorithm}
	
	\clearpage
	
	Finally, the \textit{for-loop} implements the \textit{pseudo-code}, described in Section~\ref{ch6:fuzzy_model}, to determine the tuple indices of each data set that has the largest MD. The indices are used to cross data sets ($line~24$).

	\section{Experiments}
	\label{ch6:experiments}
	In this section, we analyze the efficiency of \textit{FuzzTX} algorithm and discuss its performance results. It is important to mention that we implemented the algorithm using \texttt{Python} language in order to benefit from the language's dynamism especially when dealing with large data sets.
	
	\subsection{Source Code}
	\label{ch6:source_code}
	The \texttt{Python} source code for our \textit{FuzzTX} algorithm is available at our GitHub repository: \url{https://github.com/owuordickson/data-crossing.git}.
	
	\subsection{Computational Complexity}
	\label{ch6:complexity}
	We apply the \textit{Big-O} notation to determine the limiting behavior of the \textit{FuzzTX} algorithm as the number and size of the time-series data sets increase \cite{Vaz2017, Bae2019}. A part from the \textit{control structures} (i.e. \textit{for-loop}, \textit{while-loop} and \textit{if-statements}), we assume that the computational complexity of other statements are relatively constant and they are denoted as $\mathcal{C}$. This is because the average time taken to execute these statements is approximately equal in every repetition.
	
	The FuzzTX Algorithm~\ref{alg6:fuzztx} has: 2 \textit{for-loop} statements, 1 \textit{while-loop} statement and 2 \textit{if} statements. If we let $n$ denote the number of time-series data sets and $x$ denote the number of tuples in the data set with the largest \textit{date-time} difference, then we approximate the asymptotic time complexity as: $2nx\mathcal{C} + nx^{2}\mathcal{C}$. This is because the first \textit{for-loop} iterates through all the tuples of each data set in order to build the $T_{arr}$.
	
	
	The \textit{while-loop} approximately increments its boundaries as many times as the number of tuples of the data set with the largest \textit{date-time} difference and each time it iterates through all the tuples of each data set. Therefore, the upper bound is slightly higher than $\mathcal{O}(nx^{2})$. This implies if the number of time-series data sets are few, the computational time performance of the FuzzTX algorithm is almost proportional to the square of its input size. 
		
	\subsection{Parallel Multiprocessing}
	\label{ch6:parallel}
	We analyze the multiprocessing behavior of the \textit{FuzzTX} algorithm using the \texttt{speedup} and \texttt{parallel efficiency} performance measures. 
	
	(1) \texttt{Speedup S(n)} may be defined as: \textit{``the ratio of the execution time of a single processor to the execution time of $n$ processors''} as shown in Equation~\eqref{eqn6:speedup}. 
	
	\begin{equation}\label{eqn6:speedup}
		S(n) = \dfrac{T_{1}}{T_{n}}
	\end{equation}
	\begin{center}
		where $n$ is the number of available processors
	\end{center}
	
	(2) \texttt{Parallel efficiency E(n)} may be defined as: \textit{``the average utilization of $n$ processors''} as shown in Equation~\eqref{eqn6:parallel_efficiency} \cite{Eager1989}.
	
	\begin{equation}\label{eqn6:parallel_efficiency}
		E(n) = \dfrac{S(n)}{n}
	\end{equation}
	
	In the FuzzTX algorithm (see Pseudocode~\ref{alg6:fuzztx}, we implement parallel processing at 2 code segments: (1) the \textit{for-loop} between $lines~3-8$ and (2) the \textit{while-loop} between $lines~12-26$ since each of their steps can be executed in isolation. We record the results in Table~\ref{tab6:buoys_dataset} and we use the results to plot the \textit{speedup} and \textit{parallel efficiency} in Figure~\ref{fig6:parallel2} (a) and (b). We discuss these results in Section~\ref{ch6:computational_results}.
	
	\subsection{Data Set Description}
	\label{ch6:dataset}
	In order to test computational efficiency, the \textit{FuzzTX} algorithm was executed on 7 time-series data sets obtained from OREME's data portal that recorded meteorological observations at the Puéchabon\footnote{Puéchabon is city located in southern France} weather station between the years 2000 and 2017 \cite{Puechabon2019}. This data is licensed under a Creative Commons Attribution 4.0 License and the site is annually supported by Ecofor, Allenvi and ANAEE-F\footnote{\url{http://www.anaee-france.fr/fr/}}. For this experiment, each data set has 4 attributes and 216 tuples. We performed test runs on a 2.9 GHz Intel Core \textit{i7} MacBook Pro 2012 model, with 8 GB 1600 MHz DDR3 RAM.
	
	In order to test parallel processing efficiency, the \textit{FuzzTX} algorithm was executed on 3 time-series data sets obtained from OREMES's data portal that recorded swell sensor signals of 4 buoys near the coast of the Languedoc-Roussillon region in France between the years 2012 and 2019. The data is available at \url{https://oreme.org/observation/ltc/}. For this experiment, each data set has 30 attributes and tuples ranged 15,000, 100,000 and 200,000. The test runs were performed on a (High Performance Computing) HPC platform \textbf{Meso@LR}\footnote{\url{https://meso-lr.umontpellier.fr}}. We used one node made up of 28 cores and 128GB of RAM.

	\subsection{Experiment Results}
	\label{ch6:results}
	In this section, we present results of our experimental study on 2 data sets obtained from OREME data portal. We show the computational performance results and the parallel performance results. All the results of our test runs are available at our GitHub link: \url{https://github.com/owuordickson/meso-hpc-lr/tree/master/results/fuzztx}. We also show the results of a use case example when the crossed data set is mined for gradual patterns.
	
	\subsubsection{Computational Performance Results}
	\label{ch6:computational_results}
	In order to deduce the computational time efficiency of FuzzTX algorithm, we apply the algorithm on the Puéchabon data set.
	
	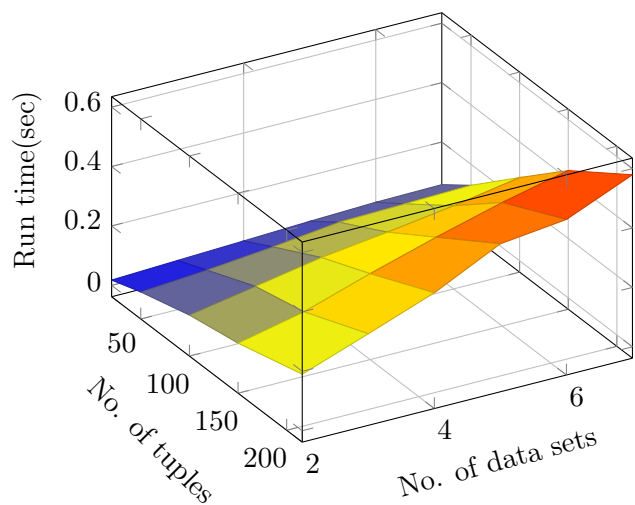
\begin{figure}[h!]
		\centering
		\texttt{Puéchabon: attributes=2-4/tuples=216/cpu cores=4}\par\medskip
		\begin{tikzpicture}
        	\begin{axis}[
        		grid=major,
        		view = {60}{40},
				3d box,
				xtick={50,100,150,200,250},
				xlabel=No. of tuples, ylabel=No. of data sets, zlabel=Run time(sec),
				ylabel style={rotate=15},
				xlabel style={rotate=-45}]
        		\addplot3[surf,mesh/ordering=y varies,mesh/cols=5,mesh/rows=6] table [x=tuples, y=datastreams, z=runtime]{results.dat};
        	\end{axis}
    	\end{tikzpicture}
		\caption{Plot of run time versus data sets' tuples verses number of data sets}
		\label{fig6:computational1}
	\end{figure}
			
	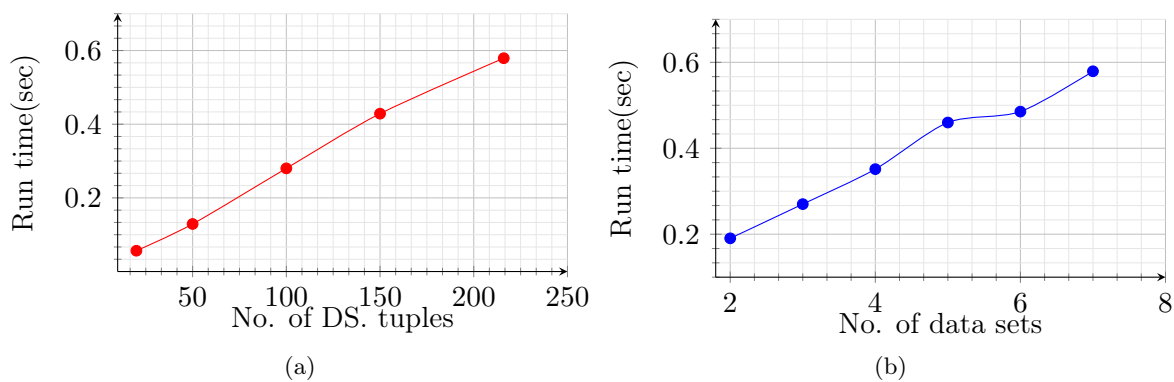
\begin{figure}[h!]
		\centering
		\texttt{Puéchabon: attributes=4/tuples=216/cpu cores=4}\par\medskip
		\subfloat[]{%
		\begin{tikzpicture}
  			\begin{axis}[
  				height=5cm, width=7.5cm,
  				grid=both,
  				grid style={line width=.1pt, draw=gray!20},
    			major grid style={line width=.2pt, draw=gray!50},
  				axis lines=middle,
  				minor tick num=5,
  				ymin=0,ymax=0.7,
  				xmin=10,xmax=250,
  				xlabel=No. of DS. tuples, ylabel=Run time(sec),
  				xlabel style={at={(axis description cs:0.5,-0.1)},anchor=north},
  				ylabel style={at={(axis description cs:-0.15,0.5)},rotate=90,anchor=south},
 				legend style={nodes={scale=0.5}},
 				legend pos=north west]
  			\addplot[smooth, mark=*, color=red] table[x=tuples, y=runtime]{results1.dat};
  			\end{axis}
  		\end{tikzpicture}}%
  		\subfloat[]{%
  		\begin{tikzpicture}
  			\begin{axis}[
  				height=5cm, width=7.5cm,
  				grid=both,
  				grid style={line width=.1pt, draw=gray!20},
    			major grid style={line width=.2pt, draw=gray!50},
  				axis lines=middle,
  				minor tick num=5,
  				ymin=0.1,ymax=0.7,
  				xmin=1.8,xmax=8,
  				xlabel=No. of data sets, ylabel=Run time(sec),
  				xlabel style={at={(axis description cs:0.5,-0.1)},anchor=north},
  				ylabel style={at={(axis description cs:-0.15,0.5)},rotate=90,anchor=south},
  				legend style={nodes={scale=0.5}},
 				legend pos=north west]
			\addplot[smooth, mark=*, color=blue] table[x=datastreams, y=runtime]{results2.dat};
  			\end{axis}
  		\end{tikzpicture}}%
		\caption{(a) Plot of run time versus data sets' tuples with number of data sets held at 7 (b) plot of run time versus data sets with number of tuples held at 216}
		\label{fig6:computational2}
	\end{figure}
	
	In Figure~\ref{fig6:computational1}, run time increases with increase in both the number of data sets and the number of the data sets' tuples. For the purpose of getting a clearer picture of the algorithm's computational performance, we plot axes shown in Figure~\ref{fig6:computational2} (a) and (b). As can be observed the growth rate of run time is almost linearly proportional to the growth rate of the data set size. This linear growth rate performance is better than the deduced performance in Section~\ref{ch6:complexity} which implied a quadratic growth rate.

\begin{figure}[h!]
		\centering
		\texttt{Buoys: attributes=30/tuples=15k-200k/cpu cores=1-28}\par\medskip
		\begin{tikzpicture}
  			\begin{axis}[
  				height=5cm, width=7.5cm,
  				grid=both,
  				grid style={line width=.1pt, draw=gray!20},
    			major grid style={line width=.2pt, draw=gray!50},
  				axis lines=middle,
  				minor tick num=5,
  				ymin=0,ymax=2400,
  				xmin=1,xmax=30,
  				xlabel=No. of cores, ylabel=Run time(sec),
  				xlabel style={at={(axis description cs:0.5,-0.1)},anchor=north},
  				ylabel style={at={(axis description cs:-0.2,0.5)},rotate=90,anchor=south},
 				legend style={nodes={scale=0.5}},
 				legend pos=north east]
  			\addplot[smooth, mark=*, color=red] table[x=Cores, y=Runtime]{fuzztx_15k.dat};
  			\addplot[smooth, mark=*, color=blue] table[x=Cores, y=Runtime]{fuzztx_100k.dat};
  			\addplot[smooth, mark=*, color=brown] table[x=Cores, y=Runtime]{fuzztx_200k.dat};
  			\addlegendentry{15k lines}
			\addlegendentry{100k lines}
			\addlegendentry{200k lines}
  			\end{axis}
  		\end{tikzpicture}
  		\caption{Plot of run time versus number of cores}
		\label{fig6:parallel1}
	\end{figure}
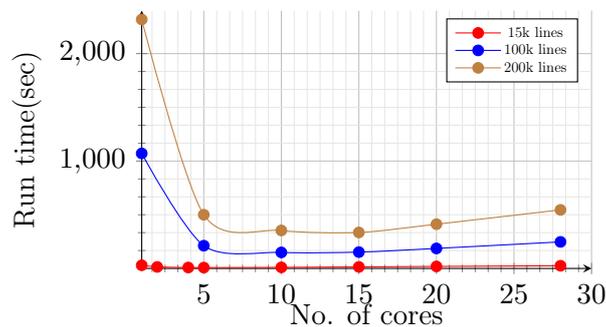

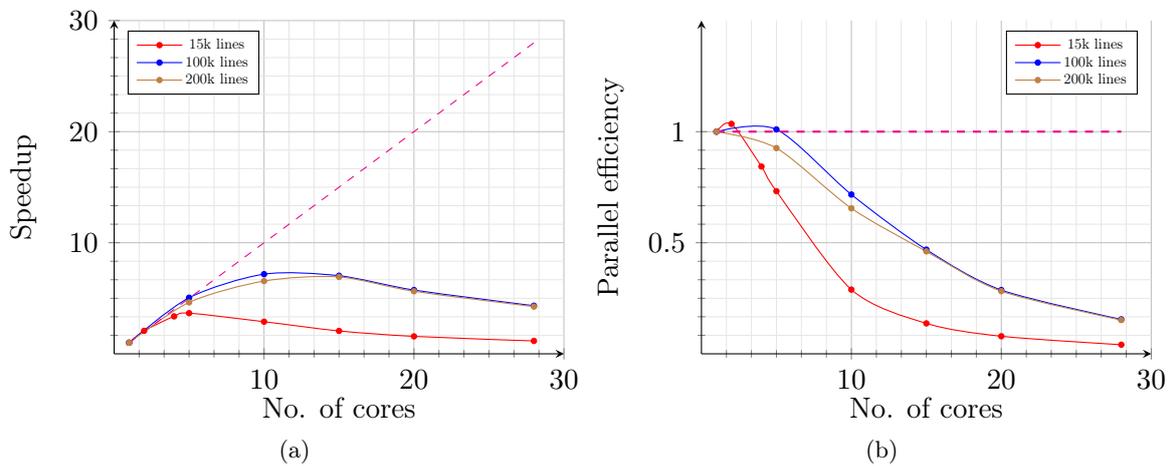
\begin{figure}[h!]
		\centering
		\texttt{Buoys: attributes=30/tuples=15k-200k/cpu cores=1-28}\par\medskip
  		\subfloat[]{%
  		\begin{tikzpicture}
  			\begin{axis}[
  				height=6cm, width=7.5cm,
  				grid=both,
  				grid style={line width=.1pt, draw=gray!20},
    			major grid style={line width=.2pt, draw=gray!50},
  				axis lines=middle,
  				minor tick num=5,
  				ymin=0,ymax=30,
  				xmin=0,xmax=30,
  				xlabel=No. of cores, ylabel=Speedup,
  				xlabel style={at={(axis description cs:0.5,-0.1)},anchor=north},
  				ylabel style={at={(axis description cs:-0.15,0.5)},rotate=90,anchor=south},
 				legend style={nodes={scale=0.5}},
 				legend pos=north west]
  			\addplot[smooth, mark=*, mark size=1pt, color=red] table[x=Cores, y=Runtime]{fuzztx_15k_sp.dat};
  			\addplot[smooth, mark=*, mark size=1pt, color=blue] table[x=Cores, y=Runtime]{fuzztx_100k_sp.dat};
  			\addplot[smooth, mark=*, mark size=1pt, color=brown] table[x=Cores, y=Runtime]{fuzztx_200k_sp.dat};
  			\addplot[dashed, color=magenta] table[x=Cores, y=Runtime]{speedup_2.dat};
  			\addlegendentry{15k lines}
			\addlegendentry{100k lines}
			\addlegendentry{200k lines}
  			\end{axis}
  		\end{tikzpicture}}%
  		\subfloat[]{%
  		\begin{tikzpicture}
  			\begin{axis}[
  				height=6cm, width=7.5cm,
  				grid=both,
  				grid style={line width=.1pt, draw=gray!20},
    			major grid style={line width=.2pt, draw=gray!50},
  				axis lines=middle,
  				minor tick num=5,
  				ymin=0,ymax=1.5,
  				xmin=0,xmax=30,
  				ytick={0,.5,1},
  				xlabel=No. of cores, ylabel=Parallel efficiency,
  				xlabel style={at={(axis description cs:0.5,-0.1)},anchor=north},
  				ylabel style={at={(axis description cs:-0.15,0.5)},rotate=90,anchor=south},
 				legend style={nodes={scale=0.5}},
 				legend pos=north east]
  			\addplot[smooth, mark=*, mark size=1pt, color=red] table[x=Cores, y=Runtime]{fuzztx_15k_sp_eff.dat};
  			\addplot[smooth, mark=*, mark size=1pt, color=blue] table[x=Cores, y=Runtime]{fuzztx_100k_sp_eff.dat};
  			\addplot[smooth, mark=*, mark size=1pt, color=brown] table[x=Cores, y=Runtime]{fuzztx_200k_sp_eff.dat};
  			\addplot[thick, dashed, color=magenta] table[x=Cores, y=Runtime]{sp_efficiency_2.dat};
  			\addlegendentry{15k lines}
			\addlegendentry{100k lines}
			\addlegendentry{200k lines}
  			\end{axis}
  		\end{tikzpicture}}%
  		\caption{(a) Plot of speed up versus number of cores (b) plot of parallel efficiency versus number of cores}
  		\label{fig6:parallel2}
  	\end{figure}

	In order to test the parallel efficiency of the FuzzTX algorithm, we applied the algorithm on the Buoys data sets and we use the results to plot Figure~\ref{fig6:parallel1} and Figure~\ref{fig6:parallel2}. Figure~\ref{fig6:parallel1} shows the overall behavior of the FuzzTX algorithm when we apply parallel processing. Generally,  run time decreases as the number of cores increase from 1 to 28.
	
  	Figure~\ref{fig6:parallel2} (a) and (b) show the \textit{speedup} and \textit{parallel efficiency} of the FuzzTX algorithm. We observe that for each data set, there is an optimum number of processors where parallel efficiency is highest. For the 15k-line data set this is approximately 2 processors; for the 100k-line and 200k-line data set this is approximately 5 processors.
  		
	\subsubsection{Use Case Example: Mining Temporal Gradual Patterns}
	\label{ch6:use_case}
\begin{figure}[h!]
		\centering
		\center{\includegraphics[width=.75\textwidth]
        {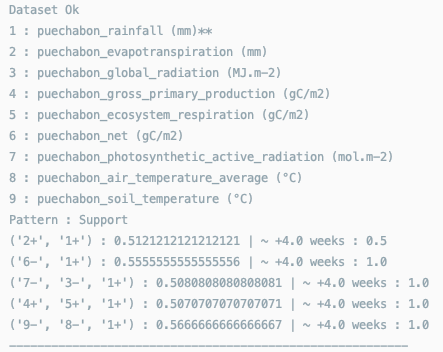}}
		\caption{A sample of the fuzzy-temporal gradual patterns extracted from crossed data}
		\label{fig6:ftgp}
	\end{figure}

	We applied the T-GRAANK algorithm proposed in Chapter~\ref{ch3} on the crossed data we obtained after applying our \textit{FuzzTX} algorithm on the time-series data sets from the Puéchabon weather station. We obtained temporal gradual patterns shown in Figure~\ref{fig6:ftgp}. For instance the pattern \textbf{$\{$(`2+',`1+') : 0.5121 | $\sim$ +4.0 weeks : 0.5$\}$} may be interpreted as: \textit{``the higher the evapotranspiration, the higher the rainfall amount, almost 4 weeks later''}.
	
	\section{Summary}
	\label{ch6:summary}
	In this chapter, we propose a fuzzy model that applies a triangular membership function to cross time-series data sets. This model is most suitable for adoption by research observatories (such as OREME) that support \textit{data lakes} which store time-series data or data streams from different sources.
		
	In order to emphasize the applicability of our model, we integrated the \textit{FuzzTX} algorithm into a \textit{Docker} implementation of the \textit{OGC SensorThings framework} to cross different data streams and extract relevant gradual patterns. We describe this implementation in Chapter~\ref{ch7}. The source code for this work is available at our Github repository: \url{https://github.com/owuordickson/cloud-api.git}.


\chapter{Cloud Integration of GP Mining Algorithms}
\label{ch7}
\minitoc \clearpage 

	\begin{chapquote}{Wernher von Braun \textit{(1912 – 1977)}}
		``Research is what I'm doing when I don't know what I'm doing''
	\end{chapquote}

	\section{Introduction}
	\label{ch7:introduction}
	In this chapter, we describe a software architecture model that integrates two algorithm implementations: (1) the fuzzy data crossing algorithm implementation described in Chapter~\ref{ch6} and (2) the temporal gradual pattern mining algorithm implementation described in Chapter~\ref{ch3} into a Cloud platform which implements \textit{OGC SensorThings} API. We build the model on top of the SensorThings API in order to exploit the time-series data sets it manages for extraction of temporal gradual patterns.

	\section{Gradual Pattern Mining Tools on Cloud}
	\label{ch7:cloud_tools}
	Scientific researchers are constantly collecting, crossing and analyzing data in order to help them understand various phenomena. For example environmental data is important for helping to understand phenomena like global warming, typhoons, rainfall patterns among others; biological data is important for helping to understand phenomena like cancer cells, Ebola virus among others. Most of such data is collected using sensors that are enabled to upload data into Cloud platforms \cite{Liang2016, Hicham2018, Joshi2018}.

	Temporal gradual pattern mining is an instance of a data analysis technique that allows for extraction of gradual correlations among attributes of a data set. For instance given a data set with attributes $\{A, B\}$, a temporal gradual pattern may take the form: \textit{``the more A, the less B almost 6 days later''}. Under those circumstances, it comes as no surprise that temporal gradual pattern mining algorithms are largely applied on time-series data sets.

	Again, this research study advocates the view that different time-series data sets may be crossed with the aim of exploiting the crossings for extraction of temporal gradual patterns. This allows for identification of interesting correlations among time-series data sets from different sources which otherwise would have been difficult to retrieve from the individual data sets \cite{Hicham2018}. For example, let us assume that a researcher has: (1) set up an IoT-based data collection station which has numerous sensors monitoring different environmental phenomena (such as temperature, humidity, wind speed) and (2) a time-series data set that records the population of birds in same geographical location.

	If the researcher uses a Cloud platform to store all the time-series data sets, then it would be very convenient as well to integrate a time-series data analysis tool into the Cloud platform. The existence of the tool in the Cloud platform saves the researcher the trouble of retrieving, cleaning and preparing the raw time-series data for analysis on an offline software. This is because the tool allows for data analysis within the same Cloud platform that stores the data. It is for such a similar reason that we propose in this chapter a software model that allows crossing of time-series data for the purpose of exploiting the crossings for temporal gradual pattern mining.

	The motivation of the study emerges from the proliferation of IoT in research institutions and with this, comes the provision of large-scale time-series data from different sources. Time-series data can also be defined as a \textit{data stream} when it becomes a potentially infinite sequence of precise recorded over a period of time \cite{Pitarch2010}. As a result of this proliferation, frameworks such as OGC SensorThings have emerged to allow for (FAIR) Findable, Accessible, Interoperable and Reusable time-series data \cite{Liang2016, Hicham2015}.

	\subsection{Proposed Software Architecture Model}
	\label{ch7:architecture}
\begin{figure}[h!]
    	\centering
		\includegraphics[width=\textwidth] {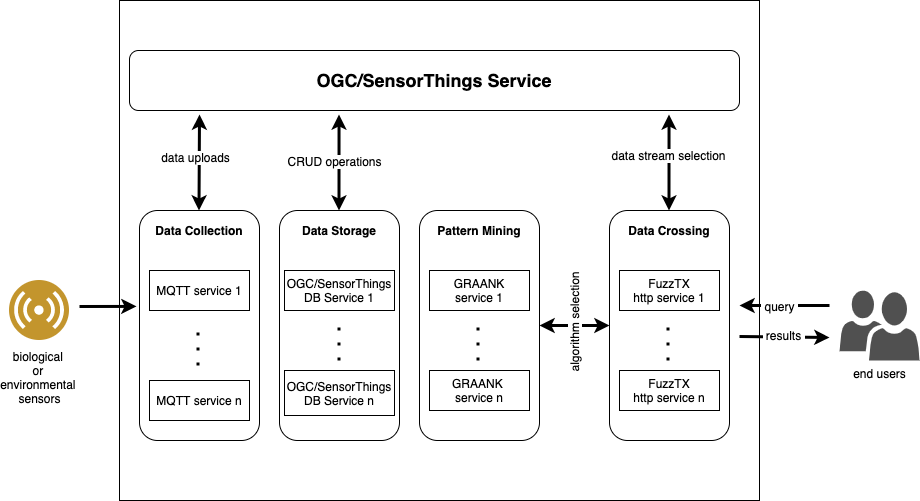}
		\caption{Proposed software model architecture for Cloud integration}
		\label{fig7:architecture}
	\end{figure}

	We propose to extend GOST\footnote{\url{https://www.gostserver.xyz}} (golang SensorThings), a certified software implementation of the OGC SensorThings API, through: (1) a \textit{Data Crossing} software component that provides a user interface and allow users to cross numerous datastreams (2) a gradual \textit{Pattern Mining} software component as illustrated in Figure~\ref{fig7:architecture}.

	The proposed OGC SensorThings software model architecture that represents our proposed model for Cloud integration is composed of 5 components, where each is implemented by a separate server computing device. The arrows illustrate the direction of communication between the software components and the type of messages exchanged that are permitted.

	The \textit{Data Crossing} software component integrates into the OGC SensorThings API, our fuzzy model for crossing time-series data (which we describe in Chapter~\ref{ch6}). This tool provides a \textit{Web} interface that allows the user to select which time-series data (\textit{data streams}) to cross. The \textit{Pattern Mining} software component integrates into the OGC SensorThings API, our ACO-GRAANK and T-GRAANK algorithm implementations (which are described in Chapter~\ref{ch3} and Chapter~\ref{ch4}). This allows for extraction of gradual patterns from the crossed data streams.

	The \textit{Data Collection} and \textit{Data Storage} software components come with the OGC SensorThings API implementation. They allow for collection of sensor data through the MQTT protocol and storage of the data into a \textit{Postgres} database. The \textit{OGC SensorThings Service} component provides hyper-text interface that allows for GET and POST requests from the other software components.

\begin{figure}[h!]
    	\centering
		\includegraphics[width=\textwidth] {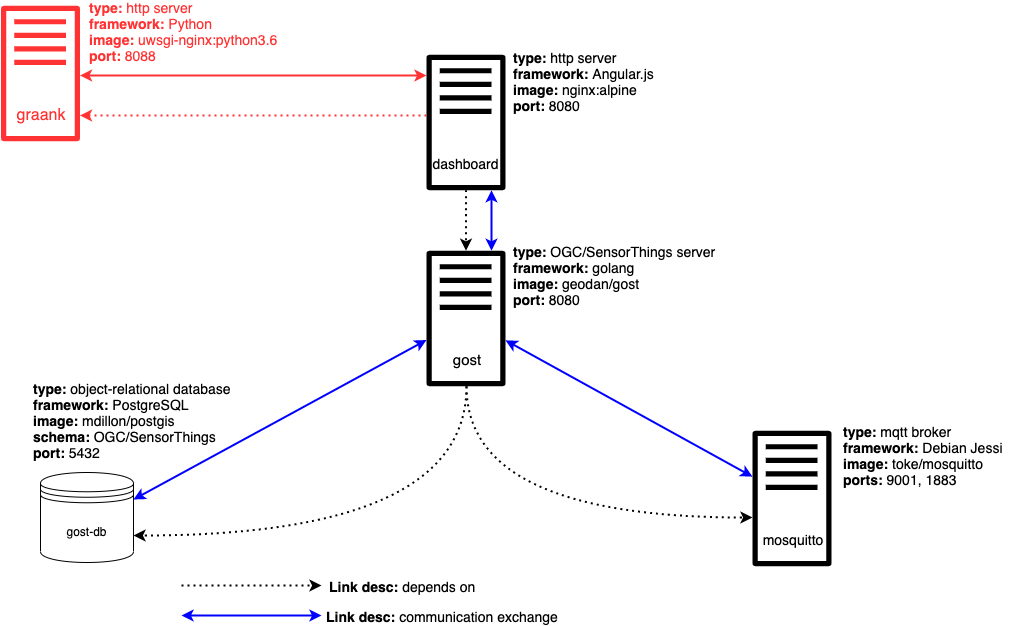}
		\caption{Docker architecture of a Cloud API for integrating GP mining algorithms into a Docker implementation of the OGC SensorThings API}
		\label{fig7:docker}
	\end{figure}

	We implement our proposed model on top of the OGC SensorThings API because, to the best of our knowledge, it stands out as the best API to interconnect IoT devices, sensor data and applications over the Cloud. As shown in Figure~\ref{fig7:docker}, the proposed software model is composed of 5 items described below.

	\textbf{(1) OGC SensorThings Service} exposes the document resources of the 8 entity sets (we explain them in Section~\ref{ch2:ogc}) for the SensorThings clients. This is the core task of the SensorThings API since every CRUD (Create, Read, Update, Delete) operation goes through it. The \textbf{GOST Server} implements the OGC SensorThings API and it provides the \textit{OGC SensorThings Service}. It is implemented using Google's \texttt{Go} language (golang).

	\textbf{(2) Data Collection Tool} implements MQTT protocol which is a lightweight publish/subscribe protocol designed for constrained devices. This tool is used to connect to sensors in order to collect data from the environment and send them to \textit{OGC SensorThings Service}. The \textbf{Mosquitto Server} provides the \textit{Data Collection} services. It is implemented on \texttt{Mosquitto} which supports an MQTT broker that allows sensors to publish their data into the system.

	\textbf{(3) Data Storage Tool} serves an object-relational database which implements the schema of the 8 OGC SensorThings sensing entities. This tool provides storage services for all the time-series data collected by the sensors. The \textbf{GOST Database Server} provides the \textit{Data Storage} services. It is implemented on top of \texttt{PostGIS} which is a spatial object-relational database software that supports \texttt{PostgreSQL} language. This server stores all the real-time sensor data collected in form of \texttt{Observation}s which belong to \texttt{Datastream}s.

	\textbf{(4) Data Crossing Tool} serves a HTTP \textbf{Dashboard Server} which enables users to interact with our system. Through this tool users are able to send HTTP requests to the \textit{OGC SensorThings Service} and the \textit{Pattern Mining tool}. Equally important, this tool integrates the fuzzy data crossing algorithm described in Chapter~\ref{ch6} into this software model. The \textbf{Dashboard server} is implemented using \texttt{Angular.js} which creates all the HTML views and it is served by \texttt{NGINX} (engine x) to enable users to interact with the system.

	\textbf{(5) Pattern Mining Tool} (implemented by \textbf{GRAANK} server) serves the gradual pattern mining algorithms which are variants of the \textit{T-GRAANK} technique described in Chapter~\ref{ch3} and are implemented in Python. This tool runs services that receive HTTP $POST$ requests from the \textit{Data Crossing tool} and responds with the extracted patterns. The \textbf{GRAANK Server} provides the \textit{Pattern Mining} services. It is implemented using \texttt{Python} and deployed with \texttt{uWSGI}\footnote{\url{https://uwsgi-docs.readthedocs.io/en/latest/}} server which enables it to accept HTTP POST requests forwarded by the \textit{Dashboard Server} and respond with extracted temporal gradual patterns.

	The \texttt{uWSGI} is a software application named after Web Server Gateway Interface (WSGI) plugin which supports the \texttt{uwsgi} protocol, and it is popularly used to serve \texttt{Python} web applications since it offers multiprocessing capabilities and efficient usage of computing resources. The \textbf{GRAANK Server} exploits these benefits to improve the performance of gradual pattern mining algorithms.

	\subsection{Benefits of Integrating Pattern Mining Tools into Cloud}
	\label{ch7:benefits}
	It is important to highlight that the \textit{pattern mining tool} enables the access of gradual pattern mining algorithms through a SaaS (Software-as-a-service) model. In a SaaS distribution model, software is centrally hosted and licensed on a subscription basis. This introduces a flexibility that spares users the agony of spending hours trying to install analysis software \cite{Joshi2018}.

\begin{figure}[h!]
    	\centering
		\includegraphics[width=\textwidth] {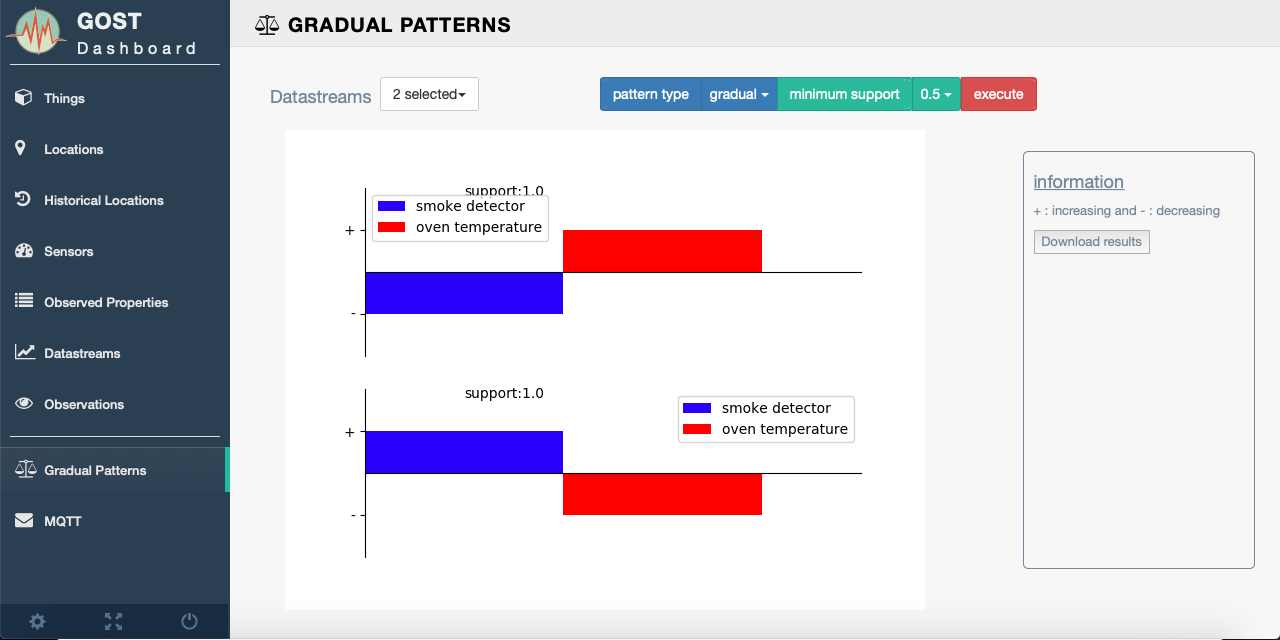}
		\caption{Screenshot of OGC SensorThings API Web interface}
		\label{fig7:ogc_screen}
	\end{figure}

	The \textbf{Data Crossing Tool} allows the user to select multiple data streams to cross them to form a single data set from which the \textbf{Pattern Mining Tool} is employed to extract temporal gradual patterns. For instance the user can cross a \textit{``temperature''} data stream with a \textit{``no. of bees''} data stream to test for a pattern like: \textit{``the higher the temperature, the higher the population of bees almost 3 days later''}. As can be noted, all these tools are executed on top of a \textit{Docker} Cloud platform meaning that the user can access the services from any terminal device that has Internet connection (i.e. computer, smartphone, tablet among others). Figure~\ref{fig7:ogc_screen} shows the web implementation of this software model.

	With regard to Chapter~\ref{ch3} the computational complexity of mining temporal gradual patterns increases with the size of the data set. For this reason, the algorithms are implemented in Python language in order to harness Python's ability to improve efficiency through multi-core parallel processing. As a result, we argue that a SaaS model is best suited for offering gradual pattern mining tools to users since they are installed on one powerful mainframe which may be accessed on a subscription basis.

	The proposed software model enables a Python implementation of these algorithms on a single server and also provides a user-friendly HTTP API for access. Again, the Cloud implementation of the proposed software model allows for a configuration setting that limits access of these tools to a subscription basis. Further, usage analytics may also be easily generated since every user request and operation is recorded in log files. These analytics may be used to improve user experience or identify and fix software bugs.

	\section{Software Implementation}
	\label{ch7:implementation}

	\subsection{Source Code}
	\label{ch7:source_code}
	The entire source code for this work is available at our GitHub repository: \url{https://github.com/owuordickson/cloud-api.git}. The \textit{GRAANK Server} image is available at our \texttt{Docker repository}: \url{https://hub.docker.com/r/owuordickson/graank:ogc}.

	\subsection{Use Case Example: OREME}
	\label{ch7:oreme}

	OREME\footnote{\url{https://oreme.org}} (Observatory of Mediterranean Research of the Environment) is a scientific observatory recognized by INSU (National Institute of Science of the Universe) and INEE (Ecology and Environment Institute) of CNRS (National Center for Scientific Research) and the IRD (French national Institute for Sustainable Development).

	OREME supports research activities that involve continuous observation of the natural environment over long periods in the Mediterranean region. It is constituted by 8 laboratories: CEFE (Functional and Evolutionary Ecology Center), UMR WMAU (Water Management, Actors, Uses), Geosciences Montpellier, HSM (HydroSciences Montpellier), ISEM (Institute of Evolution Sciences of Montpellier), LUPM (Laboratory Universes and Particles of Montpellier), MARBEC (Marine Biodiversity, Exploitation and Conservation) and UMR TETIS (Territories, Environment, Remote Sensing and Spatial Information).

	OREME complies with INSPIRE\footnote{\url{https://inspire.ec.europa.eu}} \textit{Implementing Rules} which requires that common frameworks are implemented by member research institutions in order to ensure that collected spatial data is made FAIR (Findable, Accessible, Interoperable and Reusable). Therefore, it currently implements in its data portal\footnote{\url{https://data.oreme.org}} OGC geospatial standards like WFS (Web Feature Service) and WMS (Web Map Service) which standardizes the creation, modification and exchange of geographic information over the Internet..

	However, OREME has not yet implemented data stream standards like the OGC SensorThings API into its data portal due to data entity incompatibility. One solution may involve installing IoT sensors at the data collection stages, which integrate into an MQTT instance to generate compatible data streams \cite{Grellet2017, Kotsev2018}.


	\section{Summary}
	\label{ch7:summary}
	In this chapter, we propose and develop a software architecture model that allows gradual pattern mining algorithms to be accessed over a Cloud platform. This allows for these algorithms to be applied on real-time sensor data that is collected over the Cloud through frameworks such as the OGC SensorThings. In future, we intend to implement the proposed software model into the OREME's data portal.


\chapter{Conclusions and Perspectives}
\label{ch8}
\minitoc \clearpage 
	
	\begin{chapquote}{Ecclesiastes 7:8 \textit{(KJV)}}
		``Better is the end of a thing than the beginning thereof''
	\end{chapquote}
	
	\section{Summary}
	\label{ch8:summary}
	This research study mainly addresses the aspect of mining temporal gradual patterns and the integration of gradual mining techniques into a cloud platform. We summarize the main contributions made by this study in field of knowledge discovery and data mining through points that follow:
	
	\begin{itemize}
		\item We reviewed literature concerning two existing gradual pattern mining techniques: GRITE and GRAANK. We discover that: (1) both techniques apply a \textit{breadth-first search} strategy in order to generate gradual item set candidates and (2) both strategies do not address the question of additionally extracting the temporal tendencies of gradual patterns.
		\item We introduce and propose formal definitions of temporal and fuzzy-temporal gradual item sets and temporal gradual emerging pattens.
		\item We describe an ant colony optimization technique for \textit{breath-first search}-based and \textit{depth-first search}-based approaches for mining gradual patterns. Given the computational results in, we establish that ACO-GRAANK and ACO-ParaMiner algorithms out-perform classical GRAANK and ParaMiner algorithms and, they mine fewer but high quality maximal and closed patterns respectively.
		\item We propose and describe T-GRAANK approach that is an extension of the GRAANK approach. The T-GRAANK approach allows for extraction of temporal gradual patterns through a fuzzy model. We establish that ACO-TGRAANK which is based on the ACO-GRAANK approach offers a better computational run time performance than T-GRAANK which based on the classic GRAANK technique.
		\item We propose and describe two approaches for mining temporal gradual emerging patterns: (1) an ant colony optimization approach and, (2) a border manipulation approach. We develop algorithm implementations for both approaches: TRENC and BorderT-GRAANK respectively. We compare the computational performance of both algorithm implementations and establish that TRENC is the more efficient algorithm.
		\item We propose and describe FuzzTX approach that allows for \textit{crossing} of time-series data sets from different sources through a fuzzy model. We show that through the FuzzTX approach, it is possible to cross unrelated time-series data sets and mine for temporal gradual patterns.
		\item Finally, we propose and describe a software model architecture that allows for integration into a \textit{Docker} Cloud platform that implements the OGC SensorThings API: (1) the FuzzTX data crossing algorithm implementation and (2) temporal gradual pattern mining algorithm implementations. In addition to the Cloud platform, we develop a Desktop and Web application that allows users to upload numeric data sets in \texttt{CSV} format then it may extract gradual patterns, or temporal gradual patterns, or emerging patterns. A screenshot of this application is shown in Figure~\ref{fig8:electronjs}. We provide installation instructions about all our software tools in Appendix~\ref{appA}.  
	\end{itemize}
	
	\begin{figure}[h!]
    	\centering
		\includegraphics[width=.9\textwidth] {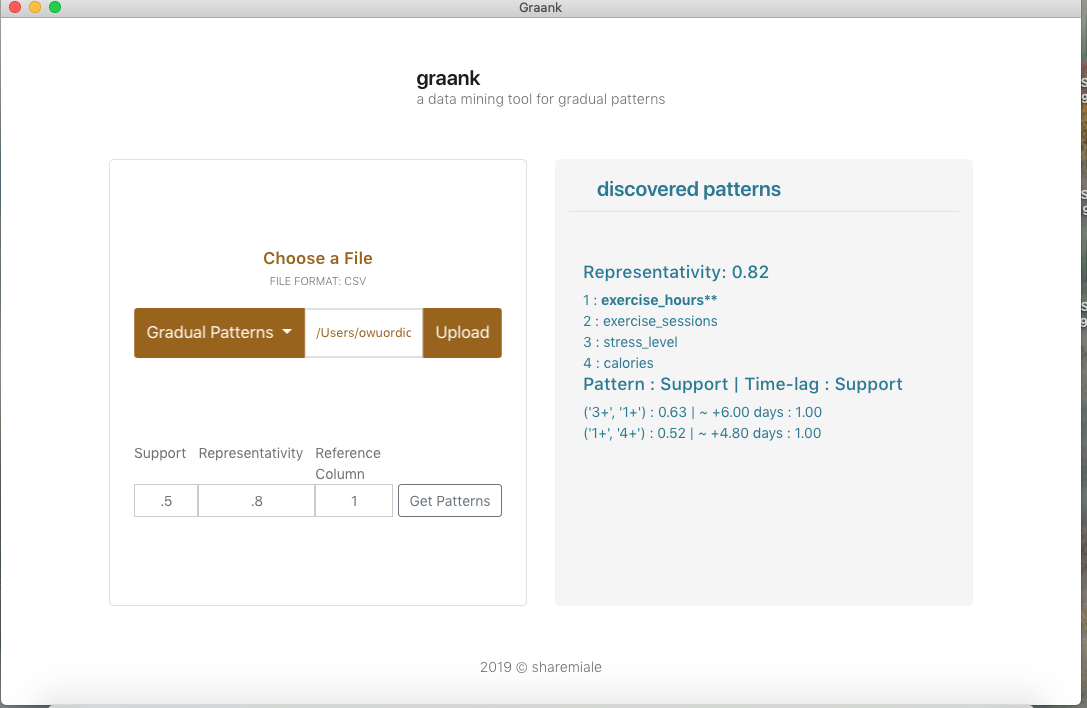}
		\caption{Screenshot of desktop application for gradual pattern mining tools}
		\label{fig8:electronjs}
	\end{figure}
	
	
	
	\section{Perspectives}
	\label{ch8:perspectives}
	In consideration of the contributions made by this work in the field of knowledge discovery and data mining, we present the following research perspectives which can be taken as future research directions.
	
	\subsection{Further Efficiency Optimization of Gradual Pattern Mining Technique}
	\label{ch8:gradual}
	In Chapter~\ref{ch4}, we propose and describe an ant colony optimization strategy for optimizing the efficiency GRAANK (GRAdual rANKing) which is a classical gradual pattern mining technique proposed by \cite{Laurent2009}. Ant colony optimization is a general purpose heuristic approach mainly used to optimize combinatory problems \cite{Dorigo1996}. For the reason that generation of gradual item set candidates may be represented as a combinatory problem (especially when dealing with data sets that have large numbers of attributes), we extend an ant colony optimization strategy to the case of gradual pattern mining in order to provide a solution to this problem.
	
	In Chapter~\ref{ch4:aco} We fully describe an ant colony optimization approach for the case of gradual pattern mining and temporal gradual patten mining, and develop an algorithms: ACO-GRAANK and ACO-TGRAANK respectively. We test and compare the computational performance of these two algorithms to their classical counterparts GRAANK and T-GRAANK (Temporal-GRAANK proposed in Chapter~\ref{ch3}). We establish that the ACO-based algorithms significantly outperform their classical counterparts in terms of computational run-time efficiency.
	
	However, we observe that as the size of the data set increases by number of tuples; the computational run-time efficiency of the ACO-based algorithms reduces. Although the situation is relatively worse in the case of the classical gradual pattern mining algorithms, it is significant enough to draw attention. It should be remembered that the frequency of a gradual pattern is determined by number of tuple pairs or size of tuple orders, and many classical gradual pattern mining techniques employ a bitmap representation for capturing tuples pairs/orders that respect a gradual pattern.
	
	As described in Chapter~\ref{ch2:grite_binary}, bitmap representation allows for the modeling of a gradual pattern as a $n\times n$ binary matrix (where $n$ is the total number of tuples in a data set). In the matrix, tuple pairs that respect a pattern are set to $1$, and tuple pairs that do not respect the pattern are set to $0$.
		
	\textit{Example 8.1.} An arbitrary data set $\mathcal{D}_{8.1}$ containing recordings of atmospheric temperature, atmospheric humidity.
	
	\begin{table}[h!]
  		\centering
    	\begin{tabular}{c c c} 
	  	\toprule
      	\textbf{id} & \textbf{temperature} & \textbf{humidity}\\
      	\hline \hline
      	r1 & 30 & .2\\
      	r2 & 28 & .4\\
      	r3 & 26 & .5\\
      	r4 & 26 & .8\\
      	\bottomrule
    	\end{tabular}
    	\caption{Sample data set $\mathcal{D}_{8.1}$}
    	\label{tab8:sample1}
	\end{table}
	
	For example given the data set in Table~\ref{tab8:sample1} (which has $4$ tuples), a gradual pattern \textit{``the lower the temperature, the higher the humidity''} (denoted as $\{ (temp,\downarrow ), (hum,\uparrow ) \} $) may be represented as a $4\times 4$ matrix as shown in Figure~\ref{fig8:matrix1}.
	
	\begin{figure}[h!]
		\centering
		\begin{tabular}{|c|c c c c|}
			\hline
			$\Rsh$ & r1 & r2 & r3 & r4\\
			\hline
			r1 & 0 & 1 & 1 & 1\\
			r2 & 0 & 0 & 1 & 1\\
			r3 & 0 & 0 & 0 & 0\\
			r4 & 0 & 0 & 0 & 0\\
			\hline
		\end{tabular}
		\caption{Binary matrix $M_{G}$ for gradual pattern $\{ (temp,\downarrow ), (hum,\uparrow ) \} $}
		\label{fig8:matrix1}
	\end{figure}
	
	Under those circumstances, it should be observed that \textit{tuple-size} of a data set directly determines the sizes of binary matrices of its patterns. Therefore, the computational complexity that is introduced by the size of binary matrices negatively affects the efficiency of the gradual pattern mining technique as the tuple-size of the data set increases.
	
	\clearpage
	
	With regard to the points given above, future works may involve proposing approaches for optimizing gradual pattern mining techniques by reducing the size of binary matrices needed to represent respective gradual patterns. One such approach may extend a heuristic technique that constructs a binary matrix which carries only significant representation knowledge of a gradual pattern using a randomly selected portion of the tuple pairs. As a result, tuple pairs with less significant representation knowledge do not form any part of the binary matrix. If a portion of few tuples are selected to successfully represent frequent gradual patterns then, the computational complexity may be reduced to a constant regardless of the size of the data set.
	
	\subsection{Memory Limitation}
	\label{ch8:memory_limitation}
	In Chapter~\ref{ch4:results}, we observe from the experiment results that ParaMiner and ACO-ParaMiner algorithms yield \textit{Memory Error} when applied on slightly large data sets. GRAANK and ACO-GRAANK algorithms perform relatively better in terms of memory usage; however, they also yield \textit{Memory Error} when applied to \texttt{BigData}. As mentioned in the previous section, these algorithms rely on a technique that represents tuple gradual relationships in the form of binary matrices. This implies that the greater the number of tuples in a data set, the larger the size of the binary matrices that is stored in primary memory. As result, this introduces a memory limitation on these algorithms (GRAANK, ParaMiner, ACO-GRAANK, ACO-ParaMiner) that undermines their computational efficiency.
	
	Concerning the memory limitation, we propose as future works a more efficient data format for dealing with huge binary matrices that may be constructed from \texttt{BigData}. Version 5 of Hierarchical Data Format \texttt{(HDF5)} is an instance of a high performing data software library that allow for management, processing and storage of \texttt{BigData}. Another advantage of HDF5 is that it allows for execution of \texttt{Map-Reduce} jobs. Map-Reduce is a program model that employed on \texttt{BigData} which performs \textit{filtering and sorting} operations on a \texttt{Map phase} and \textit{summary} operations on a \texttt{Reduce phase}. In the case of gradual pattern mining, the \texttt{Map phase} may identify all gradual relationships between tuples and the \texttt{Reduce phase} may summarize the relationships into a binary matrix with an efficient HDF5 data format.

	\subsection{Detecting Interesting Data Crossings}
	\label{ch8:crossing}	
	In Chapter~\ref{ch6}, we propose and describe an approach that extends a fuzzy model in order to cross time-series data sets from unrelated sources. It should be noted that this approach is designed to allow as input, only data sets that are manually specified by a user. Therefore, the proposed approach: (1) may require some modification to allow it to operate in an automated environment, also (2) it crosses all data sets that a user specifies without pre-testing if the resulting crossing may be of significant interest to the user. For instance, it may be the case that crossing two unrelated data sets generates a \textit{crossed data set} that does not produce any non-trivial (or interesting) pattern or knowledge when mined.	
	
	In relation to this, future works may involve extending the approach through a \textit{heuristic bot} that mines a data lake's catalog in order to identify relevant time-series data sets that may produce interesting knowledge when crossed. A data lake might contain thousands of large and unrelated data sets; therefore, identifying and crossing only interesting data sets instead of crossing all the available data sets is a prudent strategy for saving time and computational resources.

\cleardoublepage \phantomsection
\mtcaddchapter[Bibliography]
\Urlmuskip=0mu plus 1mu\relax
\bibliography{bibliography}

\renewcommand{\theHchapter}{A\arabic{chapter}}
\appendix
\cleardoublepage \phantomsection
\mtcaddchapter[Appendices]

\chapter{Publications} 
\label{appA}

	\section{Published Papers}
	\label{appA:accepted_papers}

	\begin{enumerate}
		\item \bibentry{Owuor2019}
    	\item \bibentry{Owuor2020}
    	\begin{itemize}
    		\item Parallel processing in this work has been realized with the support of the High Performance Computing Platform provided by MESO@LR 
    		\begin{sloppypar}
    	    	\url{https://meso-lr.umontpellier.fr}\footnote{\includegraphics[scale=0.4]{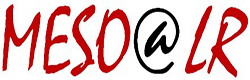}} 		
    		\end{sloppypar}
    		\begin{sloppypar}
    			\url{https://github.com/owuordickson/meso-hpc-lr.git}
    		\end{sloppypar}
    	\end{itemize}
    	\item D. Owuor, A. Laurent, and J. Orero. Extraction de motifs graduels temporels flous. \textit{2019 Rencontres francophones sur la logique floue et ses applications (LFA 2019)}. ISBN: 9782364937369. Référence: 173625.
    	\begin{sloppypar}
      		URL: \url{https://www.cepadues.com/auteur/owuor-dickson/1820.html}
    	\end{sloppypar}
     	\item D. Owuor, A. Laurent, and J. Orero. Croisements de données de l'Internet des Objets pour l'extraction de motifs graduels temporels flous. \textit{2020 Rencontres francophones sur la logique floue et ses applications (LFA 2020)}. 
	\end{enumerate}

	\section{In Progress}
	\label{appA:submitted_papers}
	
	\begin{enumerate}
		\item We submitted our work on \textit{``mining fuzzy temporal gradual emerging patterns''} to the `International Journal of Uncertainty, Fuzziness and Knowledge-Based Systems' journal in June 2020. We are awaiting feedback from the reviewing committee.
		\item We submitted our work on \textit{``ant colony optimization for mining gradual patterns''} to the `International Journal of Machine Learning and Cybernetics' in July 2020. We are awaiting feedback from the reviewing committee.
		\item The work \textit{``gradual pattern mining tools on Cloud''} will be submitted as a demo paper in an International conference and could also be submitted to the 2021 conférence Extraction et Gestion des Connaissances (EGC) to be held in Montpellier, France.
	\end{enumerate}

\chapter{Software Installations} 
\label{appB}

	\section{Installing Cloud API Framework}
	\label{appB:cloud}
	A Docker implementation of OGC/SensorThings API with gradual pattern mining tools integrated into it.
	
	\subsection{Requirement}
	\label{appB:cloud_req}
	You will be required to install \textbf{Docker}.
	
	\subsection{Installation}
	\label{appB:cloud_install}
	Download/clone into a local package from \url{https://github.com/owuordickson/cloud-api.git}. Use a command line program with the local package:
	
	\begin{lstlisting}[language=bash]
		docker-compose up
	\end{lstlisting}
	
	\subsection{Usage}
	\label{appB:cloud_usage}
	
	Launch your Browser and go to: \url{http://localhost:8080}

	\textbf{NB:} a sample IoT data set provided in $sample\_data$, follow the steps to populate your database.
	
	\newpage
	\section{Installing Web GRAANK Tool}
	\label{appB:web}
	Graank tool is a gradual pattern mining Web application implemented in \texttt{Node.js}. It allows users to:
	\begin{itemize}
		\item Extract gradual patterns, fuzzy-temporal gradual patterns and emerging gradual patterns from numeric data sets (\texttt{csv} format).
		\item Cross different time-series data sets (in \texttt{csv} format).
	\end{itemize}
	
	\subsection{Installation}
	\label{appB:web_install}
	Download/clone into a local package from \url{https://github.com/owuordickson/graank-tool-nodejs.git}. There are two options for installing this application (through a command line program with the local package):
	
	\textbf{1. (**RECOMMENDED**)} Install \texttt{Docker}, build the image from the $Dockerfile$ and create a container to run the image:

	\begin{lstlisting}[language=bash]
		cd graank-tool-nodejs-master
	 	
		docker image build -t graank:nodejs

		docker run -d graank:nodejs
	\end{lstlisting}
	
	\begin{center}
		\textbf{OR}
	\end{center}
	
	\begin{lstlisting}[language=bash]
	docker pull owuordickson/graank:nodejs-tool

	docker run -d -p 80:80 owuordickson/graank:nodejs-tool
	\end{lstlisting}

	\textbf{2.} Install \texttt{Nodejs $\&$ npm, Python3 $\&$ Pip3}, then run the application (see below):
	
	\begin{lstlisting}[language=bash]
		cd app && npm install

		sudo pip3 install scikit-fuzzy python-dateutil

		npm start
	\end{lstlisting}
	
	\subsection{Usage}
	\label{appB:web_usage}

	Launch your Browser and go to: \url{http://localhost:80} or \url{http://localhost:80/x}

	\newpage
	\section{Installing Desktop GRAANK Tool}
	\label{appB:desktop}
	Graank tool is a gradual pattern mining Desktop application implemented in \texttt{Electron.js}.
	
	\subsection{Requirements}
	\label{appB:desktop_req}
	The main algorithm in graank-tool has been implemented in \texttt{Python}, therefore you will require the following \texttt{Python} dependencies on your host machine before installing \texttt{graank-tool} software:
	
	\begin{lstlisting}[language=bash]
		install python (version => 2.7)
		
		pip install numpy
		
		pip install python-dateutil scikit-fuzzy
	\end{lstlisting}
	
	\subsection{Installation}
	\label{appB:desktop_install}
	Download installation files for different OS environments:
	
	\small MacOS: \footnotesize \url{http://sharemiale.info.ke/graank-tool/download/mac/graank-tool.dmg}
	
	\small Linux: \footnotesize \url{http://sharemiale.info.ke/graank-tool/download/debian/graank-tool_1.0.0_amd64.deb}
	
	\small Windows: \footnotesize \url{http://sharemiale.info.ke/graank-tool/download/windows/graank-tool-1.0.0-setup.exe}
	
	\subsection{Usage}
	\label{appB:desktop_usage}
	Run \texttt{graank-tool}.



\end{document}